\documentclass[apj]{emulateapj}

\newcommand{\mdot}{\mbox{$\dot M_{\rm wind}$}}
\newcommand{\lsun}{\mbox{L$_\odot$}}
\newcommand\msunyr{\rm M_{\odot}\,yr^{-1}}
\newcommand\be{\begin{equation}}
\newcommand\en{\end{equation}}

\newcommand{\kms}{\mbox{km s$^{-1}$}}

\newcommand{\ewidth}{\mbox{EW}}
\newcommand{\lineflux}{\mbox{$F_{\rm line}$}}
\newcommand{\contflux}{\mbox{$S_{\lambda}$}}
\newcommand{\wmmicron}{\mbox{\rm{W m$^{-2}$ \micron$^{-1}$}}}
\newcommand{\wmeter}{\mbox{\rm{W m$^{-2}$}}}
\newcommand{\lsmm}{\mbox{$L_{\rm submm}$}} 
\newcommand{\lbol}{\mbox{$L_{\rm bol}$}} 
\newcommand{\tbol}{\mbox{$T_{\rm bol}$}} 
\newcommand{\eup}{\mbox{$E_{\rm u}$}} 
\newcommand{\trot}{\mbox{$T_{\rm rot}$}} 
\newcommand{\tk}{\mbox{$T_{\rm K}$}} 
\newcommand{\tkin}{\mbox{$T_{\rm K}$}} 
\newcommand{\ee}[1]{\mbox{${} \times 10^{#1}$}}
\newcommand{\eten}[1]{\mbox{$10^{#1}$}}
\newcommand{\mean}[1]{\mbox{$\langle#1\rangle$}} 
\newcommand{\cmv}{\mbox{cm$^{-3}$}}

\newcommand{\water}{\mbox{H$_{2}$O}}
\newcommand{\OI}{\mbox{[\ion{O}{1}]}}
\newcommand{\CII}{\mbox{[\ion{C}{2}]}}

\newcommand{\NeII}{\mbox{[\ion{Ne}{2}]}}
\newcommand{\FeII}{\mbox{[\ion{Fe}{2}]}}

\newcommand{\Sitwo}{\mbox{[\ion{Si}{2}]}}
\newcommand{\jj}[2]{\mbox{$J = #1\rightarrow#2$}}
\newcommand{\jkkjkk}[6]{\mbox{$J_{K_{-1}K_{+1}}
                              = #1_{#2#3}\rightarrow#4_{#5#6}$}}
\newcommand{\funnyN}{\mbox{$\mathcal{N}$}}
\newcommand{\funnyNJ}{\mbox{$\mathcal{N}_{\rm J}$}}

\shorttitle{DIGIT Embedded Sources}
\shortauthors{Green et al.}

\begin{document}

\title{Embedded Protostars in the Dust, Ice, and Gas In Time (DIGIT) {\it Herschel}\footnote{{\it Herschel} is an
ESA space observatory with science instruments provided by European-led
Principal Investigator consortia and with important participation from NASA.} 
 Key Program: 
Continuum SEDs, and an Inventory of Characteristic Far-Infrared Lines from PACS Spectroscopy
}

\author{Joel D. Green \altaffilmark{1},
Neal J. Evans II\altaffilmark{1},
Jes K. J{\o}rgensen\altaffilmark{2,3},
Gregory J. Herczeg\altaffilmark{4,5},
Lars E. Kristensen\altaffilmark{6,7},
Jeong-Eun Lee\altaffilmark{8},
Odysseas Dionatos\altaffilmark{3,2,9},
Umut A. Yildiz\altaffilmark{6},
Colette Salyk\altaffilmark{10},
Gwendolyn Meeus\altaffilmark{11},
Jeroen Bouwman\altaffilmark{12},
Ruud Visser\altaffilmark{13},
Edwin A. Bergin\altaffilmark{13},
Ewine F. van Dishoeck\altaffilmark{6,5},
Michelle R. Rascati\altaffilmark{1},
Agata Karska\altaffilmark{5},
Tim A. van Kempen\altaffilmark{6,14},
Michael M. Dunham\altaffilmark{15},
Johan E. Lindberg\altaffilmark{3,2},
Davide Fedele\altaffilmark{5}, 
 \& the DIGIT Team }

\affil{
1.  The University of Texas at Austin, Department of
Astronomy, 2515 Speedway, Stop C1400,
Austin, TX 78712-1205, USA; joel@astro.as.utexas.edu \\
2.  Niels Bohr Institute, University of Copenhagen. Denmark \\
3.  Centre for Star and Planet Formation, Natural History Museum of Denmark,
University of Copenhagen, Denmark \\
4. Kavli Institute for Astronomy and Astrophysics, Peking University, Beijing, 100871, PR China \\
5. Max-Planck Institute for Extraterrestrial Physics, Postfach 1312, 85741, Garching, Germany \\
6.  Leiden Observatory, Leiden University, PO Box 9513, 2300 RA Leiden, The Netherlands \\
7. Harvard-Smithsonian Center for Astrophysics, 60 Garden St., Cambridge, MA, 02183, USA \\
8. Department of Astronomy \& Space Science, Kyung Hee University, Gyeonggi, 446-701, Korea  \\
9. University of Vienna, Department of Astronomy, T\"{u}rkenschanzstrasse 17, 1180 Vienna, Austria \\
10. National Optical Astronomy Observatory, 950 N Cherry Ave Tucson, AZ 85719, USA \\
11. Universidad Autonoma de Madrid, Dpt. Fisica Teorica, Campus Cantoblanco, Spain \\
12. Max Planck Institute for Astronomy, Heidelberg, Germany \\
13. Department of Astronomy, University of Michigan, 500 Church Street, Ann Arbor, MI 
48109-1042, USA\\
14. Joint ALMA offices, Av. Alonso de Cordova, Santiago, Chile \\
15. Dept. of Astronomy, Yale University, New Haven, CT, USA \\
 }
 
\begin{abstract}

We present 50-210 \micron\ spectral scans of 30 Class 0/I protostellar 
sources, obtained with {\it Herschel}-PACS, and 0.5-1000 \micron\ SEDs, as part of the Dust, Ice, and Gas in Time (DIGIT) Key Program. Some sources exhibit up to 75 H$_2$O lines ranging in excitation energy from 100-2000 K, 12 transitions of OH, and CO rotational lines ranging from \jj{14}{13} up to \jj{40}{39}. \OI\ is detected in all but one source in the entire sample; among the sources with detectable \OI\ are two Very Low Luminosity Objects (VeLLOs). The mean 63/145 \micron\ \OI\ flux ratio is 17.2 $\pm$ 9.2. The \OI\ 63 \micron\ line correlates with \lbol, but not with the time-averaged outflow rate derived from low-$J$ CO maps.  \CII\ emission is in general not local to the source. The sample \lbol\ increased by 1.25 (1.06) and \tbol\ decreased to 0.96 (0.96) of mean (median) values with the inclusion of the {\it Herschel} data. Most CO rotational diagrams are characterized by two optically thin components ($\mean{\funnyN} =  (0.70 \pm 1.12)\ee{49}$ total particles). $\funnyN_{\rm CO}$ correlates strongly with \lbol, but neither \trot\ nor $\funnyN_{\rm CO}$(warm)/$\funnyN_{\rm CO}$(hot) correlates with \lbol, suggesting that the total excited gas is related to the current source luminosity, but that the excitation is primarily determined by the physics of the interaction (e.g., UV-heating/shocks).  Rotational temperatures for H$_2$O ($\mean{\trot} = 194 \pm 85$ K) and OH ($\mean{\trot} =183 \pm 117$ K) are generally lower than for CO,  and much of the scatter in the observations about the best fit is attributed to differences in excitation conditions and optical depths  amongst the detected lines.

\end{abstract}

\section{Introduction}

Embedded protostars represent a transition from the core collapse within
molecular clouds to the eventual young star and protoplanetary disk system,
revealed as the surrounding envelope is accreted and/or removed.
Processes occurring in the protostellar stage of evolution determine
the conditions in the disks, including the amount and nature of material
available for planet formation in later stages of evolution.
In a picture of episodic accretion \citep[e.g.,][]{dunham10}, the disk mass at
the end of the protostellar phase may be determined by the phasing
of the last burst of accretion onto the star and the end of infall.
Tracing the physical processes in these systems will lead to a
better understanding of the constraints on planet formation.
Physical processes can be traced
through a wealth of molecular, atomic, and ionic tracers available to optical,
infrared, and millimeter-wave telescopes.

Large samples of protostars in relatively nearby ($d \leq 300$ pc) clouds
have been developed through recent surveys with the Spitzer Space
Telescope (e.g, \citealt{evans09}, Allen et al. in prep.), along with
ground-based surveys (e.g., \citealt{jorgensen09}, \citealt{dunham13}).
We have selected a sample of well-studied protostars from these
studies, including both Class 0 and Class I objects.
Class 0 and Class I sources are characterized observationally by rising spectral 
energy distributions (SEDs)
between near-infrared and mid-infrared wavelengths.

In addition to the continuum emission, 
the far-infrared/submillimeter bands contain
numerous pure rotational transitions of the CO ladder,
as well as OH and H$_2$O, and several fine structure lines, all useful tracers
of gas
content and properties.   The
transitions and collisional rates of these simple molecules are
well-understood
\citep[see, e.g.,][for a recent update on CO]{yang10,neufeld12}.  Thus these lines make
excellent diagnostics of opacity, density, temperature, and shock velocities
\citep[e.g.,][]{kaufman96,flower10} of the
gas surrounding these systems. 

ISO/LWS covered the 20-200 $\mu$m spectral region and was well-suited to study
the warm (T $>$ 100 K) region of protostellar envelopes, distinguishable from
the
ambient cloud typically probed in ground-based millimeter studies.
ISO-LWS detected gas phase H$_2$O, high-$J$ CO rotational transitions, and fine
structure emission lines toward protostars
and related sources
\citep[e.g.,][]{lorenzetti99,giannini99,ceccarelli99,lorenzetti00,giannini01,nisini02}.  These lines are
indicative of the innermost regions of the protostellar envelope, 
exposed to heating by the central object, and the outflow cavity region, 
where winds and jets may interact with the envelope and the surrounding cloud.

The Herschel Space Observatory \citep{pilbratt10} provides greater spatial 
and spectral resolution, as well as 
sensitivity, allowing study of sources with weaker continuum and
line emission.   {\it Herschel} is an ESA space-based submillimeter telescope
with a 3.5-meter primary mirror.  
The  {\it Herschel}-PACS spectrograph \citep[Photodetector Array Camera and Spectrometer,][]{poglitsch10} provides a generational 
improvement over the spectral range covered by 
ISO-LWS, and is complementary with the Infrared Spectrograph
\citep{houck04}
on the Spitzer Space Telescope 
\citep{werner04}.  {\it Spitzer}-IRS revealed a wealth of diagnostics
in the mid-infrared, at relatively low spectral (R $\sim$ 60-600) and spatial
(2--5$\arcsec$) resolution.
For example, H$_2$O and OH were detected in some protostars
\citep{watson07a} but in transitions with typically greater critical densities than those
detectable by
 {\it Herschel}.  Combined with enhanced spatial resolution, this difference in excitation 
requirements can lead to significantly 
different conclusions on the origin of the emission without the context of the  {\it Herschel} lines 
\citep{herczeg12}.  Similarly
while pure rotational lines of H$_2$ were detected with {\it Spitzer} in many regions
\citep[e.g.,][]{maret09},  CO far-infrared lines are
 often used as a tracer for the warm gas mass,
through a conversion factor to H$_2$.  
Fine structure 
lines are a third example: 
although [Fe II] and [Si II]
were detected as tracers of outflow properties with {\it Spitzer} \citep[e.g.,][]{maret09,podio12}, a conversion factor
to \OI\ 63 $\mu$m ($^3$P$_1$--$^3$P$_2$) \citep{hollenbach85,hollenbach89,hollenbach97} was used to derive outflow
rates; this \OI\ line
falls in the  {\it Herschel} bands and can be used to inform models \citep[e.g.,][]{ceccarelli97}.
 {\it Herschel}'s spatial resolution (9--40$\arcsec$) resolves
envelopes around systems within a few hundred pc.  {\it Herschel}-HIFI  \citep[Heterodyne Instrument for the Far
Infrared,][]{degraauw10} provides better than 1 
km s$^{-1}$ resolution.

In this work we present the first summary of results of the ``Dust, Ice, and Gas in Time'' (DIGIT, PI: N. Evans) Key Program survey of protostars.  We consider the sample as a whole in characteristic spectral lines, using a standardized data reduction procedure for all targets, and analyze the differences in the continuum and gas over the full sample, presenting an overview of trends.  Specific sources are detailed in individual papers; a  paper including a full inventory of lines with post-mission {\it Herschel} data calibration will follow.  In \S \ref{observations}  
we describe the sample in the context of other surveys, and describe the observations.  
In \S \ref{reduc} we present our detailed data reduction method for acquiring accurate continuum 
SEDs and linefluxes.  In \S \ref{results} we present results and statistics for the sample.
In \S \ref{rotdiagrams} we analyze rotational diagrams for all detected molecular species.
In \S \ref{correlations} we present correlations amongst the detected and derived 
parameters.  In \S \ref{emissionorigin} we consider the origin of the line emission.  We conclude in \S \ref{conclusions}.

\section{Observations}\label{observations}

The DIGIT program uses  {\it Herschel} to achieve far-infrared
spectral coverage of protostars and more evolved young stellar objects in order
to make a full inventory of their chemical
tracers.  The full DIGIT sample consists of 94 sources: 24 Herbig Ae/Be stars,
9 T Tauri stars, 30 protostars observed with PACS and HIFI spectroscopy, and
an additional
31 weak-line T Tauri stars observed with PACS imaging.
\citet{cieza13} presented the results for the weak-line
T Tauri stars; various papers are summarizing the results on disks
in the Herbig and T Tauri star spectroscopic survey
(\citealt{sturm13}, Meeus et al., Fedele et al., subm.)
In this paper we focus on the
embedded protostar sample.  We presented the first 
DIGIT observations of a protostar, DK Cha, in \citet{vankempen10}.
The 30 object sample is
drawn from previous studies (summarized by \citealt{lahuis06}), focusing on protostars
with high-quality {\it Spitzer}-IRS 5-40 $\mu$m and UV, optical,
infrared, and submillimeter complementary data.  Thus all of the sources in the
DIGIT
sample are well-studied at other wavelengths, and many are part of ongoing studies with
 {\it Herschel}
in other programs, especially in the Water in Star-Forming Regions with  {\it Herschel} 
(WISH) program (\citealt{vandishoeck11}; see also \citet{nisini10,kristensen12,karska13,wampfler13}).  Specific outflow positions that are well separated 
from the central source positions are observed by, e.g., \citet{lefloch10,vasta12,
santangelo12,benedettini12,tafalla13}.  In this work we characterize
the DIGIT sample as a whole, using a single standardized data reduction technique.
A complementary sample of protostars from a single region (Orion) can be found in 
\citet{manoj12}.

\subsection{The Sample}

The full DIGIT embedded protostellar sample consists of 30 Class 0/I targets.
Two sources (IRS44 and IRS46) were observed in a
single pointing centered on IRS46.  
These objects are selected from some of the nearest and best-studied 
molecular clouds: Taurus (140 pc; 6 targets), Ophiuchus (125 pc; 7 targets),
Perseus (230-250 pc; 7 targets), R Corona Australis (130 pc; 3 targets),
Serpens (429 pc; 2 targets), Chamaeleon (178 pc, 1 target), and 4
additional isolated cores.  The sources span two orders of magnitude in
luminosity, from 0.11 to 27.8 L$_{\odot}$, with bolometric temperatures
spanning the Class 0/I divide from 27 to 592 K.  Their \lbol/\lsmm\ ranges from 
5 to $>$ 10000, with most (18/22 with well-constrained submm data) 
falling between 10 and 1000.
 The full list of targets
appears in Table \ref{obslog} in order of observation and in
Table \ref{sourcelist} in order of right ascension.

We utilized two of the three instruments on  {\it Herschel}:
PACS and HIFI.  The
spectral resolving power ($\lambda/\Delta \lambda$)
ranges from 1000-3000 for PACS 
and up to 10$^7$ for HIFI, depending on wavelength and 
the selected observing mode.

All 30 of the sources in this work were observed with PACS
as part of DIGIT 
during Science Demonstration and Regular Science Phase Cycle 1 schedules
between December 2009 and September 2011.
Thirteen sources in the DIGIT embedded sources sample were observed with HIFI
at 557 GHz to survey the ground-state H$_2$O 1$_{10}$-1$_{01}$ line.
All of the DIGIT sources not observed with HIFI  as part of DIGIT (17 sources) were observed
in the WISH Key Program in similar, but slightly different,
observing modes.
These are discussed in \citet{kristensen12}; here we present only the
DIGIT portion of the sample (\S \ref{linestats}).  

The sample includes a few outliers from the Class 0/I split.  IRS46 appears to be an 
edge-on disk, perhaps a Stage II source with a Class I SED \citep{lahuis06}.  
GSS30-IRS1 had previously also been an edge-on disk candidate, but has been 
shown to be a protostar from submillimeter analysis \citep{vankempen09}.  Additionally, 
DK Cha is often considered a borderline Class I/II source.

WISH has also observed specific \OI\, CO, H$_2$O and OH lines with
 PACS in the line observing mode for 8 of the 30 DIGIT sources 
 \citep{vankempen10b,
 wampfler10,wampfler13}. These data sample only a small spectral window around a
 limited number of individual lines but with higher spectral sampling
 and somewhat higher sensitivity, which facilitates the detection of
 weak lines. Full PACS spectral scans have been performed within
 WISH for four Class 0 sources \citep{herczeg12,goicoechea12}.

\subsection{ {\it Herschel}-PACS} \label{pacsobs}

PACS is a
5$\times$5 array of 9.4\arcsec$\times$9.4\arcsec\ spatial pixels
(hereinafter referred
to as ``spaxels'') covering the spectral range from 50-210 $\mu$m with
$\lambda$/$\Delta\lambda$ $\sim$ 1000-3000, divided into four segments,
covering
$\lambda \sim$ 50-75, 70-105, 100-145, and 140-210 $\mu$m.
The PACS spatial resolution ranges from $\sim$ 9$\arcsec$ at the shortest
wavelengths (50 $\mu$m) to $\sim$ 18$\arcsec$ at the longest (210 $\mu$m).
The nominal pointing RMS of the telescope is 2$\arcsec$.  However,
targets observed during certain periods show consistently greater offsets
(see below).

For the DIGIT embedded sources sample we utilized the full range of PACS 
(50-210 $\mu$m) in
two linked pointed chop/nod rangescans: a blue scan covering 50-75 and
100-150 $\mu$m (SED B2A + short R1); and a red scan covering 70-105 and
140-210 $\mu$m (SED B2B + long R1). We used 6 and 4 range repetitions
respectively, for integration times
of 6853 and 9088 seconds (a total of $\sim$ 16000 seconds per target for
the entire 50-210 $\mu$m scan); excluding overhead, 50\% of the integration 
time is spent onsource and on sky.  Thus the effective on-source integration 
times are 3088 and 4180 seconds, for the blue and red scans, respectively. 
The total on-source integration time to achieve the entire 50-210 $\mu$m 
scan is then 7268 seconds.

The telescope and sky background emission was subtracted using two nod
positions 6\arcmin\ from the source in opposite directions.
The telescope chopped between the source and nod positions, cycling every
1/8 of a second in a
pre-determined pattern of on and off positions \citep{poglitsch10} during
the integration.

\subsection{ {\it Herschel}-HIFI} \label{hifiobs}

The HIFI observations targeting the H$_2$O 1$_{10}$--1$_{01}$ line at 556.9 
GHz were made with the wide-band spectrograph (WBS) and high resolution 
spectrograph (HRS) in Band 1b. The range of observed frequencies is 554.6 
to 558.6 GHz (lower sideband) and 566.6 to 570.6 GHz (upper sideband). The 
observations were executed in single point mode using position switching with 
a 30\arcmin\ offset in both right ascension and declination and the total integration 
time was 1144 seconds per observation, with an effective on-source integration time 
of 318.6 seconds. The beamsize of HIFI is diffraction limited 
at this frequency ($\sim$ 39$\arcsec$; \citealt{roelfsema12}), and the uncertainty 
in frequency is less than 100 kHz, or 0.05 km s$^{-1}$.  

The HIFI data are reduced in a manner similar to that described in 
\citet{kristensen12}. The spectra were reduced in HIPE ({\it Herschel} 
Interactive Processing Environment) 
v6.0 \citep{ott10} and exported 
to the ``CLASS'' analysis package for further reduction and analysis. The 
reduction consisted of subtracting linear baselines and averaging data 
from the H- and V-polarizations, the latter only after visual inspection of 
the data from the two polarizations. A main beam efficiency of 0.75 \citep{roelfsema12} is 
adopted to convert from antenna temperature ($T_A^*$) 
to main beam temperature ($T_{\rm MB}$), and integration over the line
yields a value for the integrated intensity ($\int T_{\rm MB}$ d$v$).
The final spectral RMS noise 
is 8--13 mK for a binsize of 0.3 km s$^{-1}$.

\section{Data Reduction with PACS} \label{reduc}

We have two primary science goals for the DIGIT embedded sources sample. First, we 
want to produce
the most reliable Spectral Energy Distributions (SED) to inform models, better constraining
source temperature and density profiles.  For this purpose, our criteria are: 
(a) the best match
between spectral modules and (b) the best consistency with photometric
measurements. Second, we want to measure linefluxes for as many gas lines
as possible to constrain excitation conditions and chemical evolution of the
gas phase emission.
For this goal,
our main criterion is the best signal-to-noise ratio in narrow wavelength
regions around lines. In the future, we hope to study weaker, broader features
in the continuum emission that may constrain refractory dust and ice
components, but the data reduction quality is not yet capable of
this last point.

After extensive tests of various reduction methods with different versions
of the HIPE program, we
found that two different methods and HIPE versions (6.0 and 8.0) performed 
best in
meeting the separate goals of achieving the most accurate SED and
the best signal-to-noise ratio for line emission. We combined the best features
of both to get well-calibrated, low-noise linefluxes, as we summarize below; the 
full description may be found in Appendix A.

First, we apply a modified pipeline based on the default data reduction script for that 
version of HIPE.  We used an oversampling rate of 2 (closely matching the raw data 
sampling rate, with 2 pixels per resolution element) in all sources except for 
our science demonstration phase target, DK Cha, for which we used an oversampling of 1.  
This target was observed in a 
slightly different setup not sufficiently sampled in wavelength to use an oversampling 
of 2; all of the other sources are Nyquist sampled.  This process yields spectral cubes for each nod position, which are then averaged to 
produce the final pipeline data product.

At the edges of each segment, 
the noise increased greatly and the overall calibration
became very erratic ($\lambda$ $<$ 53.5 $\mu$m, and 95.0 $\mu$m $<$
$\lambda$ $<$ 102.5 $\mu$m) due to decreased sensitivity and order confusion.
At the longest wavelengths of PACS ($\lambda$ $>$ 190 $\mu$m), 
echoes of spectral lines appear at twice their
actual wavelength, a side-effect of light leakage from the short wavelengths, combined 
with decreased sensitivity on the edge of the array.  Thus neither the continuum nor the 
line emission is reliable at these wavelengths, and 
detections in these ranges are not further considered in this paper; a discussion 
can be found in the Appendix of \citet{herczeg12}.

After this point, we used two different procedures, optimized to produce
the best results for continuum and lines, respectively.  This dual procedure 
was necessary to produce reliable absolute flux calibration, while also 
obtaining the best S/N on continuum and lines.
For convenience, we refer to these two reductions as the ``bgcal'' (Background
Calibrated; best for absolute flux calibration) and ``calblock'' (Calibration Block; best 
for S/N, or point-to-point flux calibration) spectra. These two processes are described in 
detail in Appendix A.1 and A.2.  In the next two
subsections,
we summarize the analysis, including the post-processing
needed to deal with the
point-source profile (PSF) and extended emission, first for continuum
and then for lines.

\subsection{Continuum Emission} \label{reduccont}

The purpose of the steps described here is to produce a spectral
energy distribution that can be compared to existing photometry
and used to calibrate linefluxes derived from the other reduction.
This process used the bgcal reduction, within HIPE 6.0.  The individual steps in
the reduction are detailed in Appendix A.3.  In summary, we calibrate linefluxes from 
regions of good signal-to-noise to a separate determination of the absolute flux level 
and spectral shape, noting and accounting for extended emission and mispointing 
effects.  In the most complicated fields, we defer full analysis to individual source 
papers, as noted below.

First, we determined the centroid position
of the continuum (using the HIPE 6.0 reduction) for two wavelength regions
in each segment, selected to avoid detectable line emission.
Using maps of these regions, we were able to interpolate a peak
position to within 0.2 spaxel widths. 
We find by this method that most of the sources have continuum centroids 
that fall between 0\farcs2 and 3.0\arcsec\ of the center coordinates, with the majority 
landing within 2\arcsec\ (0.2 spaxel widths) of the center.  For those sources, 
the pointing is consistent between the ``blue'' and ``red'' observations, independent 
observations executed at different times.  

Second, in order to produce a spectrum with the best S/N and an accurate absolute
flux density, we chose two apertures for each spectrum: a smaller aperture ($S_{\rm sm}$) 
chosen for the best S/N, 
and a larger aperture ($S_{\rm lg}$) that included all of the source flux for calibration.  To do this, we summed the flux in spaxels containing high-S/N spectra of the source 
continuum, producing a single spectrum.

In most cases, we 
simply use the central spaxel for this purpose.  For some sources we measured a
centroid offset from the center of greater
than 0.2 spaxels.  In particular, the series of DIGIT observations taken 
between 29 March and 02 April 2010 were consistently offset by 6\arcsec,
generally to the upper left of the central spaxel.
In these cases we extracted a
2$\times$2 spaxel area centered on the apparent centroid location
of the central source from the PACS continuum emission, and scaled it to
the full 25 spaxel
continuum values.  We applied this technique to DK Cha,
TMR 1, and L1527 (Table \ref{obslog}).   In the case of TMC 1A, which was slightly 
less offset, we used only the two brightest spaxels for the centroid.

Third, we determined the largest set of spaxels containing the source without
contamination from other sources in the field-of-view; in most cases this was
the
full 5$\times$5 array, but in some cases we selected a smaller set of spaxels (see 
section on extended emission, below).  We then computed the flux density summed 
over these spaxels (the total flux).

Fourth, we fitted a 2nd (or smaller) order polynomial scaling factor in wavelength, 
determined from the ratio of the flux density in the larger aperture to that in the 
smaller aperture.  Then we scaled the spectrum by this correction 
factor.

This correction is applied  
module-by-module (e.g. a separate polynomial for each range of 50-75, 70-95, 
100-145, and 145-190 $\mu$m).
After scaling $S_{\rm sm}$ to $S_{\rm lg}$, the two ``blue'' (50-75 and 70-95 $\mu$m) 
and the two ``red'' (100-145 and 145-190 $\mu$m) 
modules were generally well-aligned with each other 
(less than 5\% difference on average), but retained the S/N of $S_{\rm sm}$.  
Unfortunately, 
even after applying this entire procedure,  $S_{\rm lg}$(red) $>$ $S_{\rm lg}$(blue) in 
all cases, a systematic discontinuity at 100 $\mu$m for every source.  
This jump was not uniform; it is most problematic for faint continuum
sources IRS63, IRAS 03301, B1-a, L1014, and WL12.
Ultimately the SEDs will benefit from comparison 
to PACS photometry.

Furthermore, there are a few cases where the reduction process did not produce a smooth 
continuum across modules even within the ``blue'' or ``red'' halves.
In cases with a significant misalignment, as an optional 
sixth step we apply an additional correction.  The sources that are significantly misaligned 
between 
modules are Serpens-SMM3 and SMM4, RCrA-IRS5A, 7B,  IRS 46, L1551-IRS5,  
L1455-IRS3, and L1014.  
The Serpens and R CrA sources are in highly confused regions; an analysis of PACS 70/100/160 $\mu$m imaging can be found in \citet{sicilia13}.  Serpens-SMM3 is the most
problematic; the B2A/short R1 observation is 30\% higher than the B2B/long R1
observation, probably an effect of the complicated interaction between the
extended emission in the region, and the PSF; once again we scaled the 
B2A/short R1 spectrum up 
by 30\%.  IRS46 is complicated 
by emission from the much brighter nearby source IRS 44.
The L1551-IRS5 B2A/short R1 observation is 10\% higher than the B2B/long R1
observation, which was observed 1.5 years later.  We 
scaled up the B2B/long R1 by 10\% to match the B2A/short R1 data, for the continuum 
SEDs.  L1455-IRS3 and L1014 are only marginally detected in continuum; thus when scaling to 
the full array the noise becomes comparable to the signal, causing the scaling to 
become unreliable.  In this case we scaled the brightest module (the blue half of each 
observation) to the full array 
as usual and then scaled the red halves upward by a factor of 2 to match the blue.  
This exercise shows that in the case of very faint sources absolute calibration using the 
full array becomes unreliable.  No other additional continuum scaling was 
performed on any of the other sources in the sample after this point.  

In a few cases, we observed clear evidence of {\it multiple} emitting sources
within the PACS field-of-view.  In the case of VLA 1623-243 we detected a clear
second source to the northeast
edge of the grid (Appendix B, Fig. \ref{spatial8}), and
avoided the emission from those spaxels.  This is consistent with the emission from the 
Oph-A ridge detected in SCUBA maps \citep{wilsonc99,johnstone00}.  IRS 46 is 
well-centered, but
by design IRS 44 occupies only the southwestern 3 spaxels in the map, which 
greatly underestimates the true flux density of IRS 44 and distorts the spectral 
shape. In these cases,
we selected spaxels most likely to be attributable to the target source.  We
extracted the central source,
and scaled to the flux in the spaxels clearly attributed to that
source; in some cases
this amounted to only 60\% of the spaxels.
In the case of IRS 44 we extracted
the flux from only the southwestern 3 spaxels.  In the complicated Serpens-SMM3/4 
and the RCrA-IRS7B/7C
fields we observed significant extended diffuse emission contributing
to the flux. We therefore calibrated these to flux density per pixel and 
detailed analysis will be presented by Dionatos et al. (subm.) and Lindberg et al. (in prep).
In the case of L1448-MM, we were able to decompose the flux based on 
resolved observations at shorter wavelengths (see the section on continuum analysis for further detail, and Lee et al., subm.).

\subsubsection{Absolute flux calibration}

Comparison to previous photometry in the PACS wavelength range is
problematic. The {\it IRAS} beam was much larger, and MIPS data
often suffered from saturation on these sources.
The typical MIPS 70 $\mu$m flux density is almost always low compared to the PACS
spectra extracted from the full array by $\sim$ 20\%, except in 
the brighter half of our sample, where the MIPS value is even lower -- a
factor of two -- compared to the  {\it Herschel} data.

\begin{figure}
\begin{center}
\includegraphics[scale=0.85]{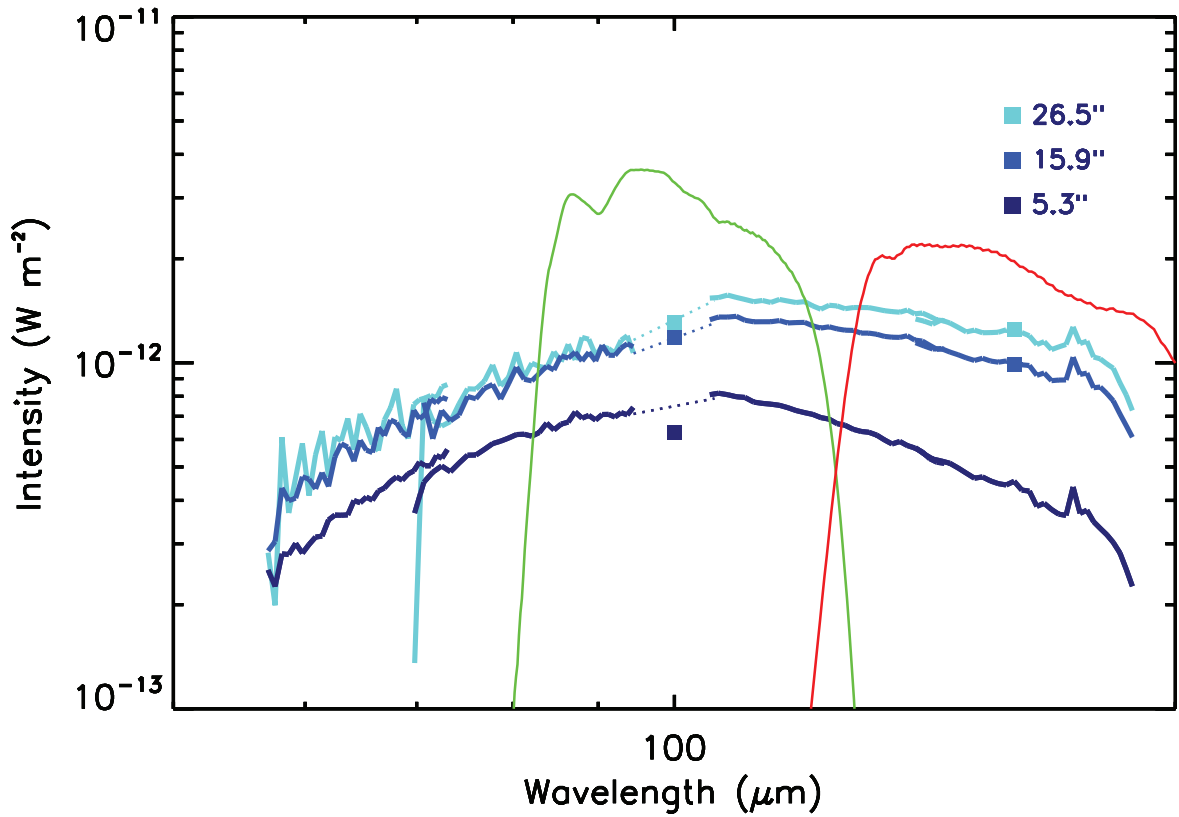}
\caption{Comparison of B335 PACS spectroscopy (rebinned to low resolution
to remove line emission) and PACS photometry (priv. comm.: A. Stutz, 2012),
at similar aperture sizes.  The PACS photometry is taken from circular
apertures of
radius 5\farcs3, 15\farcs9, and 26\farcs5, equivalent to
9\farcs4 (black-blue), 28\farcs2 (dark blue), and 47\farcs0 (light blue) squares.  
The dash lines indicate linear interpolations for the spectra at 100 $\mu$m.
These sizes are comparable to the central spaxel only,
central 3$\times$3 spaxels, and all 25 spaxels, respectively.
The agreement is good to better than 10\%, except for the central spaxel.  The central 
spectrum is higher than the PACS photometry by 30\% at 100 $\mu$m.}
\label{b335specphot}
\end{center}
\end{figure}

Instead,
we compared SEDs from this reduction method to PACS
photometry of disk sources, obtained in the GASPS (PI: W. Dent) Key Program.
These isolated point
sources provide a better photometric test than the extended, embedded sources.
We
convolved our SEDs to the transmission curve for each PACS photometric filter
and compared the photometry to the resulting spectrophotometry (Table \ref{specphot}).
We found the best agreement with HIPE 6.0 using the bgcal reduction, for which we found the following
average factors for photometry/spectrophotometry: $1.071\pm0.106$ at 70 $\mu$m,
 and $0.993\pm0.136$ at 160 $\mu$m, where the second
number is the standard deviation over 10 sources.  
Eventually, PACS photometry for our embedded sources should become available
from other programs. For now,
we were able to compare to PACS photometry for one embedded source, B335
(A. Stutz, priv. comm. 2012; \citealt{launhardt13}).
The SEDs derived from various combinations of spaxels (central spaxel only, 
central 3$\times$3 spaxels, and total from all spaxels) are compared
to PACS photometry in apertures set to match the solid angle 
in Figure \ref{b335specphot}; e.g. circular apertures of radius 5\farcs3, 15\farcs9, and 26\farcs5, 
equivalent to 9\farcs4 (black-blue), 28\farcs2 (dark blue), and 47\farcs0 (light blue) squares.  The spectral 
flux density at 100 $\mu$m was determined from a linear interpolation between the average 
flux density in the 94--95 $\mu$m, and 104--105 $\mu$m ranges.
The agreement is good to 10\%, except for 
the central spaxel, where the photometric aperture size (10\farcs6 diameter) is very close to the 
beamsize at 100 $\mu$m ($\sim$ 10\arcsec) and much less than the beamsize at 160 $\mu$m 
($\sim$ 14\arcsec).  The extent of the source is well reflected in the rise of flux density  
as we add increasing numbers of spaxels, but the images show that the 
emission continues to still larger scales. 
If this situation is general, our SEDs
will underestimate the total flux from the source.

\subsection{Line Emission} \label{reduclines}

Our reduction technique for lines utilizes HIPE 6.0 {\it and} 
HIPE v8.0.2489 with the
``calibration block'', referred to as the ``calblock'' reduction,
which multiplies the chopped spectrum by a
similar spectrum of an on-board calibration source.  It works in a 
similar fashion to the continuum procedure, but with several key differences.  The 
steps in
the reduction are detailed in Appendix A.4.

Our objective here is to measure line equivalent widths (\ewidth)
for gas phase lines,
which are spectrally unresolved, or in a few cases, perhaps slightly
resolved.
This reduction produced the lowest local (point-to-point) noise but larger
shifts between modules and an overall calibration less consistent
with photometry. Tests indicate that the equivalent width of lines
is independent of reduction. Consequently, we use the calblock reduction to
define the
equivalent width of lines and then use the bgcal reduction of the same
spectral region to convert to linefluxes. 
This process allows us to combine the best signal-to-noise on the
equivalent width from HIPE 8.0.v2489 with the best overall line flux
calibration using HIPE 6.0.

The first step is to extract linefluxes from the HIPE 8 calblock reduction using
a narrow region to define the continuum (\contflux) and calculate the
\ewidth. We
utilized a modified version of the SMART reduction package \citep{higdon04}
to fit Gaussians and first or second-order baselines to the spectra
to remove local continuum features.
The LAMDA database of lines \citep{schoier05}, supplemented by HITRAN 
\citep{rothman05}, was
used to identify the features.
The fit for the line center varies only by 0.01 $\mu$m or less for
our different extraction methods (10-20\% of a resolution element);
therefore the fit uncertainty in velocity space is $\sim$ 30-50 km s$^{-1}$.
Line centers usually lie within 50 km s$^{-1}$
of the theoretical line center from laboratory measurements.  At our level 
of precision, we do not observe believable shifts in our data; nonetheless, 
our observed shifts are still consistent with shifts observed in other datasets 
\citep[e.g.][]{karska13}.
However, the location of the
source within each slit also affects the velocity.
The uncertainty in \ewidth\ caused by the fit is quite small (S/N ratios as
high as 300 were obtained on bright
lines, particularly \OI), but for faint lines, the
uncertainty rises to 20\%.
In this treatment we do not include lines within the masked wavelength regions 
($<$ 55 $\mu$m, 95--102.5 $\mu$m, $>$ 190 $\mu$m) as they 
exhibit large calibration uncertainties.  The one exception is the 
CO  \jj{27}{26} line at 96.77 $\mu$m; the local RMS uncertainty is somewhat 
larger than, but still comparable to, the local continuum surrounding the CO  \jj{28}{27} 
line at 93.3 $\mu$m, 
and we only include this line in sources with confirmed detections of CO 
lines at shorter wavelengths.  The line does not significantly affect the derived 
source properties in the following sections.

\subsubsection{Corrections for spatial extent, and comparison to the empirical PSF}

Second, the linefluxes were corrected for PSF and extended emission.
As with the continuum, simply adding all the spaxels caused a serious loss  
of signal-to-noise and
a diminished rate of line detection, and the PSF correction
provided in the standard pipeline does not
account for extended emission.  Thus 
we do {\it not} use the PSF correction function, but instead compare the flux
over different sized apertures to determine an empirical correction for each
source.  In order to do this, we measured the lineflux, \lineflux(sm), in
the central spaxel (or a sum over a few spaxels for poorly centered sources)
that gave the best signal-to-noise on relatively strong lines.  Mathematically, 
the process is identical to the continuum, except for lineflux rather than flux 
density.

Additionally, 
we correct the fluxes by the ratio between the bgcal and calblock local 
continuum for each individual line.
Because the spatial distribution of line emission could in principle
differ from that of the continuum, we did not simply scale the value
in the central spaxel by the continuum ratios.
To determine the total linefluxes, \lineflux(lg), we applied a polynomial 
correction, developed by comparing the spectral lineflux
detected in the central spaxel compared to that in the surrounding spaxels
for strong lines -- mostly CO lines, but also including strong H$_2$O lines, a technique 
developed by \citet{karska13}.  
We did not include the 63 and 145 $\mu$m \OI\ lines in this correction, because they  
were frequently extended compared to these species and to the PSF 
(see Section \ref{reduccont}).
Weaker lines of all species were then scaled with the same factors.

This
method assumes that weak and strong lines have similar spatial distributions.
In Figure \ref{extendedlines}, we plot linefluxes of all
species measured from 3$\times$3 spaxel regions vs. linefluxes contained within
just the central spaxel, for four 
well-pointed sources.  For each source we fit a constant ratio from
50-100 $\mu$m, and a first order polynomial from 100-200 $\mu$m, a simplified
fitted version of the PSF correction, to the bright CO and H$_2$O lines.  For this 
exercise we set a minimum correction factor of 1.4 to the 50-100 $\mu$m fit, 
approximately the PSF correction.  Although 
additional lines are plotted, they are not considered in the fit.  In the first two cases 
(L1489 and Elias 29), 
the pipeline-provided PSF correction (dashed line) is very similar to the analytical 
correction.  In Elias 29, the \OI\  (purple)
emission exceeds the PSF by a significant margin, indicating that it is 
extended compared to the PSF.  In the third case, BHR71, we find that all line ratios  
significantly exceed the PSF correction, suggesting a substantial contribution from 
extended and/or diffuse emission in the field-of-view.  In GSS30-IRS1, we find that the 
long wavelength line emission exceeds the PSF, as does the continuum; at short wavelengths the \OI\ line is far more extended than the molecular emission or the continuum (see below).  This 
may be associated with an HCO$^+$ flow extending to the NE from the source position \citep{jorgensen09}.

In cases for which we lack sufficient strong lines to determine the correction, we 
use a linear approximation to the PSF correction.  This was applied to IRAS 03245, 
IRAS 03301, IRAM 04191, 
IRS 46, IRS 63, L1014, and L1455-IRS3.  We also correct IRS 44 with the PSF, but note that 
the final derived flux is still close to an order of magnitude too low, judging by the 
{\it Spitzer} spectra and long wavelength photometry.  This is due to the position of the 
source just off the edge of the PACS footprint; we do not fully see the source peak and lose substantial flux.

\section{Results} \label{results}

\subsection{Detection Limits}\label{limits}

We measured the continuum RMS in selected
(line-free) portions of the spectrum.
The measured RMS uncertainties are dominated by three factors.  Second
order (B2A and B2B; 50-75 and 70-105 $\mu$m, respectively) spectra have
$\sim$ sixfold higher noise for a constant spectral width, 
at the given integration time/repetition settings.
Figure \ref{unc1} shows the RMS uncertainties vs. wavelength for a line-rich
source (Elias 29) separated by order, for both CO and H$_2$O lines, using fitted Gaussian 
linewidths.  
The uncertainties in flux are independent of species (unsurprisingly)
and broadly decrease with increasing
wavelength, discontinuously at the order boundaries.
Portions of the spectra within 5 $\mu$m
of the order edges have far higher uncertainties.
Finally, we show the RMS uncertainties for
the 63 $\mu$m \OI\ line across the full DIGIT sample, against the
local continuum flux.  Above a certain continuum strength -- greater than 
$\sim$ 3 $\ee{-16}$ Wm$^{-2}$ at 63 $\mu$m --
the RMS uncertainty rises roughly linearly with 
continuum flux, reaching 70 $\times$ 10$^{-18}$ W m$^{-2}$ at a continuum flux 
of 4000 $\times$ 10$^{-18}$ W m$^{-2}$.  Thus the S/N of lines and continuum 
seem to reach a maximum at $\sim$ 60, compared to the continuum RMS noise.

\subsection{Continuum Results: SEDs}\label{seds}

Analysis of the continuum yields several interesting comparisons.  In brief, 
the classification of sources is altered only slightly, but systematically, by the 
inclusion of the Herschel data. 
Figures \ref{seds1} to \ref{seds5} show the SEDs
of our sample over the entire wavelength range with available data.
We plot available photometry for comparison, in the following
order of precedence: PACS photometry from other programs, {\it Spitzer}-MIPS
photometry (with arrows for lower limits due to saturation), ISO
photometry, and IRAS photometry.  Where available, we include
{\it Spitzer}-IRS spectroscopy from the ``IRS\_Disks'' Instrument Team
\citep{furlan06} and ``Cores to Disks'' Legacy \citep{lahuis06, boogert08,lahuis10}
programs.  In the case of L1448-MM, {\it Spitzer}-IRAC/IRS and submillimeter 
data were used to determine the flux ratio between the two components (separated by 
8\arcsec, comparable to the PACS beam at the shortest wavelength; \citealt{jorgensen06}).  Extracting a 
separate spectrum from each spaxel in PACS, and from IRS maps, we present in 
Figure \ref{l1448ab} the decomposed SEDs of L1448-MM(A) or (C), and L1448-MM(B) 
or (S).  (S), the southern source, appears to be fainter and bluer than (C).  The 
separated sources are discussed in detail in Lee et al. (subm.).  
For the remainder of our analysis in this paper, 
we consider the combined SED.

Using our spectral data, along with photometric data for wavelengths
not covered by  {\it Herschel}, we have recalculated the bolometric luminosities
(\lbol) and temperatures (\tbol) for the sample, using standard definitions
\citep{myers93} and the calculation algorithm from \citet{dunham10}.  In summary, \lbol\ is the luminosity integrated
over the entire SED, while \tbol\ is the temperature of a blackbody with a
flux-weighted mean frequency equal to that of the observed SED.  The weights of the
points are proportional to the fraction of the SED they cover 
(``trapezoidal'' weighting); thus individual points in the PACS and 
IRS spectra receive little weight relative to the photometric points 
in sparsely covered regions of the SED.

The dashed lines in the SED plots show a blackbody at \tbol; these
resemble the full SED only for sources of low \tbol\ with little
near-infrared emission. Of course, \tbol\ is not meant to fit the SED,
but only to characterize it with a single parameter.
The values of \lbol\ and \tbol\ with and without  {\it Herschel} data are given in Table
\ref{boltable}.
The mean, median, and standard deviation of \lbol(wHer) / \lbol(w/oHer) are
1.25, 1.06, and 0.66, after removing those five with multiple known emitting
sources
(RCrA-IRS5A, 7B, and 7C, Serpens-SMM3 and SMM4).
The mean, median, and standard deviation of \tbol(wHer) / \tbol(w/oHer) for the sample 
are 0.96, 0.96, and 0.21.
The sample means now are $\mean{\lbol} = 6.1$ \lsun\
and $\mean{\tbol} = 167$ K.
Figure \ref{lbolnew} shows the new distribution in \lbol-\tbol\ space,
plotted alongside the ``without  {\it Herschel}'' distribution. The sample still represents
sources spanning two orders of magnitude in bolometric temperature, and 2.5 
orders of magnitude in bolometric luminosity.
Using the standard 70 K Class 0/I boundary \citep{chen95},
we now have 9 Class 0 sources and 16 Class I sources (previously 
8 and 17, respectively) (Table \ref{boltable}), with B1-c and BHR71 shifting 
across the boundary from Class I to 0, and L1014 shifting 
from Class 0 to I.

There is some uncertainty in the value of \tbol\, which is affected by the data at wavelengths 
shorter than 
PACS wavelengths, because the short-wavelength data are more extinguished (characterized broadly in similar sources by \citealt{dunham13}), and vary over multiple epochs of observation.  Thus different values for the near-IR/optical photometry strongly affect \tbol\, but not \lbol.  In the case of L1455-IRS3, literature values for $\lambda < 70 \mu$m were an order of magnitude too high.  Investigation revealed these data to be associated with stronger nearby sources, and thus we exclude them from the SED. Despite these uncertainties, our measurements for \tbol\ and \lbol\ are generally consistent 
with \citet{karska13}.  
There are six sources that show substantial ($>$ 20\%) shifts in this space.
GSS30-IRS1, IRS46, L1014, L1455-IRS3,  all show substantial decrease 
in \tbol\ and increase in \lbol\, while VLA1623 shows the reverse effect.
Of these five sources,  VLA1623 is the only source that falls on the Class 0 side 
of the \tbol\ line, while the other four fall on the Class I side.  IRS44 shows the reverse 
trend of the four Class I sources; this is due 
to its position on the edge of the array footprint, as noted earlier, missing a large amount of 
flux; the IRS 44 PACS flux is too low by a factor of a few even after applying 
the typical PSF correction.  Thus we conclude that overall the PACS data is 
substantially increasing 
the derived luminosity in bluer sources for which it samples the peak 
and longer wavelengths.  {\it Spitzer}-MIPS and IRAS photometry were 
previously used to sample this region; the MIPS data is often close to saturation 
in these sources.  Without the PACS data, one will consistently derive lower
\lbol, and higher \tbol, in Class I sources.

Another diagnostic of evolutionary state is  \lbol/\lsmm.  By this metric, a source with 
\lbol/\lsmm\ $<$ 200 is Class 0.  Using this criteria, our  
original sample consisted of 16 Class 0 and 9 Class I sources, rather than 8 
and 17, respectively, even when including three sources (BHR71, IRS 44, and IRS46) 
lacking submm under 
Class I.  This total was unchanged by the addition of  {\it Herschel}-PACS data.  
The Class 0 
sources under this definition include all of the Class 0 sources 
from the \tbol\ diagnostic (except BHR71, which lacked submillimeter 
data to be classified) and 9 additional sources identified as Class I 
by \tbol.
The submillimeter luminosity diagnostic
is also stable with respect to what data are included, as long
as sufficient submm data are available, and evolutionary models
indicate that it is a better guide to the evolutionary stage
\citep{young05}, 
though both indicators become less reliable in models with episodic
accretion \citep{dunham10}.  However, we chose to focus on the \tbol\ classification for 
two reasons: first, for straightforward comparison to the existing work \citep[e.g.][]{manoj12,karska13}, and second, as we lack sufficient data to classify all of our sources by 
the \lbol/\lsmm\ method.
%
%
%

\subsection{Lines: Atomic and Molecular Emission}\label{lines}

\subsubsection{Detection Statistics}\label{linestats}

In general, the PACS spectra are very rich in line emission, including 
\OI, CO, OH, and H$_2$O emission lines, and absorption only in two sources 
(in OH 119 $\mu$m) and diffuse \CII.
Spectra characteristic of three kinds of sources, classified by their 
highest $J$ CO line observed, are shown
after removal of the continuum (Fig. \ref{quiescent} -- Fig. \ref{hot}).  Although the sources are classified by their 
CO emission, there are aspects of other lines common to each subset.  
L1014 (Fig. \ref{quiescent})
has almost no line emission, with only \OI\ at 63 \micron\
detected.  The lack of CO is a significant non-detection; by contrast, 
the faintest source in our sample (IRAM04191)
does show CO emission.
L1551-IRS5 
(Fig. \ref{warm}) -- one of the most luminous sources in our sample -- 
shows more lines, including 15 
CO lines, 13 \water\ lines and the two lowest energy OH doublets in the 
3/2--3/2 ladder at 84 and 119 $\mu$m; 
unusually, the lowest excitation OH doublet 
at 119 $\mu$m is in absorption.
Finally, Elias 29 (Fig. \ref{hot}) shows
the richest spectrum in the sample,
with 25 CO lines (ranging from \jj{13}{12} up to CO \jj{38}{37}; $\eup\ = E_{\rm upper}/k$, 
expressed in Kelvin, of 503 -- 4080 K), 18 OH lines (in 8 resolved pairs from 
two different ladders, 
and 4 blended or cross-ladder 
transitions (\eup\ $=$ 121 -- 875 K), 
and 74 lines of H$_2$O including one 
blended pair (\eup\ $=$ 114 -- 1729 K).
No atomic species other than
\OI\ ($^3$P$_1$--$^3$P$_2$ and $^3$P$_0$--$^3$P$_1$) and \CII\ 
($^2$P$_{3/2}$--$^2$P$_{1/2}$) and no molecular 
species other than CO, OH,
and \water\ have so far been identified in any of these spectra.

The measured linefluxes for seven characteristic lines 
are shown in Table \ref{fluxtable}, 
including corrections for 
extended emission (see \S \ref{linestats}, below).  
A full inventory of lines will appear in a later paper 
including the final post-mission {\it Herschel} calibration dataset.
The lines selected as characteristic are as follows:
\OI\ at 63.18 \micron, CO \jj{29}{28} at 90.16 \micron, 
CO \jj{16}{15} at 162.81 \micron, 
OH 3/2--3/2 ($7/2+ \rightarrow 5/2-$) at 84.61 \micron, 
OH 3/2--3/2 ($9/2+ \rightarrow 7/2-$) at 65.28 \micron,
ortho-\water\ ($2_{12} \rightarrow 1_{01}$) at 179.53 \micron,
and ortho-\water\ ($3_{30} \rightarrow 2_{21}$) at 66.44 \micron. 
These lines were chosen
to represent species and/or excitation above the ground state,
quantified by \eup, and
roughly characterized as ``cool" ($\eup < 300$ K), 
``warm" ($\eup \sim 500$ K), and 
``hot" ($\eup > 1000$ K).  
Although both OH lines are doublets, 
the individual lines in the selected transitions are cleanly resolved at the
resolution of PACS.  
We avoided the 119 $\mu$m doublet because it is complicated in at least two
sources (L1551-IRS5 and B1-c) by apparent foreground absorption.

All  30 sources have at least one detected line, 
including the two VeLLOs (Very Low Luminosity
Objects -- L1014 and IRAM04191).  The most universal  
is the \OI\ line at 63 \micron, detected
in 29 of 30 sources (only B1-c lacks a detection of this line).
\OI\ at 145 \micron\ is also detected in 24 of these 29 sources.
The \CII\ emission line at 158 \micron\ is also widely seen, and in some cases 
(DK Cha, IRAM04191, and TMC1) is spatially coincident with the local continuum 
and with the \OI\ emission.  
More often the peak emission is offset, and in several cases appears  
in absorption, indicating stronger emission in the reference
position. When seen in emission,
it shows little correlation with the continuum or other lines.
We conclude that it mostly arises from the general cloud and is not related
to the source. We do not consider it further in this paper.

Across the full sample, the detection of ``warm" CO (\jj{16}{15}; $\eup = 752$ K)
is nearly universal, appearing in 27 of 30 sources (including IRAM04191, but 
not the slightly more luminous L1014).
Most of these (22/27) also show ``hot" CO (\jj{29}{28}; $\eup = 2400$ K).
Similarly, the ``cool'' o-\water\ \jkkjkk221101\
line at  179.5 $\mu$m ($\eup  = 114$ K) is detected in 23 of 30 sources,
while the ``warm'' o-\water\ line at 66.5 \micron\
($\eup = 411$ K) is detected in 14/23 of these.
``Cool'' OH 3/2--3/2 ($7/2+ \rightarrow 5/2-$) at 84.61 $\mu$m 
($\eup = 291$ K)
appears in 27 of 30 sources, whereas ``warm'' OH 
at 65.28 \micron\ ($\eup$ = 511) appears in 15/27 of these.
When higher excitation lines of a given species appear, the lower excitation lines
are always present.
OH and H$_2$O emission are both commonly detected but not always in
the same source:
23 of 27 sources with OH also have H$_2$O emission at 179 $\mu$m, but
L1551-IRS5, IRAM04191, L1455-IRS3, and IRS 63 all show some
OH emission but no H$_2$O; B1-c is the only source with H$_2$O
emission but no detected OH emission.  Note that we do not see a ``cool'' 
component of CO with PACS; we reserve the term for the observed $\sim$ 100 K 
component with SPIRE and HIFI (\citealt{yildiz12}, \citealt{goicoechea12}, 
Kristensen et al., in prep., Green et al., in prep.).

The DIGIT HIFI data add spectrally resolved measurements of the emission in
a low excitation line of \water: \jkkjkk110101\ at 557 GHz ($\eup\ = 60$ K).
In Figure \ref{hifi} we show the 
line profiles from each source.  We detect H$_2$O in 9 of 13 sources, 3 of
which show an inverse P Cygni profile.  The sources without line detections
are IRAM04191, L1014, IRAS03301, and L1455-IRS3.  Additionally, WL12
and IRAS03245 are marginal detections.  IRAM04191, L1014, and L1455-IRS3 
are three of the faintest sources in our sample.  However IRAS03245 
is considerably more luminous than IRS46 and B1-a, both of which show strong 
detections.  The integrated intensities
are tabulated in the last column of Table \ref{fluxtable}.  
Eight of 9 sources with 557 GHz H$_2$O
detections also show the H$_2$O 179 $\mu$m line, the sole exception being
IRS46, which suffers from confusion with IRS44.  It is possible that the H$_2$O
in the IRS46 HIFI data is related to IRS44; the PACS map suggests that IRS44 is
the main source of molecular emission in the region.   
IRAS03301 has a rich PACS H$_2$O spectrum, but a very weak 179 $\mu$m 
H$_2$O line, and no detection with HIFI.  The reverse is slightly more common: 
 L1551-IRS5 and IRS63 both 
show detections with HIFI \citep{kristensen12}-- the latter very faint -- but no H$_2$O 
179 $\mu$m emission 
with PACS in our sample.

\subsubsection{Spatial Extent of Continuum and Line Emission}\label{lineextent}

The PACS data were taken as a single footprint, so the data are not
designed for detailed analysis of spatial structure. We discuss here
some simple metrics regarding spatial extent and discuss a few examples
of interesting structure.

The simplest metric of spatial extent is the ratio of
the emission from the central 3$\times$3 spaxels 
to the emission from the central (highest S/N) spaxel.
We have used this metric to scale the spectrum from the central spaxel
(\S \ref{reduclines}), and by comparing this ratio to that of
a point source, we can roughly characterize the extent of emission.

We characterize source extent across the sample with this metric, 
for both continuum and 
the characteristic lines, in Tables \ref{extended} and \ref{extended2}.  
Signs of extended emission are seen in both line and continuum 
in some sources, and the line emission is not always coupled to the 
continuum, but many sources are essentially pointlike at {\it Herschel} wavelengths.
To simplify the comparison, we consider only the 15 sources
in our sample that are
well-centered (we exclude DK Cha, TMR1, TMC1A, L1527, and IRS 44), lack clear
evidence for multiple, spatially separated emission peaks at the resolution
of  {\it Herschel} (we exclude Serpens-SMM3/4, RCrA-IRS5A/7B/7C, L1448-MM, and IRS
46),
and have sufficient flux density to be detectable in continuum at all PACS
wavelengths (we exclude L1455-IRS3, L1014, and IRAM 04191).  DK Cha is 
clearly extended despite the mispointing. We restrict further analysis of the
extent of line emission to the 15 sources with
sufficient S/N, and to lines with clear detections in the 3$\times$3 spectrum.

The PACS PSF is larger at longer wavelengths.  Thus a source must be more extended to 
exceed the PSF at longer wavelengths and appear extended in our data.
The ratio of the continuum flux density in the full  3$\times$3 spaxel array 
to the emission from the central spaxel averages $1.83 \pm 0.40$ (median 1.79) 
at 63 $\mu$m, the average being higher by a factor of 1.28 than the 
PSF correction factor of 1.43. At 163 $\mu$m the ratio averages $3.09 \pm 1.06$ 
(median 2.55), with the average a factor of 1.51 higher than the PSF 
correction factor of 2.04. On average (and median) our sources are 
marginally extended, and more extended at the longer wavelengths.  However, the 
lower median values compared to the mean values indicate that the extended 
emission is dominated by a few sources.  In Tables  \ref{extended} and \ref{extended2}, 
we only include formal errors in the local continuum RMS; the overall calibration 
uncertainties may be much higher, including photometric calibration and spaxel-to-spaxel 
variation.  If 
we assume the uncertainty in the 3$\times$3 and the central spaxel continuum 
is dominated by the photometric uncertainty ($\sim$ 20\%), then adding this to the ratio in quadrature, we would expect the ratio to be uncertain by 
$\sim$ 28\%.   The continuum ratios in IRAS03245, IRAS03301, B1-a, BHR71, GSS30, VLA1623, 
and Elias 29 significantly exceed the PSF correction (by more than 28\%), at most sampled wavelengths with sufficient signal-to-noise.  In contrast, B1-c, L1489, L1551-IRS5, 
TMC1, WL12, IRS63, B335, and L1157 are consistent with a point source at most 
wavelengths (within the DIGIT sample, we are characterizing the compact core; 
notably L1157 contains an extended outflow on arcmin scales; \citealt{nisini10}).  
There are no objects that show a ratio significantly smaller than that 
of a point source.

There is no obvious distinction between the sources in these 
two groups (e.g., \lbol, \tbol, class/stage, or source distance).
This measure will be a bit misleading as our method characterizes an extended 
source with an equally strong central peak as ``pointlike''; PACS imaging suggests 
this may be the case for B335 in particular (A. Stutz, priv. comm.; \citealt{launhardt13}).
Additionally, for a given source, there can be a trend with wavelength.
WL12 is a bit of an outlier in that it shows greatly extended continuum at 
$\lambda$ $>$ 150 $\mu$m but is 
otherwise pointlike.
The trend to larger sizes, relative to a point source, can also be seen
in Figure \ref{gss30cont} for GSS30-IRS1, where a trend toward 
displacement of the emission centroid toward the north-east with
increasing wavelength can be seen.
H-band VLT images \citep{chen07} also show extended
emission in this direction.

%

The extent of the line emission need not be coupled to continuum.  
Tables \ref{extended} and \ref{extended2} also give the ratios for the characteristic
lines.  For the CO \jj{16}{15} line, the ratio averages $2.38 \pm 0.60$ 
(median 2.24), a factor of 1.17 above the PSF correction factor of 2.04, 
smaller than the value for the adjacent continuum at 
163 \micron. In contrast, the ratio for the \OI\ 63 $\mu$m lineflux 
averages $2.25 \pm 1.15$ (median 2.02), 1.57 times the PSF correction of 1.43.  
Thus the \OI\ line is more extended, on average, than the adjacent continuum.  
The CO \jj{29}{28} line ratio averages 1.56 (median 1.51, PSF correction: 1.45), 
and is thus essentially pointlike where detected.  The 
H$_2$O 66.5 $\mu$m line ratio is measured in four sources, two of which appear 
to be extended (B1-a, L1157).  There are a few that appear {\it less} extended than 
the PSF; in each case the line is only faintly detected above the noise in the 
extended aperture, 
but is strongly peaked in the central spaxel -- most notably the CO, but not the \OI, 
in L1551-IRS5.  The \OI\ partially tracks extended outflow emission 
around L1551-IRS5, while the CO is more compact.
Again, in the ensemble, the line emission appears to be slightly extended, 
particularly in the \OI\ line; this does not necessarily correlate 
with extended continuum emission.

In a minority of our sample we can discern significant spatial
differences between \OI\ and the molecular species. For example, 
GSS30-IRS1 (Figure \ref{gss30lines}) shows compact, well-centered
emission for all the lines except \OI, which is extended and
peaks toward the northeast, similar to but more strongly than the
continuum at long wavelengths (Figure \ref{gss30cont}).
Similarly, in B1-a, the \OI\ emission peaks
$\sim 5\arcsec$ east of center.
The shift is particularly convincing as it appears
in both \OI\ lines in both GSS30 and B1-a.

Additional examples of spatially extended emission appear in Appendix B.
The resolution of multiple sources in the L1448-MM region is apparent: the 
continuum at all wavelengths 
is relatively compact and centered, but the \OI\ emission tracks the chain of 
sources to the 
south.  Additionally, the lower excitation lines of CO and H$_2$O are centered 
further south, 
while the higher excitation lines (including both OH lines) track the continuum.  
IRAS03245 
shows an extended wing of \OI, CO and H$_2$O to the southwest; H$_2$O 
is nearly absent from the source position.  B1-a 
shows a southwest extension in long wavelength continuum that is not replicated 
in the line emission.  B1-c shows extended bipolar emission in CO along the 
SE-NW axis, but no accompanying \OI.  
VLA1623 shows \OI\ in a strong east-west band (without any particular peak on 
the source) and extended long-wavelength continuum emission to the NE.

In summary, the majority of the DIGIT sources show only limited variation in spatial extent within the PACS footprint, in line or continuum.  This does not necessarily hold across all samples, or those with denser spatial sampling; \citet{karska13} find that most of their extended sources are the Class 0 objects; we do not see a particular trend in the DIGIT sample.

\subsubsection{Linefluxes and Luminosities}\label{linefluxes}

The measured fluxes and uncertainties for the eight characteristic lines identified in
\S \ref{linestats} are shown in Table \ref{fluxtable}, 
including corrections for 
extended emission.  We do not include fluxes 
for sources in confused regions; refer to Dionatos et al. (subm.; Serpens SMM 3/4) 
and Lindberg et al. (in prep.; RCrA-IRS5A, 7B, and 7C) for detailed studies of these 
sources.  We include in our statistics the composite L1448-MM spectrum. 
The uncertainties are calculated as the residual from a Gaussian plus first-order baseline 
fit to line-free sections of the local continuum for each line.
For undetected lines we provide 3$\sigma$ 
upper limits derived from local continuum, using a linewidth equal to the 
average over the detected lines near that wavelength in the sample.  The 
average linewidths are: 0.0454 $\mu$m (CO \jj{29}{28}  and OH 84.60 $\mu$m), 
0.0468 $\mu$m (OH 65.28 $\mu$m and H$_2$O 66.45 $\mu$m), and 0.1232 $\mu$m 
(CO \jj{16}{15} and H$_2$O 179 $\mu$m).
A full inventory of all lines across the sample will appear in a followup paper following a reprocessing with post-mission calibration.

Figure \ref{hist1} shows histograms of these linefluxes.  In each
we show a typical value for the rms noise (dashed vertical lines).
The distributions of linefluxes differ among the lines.
The \OI\ 63.2 $\mu$m, CO \jj{16}{15}, and perhaps the \water\ 179.5
\micron\ linefluxes appear to peak above the detection limit,
while the CO \jj{29}{28}, and \water\ 66.5 \micron\ linefluxes 
peak near the detection limit,
suggesting that deeper integrations would see lines in more sources (as 
confirmed by \citealt{karska13}).
In all cases, the linefluxes spread over 1-2 orders of magnitude above the
detection limit (Figure \ref{histsn}).

\subsubsection{Non-Detection of $^{13}$CO in PACS}

The highest linefluxes are found in the lowest $J$ $^{12}$CO  lines detectable with 
PACS; the highest of the 15 well-centered sources is BHR71, with CO \jj{15}{14}  
(173.63 $\mu$m)
\lineflux $=$ 7.9 $\times$ 10$^{-16}$ W m$^{-2}$.  The laboratory wavelength of 
$^{13}$CO \jj{15}{14} is at 181.6 $\mu$m.  Using the local continuum we set 
a 3$\sigma$ upper limit to the flux of the $^{13}$CO line at 1.0 $\times$ 
10$^{-16}$ W m$^{-2}$, 
a factor of $\sim$ 7.8 below the $^{12}$CO flux.  Assuming an interstellar 
abundance ratio of 70, and an equal emitting area, we
estimate an upper limit to the optical depth $\tau$ of $\sim$ 9.0. 
This upper limit is improved if we consider 
line ``stacking''.  If we assume the same warm 
\trot\ component for the $^{13}$CO and $^{12}$CO gas, we 
can add the statistical weight of all non-detections of $^{13}$CO in the PACS range, 
in comparison to the detections of $^{12}$CO -- 
excluding cases in which either line falls within the trimmed regions.  
For this dataset (22 lines in total), 
using a weight inversely proportional to the RMS noise in the local continuum for each line 
\citep{fedele12}, we derive an upper limit on the 
$^{12}$CO/ $^{13}$CO ratio of 13.7, or an 
upper limit to $\tau$ of 5.1.  
This is consistent with the models of \citet{visser12}, where the $^{12}$CO emission 
probed with PACS is optically thin.

\section{Rotational Diagrams Across the Sample}\label{rotdiagrams}

\subsection{Rotational diagrams for H$_2$O and OH}

The simplest way to analyze multiple molecular rotational transitions is the
rotational diagram, introduced by \citet{linke79}.  A detailed description 
can be found in, e.g., \citet{goldsmith99}, and we give a full description
of our method in Appendix C, with a brief overview here of the assumptions
required for each step.
We use the subscript $J$ as shorthand for the full state descriptor
for the more complex species characterized by multiple quantum numbers.
If the emission lines are optically thin, 
the number of molecules per degenerate sub-level 
in the upper state of the transition, $\funnyNJ/g_J$,
is proportional to the lineflux. If the resulting plot of the logarithm
of $\funnyNJ/g_J$ versus excitation energy of the upper state (in units
of K) can be fit by a straight line, one can assign a rotational 
temperature (\trot). Finally, if all the levels are in LTE, one
can set $\trot = \tk$, but this last step is not necessarily implied
by a good fit to a single \trot. In LTE, the total number of molecules
at that value of \tk\ (denoted by \funnyN)
can be calculated by the y-intercept of the fit and
the value of the partition function at \tk. If the levels are not in
LTE, one can still compute \funnyN, but the partition function may
be incorrect. The values of \funnyN\ discussed below carry this caveat.

A characteristic rotational diagram for each detected molecular species
(H$_2$O, OH, CO) in the source Elias 29 is shown in Figure \ref{rotdiag}.  
While a line can be fitted to the \water\ and OH diagrams, the scatter
is much larger than the intrinsic uncertainties; we include lineflux uncertainties 
in our rotational diagram fits. This scatter is probably
related to the existence of sub-thermal excitation in some lines, along with
optical depth effects (see, e.g., \citealt{herczeg12}).
There are 18 sources with enough 
detected H$_2$O lines to fit a single temperature component.
The fits for \trot\ for \water\ range from 76 to 379 K, with
$\mean{\trot} = 194 \pm 85$ K, and ranging from 80 to 520 K, with a 
mean of 183 $\pm$ 117 K, for OH.
Accurate measurement of the OH linefluxes is hampered by line-blending as most 
of the transitions are only marginally resolved by {\it Herschel}.  
The OH excitation itself is complicated by radiative (IR) pumping via the 
cross-ladder transitions, and
subthermal excitation (\citealt{wampfler13}, Lee et al., subm.).  Additionally, 
B1-c appears to have \funnyN(H$_2$O) six times greater than the 
next highest in the sample; B1-c is the only source without a detectable \OI\ line, 
and one of only two sources showing the OH 119 $\mu$m doublet in absorption.
This is consistent with its extremely strong 557 GHz H$_2$O emission.
Since both OH and H$_2$O molecules require more complex analysis 
beyond the scope of
this paper, we restrict further discussion of rotation diagrams to CO.

\subsection{Rotational Diagrams for CO}

CO is the simplest case to analyze because the critical density,
\eup, and detected frequency all increase monotonically
as a function of upper state $J$.  We present all CO rotational diagrams for the 
sources with detectable CO in Figures \ref{rotdiag}, \ref{rotdiag1}, and \ref{rotdiag2}.
The CO rotational diagrams generally show a positive curvature, 
for those 
sources with sufficient detected lines and low relative scatter; this may be indicative 
of a smooth continuum of excitation temperatures (see also \citealt{neufeld12,manoj12}).
A simpler model 
presumes that the CO at certain ranges of upper state $J$ is dominated by at most a 
pair of temperatures.  We assume that the CO \jj{23}{22} line is a blend with H$_2$O, 
and the CO \jj{31}{30} line is a blend with OH, and ignore both in our fits.
For statistical analyses of CO, we chose a single breakpoint,
at $\eup = 1800$ K (as in \citealt{manoj12}) and fitted separate power laws to transitions 
with \eup\
below (``warm'';  $J < 25 $) and above (``hot''; $J  > 25$) the breakpoint.
Fits are provided in Table \ref{rotfits} for sources with sufficient data.  We fitted 
two temperatures only
for sources with at least four detected CO lines with \eup\ above
the breakpoint, as we found the fits to be unstable with fewer lines.
For sources with fewer CO lines, we
fit only a single component, to the lines below 1800 K.
There are 22 sources with sufficient CO lines to fit a warm component, 
10 of which include a hot component in addition; only 6 of those with hot 
components are statistically well-constrained (ie. are fit with \funnyN(hot) and \trot(hot)
greater than their derived uncertainties).  
If the lower excitation lines become optically thick,
the state populations derived from the optically thin approximation
would exhibit curvature and diverge from the fitted line.
If the higher excitation lines are sub-thermally populated, the derived
state populations may fall below the fitted line (\citealt{goldsmith99}, Figure 6). 
The direct correspondence between \eup\ and the wavelength 
of the transition simplifies 
the analysis, in the case of CO; neither of these effects is seen
in CO rotation diagrams for this sample.
It should be noted that if {\it all} 
lines are sub-thermally excited (for very low densities), the curvature becomes positive 
and the state populations may fall above the fitted line \citep{goldsmith99,neufeld12}.


\subsubsection{Statistics and trends across the sample}

We consider the fits to rotation diagrams
across the sample.  In summary, while we fit both warm and hot 
components to the CO diagrams, we never observe the hot component in 
absence of the warm, and both components 
scale in proportion to the 
bolometric luminosity of the system.
Figure \ref{hist2} shows the distribution of \trot\ for the warm
and hot components of CO, along with the distributions of
\funnyN\ for each component.
The values of \trot(warm)
range from 200 to 589 K, with $\mean{\trot} = 354\pm 92$ K.
Lower \trot\ would
be difficult to distinguish with the values of \eup\ for the lines we observe.
Further exploration of the lower \trot\ region awaits data on lower
excitation transitions.
The distribution of hot components has $\mean{\trot} = 932\pm 217$ K for the 
six sources with statistically significant 
fits, similar temperatures 
to those found by fitting the noise in Elias 29, but with much higher \funnyN.  
Typical errors for \trot(warm) are of order 100 K, but 
the hot component is much less constrained. The sources with \trot(hot) $\lesssim$ 1000 K 
are reasonably well constrained ($\pm$ 300 K), but all sources with \trot(hot) $>$ 1030 K 
are derived from highly uncertain fits ($\pm$ 1000 K or greater) and are 
consistent with temperatures $\lesssim$ 1000 K.  Thus there is little 
evidence of a distinct component above 1000 K.
For the twelve sources with a warm component but
with no detected hot component,
the \trot(warm) values average 339 $\pm$ 98 K, very similar to the 373 $\pm$ 39 K average 
for the ten sources with both components, although with greater volatility due to increased 
uncertainties.
All CO lines are at least 3$\sigma$ detections.

Of the 10 sources with fitted 
hot components, the ratio of molecules in the warm component is 5-10 times 
that of the hot component, but in 2 cases (B335 and L1448-MM), we 
find a relatively small hot component.  B335 is a likely outlier because 
we detect only six CO lines in the hot component, leading to very uncertain fits.
This is not the case for L1448-MM, which has one of the richest CO spectra in the sample; 
this may reflect a larger emitting region of warm gas compared to the hot gas, consistent with 
the source line map shown in Appendix B of this work.  
L1551-IRS5 has a higher warm \trot(CO) than the rest of the sample, 
at 589 $\pm$ 335 K; it also has a relatively low line-to-continuum ratio and shows little emission 
from molecules other than CO.  

Across the sample, the total number
of CO molecules in warm and hot components, \funnyN\ 
(excluding the R CrA and Serpens 
sources and 
those sources with no detectable CO component) 
ranges from 0.02 - 4.56 $\times$ 10$^{49}$ molecules,
with $\mean{\funnyN} = (0.70\pm1.12) \times 10^{49}$.

\subsubsection{Caveats to the two-temperature fits}

Fitting two components with a breakpoint to CO rotation diagrams
is a common approach \citep[e.g.,][]{vankempen10,manoj12}, but the results
may be affected by relative flux calibration uncertainties,
the selection of the breakpoint, or uncertainties in the fluxes
of fainter lines.
A scaling factor adjusting the absolute flux of all lines will increase the derived
\funnyN\  for a given \trot, while a relative scaling factor
between modules may affect the derived temperatures.
The $\eup = 1800$ K breakpoint corresponds to the transition between the
second and first orders, thereby avoiding effects of misalignment
of orders on the derived values of a single component, although the ratio of the 
\funnyN\ of the two components would be affected.
However the analysis is not qualitatively affected by moving the
breakpoint by $\le 2$ transitions in either direction.
Shifting the breakpoint to a lower excitation will cause both temperatures to
decrease, and both \funnyN\ to increase (Table \ref{breakpoint}).

As noted above, the values of \trot\ need not correspond to kinetic
temperatures.  Moreover,
the interpretation of rotation diagrams in terms of two distinct
regimes of \trot\ is not unique. The broken power law can also
be fitted by a polynomial with terms up to second order. The
two components can also be described by 
a positive second order term (curvature upward to lower $J$).
More than 2 temperature components or a power-law distribution of
temperatures
will produce a distribution with a positive second derivative
(e.g., \citealt{goldsmith99}, \citealt{neufeld12}, \citealt{manoj12}, J{\o}rgensen et al., in prep.).
Heating and cooling across a shock
produces a distribution of temperatures \citep[][Dionatos et al., in subm.]{flower10}.  
Figure 13 in \citet{karska13} contains the CO 
\trot\ values from the full shock models from \citet{kaufman96}.

Even more interesting effects can be seen if the lines are not in LTE.
In particular, \citet{neufeld12} has shown that, at quite low
($n < \eten5$ \cmv) densities, but very high \tkin\ ($\sim$ 2000 K),
a positive curvature can appear in the rotation diagram because of
the behavior of the collision rates and the excitation balance.
Figure \ref{neufeld} shows the fraction of CO molecules in each state $J$,
normalized to
$\mathcal{N}$, for B335, overplotted on the models from Figure 5 from
\citet{neufeld12}.
The best fit density appears to vary from 10$^5$ to 10$^6$ \cmv\
over the full range of $J$, suggesting that the
1000 K (isothermal) optically thin model will not match the CO linefluxes for
$J \gtrsim 23$.  However, if we increase \tkin\ to 3000 K,
a good fit to all the CO lines in B335 can be obtained for a surprisingly
low density, $n = \eten4$ \cmv\ (Fig. \ref{neufeld}).
Deciding whether such an extreme solution is physically reasonable
is examined in other papers which study individual sources in this
sample.
\section{Correlations}\label{correlations}

We consider here a few correlations across the sample.  A Pearson's  
$r > 0.63$ is considered significant at the 3$\sigma$ level, for our sample of N$=$24 
(except where noted below); 
we use this metric for straightforward comparison to \citet{karska13}.  We use 
formal errors for each line, calculated from the residual RMS after fitting a 
Gaussian and 1st-order polynomial to the local continuum, added in quadrature to 
an overall calibration uncertainty floor of 10\%.  As an aside, we also considered 
all of the above correlations using the \lsmm\ 
classification for Class 0/I sources; qualitatively the results were unchanged.
In general, we found that \trot\ for each species was generally uncorrelated with 
any other properties, while \funnyN\ and \lbol\ tracked each other fairly well (but 
with significant scatter in the cases of OH and H$_2$O).

A tight correlation exists between the $\mathcal{N}_{\rm CO}$ and 
CO \jj{16}{15} line luminosity ($r=0.88$), 
indicating that a single line of CO will trace the warm component well
(Figure \ref{co1615}).
The warm component is also a strong predictor for $\funnyN_{\rm CO}$(hot), 
as the CO \jj{16}{15} and CO \jj{29}{28}
line luminosities correlate very well ($r=0.93$).
The other  molecular species correlate with the warm CO.
$\mathcal{N}_{\rm H_2O}$ correlates with $\mathcal{N}_{\rm CO}$ ($r = 0.86$), 
suggesting a similar abundance
ratio between CO and H$_2$O in all sources, although there is 
more scatter in the Class 0 sources.  
OH 84.60 $\mu$m ($r=0.72$, N$=$20) correlates as well.  In the case of 
OH 65.28 $\mu$m ($r=0.53$, N$=$12) we do not show a statistically 
significant correlation; this is the only case where we see a striking 
split between Class 0 and I sources, but attribute this to small sample size.
The total number of molecules shows no significant correlation
with \trot\ (Figure \ref{trotcorr}), either for CO (r$=$-0.13) 
or H$_2$O (r$=$-0.37), primarily due to the large uncertainty in \trot.

We also consider the correlations with properties of the SED derived 
from ancillary data.
The total number of molecules (\funnyN)  correlates with 
\lbol\ (Figure \ref{lbol_ntot}) for CO  
(warm component: $r = 0.61$, $N=22$; hot: $r=0.85$, $N=10$) but 
not very well for H$_2$O ($r = 0.67$, $N=18$).  The ratio of the hot 
to warm component is uncorrelated with \lbol\ ($r=0.28$, $N=7$), for 
those sources with robust hot components.

While the molecular emissions correlate with each other and with
\lbol\ (as noted in \citealt{manoj12}), the excitation (characterized by \trot) 
differs substantially among the molecules.
Together with the lack of correlation with \trot, these correlations
suggest that the emission is driven by a process related to the
luminosity of the source, but that the excitation (\trot) is determined
by the physics of the interaction, such as the properties of a shock front, 
and the much greater critical densities for H$_2$O and OH, compared to CO.  

\subsection{Comparison with the WISH and HOPS samples}

\citet{karska13} consider a larger sample from the WISH survey; 
here we compare results.   On most statistical trends we find close agreement.  
We detect H$_2$O 179.5 $\mu$m in most sources, and find that about 50\% of 
those also show more highly excited water.

We detect 
CO up to $\sim$ \jj{40}{39} in the brightest sources, although 
\citet{karska13} detect even higher-order transitions due to their increased 
sensitivity.  We also noted the strong correlation between the detection of 179 
$\mu$m H$_2$O and relatively low-J CO lines as seen by them.  We find a very similar warm 
temperature component in the rotational diagrams with perhaps a slightly 
higher \trot(hot), but within the errors in agreement with the \citet{karska13} result.
We have noted the lack of 
distinction between Class 0 and I sources in the DIGIT sample, a somewhat
different result from that of \citet{karska13}, based 
on a more complete comparison of cooling by various species.  

\citet{manoj12} analyze the HOPS ({\it Herschel} Orion Protostar Survey; PI: T. Megeath) 
sample of protostars in Orion.  Their sample includes 21 protostars ranging 1-200 \lbol, 
spanning a similar dynamic range but approximately an order of magnitude brighter 
than the DIGIT sample.
They find similar excitation for CO, but also detect higher transitions, ranging 
up to \jj{46}{45}. Using the same breakpoint 
in their rotation diagrams, they find a median \trot(warm) of 315 K (354 K for DIGIT), and 
\trot(hot) ranging from 703-924 K (690-1290 K for DIGIT under similar assumptions).  
Although they consider more luminous sources compared to DIGIT, the HOPS team 
finds a similar lack of 
correlation between \trot\ and \lbol.  In combination with our results, we suggest that there 
is no trend between \trot\ and \lbol\ over 3 orders of magnitude in \lbol\ (from 0.1--200 \lsun).
\citet{manoj12} posit subthermal excitation using the prescription from \citet{neufeld12}.  
Although the DIGIT sources are nearer to us than the HOPS sources and thus 
closer to being spatially resolved, 
the DIGIT CO results are consistent with both the subthermal low density/high temperature 
solution and the thermal/low temperature (close to \trot) solution.

\subsection{\OI\ 63 \micron\ Emission as a Wind Indicator}\label{oiinterp}

\citet{hollenbach89} suggested that emission of \OI\ would be a good
tracer of the current mass loss rate in a stellar wind.
 In Figure \ref{lbol_o1}, we plot the \OI\ 63 \micron\ line
luminosity versus \lbol; it shows a significant correlation ($r=0.85$). 
We also plot the mass loss rate (\mdot) from the
equation in \citet{hollenbach89} versus the time-averaged mass loss
rate (\mean{\mdot}) derived from maps of low $J$ CO lines 
collected by \citet{yildiz12} and \citet{kristensen12}. 
Given the smaller number of sources with suitable data, the correlation
is not significant. Furthermore, the relation of \citet{hollenbach89} only
comes close to reproducing observed fluxes 
if an unrealistically high
wind velocity of 1000 \kms\ is assumed.

These data suggest that \OI\ 63 \micron\ is not a particularly
good quantitative measure of the time averaged mass loss rate in
these sources. However, the measurements of \mean{\mdot} from CO
maps are not uniformly derived in the literature, so this question
should be reexamined when a consistent data set is available.
The mean \OI\ 63/145 $\mu$m lineflux ratio is $10.6\pm 6.5$;
shock models predict ratios of 14--20, higher than
typical in our sample.  

\section{The Origin of the Line Emission}\label{emissionorigin}

Since the line emission from many of the sources in this sample
is distributed in a way consistent with a point source, we need to
check if it could arise in a disk.
Although some disks do show
many of the CO and OH lines detected in the embedded sample (\citealt{sturm10}, 
\citealt{fedele12}, Meeus et al., Fedele et al., subm.,  Salyk et al., in prep) most are not
detected in gas lines other than \OI\ and \CII.
Detections of H$_2$O are particularly difficult in
these objects (mostly Herbig Ae/Be stars), although some features have been identified from deep PACS or HIFI exposures \citep[e.g.][]{fedele12}.  
It is unlikely that the circumstellar disk dominates the emission
in  {\it Herschel} beams because of its small spatial extent.
We can be more quantitative by comparing to disk samples.
The Herbig/T Tauri sample portion of DIGIT is discussed in detail in
forthcoming papers (Meeus et al., Fedele et al., in prep.); we note here that the
embedded sources are
much more luminous in gas features than disk sources.
The \OI\ lineflux in the disks sample ranges up to a few $\times$ 10$^{-5}$
\lsun\, roughly the same as the weakest \OI\ line amongst the embedded sources.
Similarly the CO \jj{16}{15} line luminosity ranges
up to 10$^{-4}$ \lsun\ in HD 100546 (Meeus et al., in prep.); the range 
in the embedded sources is between 10$^{-3}$ and 10$^{-5}$ \lsun.

The closest comparison to a disk source in our embedded sample is IRS46, itself a 
 candidate edge-on disk \citep{vankempen09}.  It is the second-lowest luminosity source with a detected warm component in our sample (ahead of only IRAM 04191), and it has the lowest inferred \funnyN(warm) -- comparable to that of HD 100546.  However, HD 100546 is a Herbig Ae/Be star with a bolometric luminosity over two orders of magnitude greater.  The CO line luminosity in IRS46 is low, but follows the same relationship with source luminosity as the rest of the sample.  From the line luminosity, it is consistent with disk sources, particularly T Tauri stars (Meeus et al, subm.).   
Thus it is possible that the disk emission may contribute to the
CO or \OI\ emission in the faint sources, but it is unlikely to affect
our overall statistics.   
It is clear from these comparisons that the
circumstellar disk is not a significant contributor to the emission from
embedded objects
in the PACS bands.

\citet{lahuis10} provide an overview of many of these sources using the IRS spectra.  The source sizes they derive would be unresolved by {\it Herschel}, with the possible exception of VLA1623.  They note detections of extended \FeII\ only in L1448 and Serpens-SMM3/4 -- it is not detected in IRAS03245, L1455-IRS3, B1-a, B1-c, GSS30, VLA 1623, WL12, IRS63, IRS46, or the RCrA sources.  VLA1623 is detected in \Sitwo\ extended emission.  There is no obvious correlation between these sources in \FeII\ or \Sitwo\ and their \OI\ emission strength, which in principle are linked as shock tracers \citep{hollenbach89}.  \NeII, which is associated with the source in IRAS 03301, B1-a, and the RCrA sources, is also not well-correlated with \OI.  Hot H$_2$ is found in IRAS 03425, B1-a, B1-c, WL12, and RCrA-IRS7A/7B; the first four sources are compact in the Herschel data.  ``Warm'' (in this case, T $\sim$ 100 K) H$_2$ is much more prominent, found in all except IRAS03301, B1-c, WL12, IRS63, RCrA-IRS5, and IRS46; all of these sources except RCrA-IRS5 have have relatively low \lbol. With the possible exception of Elias 29, we do not see evidnce for a colder component ($<$ 250 K) in PACS CO; observations with SPIRE will test the presence of this component (Green et al., in prep.).  In this source, the warm component fit underestimates the CO \jj{14}{13} through \jj{16}{15} line intensities, hinting at a cooler component.  The 500-1500 K ``hot'' temperature is only reported for a few sources, including B1-a at 615 K (which does not show a hot component in CO).  In summary, the CO and H$_2$ do not seem closely linked across the full sample; neither are \OI\ and mid-IR fine structure lines.

The emission lines in embedded sources may arise from multiple regions of differing density 
and temperature; some evidence for this is found in resolved line profiles showing 
multiple velocity components for H$_2$O and CO, 
in high spectral resolution observations from HIFI (H$_2$O 557 GHz, 
CO \jj{16}{15} and \jj{10}{9}; \citealt{kristensen11,santangelo12,vasta12,nisini13}, Yildiz et 
al., subm.) 
and ground-based instruments
such as APEX \citep[in CO \jj{6}{5};][]{vankempen09,yildiz12}.
The hot CO and nearly all H$_2$O
emission may be excited by the shocks along the outflow cavity
\citep{vankempen10b, visser12},
or inside the mostly evacuated outflow cavity \citep{manoj12}.  In recent models,
the UV-heating from the central protostar contributes significantly to the PACS CO 
lines, particularly in more evolved sources \citep{visser12}.  However, 
UV-heating seems to play only a minor role in the
PACS H$_2$O lines, compared with heating from outflow-driven shocks \citep{manoj12}.  

\section{Conclusions}\label{conclusions}

We have presented
a statistical analysis of the full sample of 30 Class 0/I protostars from the
DIGIT key project using  {\it Herschel}-HIFI and PACS spectroscopy,
utilizing improvements in calibration and sensitivity from the data pipeline in
the first
two years of the mission.

\begin{itemize}

\item We present 0.5-1000 $\mu$m SEDs of all 30 sources in our sample, and refine
calculations
of bolometric temperature and luminosity for each source.  The addition of the
PACS data has a modest effect on the classification of sources by
bolometric temperature in our sample, shifting the number (of the 25/30 sources 
with good continuum detections) of Class 0 and I
sources from 8 and 17, respectively, to 9 and 16, respectively.  Measured by the 
metric of \lbol/\lsmm, we find 16 Class 0 and 9 Class I sources,  
unaffected by the addition of the PACS data.
We do find substantial changes in \lbol\ or \tbol\
in several sources, but usually they are caused by complicated emitting
regions or sparse data in the infrared and millimeter regimes prior to the
addition of the  {\it Herschel} data.  In the relatively isolated sources, the
 {\it Herschel} spectrum did not significantly alter the
bolometric temperature or luminosity.

\item We detect over 100 rotational transitions of CO, OH,
and H$_2$O, as well as \OI\ $^3$P$_1$--$^3$P$_2$ and 
$^3$P$_0$--$^3$P$_1$,
associated with the sources.
Diffuse \CII\ ($^2$P$_{3/2}$--$^2$P$_{1/2}$) is also detected, but its connection to the sources is
weak.  No other species have been identified.

\item 
All 30 sources show detectable line emission.
We select characteristic ``cool'', ``warm",  and ``hot'' 
excitation energy lines for each of CO, OH, and H$_2$O.
The ``warm'' CO line is detected in
27/30 sources, while the ``hot'' CO line is detected in 22 of these.  
The ``cool'' H$_2$O line is detected in 24/30 sources, 
while the ``warm'' line is detected in 14 of these, 
although not entirely the same sources as for CO.  The ``cool'' OH 
line is detected in 27/30 sources, while the ``warm'' OH line 
appears in 15 of these.  
The lowest excitation OH line at 119 $\mu$m is detected in absorption 
toward two sources; in those sources the ``cool'' OH line is in emission.
The 557 GHz H$_2$O line is detected in 9 of 13 HIFI observations,
strongly corresponding with the ``cool'' excitation PACS H$_2$O.
\OI\ 63 $\mu$m is detected in all but one source (B1-c).

\item In many sources, \OI\ is of greater spatial extent than CO.  
However, even among
less obviously confused source regions there are sometimes signs
 of multiple or extended emission sources. The positional uncertainty of line and continuum measurements is between
0.3 and 3$\arcsec$ for most sources, comparable to reported pointing
uncertainties. However, some sources were substantially ($> 3 \arcsec$) mispointed.

\item We constructed rotational diagrams for all CO, OH, and H$_2$O detections.
Using a breakpoint of $\eup = 1800$ K for a two-temperature fit to the
CO, and a single temperature
fit for OH and H$_2$O, we calculate rotational temperatures and total number of
molecules for each source.  The rotational temperatures for CO
have a mean value, $\mean{\trot} = 354 \pm 92$ K for the ``warm''
component, and $\mean{\trot} = 932 \pm 217$ K for the ``hot'' component,
while the rotational temperatures for H$_2$O and OH are much lower, 
averaging $\mean{\trot} = 194 \pm 85$ K and 183 $\pm$ 117 K, respectively.  
The scatter in the H$_2$O and OH is partially attributable to inhomogeneous 
excitation conditions amongst the fitted lines.
The average total number of CO
molecules in warm and hot components, $\mean{\funnyN} =  (0.70 \pm 1.12)\ee{49}$.

\item
The line luminosities of \water\ and OH correlate  with those of CO. 
The total number of molecules in warm and hot CO correlate with \lbol.
The excitation, measured by \trot\ or by ratios of the number of
CO molecules in warm and hot components does not correlate with \lbol.
These facts suggest that the amount of excited gas is related to
the current source luminosity, but that the excitation is primarily
determined by the physics of the interaction (e.g., UV-heating or a shock).
No differences between Class 0 and I sources (defined either by \tbol, or \lbol/\lsmm)
are apparent in these data.

\item The \OI\ 63.2 $\mu$m lineflux correlates significantly ($r=0.85$)
with \lbol. If used to calculate a current mass loss
rate in the wind, the agreement with time averaged mass loss 
rates from low-$J$ CO maps is poor.

\item There is no discernible link between {\it Spitzer} and {\it Herschel} tracers of 
outflowing gas (H$_2$ and CO, respectively) or fine structure emission (\OI\ and \FeII/\Sitwo, respectively).

\end{itemize}

\acknowledgements

Support for this work, part of the  {\it Herschel} Open Time Key
Project Program, was provided by NASA through an award issued by the Jet
Propulsion Laboratory, California Institute of Technology.  The research of JKJ, OD, 
and JL was supported by a Junior Group Leader Fellowship from the Lundbeck 
Foundation and a grant from the Instrument Center for Danish Astrophysics.  The 
research at the Centre for Star and Planet Formation is supported by the Danish 
National Research Foundation and the University of Copenhagen's programme 
of excellence.
The research of J.-E. L. is supported by Basic Science Research Program 
through the National Research
Foundation of Korea (NRF) funded by the Ministry of Education, Science 
and Technology (No.
2012-0002330 and No. 2012-044689).  JDG would like to acknowledge numerous helpful discussions 
with Isa Oliveira, Augusto Carballido, John Lacy, Dan Jaffe, and Emma Yu.  We thank the anonymous referee for extensive comments that greatly improved the manuscript.

\bibliographystyle{apj}

\clearpage

\begin{deluxetable}{l l l r r r r}
\tabletypesize{\scriptsize}
\tablecaption{Observing Log \label{obslog}}
\tablewidth{0pt}
\tablehead{
\colhead{Source} & \colhead{Other Name} & \colhead{OBSID (Blue$^1$, Red$^1$ SEDs)} &
\colhead{Date Obs.} & \colhead{HIFI ObsID} &
 \colhead{Notes}
 }
\startdata
DK Cha & IRAS 12496-7650 & 1342188039, 1342188040 & 10 Dec 2009 & --- & off-center \\
L1551-IRS5 & & 1342192805 \& & 26 Mar 2010 \& & 
--- & \\
 & & 1342229711 &  24 Sep 2011 & 
--- & \\
L1527 & IRAS 04368+2557 & 1342192981, 1342192982 &  29 Mar 2010 &  --- & off-center \\
TMR 1 & IRAS 04361+2547 & 1342192985, 1342192986 & 29 Mar 2010 &  --- & off-center \\
TMC 1A & IRAS 04362+2535 & 1342192987, 1342192988 & 29 Mar 2010 &  --- & off-center \\
Serpens-SMM3 & & 1342193216, 1342193214 & 02 Apr 2010 &  --- & mult.
sources \\
Serpens-SMM4 & & 1342193217, 1342193215 & 02 Apr 2010 &  --- & mult.
sources \\
L1455-IRS3 & IRAS 03249+2957 & 1342204122, 1342204123 & 08 Sep 2010 &  1342202056 & \\
RCrA-IRS7C & & 1342206990, 1342206989 & 23 Oct 2010 &  1342215843 &
mult. sources \\
RCrA-IRS5A & & 1342207806, 1342207805 & 01 Nov 2010 &  --- & mult.
sources \\
RCrA-IRS7B & & 1342207807, 1342207808 & 02 Nov 2010 &  1342215842 &
mult. sources \\
B335 & & 1342208889, 1342208888 & 13 Nov 2010 &  --- &  \\
L1157 & & 1342208909, 1342208908 & 13 Nov 2010 &  --- & \\
L1014 & & 1342208911, 1342208912 & 14 Nov 2010 &  1342196408 & \\
BHR 71 & & 1342212230, 1342212231 & 01 Jan 2011 &  --- &  \\
L1448-MM & & 1342213683, 1342214675 & 02 \& 22 Feb 2011 &  --- & \\
VLA 1623-243 & & 1342213918, 1342213917 & 06 Feb 2011 &  1342205298 & \\
IRAS 03245+3002 & L1455-IRS1 & 1342214677, 1342214676 & 23 Feb 2011 &  1342202057 & \\
B1-c & & 1342216213, 1342216214 & 04 Mar 2011 &  1342203204 & \\
L1489 & IRAS 04016+2610 & 1342216216, 1342216215 & 05 Mar 2011 &  --- & \\
IRAS 03301+3111 & Perseus Bolo76 & 1342215668, 1342216181 & 09 \& 16 Mar 2011 &  1342202070 & \\
GSS30-IRS1 & & 1342215678, 1342215679 & 10 Mar 2011 &  --- & \\
B1-a & & 1342216182, 1342216183 & 16 Mar 2011 &  1342203203 & \\
IRAM 04191+1522 & & 1342216654, 1342216655 & 23 Mar 2011 &  1342203192 & \\
TMC 1 & IRAS 04381+2540 & 1342225803, 1342225804 & 06 Aug 2011 &  --- & \\
WL12 & & 1342228187, 1342228188 & 05 Sep 2011 &  1342205299 & \\
IRS 63 & & 1342228473, 1342228472 & 11 Sep 2011 &  --- & \\
IRS 46 & & 1342228474, 1342228475 & 11 Sep 2011 &  1342205301 & mult.
sources  \\
IRS 44$^2$ & & 1342228474, 1342228475 & 11 Sep 2011 &  1342205300 &
mult. sources \\
Elias 29 & & 1342228519, 1342228520 & 11 Sep 2011 &  --- & \\
\enddata
\tablecomments{Observations log for protostellar sources discussed in this
work.  If the centroid of the continuum source is more than 3$\arcsec$ distant
from
the center spaxel (and commanded coordinates), we note this as
``off-center''.\\
$^1$Blue SEDs are simultaneous 50-75 and 100-150 $\mu$m observations; Red SEDs
are simultaneous 70-105 and 140-210 $\mu$m observations. \\
$^2$IRS 44 was observed in the same exposure as IRS 46, but centered 2 spaxels
away from IRS 46.}
\end{deluxetable}

\begin{deluxetable}{l r r r r r r}
\tabletypesize{\scriptsize}
\tablecaption{Source List \label{sourcelist}}
\tablewidth{0pt}
\tablehead{
\colhead{Source} & \colhead{Cloud} &
\colhead{Dist. (pc)} & \colhead{RA (J2000)} & \colhead{Dec} & \colhead{Pos. Ref.} & \colhead{Dist. Ref.} 
}
\startdata
L1448-MM & Per & 232$^1$ & 03h25m38.9s & +30d44m05.4s & PROSAC & HI \\
IRAS 03245+3002 & Per & 250 & 03h27m39.1s & +30d13m03.1s & c2d & \\
L1455-IRS3 & Per & 250 & 03h28m00.4s & +30d08m01.3s & c2d & \\
IRAS 03301+3111 &  Per & 250 & 03h33m12.8s & +31d21m24.2s & c2d & \\
B1-a &  Per & 250 & 03h33m16.7s & +31d07m55.2s & c2d & \\
B1-c & Per & 250 & 03h33m17.9s & +31d09m31.9s & c2d & \\
L1489 &  Tau & 140 & 04h04m42.9s & +26d18m56.3s & B & \\
IRAM 04191+1522 & Tau & 140 & 04h21m56.9s & +15d29m45.9s & D & \\
L1551-IRS5 & Tau & 140 & 04h31m34.1s & +18d08m04.9s & 2MASS & \\
L1527 & Tau & 140 & 04h39m53.9s & +26d03m09.8s & PROSAC & \\
TMR 1 & Tau & 140 & 04h39m13.9s & +25d53m20.6s & H & \\
TMC 1A & Tau & 140 & 04h39m35.0s & +25d41m45.5s & H & \\
TMC 1 & Tau & 140 & 04h41m12.7s & +25d46m35.9s & A & \\
BHR 71 & Core & 178 & 12h01m36.3s & --65d08m53.0s & c2d & \\
DK Cha & Cha & 178 & 12h53m17.2s & --77d07m10.7s & c2d &  \\
GSS30-IRS1 & Oph & 125 & 16h26m21.4s & --24d23m04.3s & c2d & \\
VLA 1623-243 &  Oph & 125 & 16h26m26.4s & --24d24m30.0s & c2d & \\
WL12 &  Oph & 125 & 16h26m44.2s & --24d34m48.4s & c2d & \\
Elias 29 & Oph & 125 & 16h27m09.4s & --24d37m18.6s & c2d & \\
IRS 46 &  Oph & 125 & 16h27m29.4s & --24d39m16.1s &  c2d & \\
IRS 44 &  Oph & 125 & as IRS46$^2$ & as IRS46$^2$ & c2d & \\
IRS 63 & Oph & 125 & 16h31m35.6s & --24d01m29.3s & c2d & \\
Serpens-SMM3 & Ser & 429$^3$ & 18h29m59.3s & +01d14m01.7s & c2d & DZ \\
Serpens-SMM4 & Ser & 429$^3$ & 18h29m56.7s & +01d13m17.2s & c2d & DZ \\
RCrA-IRS5A & CrA & 130$^4$ & 19h01m48.1s & --35d57m22.7s & N & NE \\
RCrA-IRS7C & CrA & 130$^4$ & 19h01m55.3s & --36d57m17.0s & L & NE \\
RCrA-IRS7B & CrA & 130$^4$ & 19h01m56.4s & --36d57m28.3s & L & NE \\
B335 & Core & 106$^5$ & 19h37m00.9s & +07d34m09.7s & PROSAC & O \\
L1157 &  Core & 325 & 20h39m06.3s & +68d02m16.0s & PROSAC \\
L1014 &  Core & 200 & 21h24m07.5s & +49d59m09.0s & Y \\
\enddata
\tablecomments{ ~List of protostellar sources discussed in this
work by region, sorted by RA.  Coordinate reference code: D $=$ \citet{dunham06}; 
Y $=$ \citet{young04};
L $=$ \citet{lindberg11}; N $=$ \citet{nisini05}; H $=$ \citet{haisch04};
HI $=$ \citet{hirota11}; DZ $=$ \citet{dzib10,dzib11}; O $=$ \citet{olofsson09};
NE $=$ \citet{neuhauser08};
A $=$ \citet{apai05}; B $=$ \citet{brinch07}; c2d $=$ \citet{evans09,lahuis10};
PROSAC $=$ \citet{jorgensen09} \\
$^1$The distance to L1448-MM is reported as 232 pc by \citet{hirota11}. \\
$^2$IRS 44 was observed in the same exposure as IRS 46, but centered 2 spaxels
away from IRS 46.  \\
$^3$The distance to Serpens was revised upward from 260 pc to 429 $\pm$ 2 pc
in \citet{dzib10,dzib11}. \\
$^4$The distance to RCrA was revised downward from 170 pc to 130 pc in 
\citet{neuhauser08}. \\
$^5$The distance to B335 was revised downward from 250 pc to 106 pc in
\citep{olofsson09}}.
\end{deluxetable}

\begin{deluxetable}{l r r}
\tabletypesize{\scriptsize}
\tablecaption{PACS photometric/spectroscopic flux calibration \label{specphot}}
\tablewidth{0pt}
\tablehead{
\colhead{Source} & \colhead{Flux ratio (70 $\mu$m)} & \colhead{Flux ratio (160 $\mu$m)}
 }
\startdata
  HD 163296   &  1.042 &	0.842 \\
  HD 135344    & 0.982 &	0.873 \\
  HD 104237     & 1.017 &	1.179 \\
  HD 139614    & 1.120 &	0.992 \\
  HD 142527   &  1.283 &	1.263 \\
   HD 36112   &  1.149	 & 0.893 \\
  HD 100453   &  1.166 &	1.000 \\
  HD 179218   &  0.991 &	0.929 \\
     IRS 48   &  1.039 &	1.040 \\
  HD 100546  &   0.930 &	0.925 \\
 & & \\
  MEAN      &   1.071 &	0.993 \\
  STDDEV     &    0.106 & 0.136 \\
\enddata
\tablecomments{Absolute flux calibration from PACS spectroscopy compared to PACS 
photometry for pointlike Herbig Ae/Be stars in the DIGIT and GASPS disks samples, 
respectively.  The ratios are the photometric value divided by the spectroscopic value, 
using HIPE 6.0 (``bgcal'') calibration.
}
\end{deluxetable}

\begin{deluxetable}{l r r r r r}
\tabletypesize{\scriptsize}
\tablecaption{Bolometric Luminosities and Temperatures \label{boltable}}
\tablewidth{0pt}
\tablehead{
\colhead{Source} & \colhead{$L_{\rm bol}$ (wHer)} & \colhead{$T_{\rm bol}$ (wHer)} &
	\colhead{$L_{\rm bol}$ (w/oHer)} & \colhead{$T_{\rm bol}$ (w/oHer)} & \colhead{$L_{\rm bol}$/$L_{\rm submm}$}}
\startdata
& L$_{\odot}$ & K & L$_{\odot}$ & K & \\
\hline \\
L1448-MM & 8.4 & 47 & 4.4 & 69 & 38 \\
IRAS 03245+3002 & 5.7 & 46 & 4.2 & 59 & 64  \\
L1455-IRS3 & 0.32 & 204 & 0.3 & 221 & 18  \\
IRAS 03301+3111 & 3.6 & 382 & 3.2 & 438 & 903 \\
B1-a  & 2.4 & 88 & 1.5 & 132 & 39 \\
B1-c  & 4.5 & 52 & 1.8 & 76 & 14 \\
L1489  & 3.5 & 226 & 3.7 & 238 & 259 \\
IRAM 04191+1522  & 0.11 & 31 & 0.11 & 24 & 5 \\
L1551-IRS5 & 24.5 & 106 & 28 & 97 & 107 \\
L1527 & 1.7 & 67 & 2.0 & 36 & 45 \\
TMR 1 &  2.6  & 140 & 3.7 & 144 & 734 \\
TMC 1A & 2.5 & 164 & 2.2 & 172 & 143 \\
TMC 1 & 0.66 & 171  & 0.7 & 139 & 114 \\
BHR71  & 9.7 & 55 & 7.1 & 84 & --- \\
DK Cha & 27.8 & 592 & 27.7 & 565 & $>$ 10000 \\
GSS30-IRS1  & 14.5 & 138 & 4.4 & 254 & 5150  \\
VLA1623-243 & 2.6 & 35 & 1.0 & 35 & 30 \\
WL12 & 1.6 & 238 & 2.6 & 440 & 145 \\
Elias 29 & 20.1 & 387 & 4.8 & 383 & 4215 \\
IRS 46 & 0.5 & 352 & 0.8 & 633 & --- \\
IRS 44 & 5.1 & 213 & --- & ---  & --- \\
IRS 63 & 1.5 & 287 & 1.6 & 580 & 33  \\
B335 & 0.68 & 39 & 0.54 & 37 & 27 \\
L1157 & 6.5 & 40 & 6.0 & 42 & 44 \\
L1014  & 0.26 & 81 & 0.21 & 64 & 14 \\
\enddata
\tablecomments{Recalculated bolometric luminosity and temperature for the DIGIT embedded 
sources sample, including  {\it Herschel}-PACS data in the fit.  The bolometric luminosities (both old 
and new) for B335 and L1448-MM are adjusted to the new distances. The ratio of 
$L_{\rm bol}$/$L_{\rm submm}$ is derived using a submillimeter cutoff wavelength of 350 $\mu$m.
Values for confused regions (Serpens and R CrA) are presented in subsequent papers, 
as noted in the text.
}
\end{deluxetable}

\begin{deluxetable}{l r r r r r r r r}
\tabletypesize{\scriptsize}
\tablecaption{Linefluxes \label{fluxtable}}
\tablewidth{0pt}
\tablehead{
\colhead{Source} & \colhead{CO \jj{29}{28}}
& \colhead{CO \jj{16}{15}} & \colhead{OH} &\colhead{OH}
&\colhead{H$_2$O} & \colhead{H$_2$O} & \colhead{ [O I]} & \colhead{H$_2$O 557
GHz}
}
\startdata
 & 90.16 $\mu$m & 162.83 $\mu$m & 65.28 $\mu$m & 84.61$\mu$m & 66.45 $\mu$m &
 179.53 $\mu$m & 63.18 $\mu$m & $\int T_{\rm MB}$ d$v$\\
  \hline \\
 & 10$^{-18}$ W m$^{-2}$ &  & & & &  & &  K km s$^{-1}$ \\
 \hline \\
L1448-MM & 197 $\pm$ 26 & 574 $\pm$ 12 & 163 $\pm$ 14 & 232 $\pm$ 16 & 376 $\pm$ 31 & 2147 $\pm$ 71 &  760 $\pm$ 33 &  \\
IRAS 03245 & 20 $\pm$ 6 & 50 $\pm$ 5 & $<$ 49 & 47 $\pm$ 4 & $<$ 28 & 17 $\pm$ 1 &  96 $\pm$ 9 & 0.50 $\pm$ 0.02 \\
L1455-IRS3 & $<$ 25 & $<$ 15 & $<$ 22 & 33 $\pm$ 10 & $<$ 20 & $<$ 18  & 17 $\pm$ 3 & $<$ 0.08 \\
IRAS 03301 & 35 $\pm$ 10 & 38 $\pm$ 5 & 145 $\pm$ 33 & 146 $\pm$ 5 & 57 $\pm$ 14 & 29 $\pm$ 6 &  324 $\pm$ 14 & $<$ 0.08 \\
B1-a & $<$ 33 & 51 $\pm$ 6 &  73 $\pm$ 7 & 81 $\pm$ 6 & 29 $\pm$ 4 & 70  $\pm$ 12 & 298 $\pm$ 12 & 1.51 $\pm$ 0.02 \\
B1-c & $<$ 63 & 85 $\pm$ 10 & $<$ 74 & $<$ 74 &  $<$ 68 & 334 $\pm$ 11 & $<$ 116  & 5.41 $\pm$ 0.05 \\
L1489 & 24 $\pm$ 3 & 76 $\pm$ 6 & 67 $\pm$ 8 & 63 $\pm$ 8 &  59 $\pm$ 12 & 72 $\pm$ 7 & 405 $\pm$ 15 & \\
IRAM 04191 & $<$ 27 & 19 $\pm$ 4 & $<$ 25 & 32 $\pm$ 6 & $<$ 33 & $<$ 10 & 75 $\pm$ 9 & $<$ 0.08 \\
L1551-IRS5 & 61 $\pm$ 11 & 138 $\pm$ 7 & $<$ 165 & 164 $\pm$ 36 & $<$ 93 & $<$ 32 &  7194 $\pm$ 110 &  \\
L1527 & $<$ 25 & 30 $\pm$ 6 & $<$ 28 & 59 $\pm$ 15 & $<$ 27 & 51 $\pm$ 5 & 483 $\pm$ 27 & \\
TMR1 & 58 $\pm$ 15 & 137 $\pm$ 5 & 129 $\pm$ 16 & 141 $\pm$ 11 & 134 $\pm$ 26 & 36 $\pm$ 5 &  594 $\pm$ 35 & \\
TMC1A & 37 $\pm$ 7 & 37 $\pm$ 4 & 67 $\pm$ 17 & 60 $\pm$ 9 & $<$ 12 & 14 $\pm$ 2 & 682  $\pm$ 19 & \\
TMC1 & 45 $\pm$ 4 & 69 $\pm$ 7 & 104 $\pm$ 15 & 101 $\pm$ 11 & 20 $\pm$ 4 & 33 $\pm$ 5 &  837 $\pm$ 14 &  \\
BHR71 & 319 $\pm$ 16 & 624 $\pm$ 21 & 89  $\pm$ 9 & 116 $\pm$ 20 &  $<$ 63 & 447 $\pm$ 14 & 2350 $\pm$ 77 &  \\
DKCha & 280 $\pm$ 11 & 440 $\pm$ 23 & 269 $\pm$ 24 & 321 $\pm$ 16 & $<$ 34 & 106 $\pm$ 6 &  2702 $\pm$ 53 & \\
GSS30 & 164 $\pm$ 11 & 374 $\pm$ 9 & 224 $\pm$ 15 & 229 $\pm$ 9 & 362 $\pm$ 13 & 234 $\pm$  15 &  1720 $\pm$ 22 & \\
VLA1623 & 94 $\pm$ 18 & 185 $\pm$ 16 & $<$ 41 & 105 $\pm$ 6 & $<$ 39 & 123 $\pm$ 10 &  658 $\pm$ 41 & 5.71 $\pm$ 0.04 \\
WL12 & 38 $\pm$ 5 & 85 $\pm$ 3 & 59 $\pm$ 2 & 95 $\pm$ 7 & 52 $\pm$ 5 & 27 $\pm$ 5 &  441 $\pm$ 14 & 0.20 $\pm$ 0.01 \\
Elias29 & 199 $\pm$ 9 & 535 $\pm$ 6 & 457 $\pm$ 5 & 615 $\pm$ 13 & 355 $\pm$ 15 & 185 $\pm$ 5 &  1787 $\pm$ 33 & \\
IRS46 & $<$ 24 & 17 $\pm$ 3 & $<$ 29 & $<$ 30 & $<$ 25 & $<$ 13 & 60 $\pm$ 13 & 1.26 $\pm$ 0.02 \\
IRS44 & 77 $\pm$ 23 & 473 $\pm$ 9 & $<$ 118 & 175 $\pm$ 31 & 172 $\pm$ 30 & 232 $\pm$ 13 &  887 $\pm$ 52 & 3.21 $\pm$ 0.03 \\
IRS63 & $<$ 28 & $<$ 12 & $<$ 24 & 17 $\pm$ 3 & $<$ 24 & $<$ 13 & 190 $\pm$ 12 &  \\
B335 & 24 $\pm$ 4 & 131 $\pm$ 4 & $<$ 31 & 48 $\pm$ 5 & $<$ 27 & 80 $\pm$ 5 &  381 $\pm$ 15 &  \\
L1157 & 48 $\pm$ 6 & 141 $\pm$ 6 & $<$ 34 & 72 $\pm$ 12 &  24 $\pm$ 6 & 187 $\pm$ 4 & 532 $\pm$ 15 & \\
L1014 & $<$ 25 & $<$ 18 & $<$ 29 & $<$ 29 & $<$ 27 & $<$ 20 & 123 $\pm$ 13 & $<$ 0.08 \\
\enddata
\tablecomments{Census of selected lines
in the sample, corrected for extended emission: CO \jj{29}{28}, representing
the hot component; CO \jj{16}{15},
representing the warm component; OH 84.61 $\mu$m; H$_2$O 179.53 $\mu$m,
representing low-lying states of their respective species; OH 65.28 $\mu$m,
H$_2$O 66.45 $\mu$m, representing higher states; [O I], tracing excited gas in
the disk, envelope, and outflow.  3$\sigma$ upper limits are given  
for non-detected lines, accounting for the PSF correction.  The final column lists 
integrated intensities for the 557 GHz H$_2$O line for sources
observed with HIFI in the DIGIT embedded sources sample; for HIFI non-detections, 
we provide 3$\sigma$ upper limits assuming a width of 20 km s$^{-1}$. \\
}
\end{deluxetable}

\begin{deluxetable}{l r r r r}
\tabletypesize{\scriptsize}
\tablecaption{Line and Continuum Spatial Extent \label{extended}}

\tablehead{
\colhead{Source} & \colhead{CO} & \colhead{CO} &  \colhead{OH} & \colhead{OH}}
\startdata
 &
\jj{29}{28} & \jj{16}{15} & 65.28 $\mu$m &
84.61 $\mu$m \\
\hline \\
PSF & 1.45 & 2.04 & 1.43 & 1.45 \\
IRAS 03245 & --- & 2.24 $\pm$ 0.28 (2.42 $\pm$ 0.03) & ---  & --- \\
IRAS 03301 & --- & --- & 1.36 $\pm$ 0.35 (2.96 $\pm$ 0.75) & 1.33 $\pm$ 0.19 (1.20 $\pm$ 0.41) \\
B1-a  & --- & 2.44 $\pm$ 0.45 (3.67 $\pm$ 0.17) & 1.77 $\pm$ 0.41 (1.78 $\pm$ 0.11) & 1.29 $\pm$ 0.18 (1.38 $\pm$ 0.24) \\
B1-c & --- & 3.50 $\pm$ 0.47 (2.06 $\pm$ 0.03) & --- & --- \\
L1489 & --- & 1.91 $\pm$ 0.23 (2.10 $\pm$ 0.04) & 2.24 $\pm$ 0.45 (1.45 $\pm$ 0.03) & 1.37 $\pm$ 0.44 (1.48 $\pm$ 0.04) \\
L1551-IRS5 & 1.19 $\pm$ 0.35 (1.80 $\pm$ 0.01) & 1.62 $\pm$ 0.19 (2.55 $\pm$ 0.01) & --- & 1.33 $\pm$ 0.39 (1.58 $\pm$ 0.01) \\
TMC 1 & --- & 2.25 $\pm$ 0.18 (2.36 $\pm$ 0.17) & ---  & 0.96 $\pm$ 0.28 (1.48 $\pm$ 0.20) \\
BHR71 & 2.09 $\pm$ 0.31 (2.63 $\pm$ 0.02) & 3.15 $\pm$ 0.14 (3.59 $\pm$ 0.03) & --- & 1.18 $\pm$ 0.47 (2.56 $\pm$ 0.02) \\
GSS30-IRS1 & 1.67 $\pm$ 0.19 (2.89 $\pm$ 0.02) & 2.51 $\pm$ 0.08 (5.34 $\pm$ 0.06) & 1.71 $\pm$ 0.19 (2.23 $\pm$ 0.01) & 1.58 $\pm$ 0.15 (2.80 $\pm$ 0.22) \\
VLA1623 & --- & 3.36 $\pm$ 0.32 (4.30 $\pm$ 0.05) & --- & 2.48 $\pm$ 0.35 (2.74 $\pm$ 0.03) \\
WL12 & --- & 1.99 $\pm$ 0.17 (3.60 $\pm$ 0.14) & --- & 1.29 $\pm$ 0.25 (1.41 $\pm$ 0.13) \\
Elias 29 & 1.29 $\pm$ 0.13 (2.25 $\pm$ 0.02) & 2.14 $\pm$ 0.03 (3.98 $\pm$ 0.04) & 1.42 $\pm$ 0.06 (1.82 $\pm$ 0.01) & 1.42 $\pm$ 0.08 (2.16 $\pm$ 0.38)  \\
IRS 63 & --- & --- & --- & --- \\
B335 & --- & 1.88 $\pm$ 0.14 (2.30 $\pm$ 0.03) & --- & --- \\
L1157 & --- & 1.90 $\pm$ 0.14 (1.93 $\pm$ 0.03) & --- & --- \\
\enddata
\tablecomments{Comparison of spatial extent for lines and local continuum for
the 15
well-centered sources with no known source confusion at the resolution of
 {\it Herschel}.  Each entry
is a pair of values: the ratio of the lineflux in the central 3$\times$3
spaxels over the
lineflux in the central spaxel, and in parentheses the ratio of the average
continuum flux density 
in the central 3$\times$3 spaxels over the continuum flux in the central
spaxel, in a
selected range of wavelengths near the line. 
 Formal errors are provided for each line ratio, 
but systematic calibration uncertainties and spaxel-to-spaxel variation are not included.}
\end{deluxetable}

\begin{deluxetable}{l r r r}
\tabletypesize{\scriptsize}
\tablecaption{Line and Continuum Spatial Extent \label{extended2}}
\tablewidth{0pt}
\tablehead{
\colhead{Source} & \colhead{H$_2$O} & \colhead{H$_2$O} &  \colhead{[O I]}}
\startdata
 &
  66.45 $\mu$m & 
179.53 $\mu$m &  63.18 $\mu$m \\
\hline \\
PSF & 1.43 & 2.27 & 1.43 \\
IRAS 03245 & --- & --- & 2.95 $\pm$ 0.45 (1.47 $\pm$ 0.04) \\
IRAS 03301 & --- & --- & 1.37 $\pm$ 0.10 (2.23 $\pm$ 0.29) \\
B1-a  & 4.23 $\pm$ 1.26 (1.69 $\pm$ 0.80) & 2.99 $\pm$ 0.63 (4.68 $\pm$ 0.41) & 2.86 $\pm$  0.16 (1.81 $\pm$ 0.15) \\
B1-c & --- & 3.44 $\pm$ 0.13 (2.27 $\pm$ 0.03) & --- \\
L1489 & --- & 1.81 $\pm$ 0.27 (2.25 $\pm$ 0.06)  & 1.55 $\pm$ 0.13 (1.46 $\pm$ 0.04) \\
L1551-IRS5 & --- & --- & 1.88 $\pm$ 0.05 (1.51 $\pm$ 0.02) \\
TMC 1 & --- & 3.19 $\pm$ 0.70 (2.44 $\pm$ 0.21) & 2.04 $\pm$ 0.08 (1.39 $\pm$ 0.19) \\
BHR71 & --- & 4.95 $\pm$ 0.20 (3.75 $\pm$ 0.03) & 2.02 $\pm$ 0.09 (2.30 $\pm$ 0.06) \\
GSS30-IRS1 & 1.23 $\pm$ 0.08 (2.26 $\pm$ 0.21) & 2.36 $\pm$ 0.16 (5.60 $\pm$ 0.15) & 3.03 $\pm$ 0.07 (2.14 $\pm$ 0.02) \\
VLA1623 & --- & 3.36 $\pm$ 0.24 (4.70 $\pm$ 0.04) & 5.67 $\pm$ 0.38 (2.78 $\pm$ 0.17) \\
WL12 & --- & ---  & 1.17 $\pm$ 0.10 (1.60 $\pm$ 0.11)  \\
Elias 29 & 1.40 $\pm$ 0.12 (1.84 $\pm$ 0.25) & 2.00 $\pm$ 0.09 (4.40 $\pm$ 0.06)  & 1.69 $\pm$ 0.04 (1.79 $\pm$ 0.02) \\
IRS 63 & --- & --- & 1.35 $\pm$ 0.23 (1.75 $\pm$ 0.08) \\
B335 & --- & 2.02 $\pm$ 0.18 (2.46 $\pm$ 0.04) & 2.02 $\pm$ 0.16 (1.81 $\pm$ 0.11) \\
L1157 & 4.23 $\pm$ 1.60 (1.61 $\pm$ 0.11) & 2.36 $\pm$ 0.09 (2.05 $\pm$ 0.03) & 1.89 $\pm$ 0.09 (1.62 $\pm$ 0.11) \\
\enddata
\tablecomments{Comparison of spatial extent for lines and local continuum for
the 15
well-centered sources with no known source confusion at the resolution of
 {\it Herschel}.  Each entry
is a pair of values: the ratio of the lineflux in the central 3$\times$3
spaxels over the
lineflux in the central spaxel, and in parentheses the ratio of the average
continuum flux density 
in the central 3$\times$3 spaxels over the continuum flux in the central
spaxel, in a
selected range of wavelengths near the line.
Formal errors are provided for each line ratio, 
but systematic calibration uncertainties and spaxel-to-spaxel variation are not included. }
\end{deluxetable}

\begin{deluxetable}{l l l r r}
\tabletypesize{\scriptsize}
\tablecaption{CO Rotational Temperatures and Total Number of Molecules \label{rotfits}}
\tablewidth{0pt}
\tablehead{
\colhead{Source} & \colhead{\trot (Warm)} & \colhead{\trot (Hot)} &
\colhead{\funnyN(Warm)} & \colhead{\funnyN(Hot)}}
\startdata
 & K & K & 10$^{49}$ & 10$^{47}$ \\
 \hline \\
L1448-MM & 315 $\pm$ 78 & 942 $\pm$ 288 & 4.37 $\pm$ 2.18 & 19.0 $\pm$ 8.9 \\
IRAS 03245 & 348 $\pm$ 117 & --- & 0.31 $\pm$ 0.15 & --- \\
L1455-IRS3 & --- & --- & --- & --- \\
IRAS 03301 & 336 $\pm$ 89 & --- & 0.18 $\pm$ 0.10 & --- \\
B1-a & 245 $\pm$ 76 & --- & 0.50 $\pm$ 0.19 & --- \\
B1-c & 277 $\pm$ 150 & --- & 0.64 $\pm$ 0.19 & --- \\
L1489 & 409 $\pm$ 133 & 1177 $\pm$ 3846 & 0.11 $\pm$ 0.06 & 1.09 $\pm$ 0.04 \\
IRAM 04191 & 200 $\pm$ 96 & --- & 0.11 $\pm$ 0.19 & --- \\
L1551-IRS5 & 589 $\pm$ 335 & --- & 0.16 $\pm$ 0.09 & --- \\
L1527 & 416 $\pm$ 219 & --- & 0.04 $\pm$ 0.02 & --- \\
TMR1 & 392 $\pm$ 121 & 1016 $\pm$ 1163 & 0.17 $\pm$ 0.09 & 2.56 $\pm$ 0.42 \\
TMC1A & 351 $\pm$ 98 & --- & 0.08 $\pm$ 0.04 & --- \\
TMC1 & 381 $\pm$ 114 & 1290 $\pm$ 979 & 0.10 $\pm$ 0.05 & 1.13 $\pm$ 0.44 \\
BHR71 & 410 $\pm$ 133 & 693 $\pm$ 298 & 1.30 $\pm$ 0.70 & 33.9 $\pm$ 8.5 \\
DK Cha & 389 $\pm$ 119 & 1030 $\pm$ 471 & 0.91 $\pm$ 0.49  & 13.0 $\pm$ 5.0 \\
GSS30 & 336 $\pm$ 89 & 838 $\pm$ 632 & 0.50 $\pm$ 0.25 & 6.74 $\pm$ 1.24 \\
VLA1623 & 306 $\pm$ 74 & --- & 0.35 $\pm$ 0.18  & --- \\
WL12 & 369 $\pm$ 108 & --- &  0.10 $\pm$ 0.05 & --- \\
Elias29 & 423 $\pm$ 63 & 690 $\pm$ 129 & 0.401 $\pm$ 0.269 & 12.3 $\pm$ 6.0 \\
IRS46 & 345 $\pm$ 260 & --- & 0.02 $\pm$ 0.005 & --- \\
IRS44 & 291 $\pm$ 66 & --- & 0.71 $\pm$ 0.36 & --- \\
IRS63 & --- & --- & --- & --- \\
B335 & 316 $\pm$ 78 & 4513 $\pm$ 11497 & 0.124 $\pm$ 0.063 & 0.54 $\pm$ 0.20 \\
L1157 & 360 $\pm$ 103 & 716 $\pm$ 410 & 0.20 $\pm$ 0.11 & 3.66 $\pm$ 0.69 \\
L1014 & --- & --- & --- & --- \\
\enddata
\tablecomments{Rotational diagram fits to the sample.  Sources
with $>$ 
four CO lines of excitation energy $>$ 1800 K are fitted with two
temperatures
``warm'' and ``hot'' with a breakpoint at 1800 K.   Three sources (IRS 63,
L1014, and
L1455-IRS3) show little or no detectable CO emission and cannot be analyzed
with a
rotational diagram.  Of the remaining 27 sources, we fit ``hot'' components to
10 of them, of which 6 produce a statistically significant fit.  \\
}
\end{deluxetable}

\begin{deluxetable}{l r r r r }
\tabletypesize{\scriptsize}
\tablecaption{Breakpoint in CO Rotational Diagrams \label{breakpoint}}
\tablewidth{0pt}
\tablehead{
\colhead{Breakpoint} & \colhead{\funnyN(warm)}  & \colhead{\funnyN(hot)} &
	\colhead{\trot(warm)} & \colhead{\trot(hot)}}
\startdata
K & $\times$ 10$^{49}$ & $\times$ 10$^{47}$ & K & K \\
\hline \\
2300 & 1.22 & 35.8 & 433 & 684 \\
2100 & 1.22 & 30.1 & 431 & 718 \\
1900 & 1.30 & 33.9 & 410 & 693 \\
1700 & 1.44 & 43.3 & 378 & 648 \\
1600 & 1.59 & 45.6 & 352 & 639 \\
\enddata
\tablecomments{The effect of shifting the breakpoint between the ``warm'' and
``hot''
temperature components in the CO rotational diagram for BHR71.  The 1800 K breakpoint is equivalent to the 1900 K entry (there are no transitions between 1800 and 1900 K as we exclude the CO \jj{26}{25} line in this work).}
\end{deluxetable}

\begin{figure}
\includegraphics[scale=0.45]{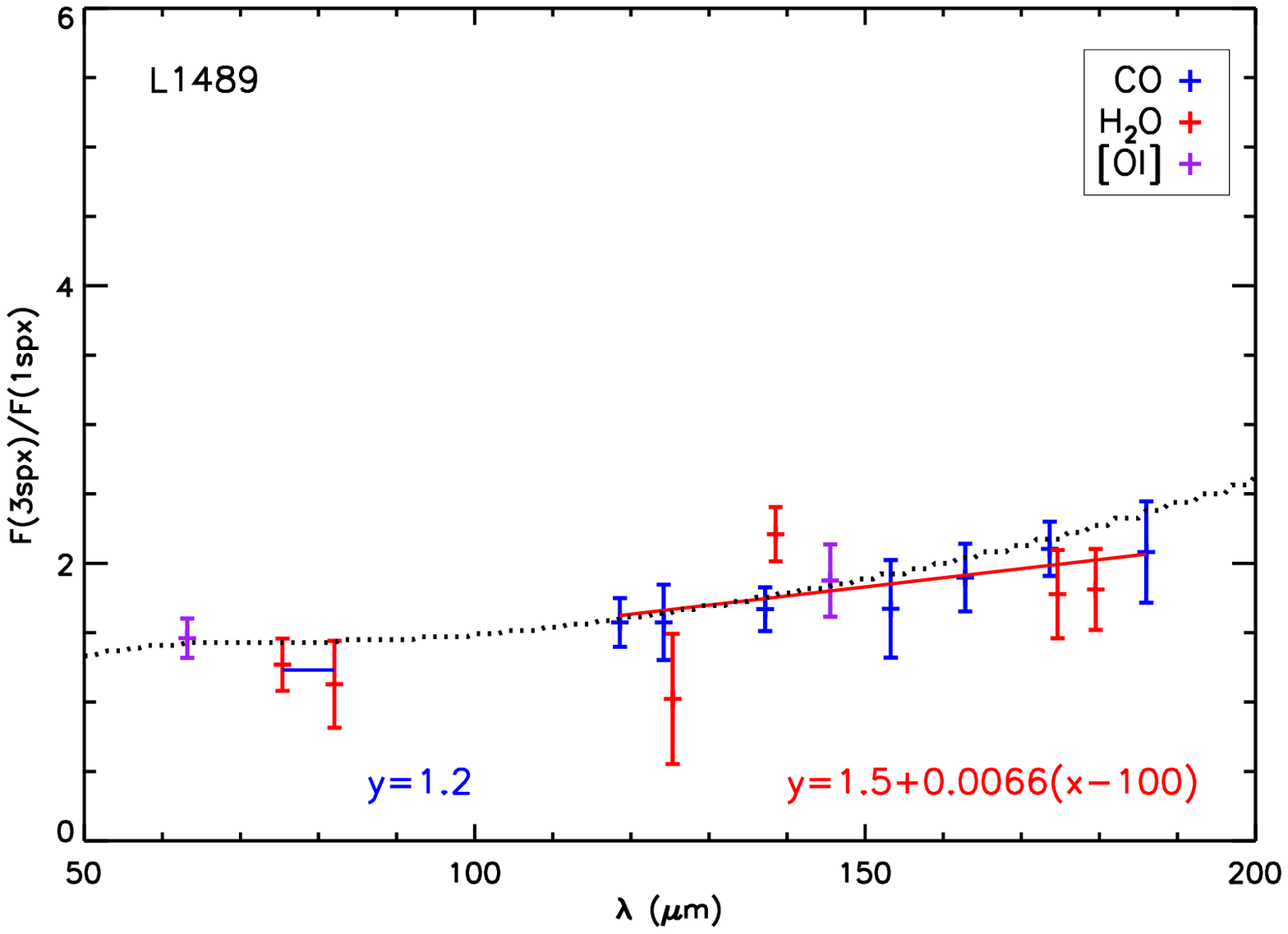}
\includegraphics[scale=0.45]{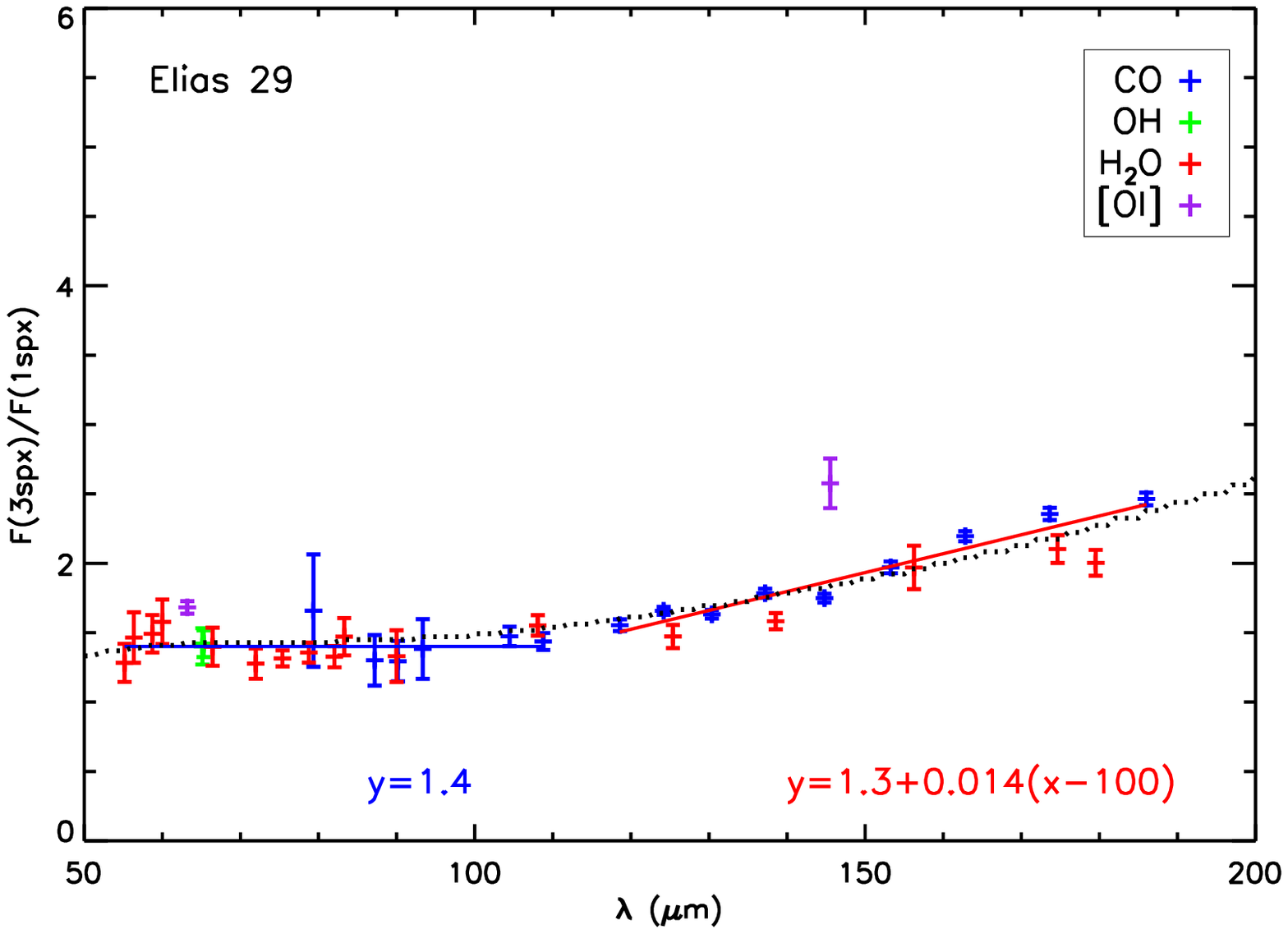}
\includegraphics[scale=0.45]{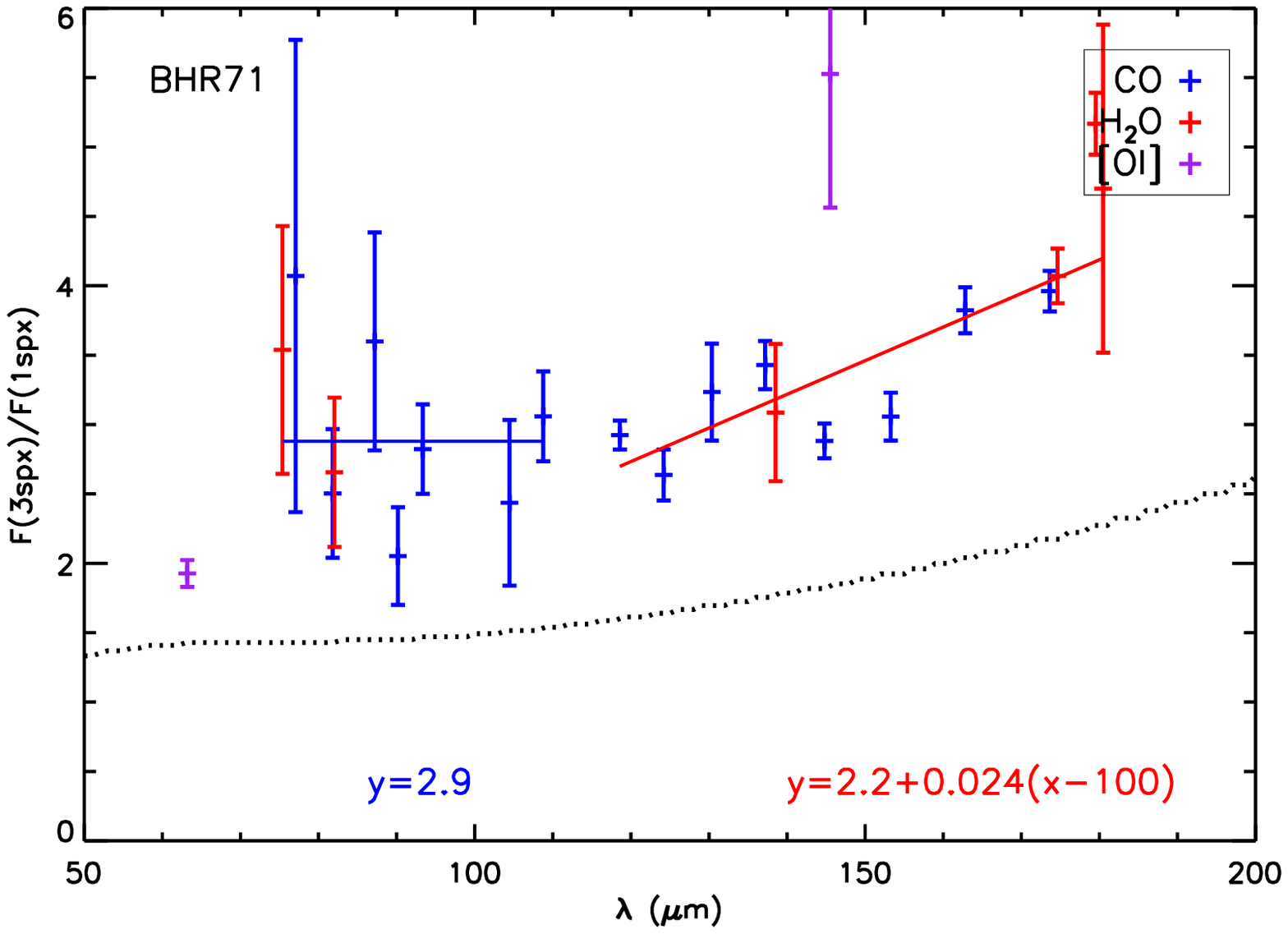}
\includegraphics[scale=0.45]{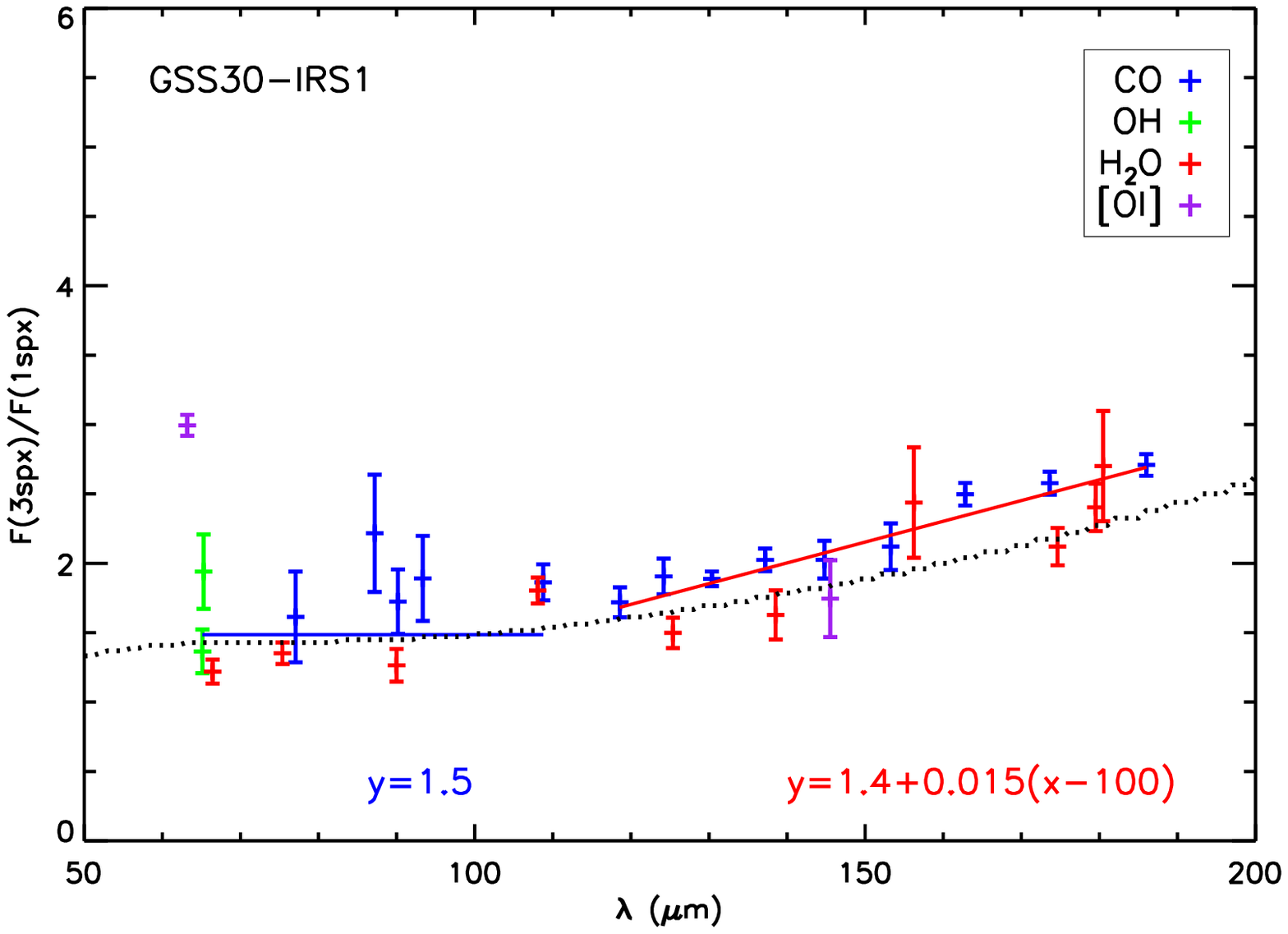}
\caption{Ratio of the flux of various lines in the 3$\times$3
aperture to the flux from the central spaxel only is plotted versus
wavelength, for L1489 (top left),  Elias 29 (top right), BHR71 (bottom left), 
and GSS30 (bottom right).  
A constant ratio is fitted to all lines between 50 and
120 $\mu$m (solid blue line), and a first-order polynomial to lines between 
120 and 200 $\mu$m (solid red line).  For comparison we plot the pipeline 
aperture correction function for the PSF (black dashed curve).
For this analysis we have selected CO (red) and H$_2$O (blue) 
lines only in selected regions of high S/N, 
ignored blended lines, and {\it excluded} OH (green), \OI\ (purple) and 
\CII\ lines (not shown).  
In L1489, applying the PSF correction would roughly
predict the ratio of 3$\times$3 to the central spaxel, for all detected
species.  In Elias 29, the \OI\ emission is extended
in comparison to the CO and H$_2$O lines.  In BHR71, all lines appear 
extended, evidence of multiple or diffuse source contributions.  In GSS30, 
the line emission marginally exceeds the PSF overall, with notable differences 
between species.}
\label{extendedlines}
\end{figure}

\begin{figure*}
\begin{center}

\includegraphics[scale=0.37]{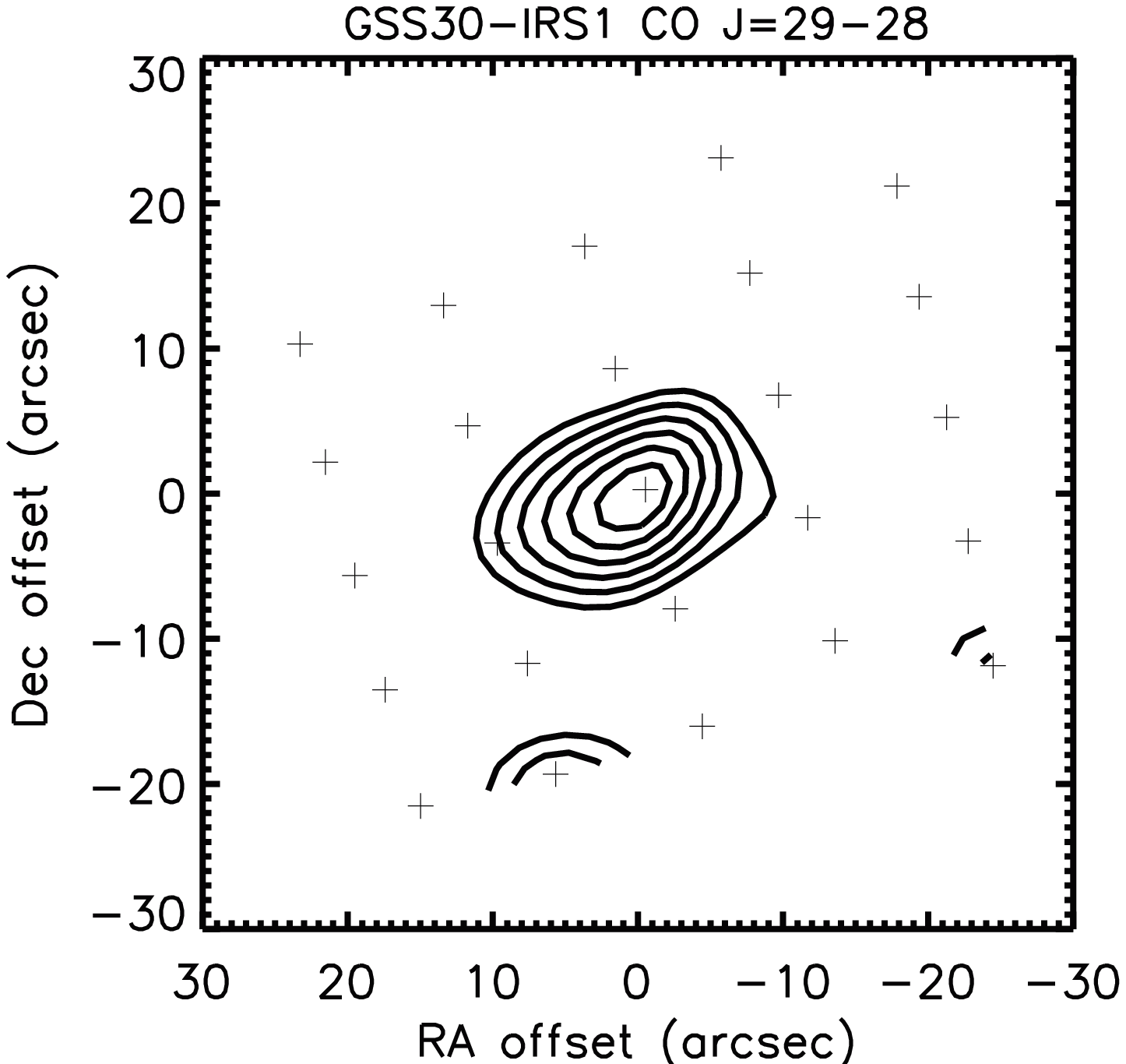}
\includegraphics[scale=0.37]{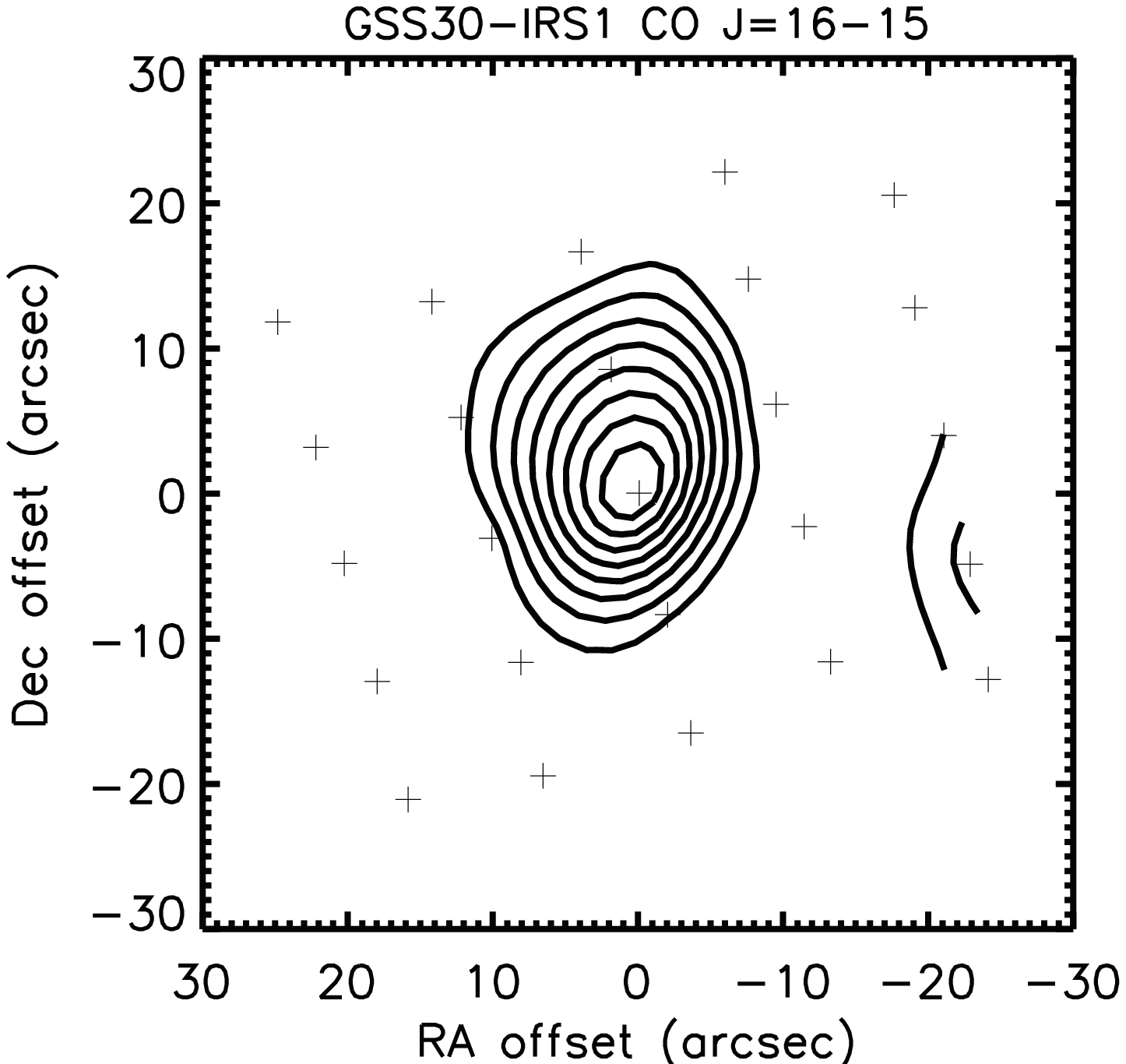}
\includegraphics[scale=0.37]{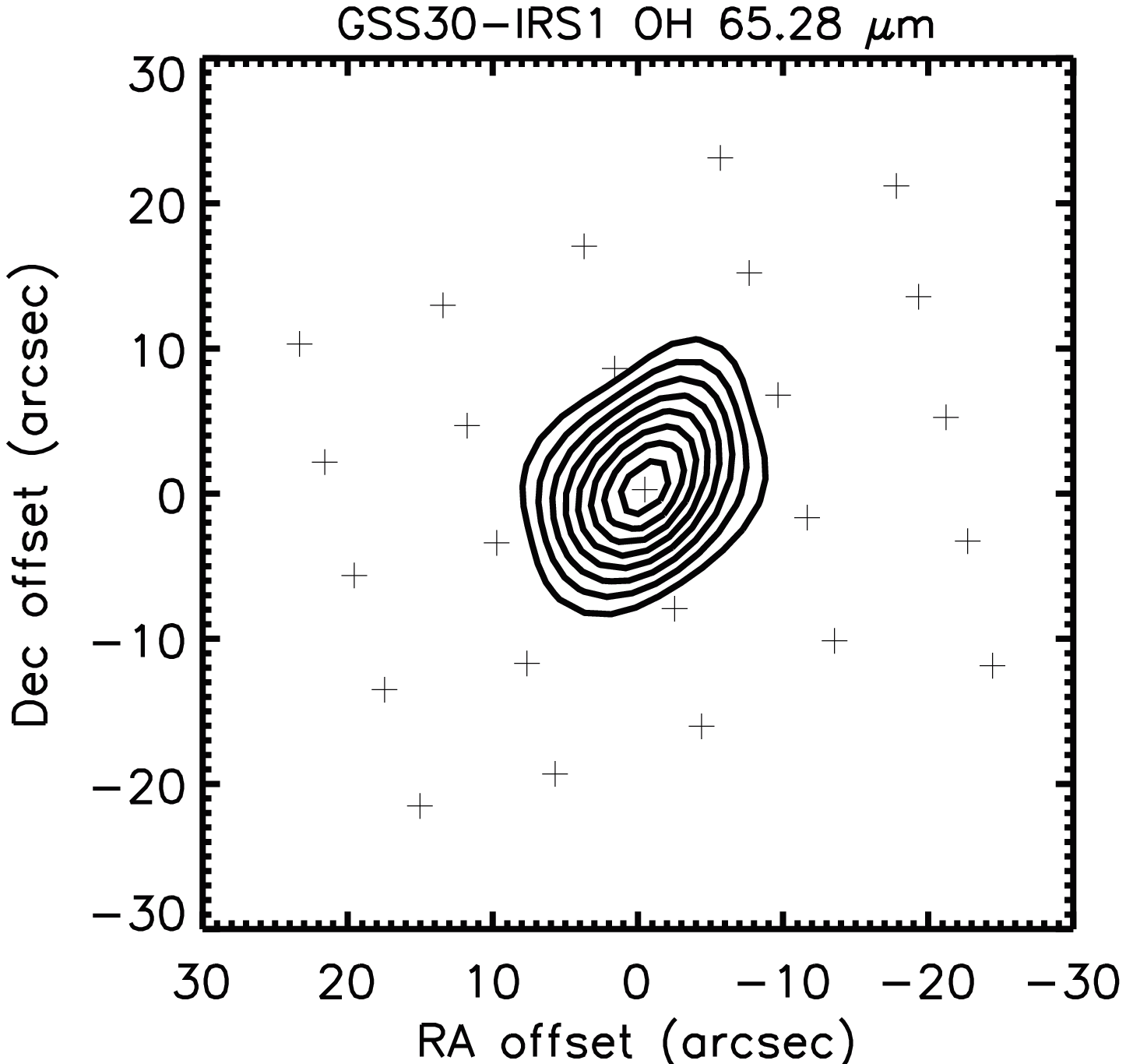}
\includegraphics[scale=0.37]{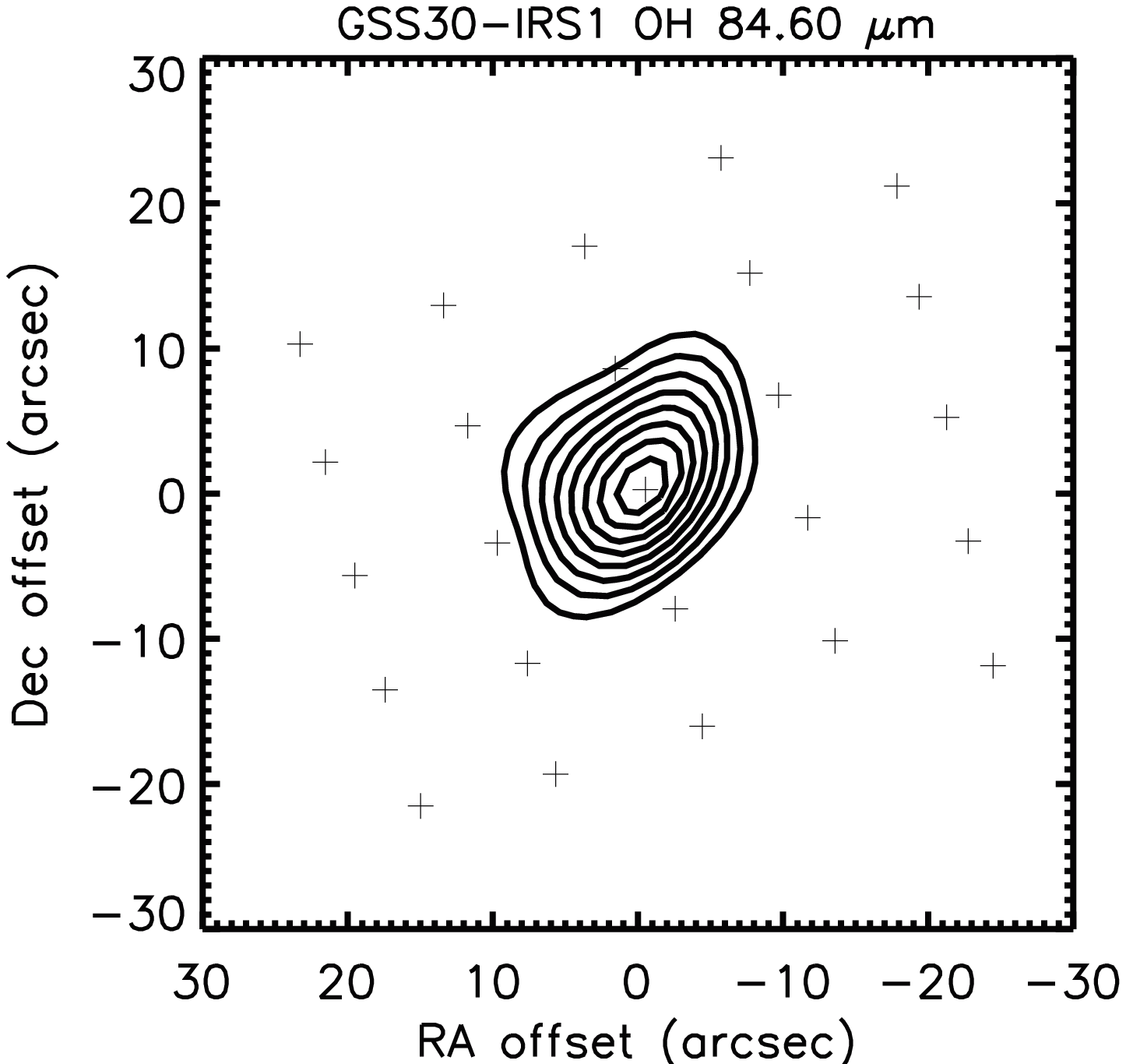}
\includegraphics[scale=0.37]{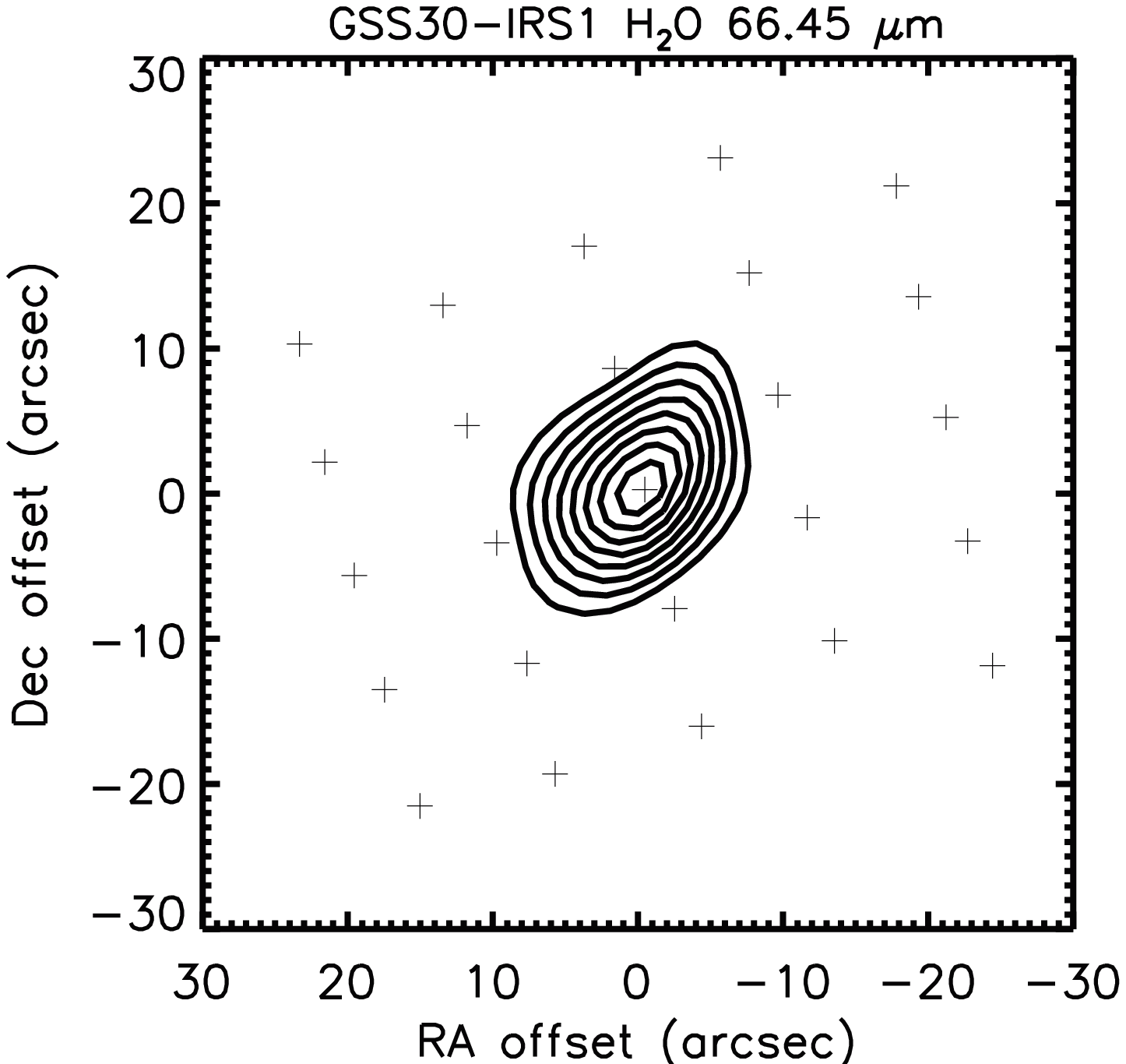}
\includegraphics[scale=0.37]{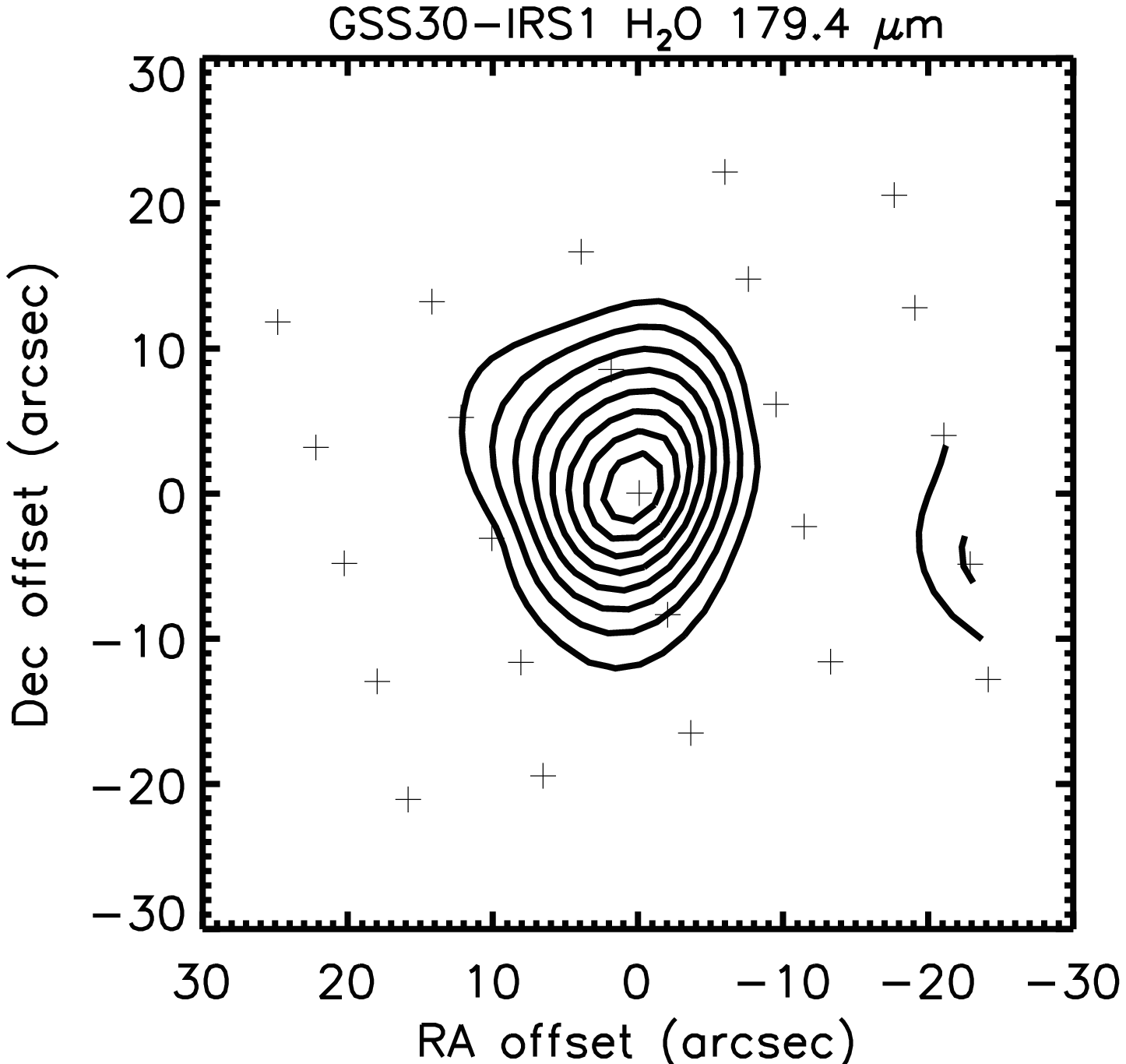}
\includegraphics[scale=0.37]{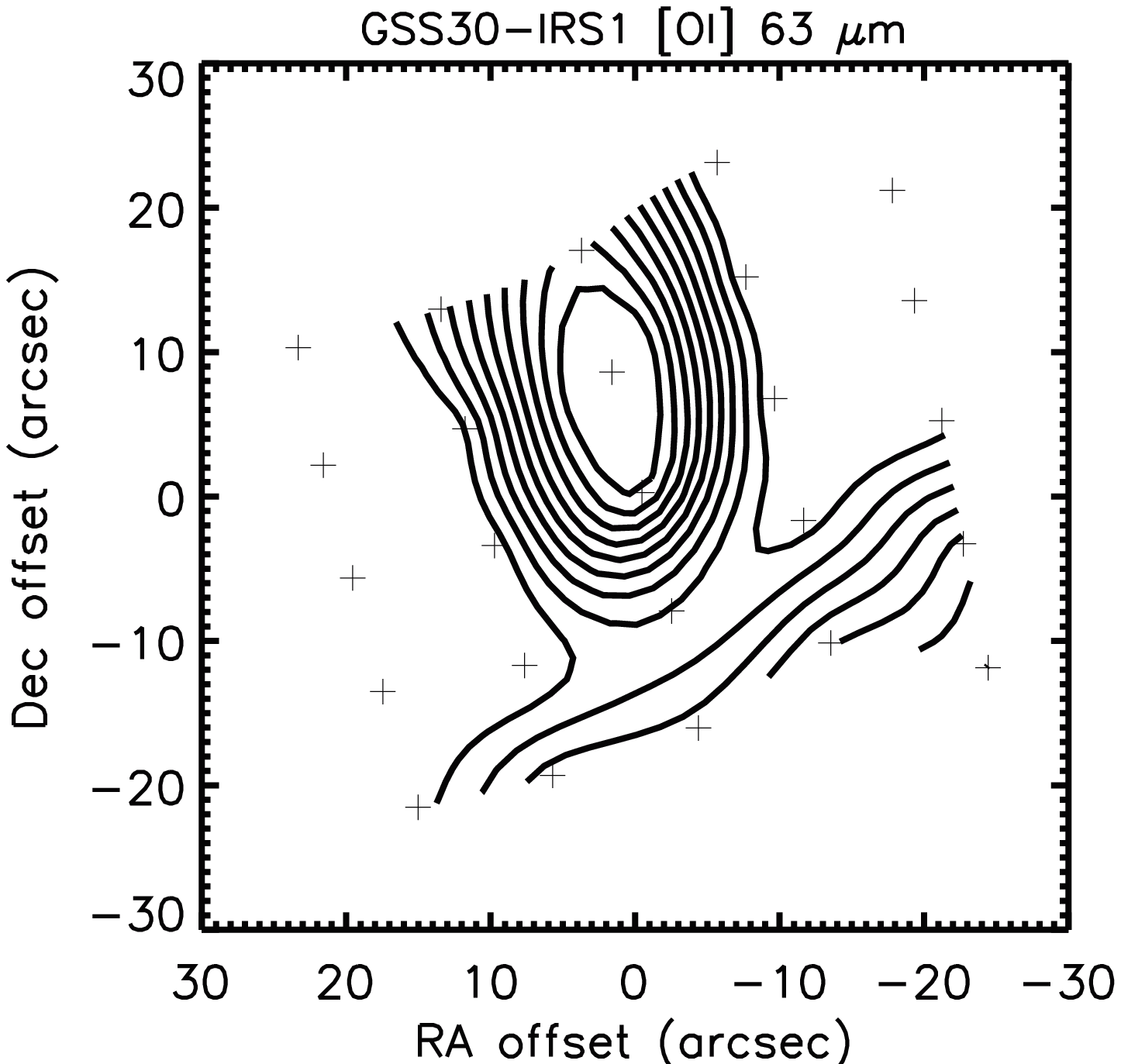}
\caption{Contour maps of archetypal lines in GSS30-IRS1.  The ``+'' indicate the positions
of the PACS spaxels; the lines are identified above. The contours are in increments of 
10\% of the central spaxel value, but cut off below $3 \sigma$.  
The left column shows contours for the high energy 
line tracers in the sample, while the right column 
shows contours for the low energy line tracers.}
\label{gss30lines}
\end{center}
\end{figure*}

\begin{figure*}
\begin{center}
\includegraphics[scale=0.37]{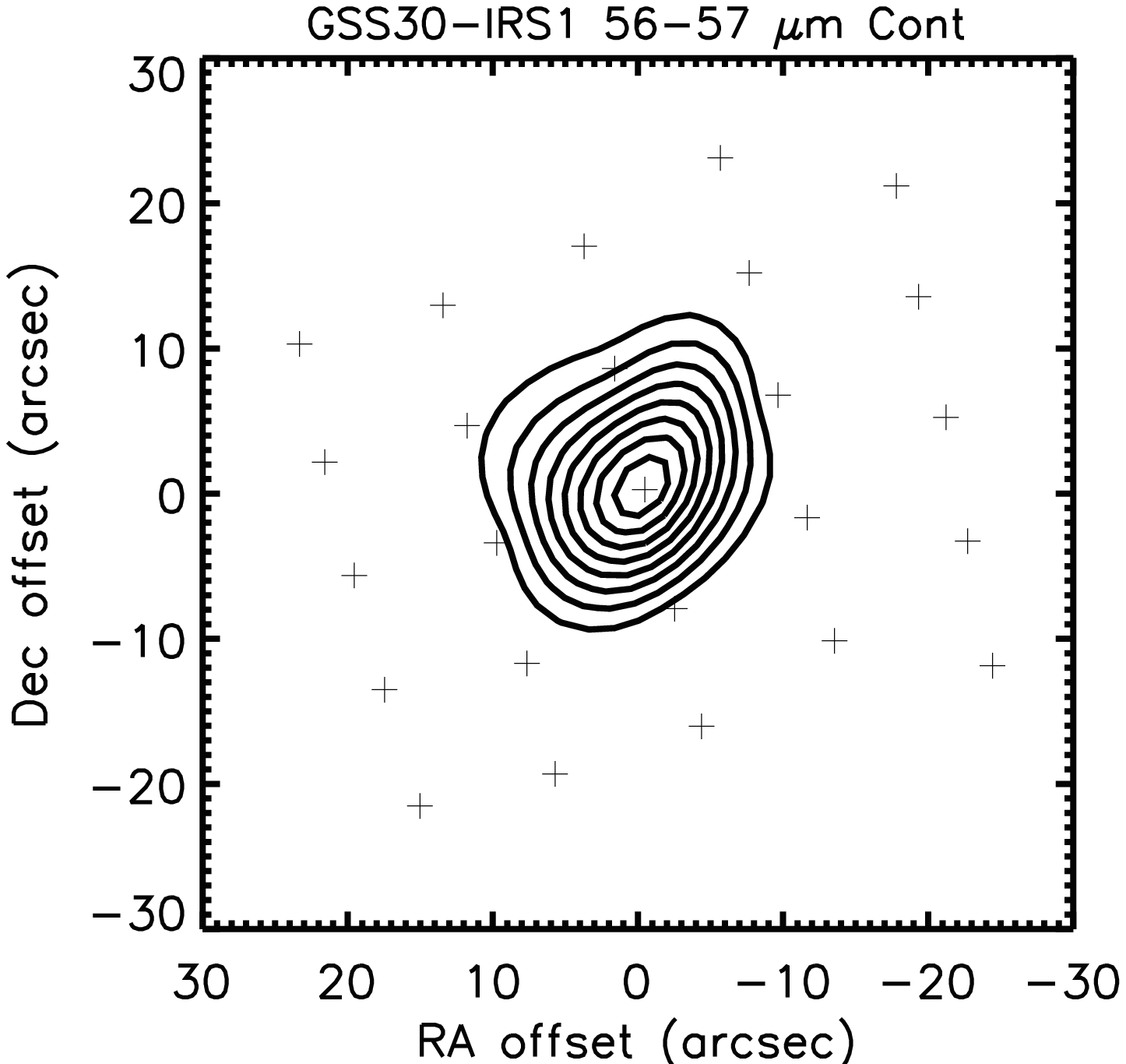}
\includegraphics[scale=0.37]{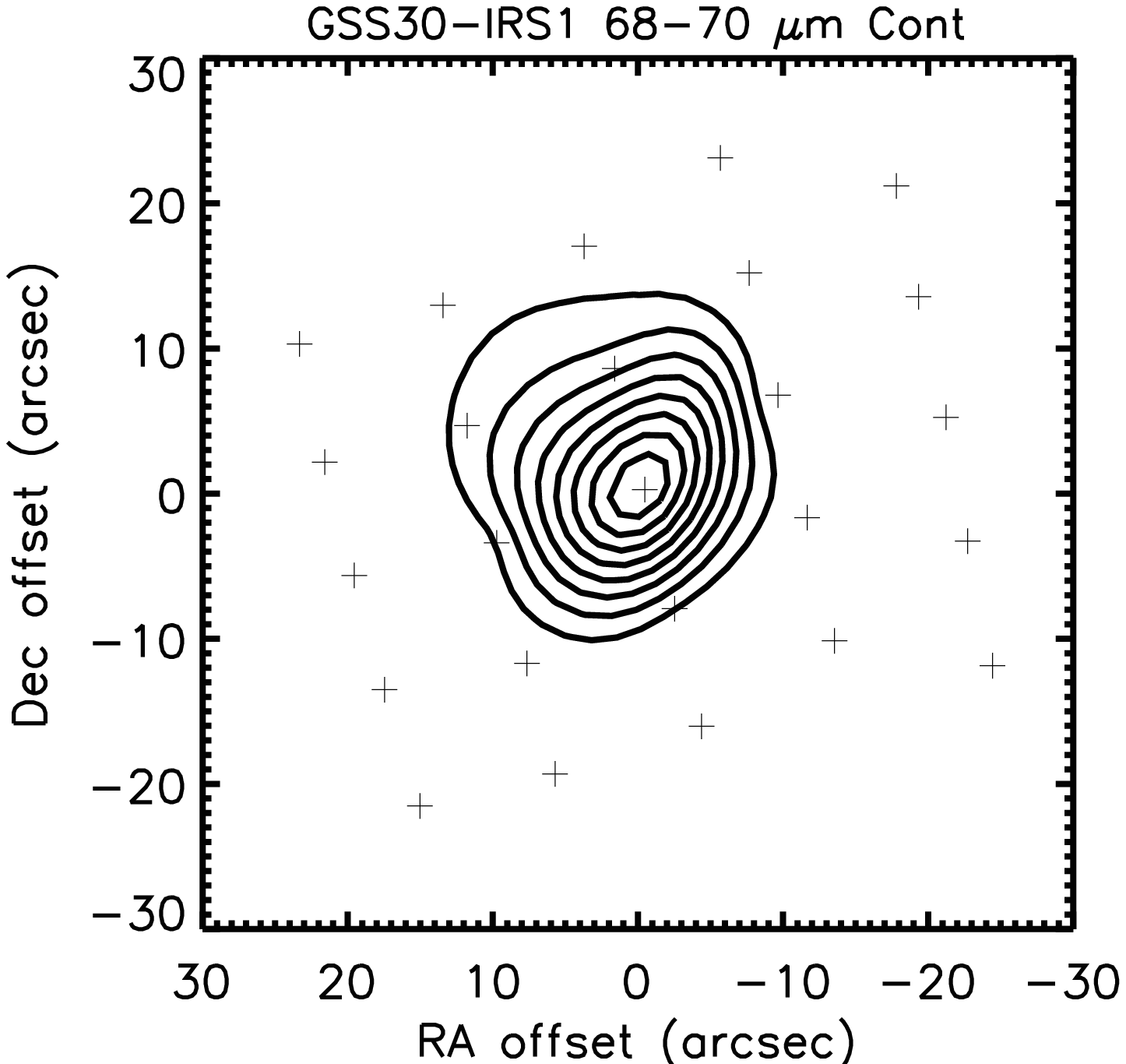}
\includegraphics[scale=0.37]{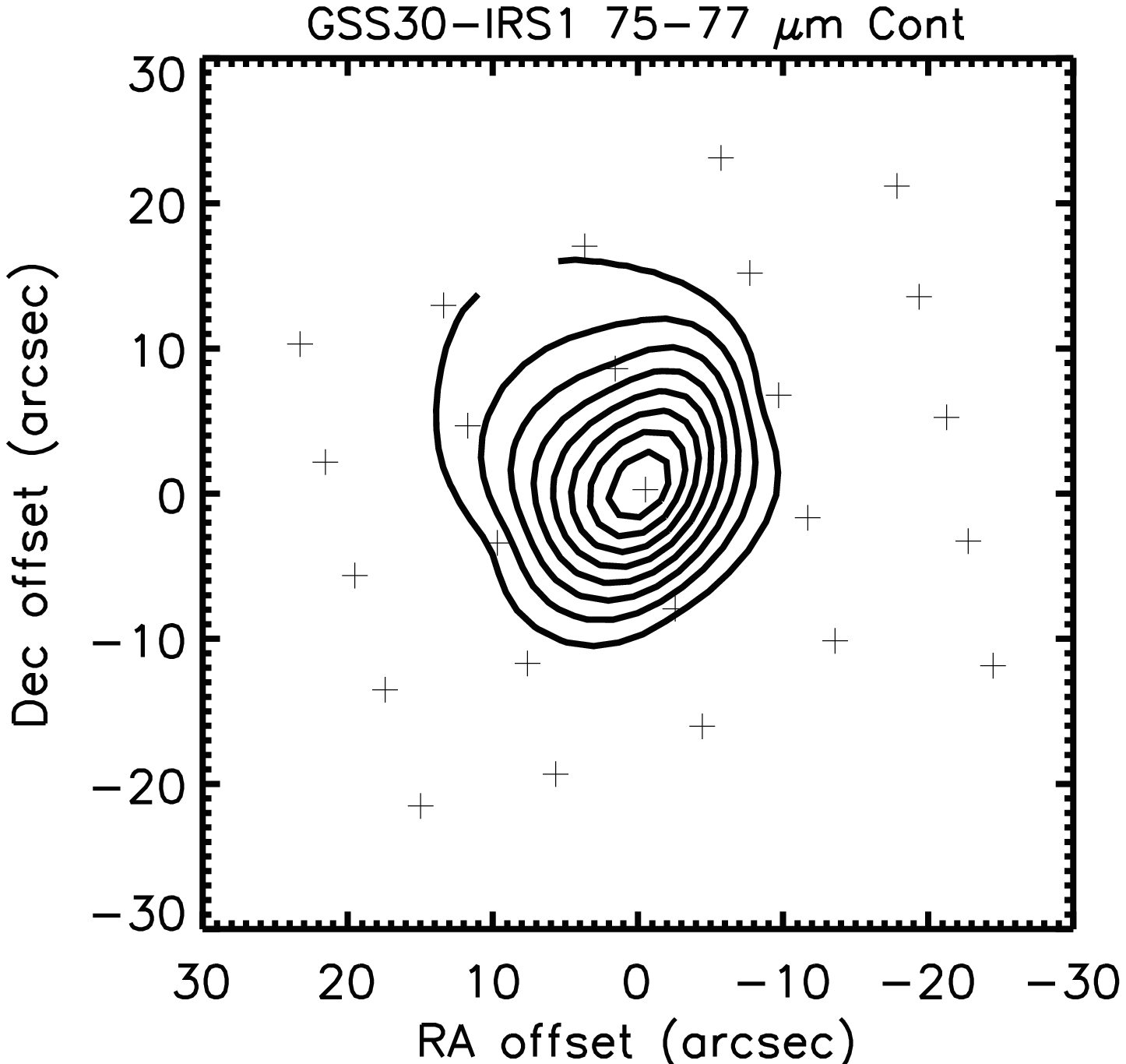}
\includegraphics[scale=0.37]{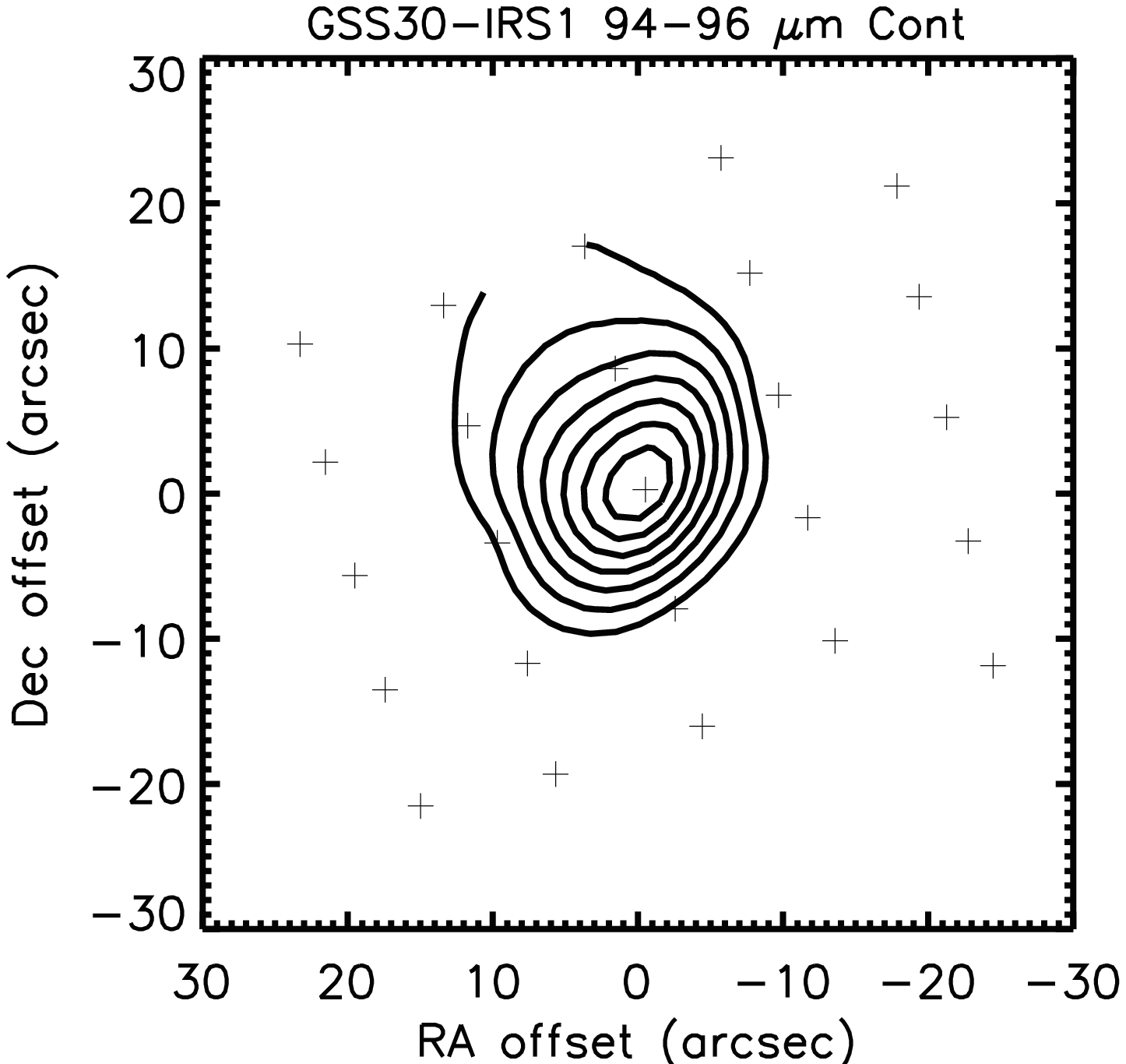}
\includegraphics[scale=0.37]{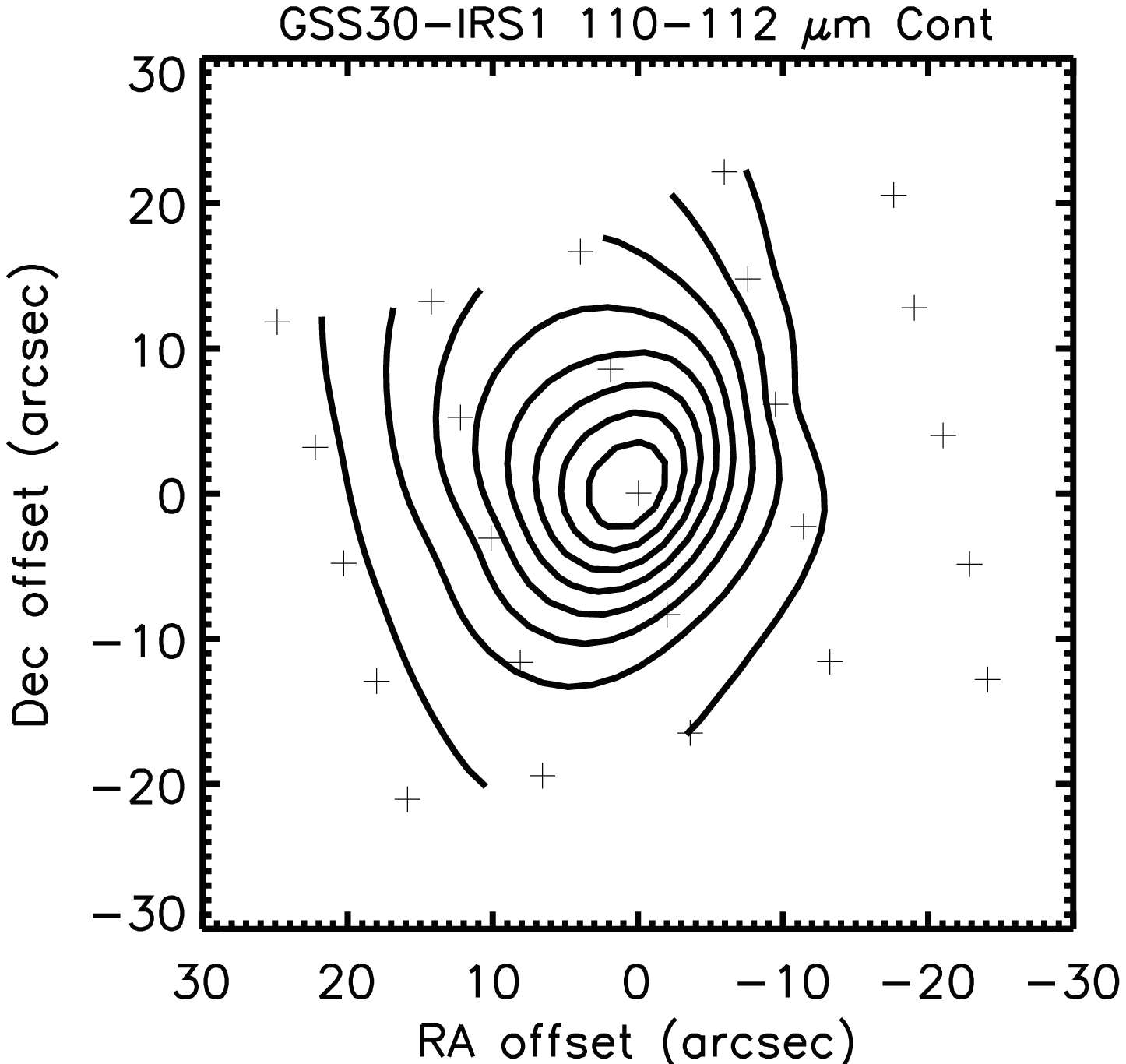}
\includegraphics[scale=0.37]{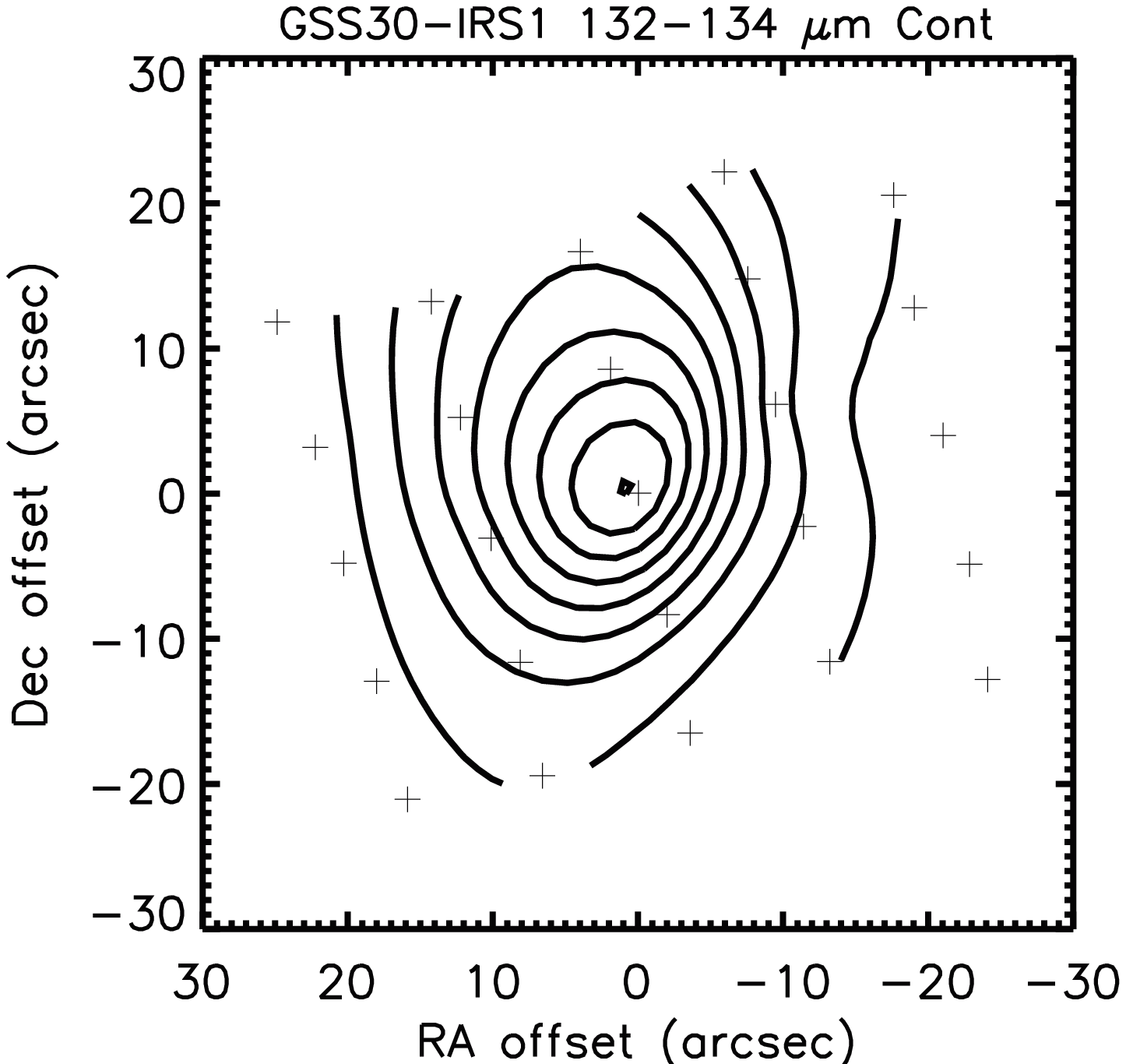}
\includegraphics[scale=0.37]{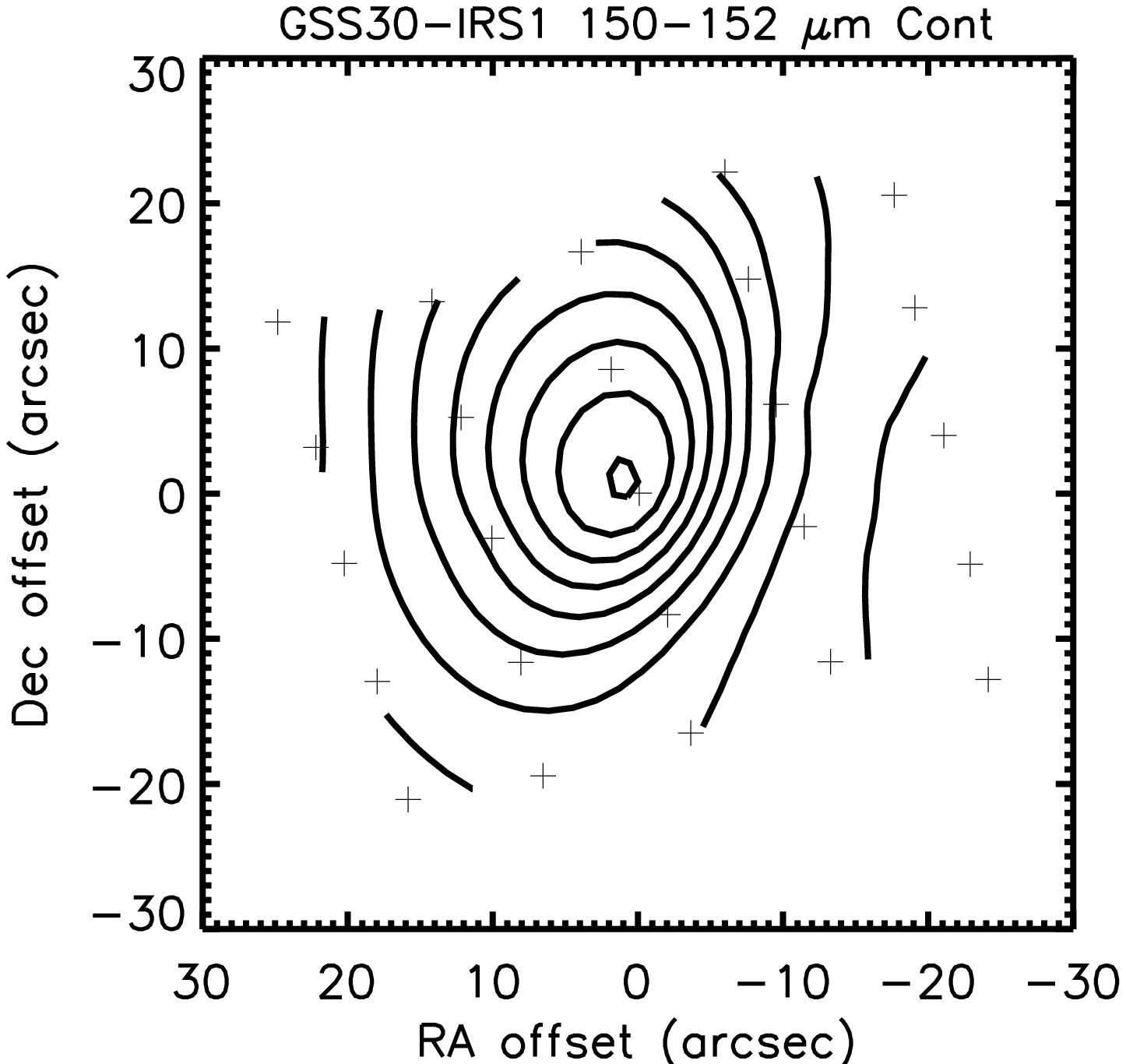}
\includegraphics[scale=0.37]{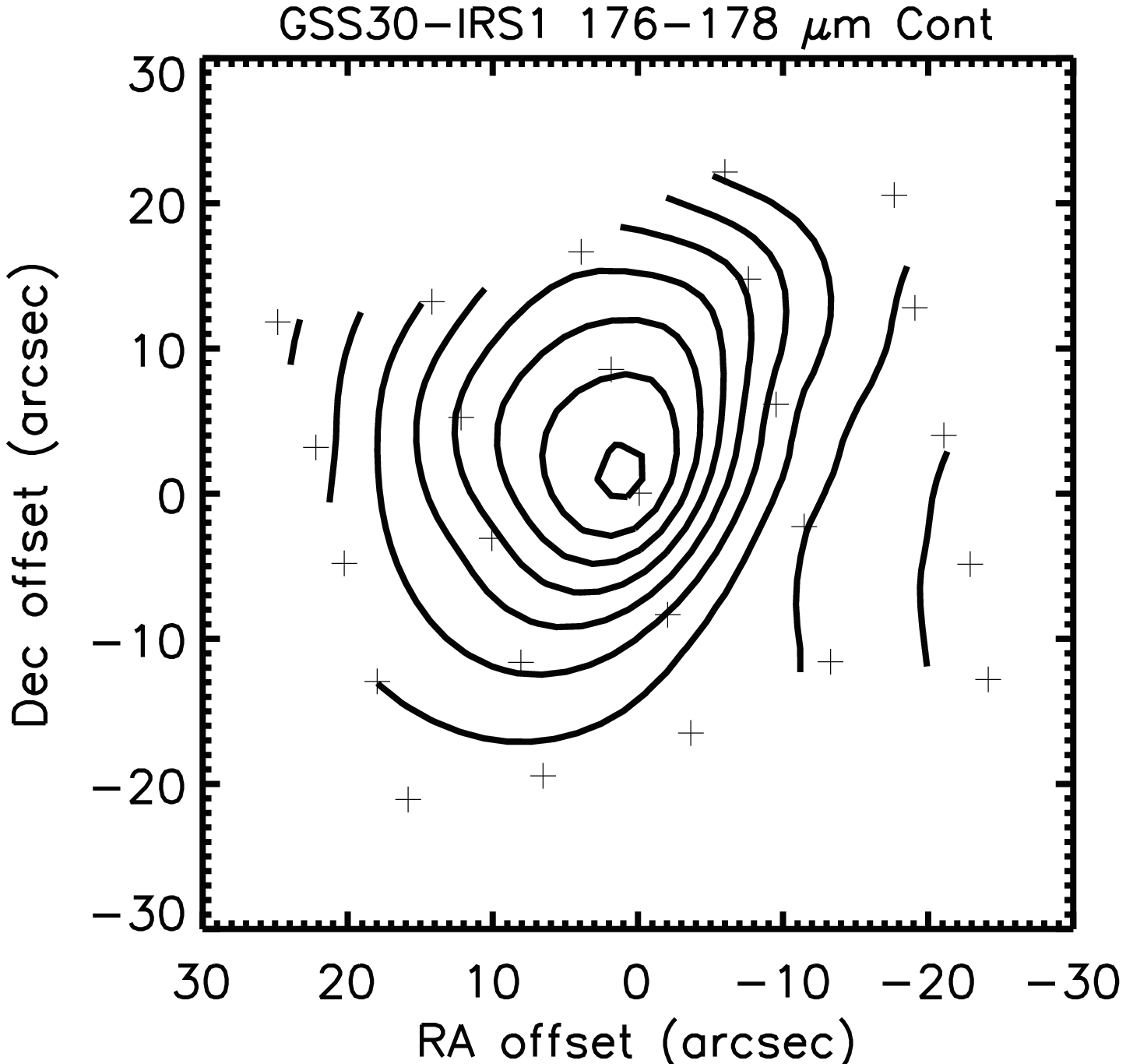}
\caption{Maps of continuum bands in GSS30-IRS1.  The ``+'' indicate the positions
of the PACS spaxels.  The contours are in increments of 
10\% of the central spaxel value, but cut off below the $3 \sigma$ level.  
The bands are, in order ($\mu$m): 
56--57, 68--70, 75--77, 94--96, 110--112, 132--134, 150--152, and 176--178.}
\label{gss30cont}
\end{center}
\end{figure*}

\begin{figure}
\begin{center}
\includegraphics[scale=0.43] {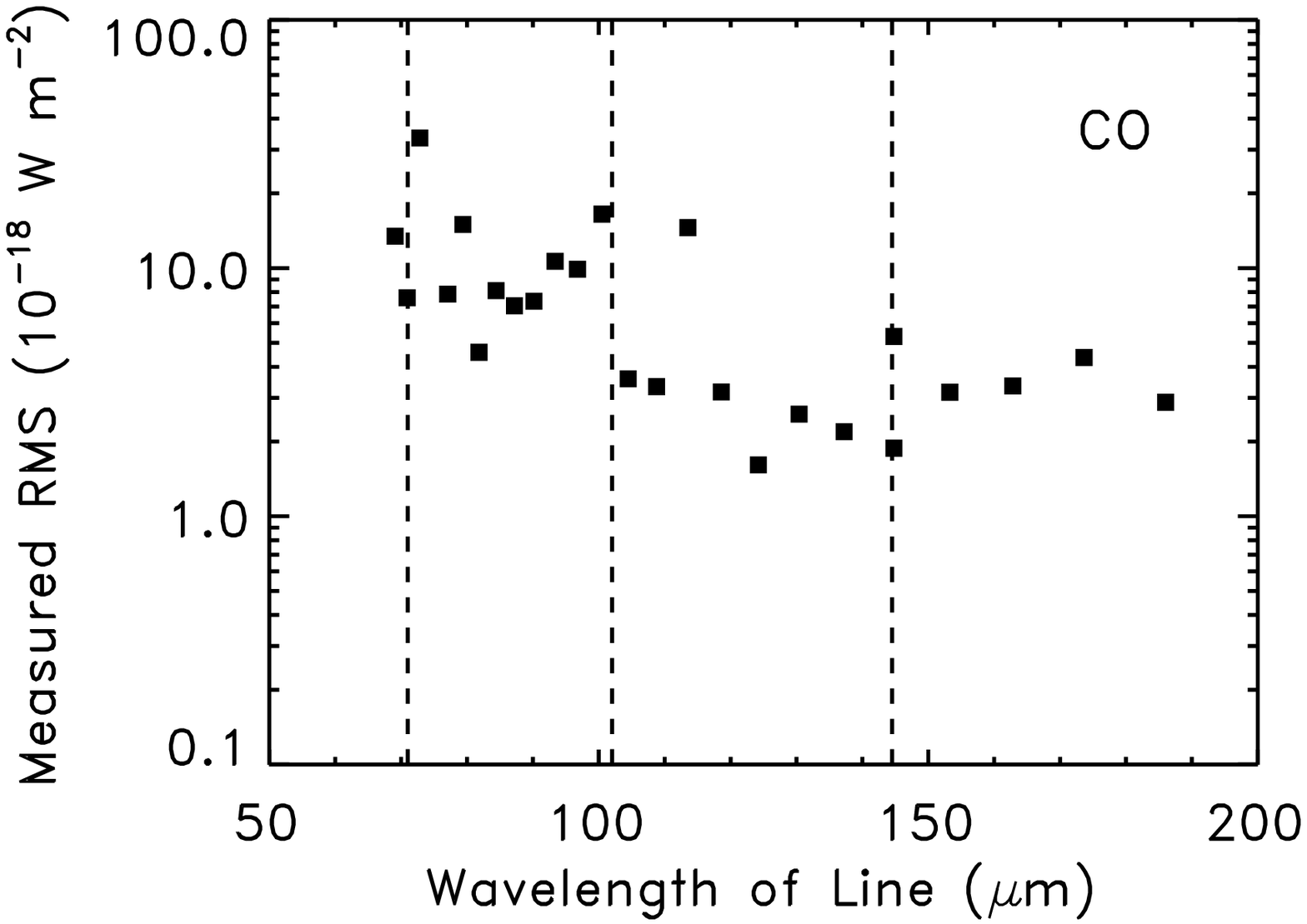}
\includegraphics[scale=0.43] {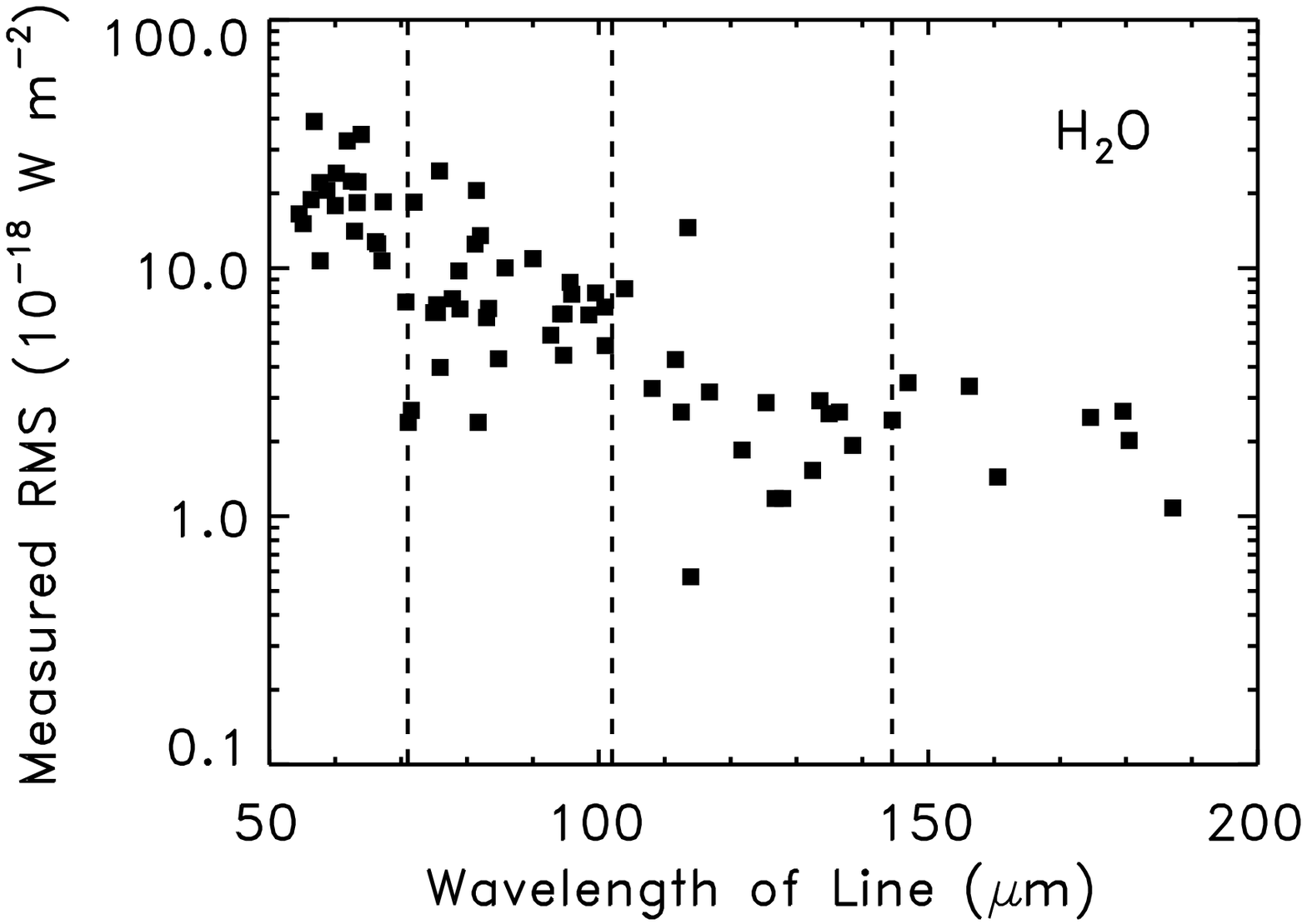}
\includegraphics[scale=0.43]{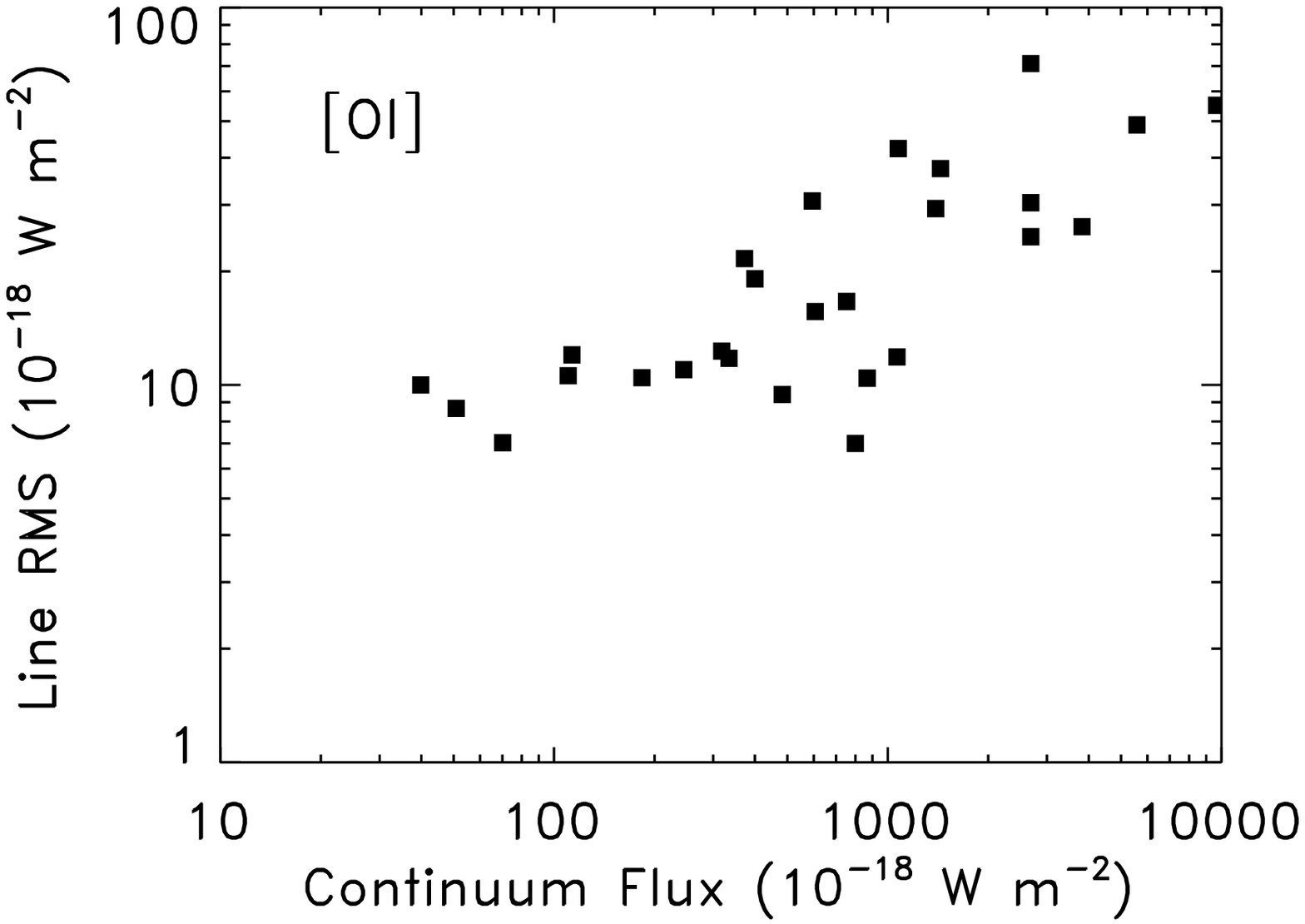}
\caption{RMS uncertainties of measured CO ({\bf top left}) and H$_2$O ({\bf top right}) 
lines for Elias 29.
The dashed lines indicate transitions between the B2A, B2B, and R1 (short and
long
wavelength) modules.  In cases for which a significantly different RMS
uncertainty is
measured for lines in the overlap regions, we
choose the higher S/N measurement.  The point at 113 $\mu$m is a blend of CO
and
H$_2$O, and appears consistently high in all sources, even those without
detectable
H$_2$O emission.  The uncertainties generally decrease to the 100 $\mu$m
breakpoint, and then flatten to a lower single value ($\sim$ 3
$\times$
10$^{-18}$ W m$^{-2}$) for $\lambda$ $>$ 100 $\mu$m.
{\bf Bottom:} RMS uncertainties measured from the \OI\
63 $\mu$m line for each source, vs. the flux of the surrounding continuum
for the same measured linewidth.  Although the uncertainty is flat at low
continuum
flux, the uncertainty rises steadily at fluxes $>$ 3 $\ee{-16}$ Wm$^{-2}$, with
a
corresponding rise in the detection limit. }
\label{unc1}
\end{center}
\end{figure}

\begin{figure}
\begin{center}
\includegraphics[scale=0.45]{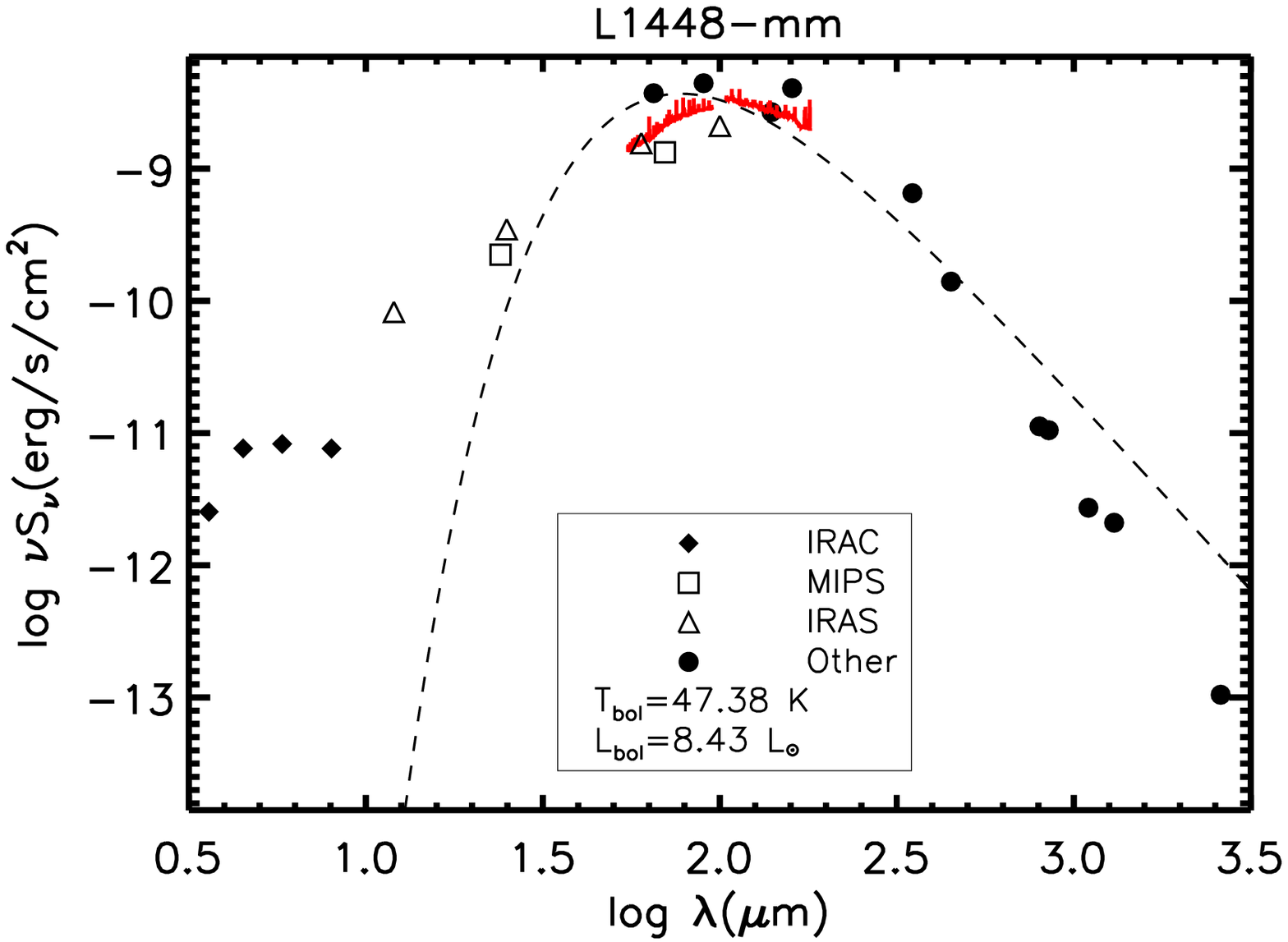}
\includegraphics[scale=0.45]{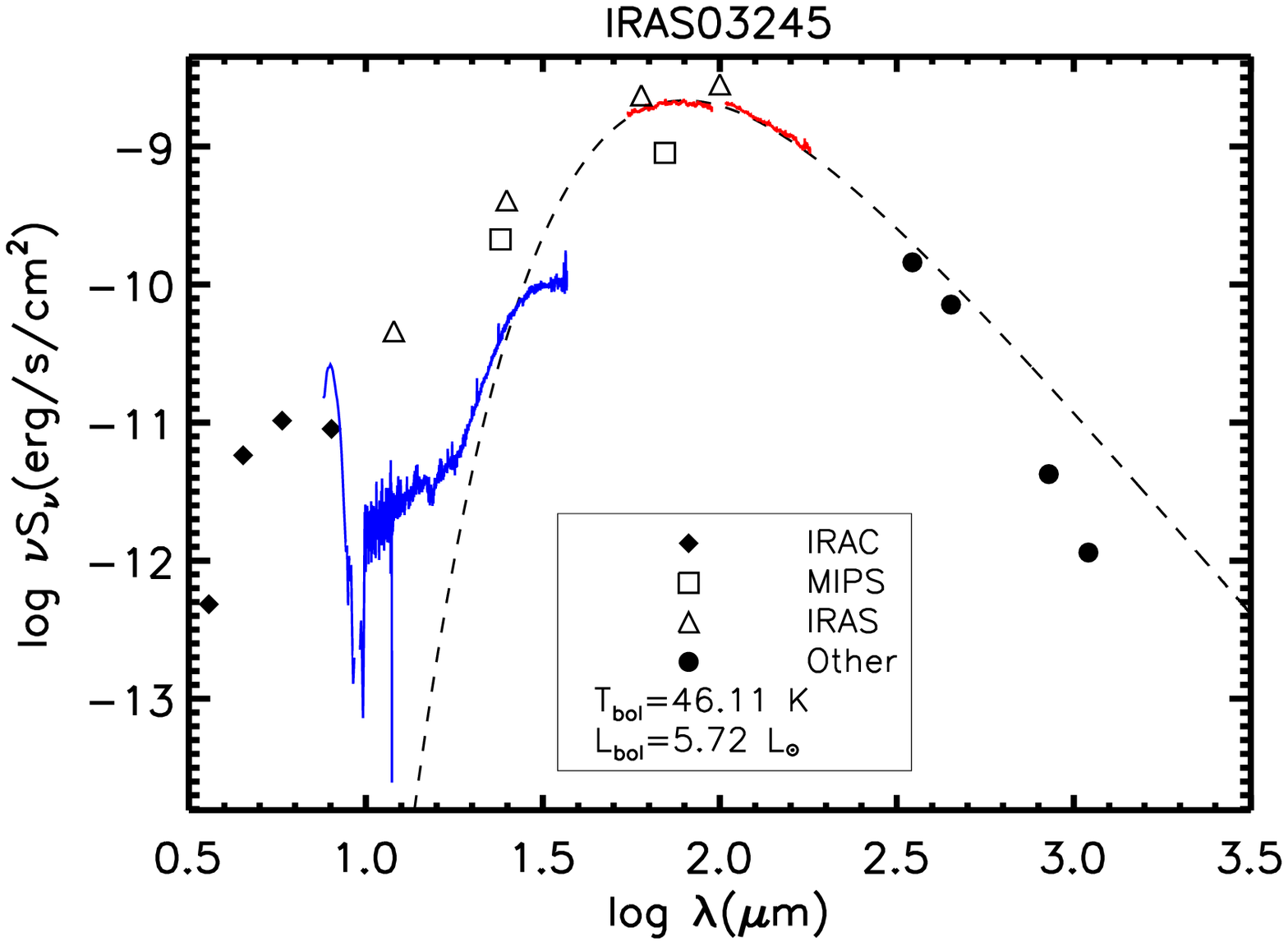}
\includegraphics[scale=0.45]{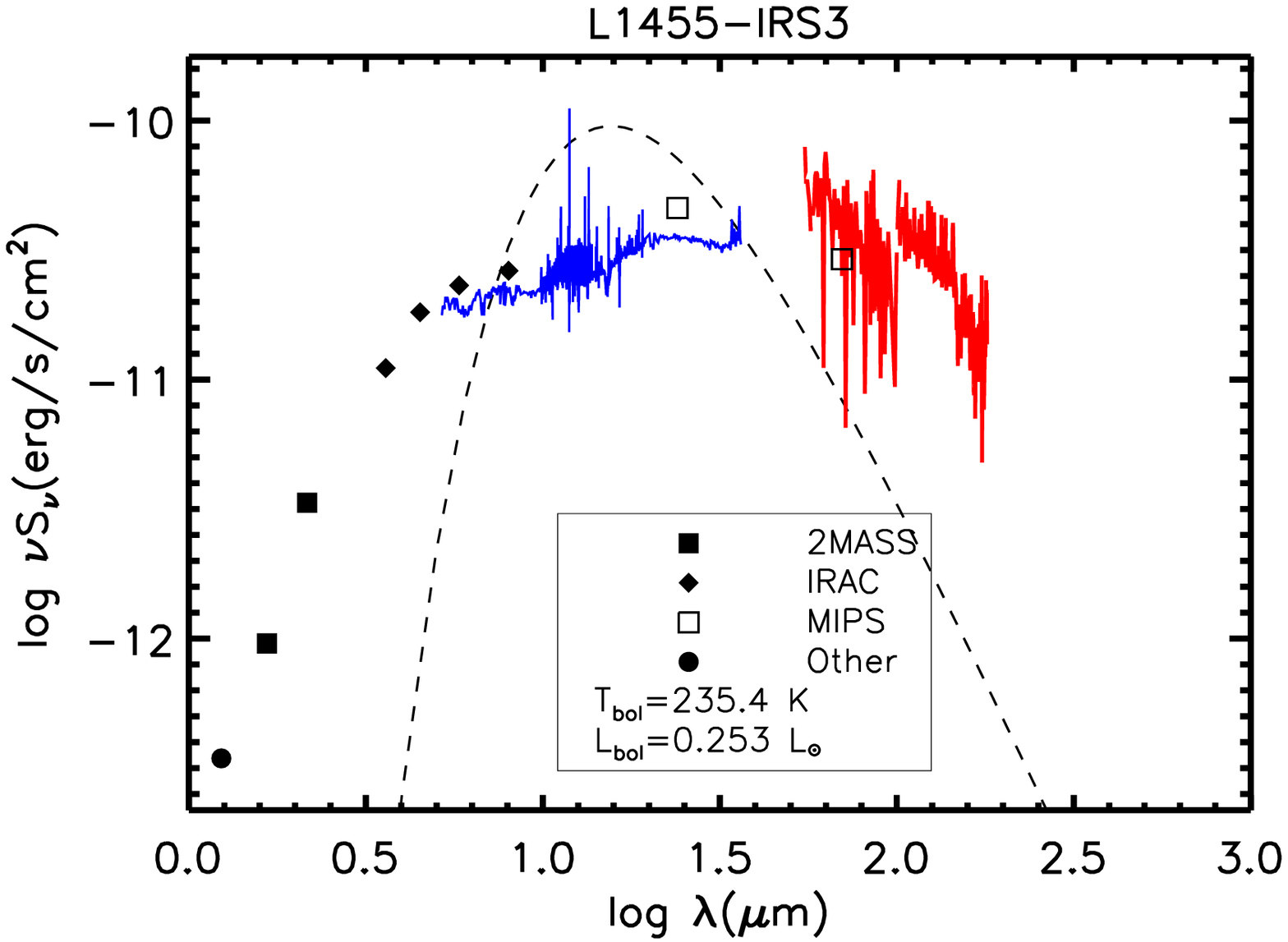}
\includegraphics[scale=0.45]{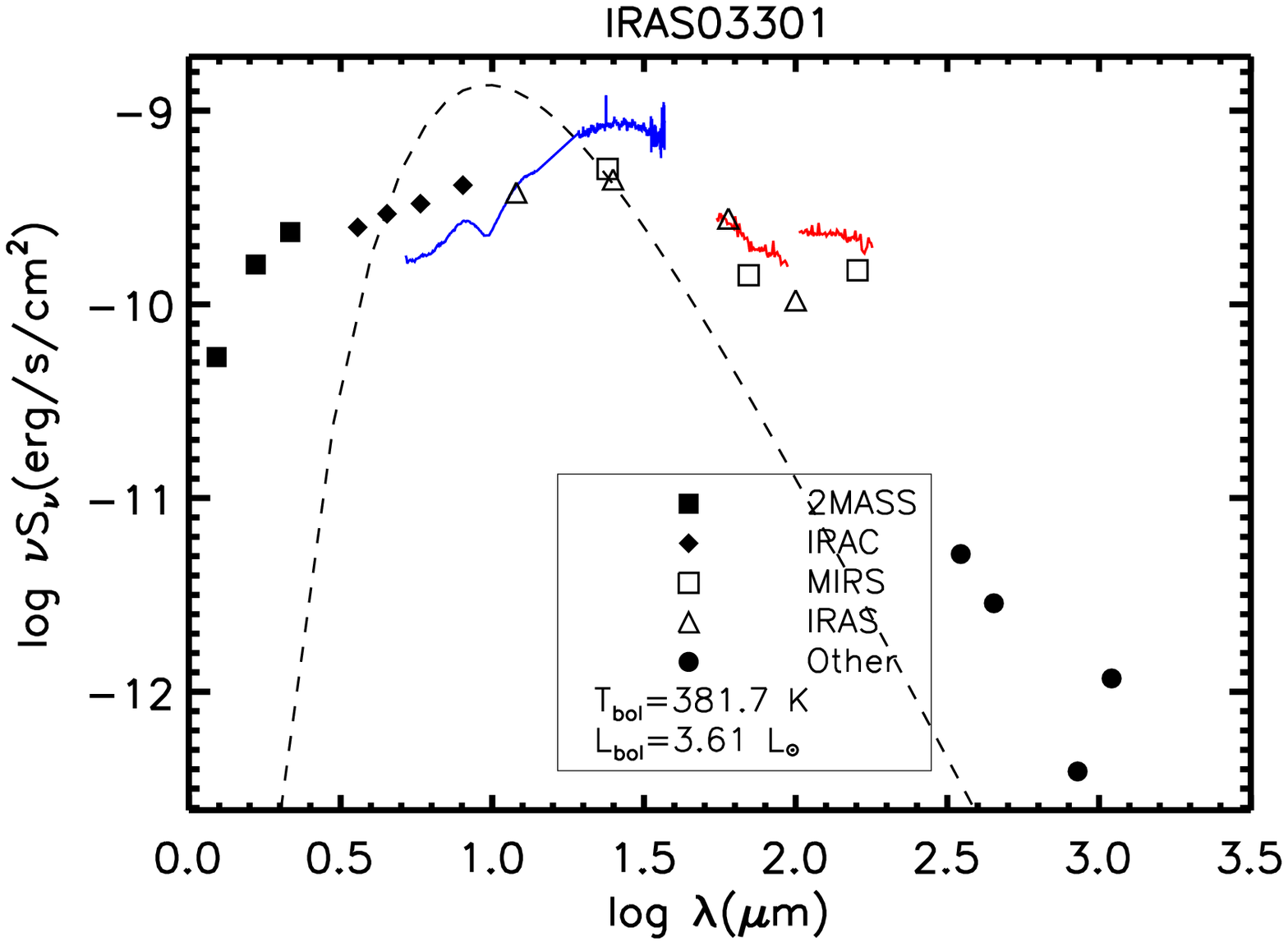}
\includegraphics[scale=0.45]{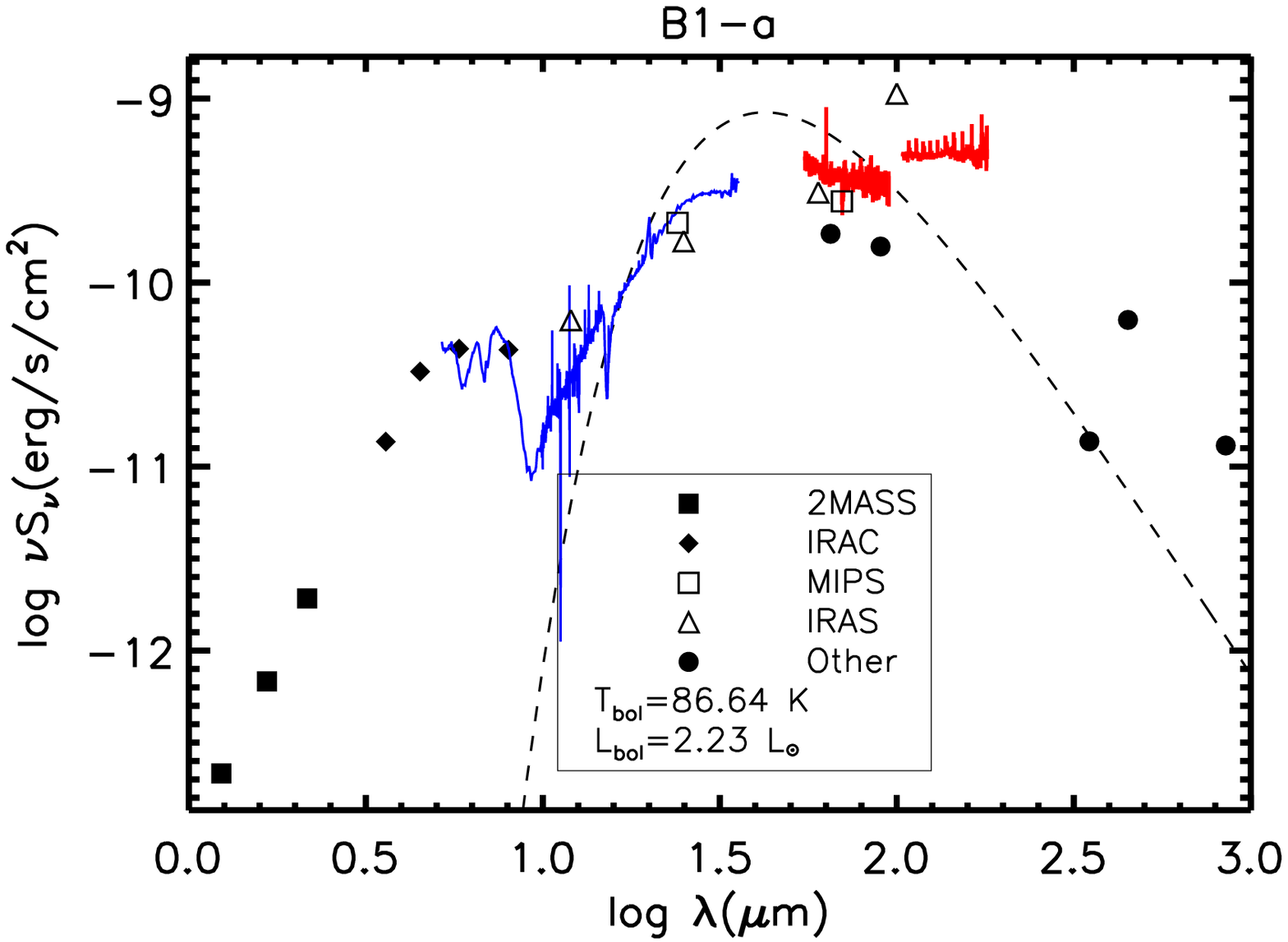}
\includegraphics[scale=0.45]{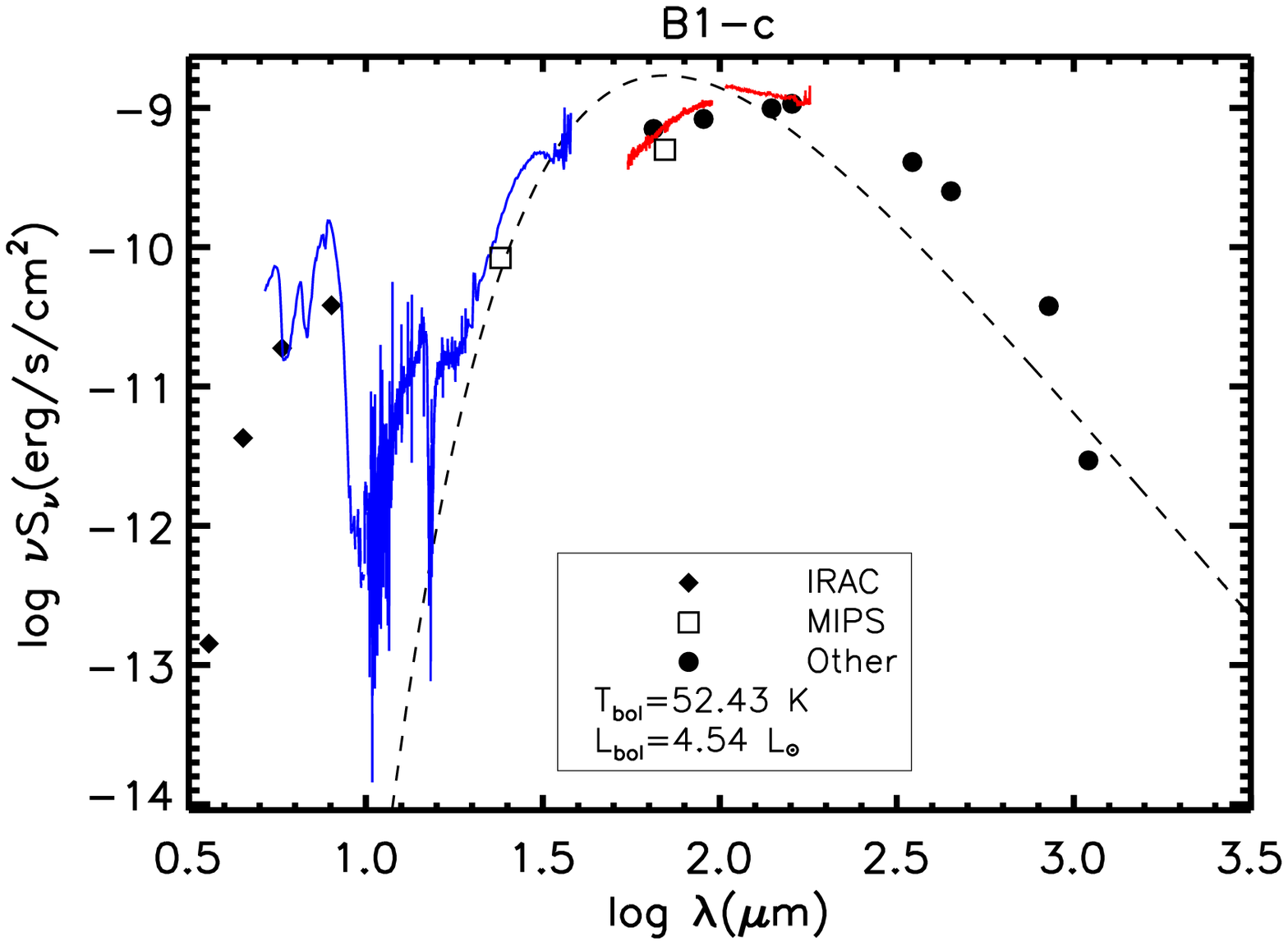}
\caption{0.5-1000 $\mu$m SEDs.
The IR photometric points are primarily from 2MASS (JHK; filled squares), {\it Spitzer}-IRAC 
(3.6, 4.5, 5.8, 8.0 $\mu$m; filled diamonds), IRAS (12, 25, and 100 $\mu$m; open triangles),
ISO (10, 13, 15, 20, 25 $\mu$m; open diamonds), and {\it Spitzer}-MIPS (24, 70 $\mu$m; open squares).  The MIPS 70 $\mu$m point is frequently 
saturated and falls below the PACS spectrum.  The submm are from 
SHARC-II \citep[350 $\mu$m][]{wu07}, BOLOCAM \citep[1.1, 1.2 mm][]{enoch06}, SCUBA 
\citep[450, 850 $\mu$m][]{difrancesco08,visser02}, AKARI (9, 18 $\mu$m; IRSA catalogue), IRAM 
\citep[1.3 mm][]{motte01}, and SEST (1.2, 1.3 mm; various).  The blue spectrum is 10-36 $\mu$m {\it Spitzer}-IRS, from the ``c2d'' legacy program \citep{lahuis06,enoch09} and the ``IRS\_Disks'' GTO program \citep{furlan06}.
For cases in which the SH (10-20 $\mu$m) and LH (20-36 $\mu$m) spectra do not match in flux, 
we have scaled the SH to the larger aperture LH.  The red spectrum is the PACS data (this work).  The dashed line is a blackbody spectrum for the measured values of \lbol\ and \tbol\ (this work).}
\label{seds1}
\end{center}
\end{figure}

\begin{figure}
\begin{center}
\includegraphics[scale=0.45]{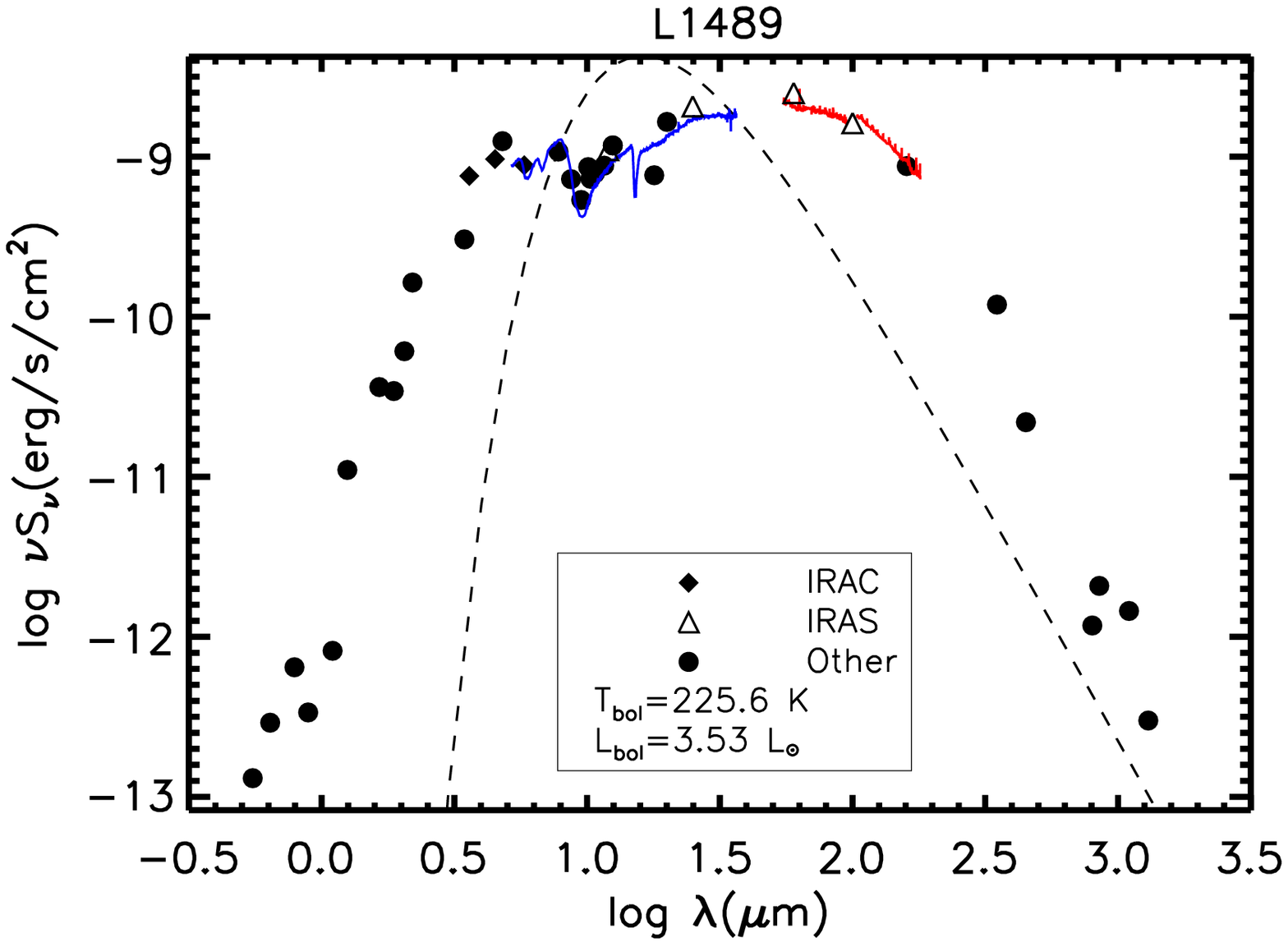}
\includegraphics[scale=0.45]{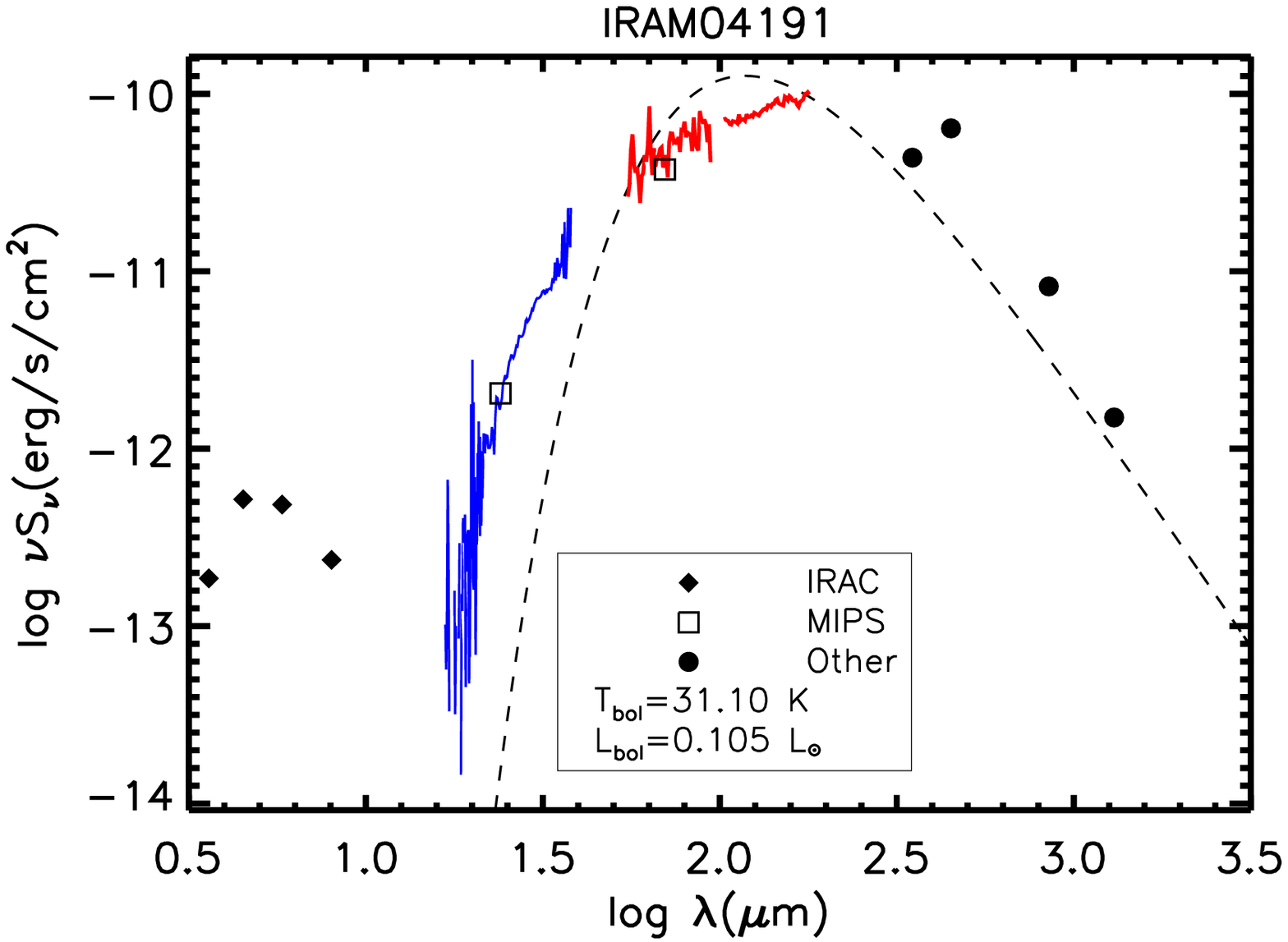}
\includegraphics[scale=0.45]{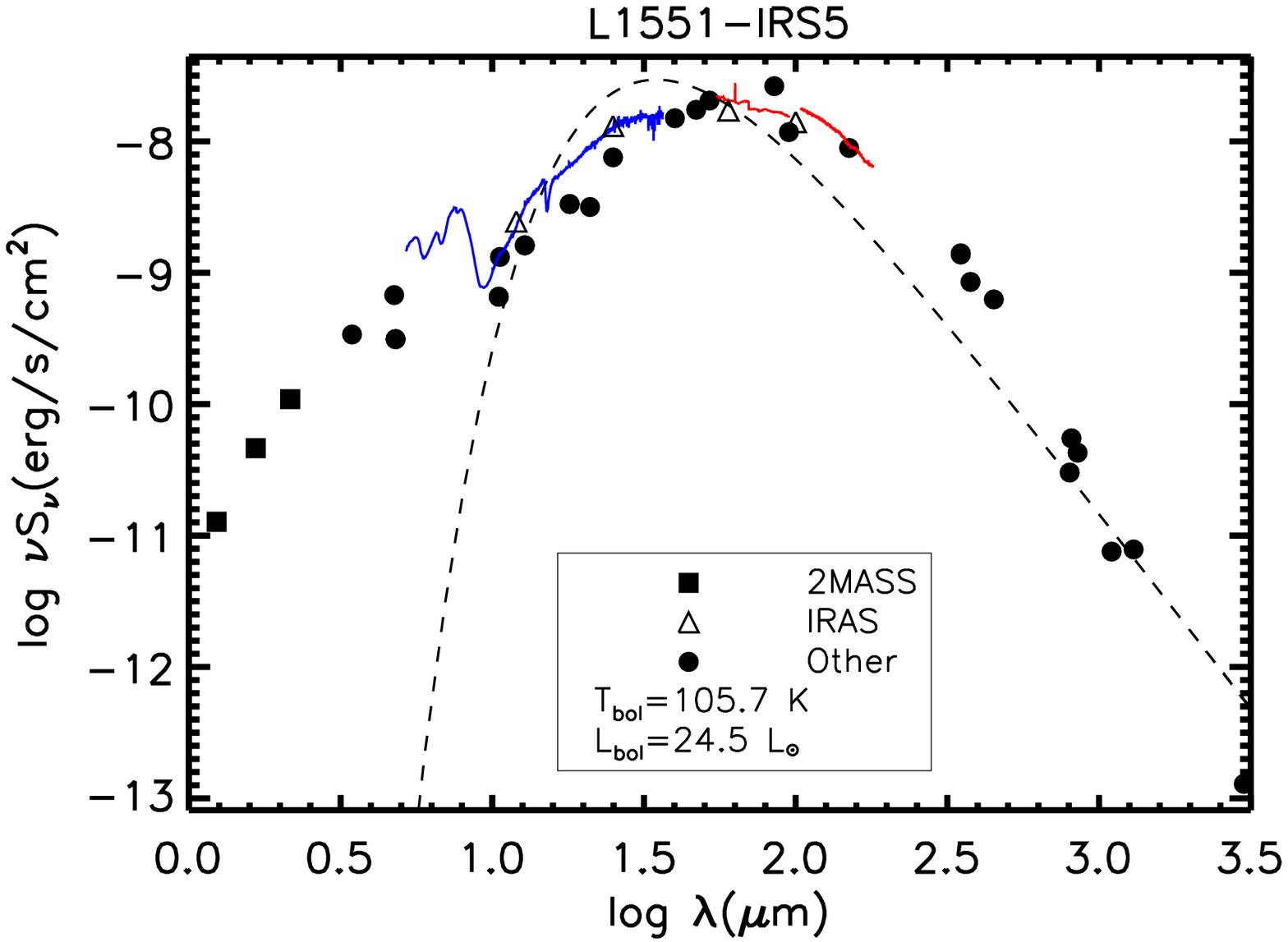}
\includegraphics[scale=0.45]{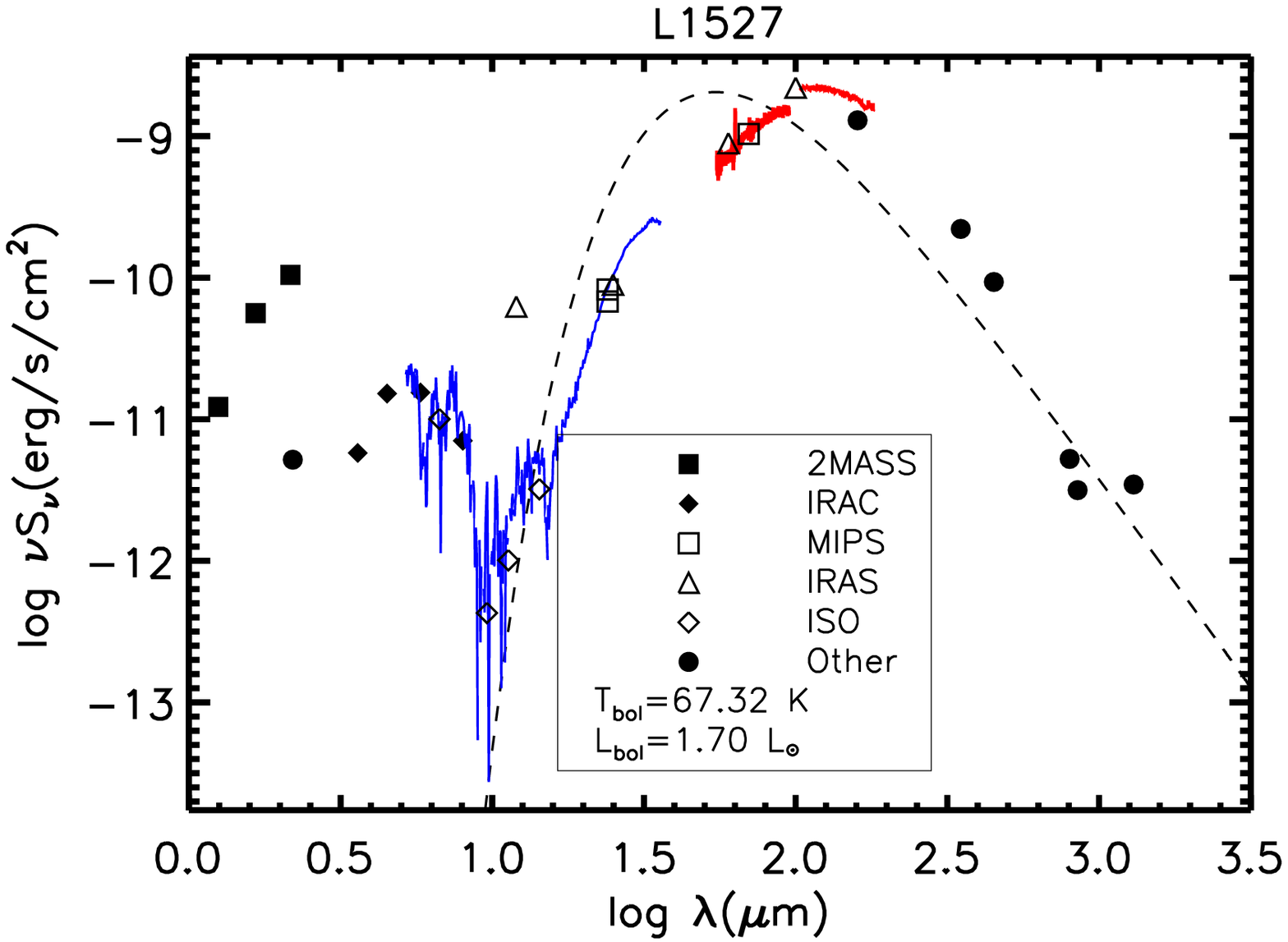}
\includegraphics[scale=0.45]{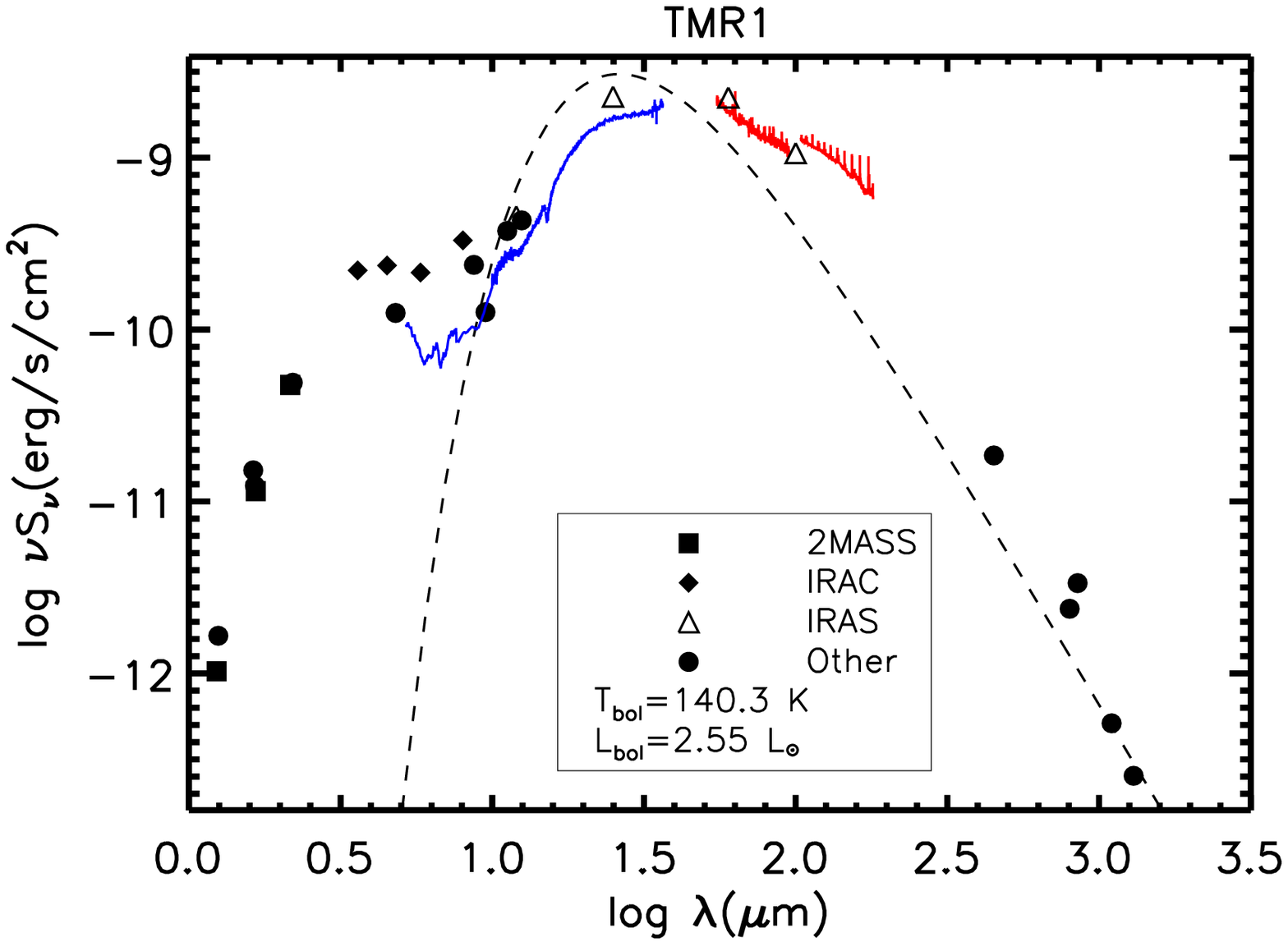}
\includegraphics[scale=0.45]{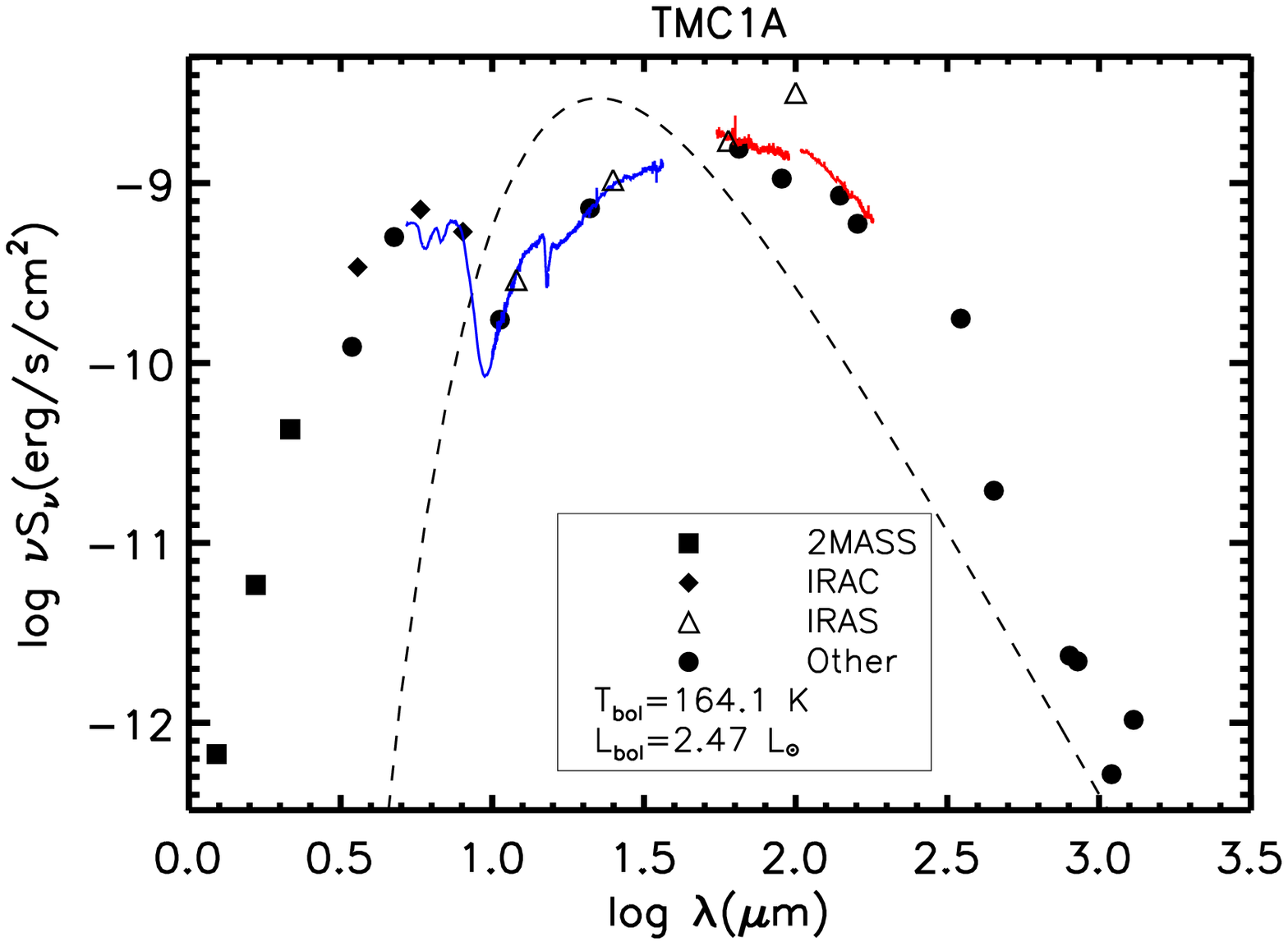}
\caption{0.5-1000 $\mu$m SEDs, as in Figure \ref{seds1}.}
\label{seds2}
\end{center}
\end{figure}

\begin{figure}
\begin{center}
\includegraphics[scale=0.45]{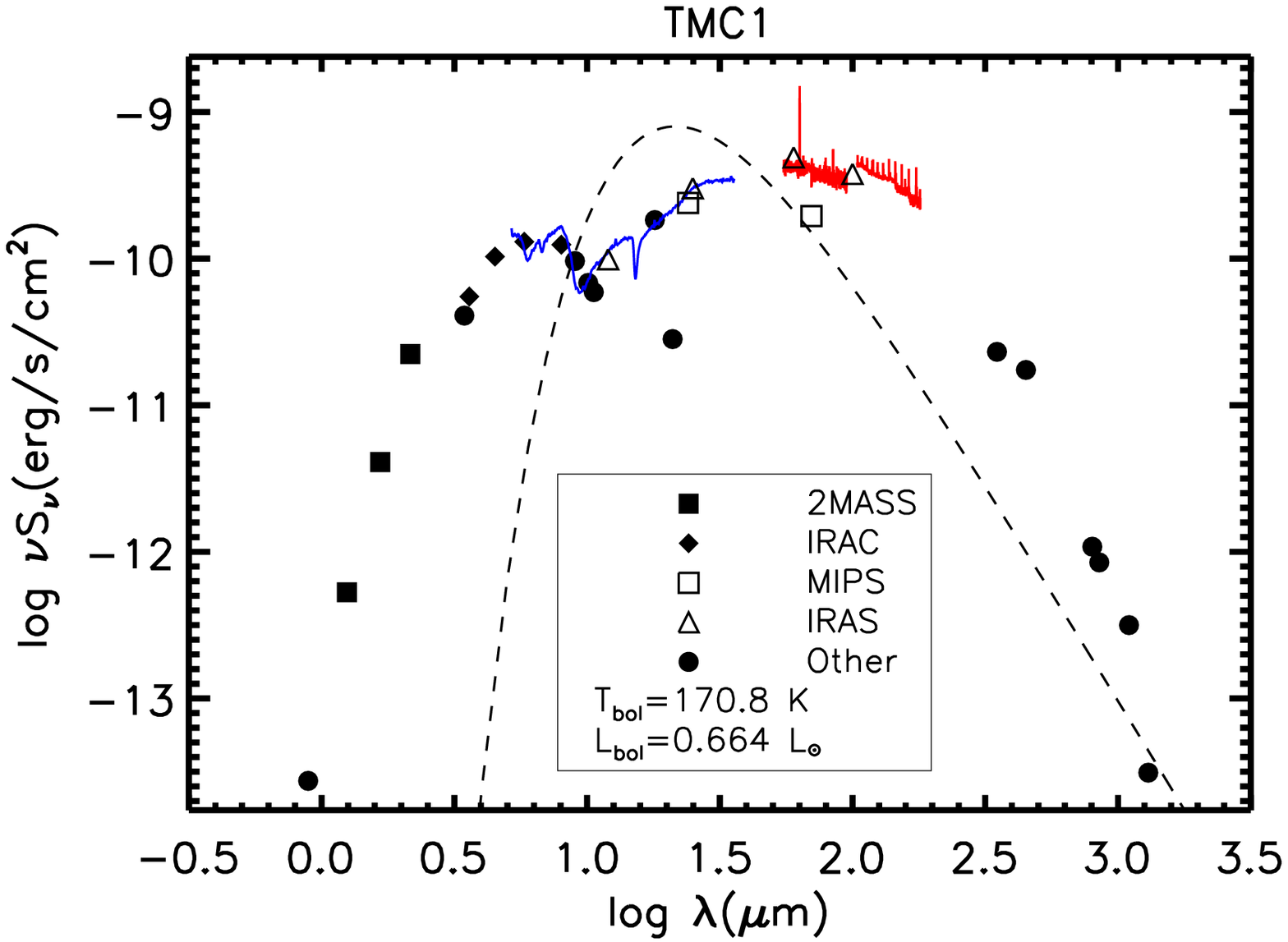}
\includegraphics[scale=0.45]{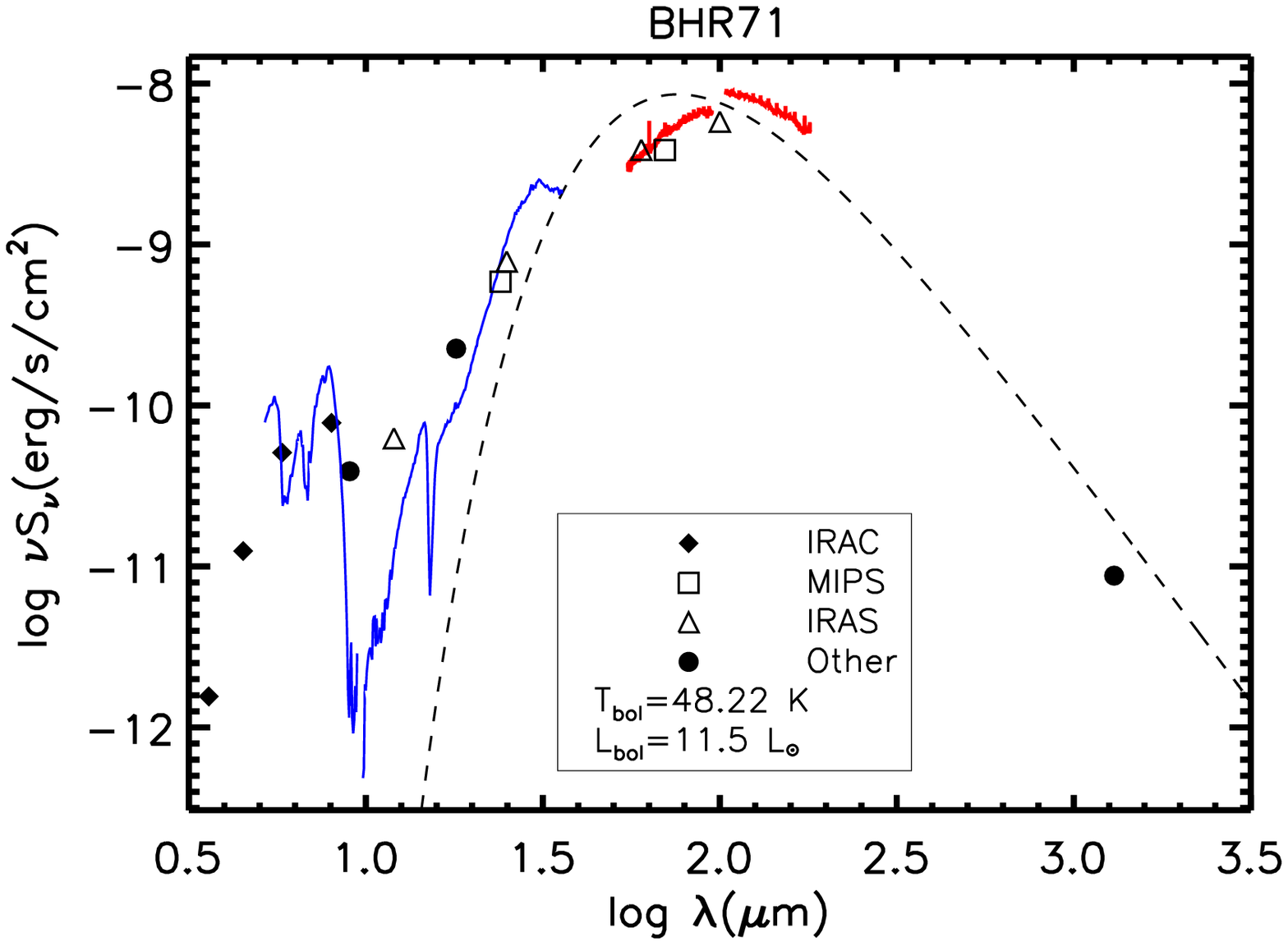}
\includegraphics[scale=0.45]{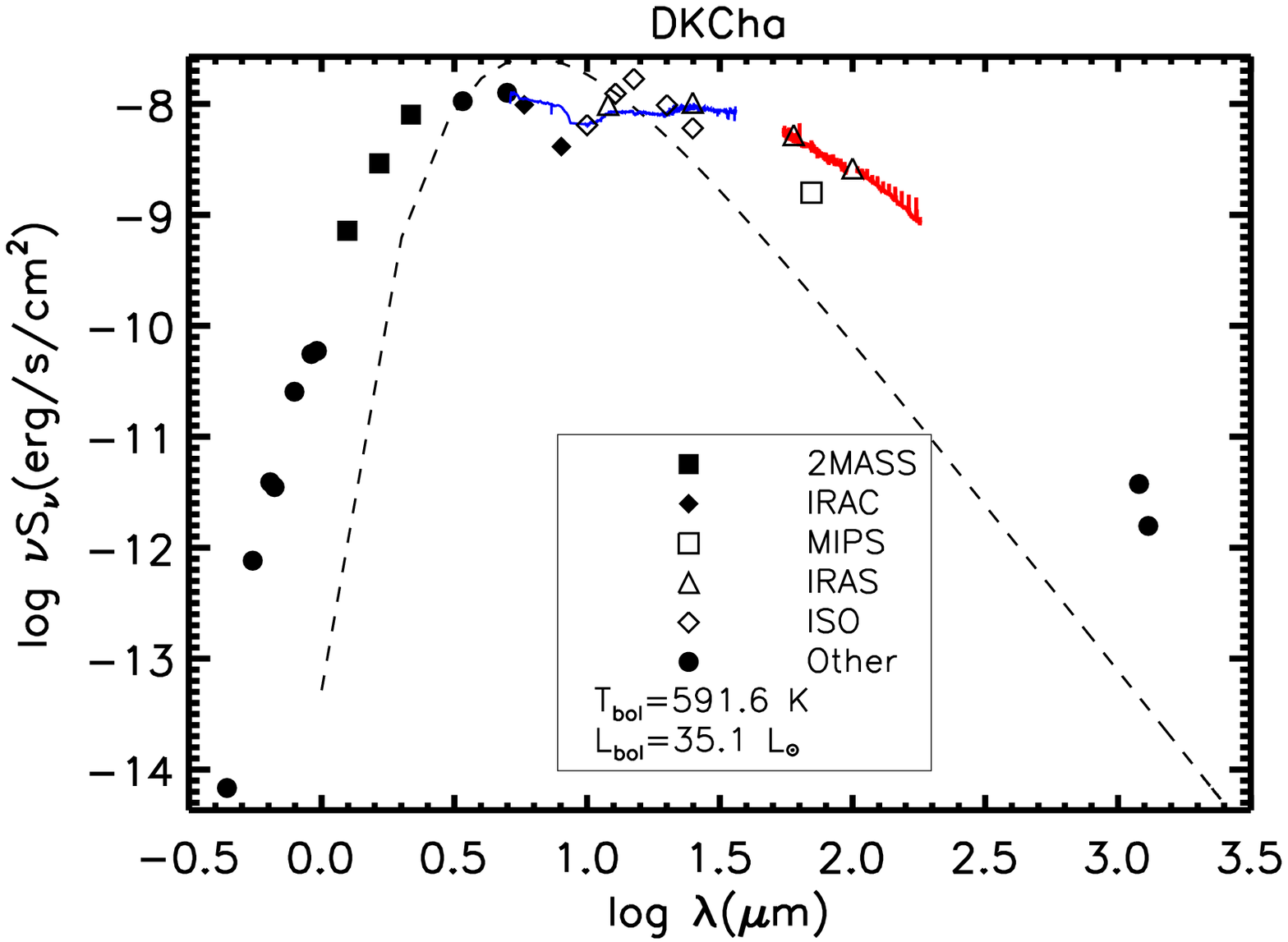}
\includegraphics[scale=0.45]{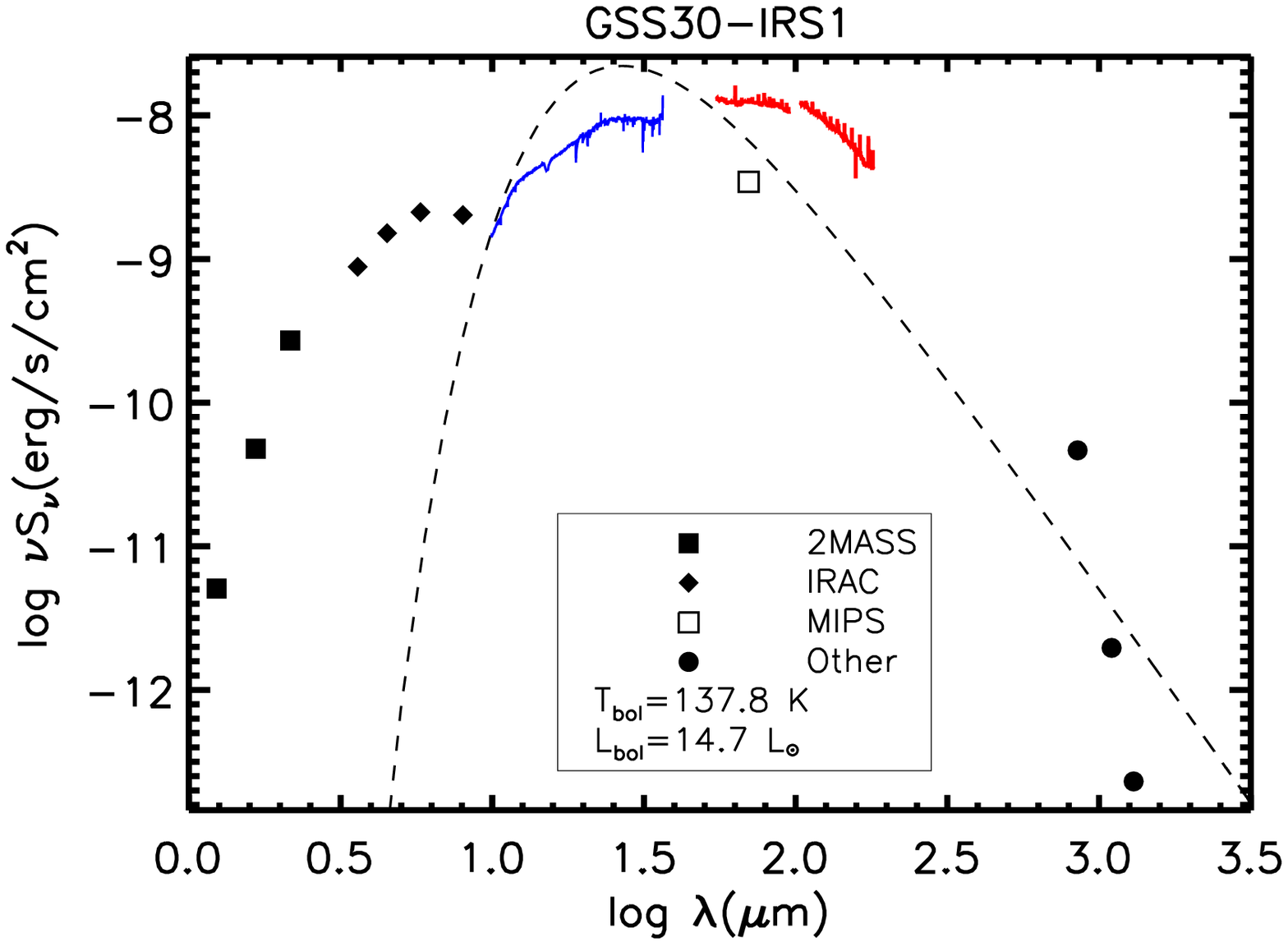}
\includegraphics[scale=0.45]{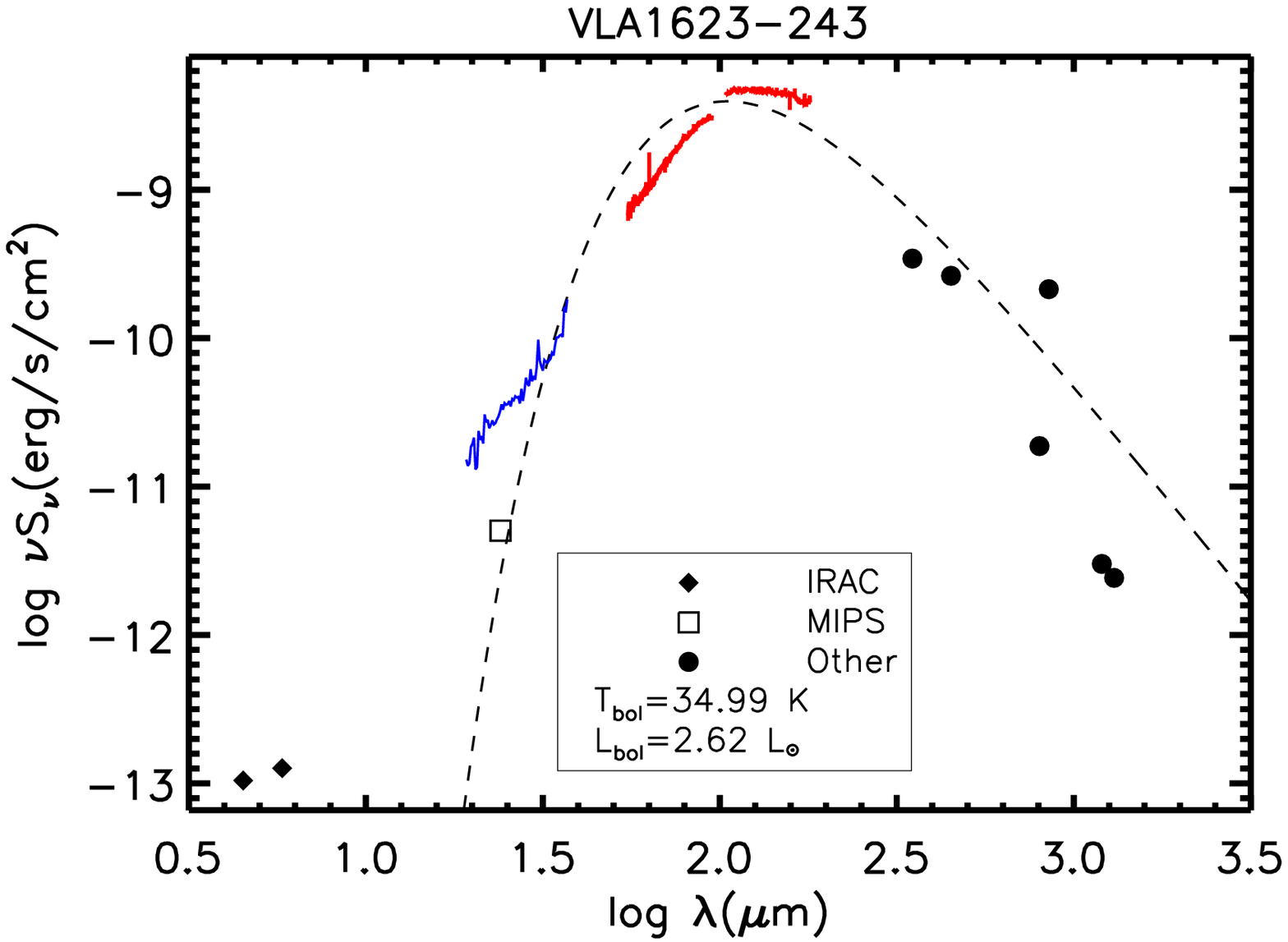}
\includegraphics[scale=0.45]{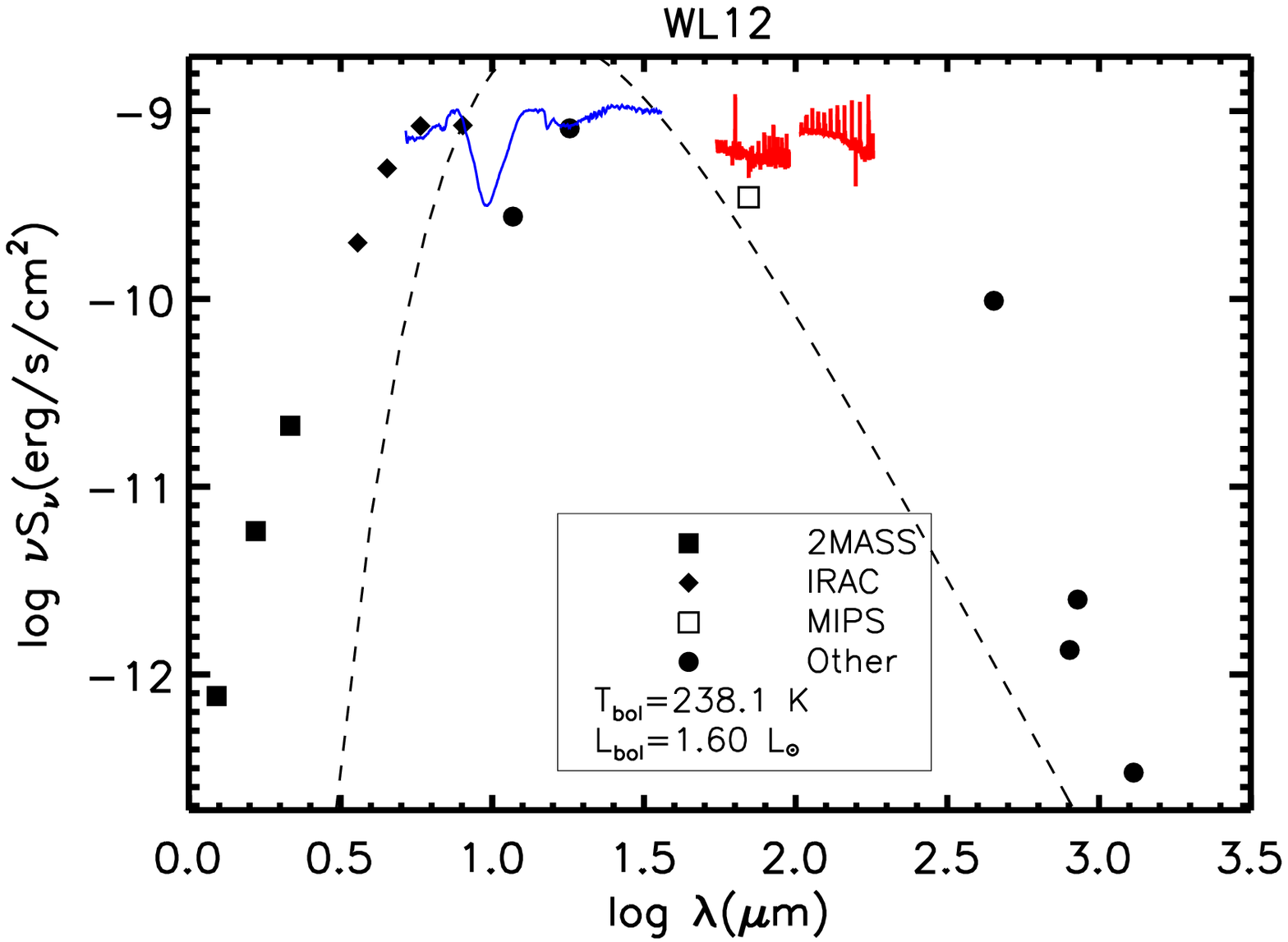}
\caption{0.5-1000 $\mu$m SEDs, as in Figure \ref{seds1}.}
\label{seds3}
\end{center}
\end{figure}

\begin{figure}
\begin{center}
\includegraphics[scale=0.45]{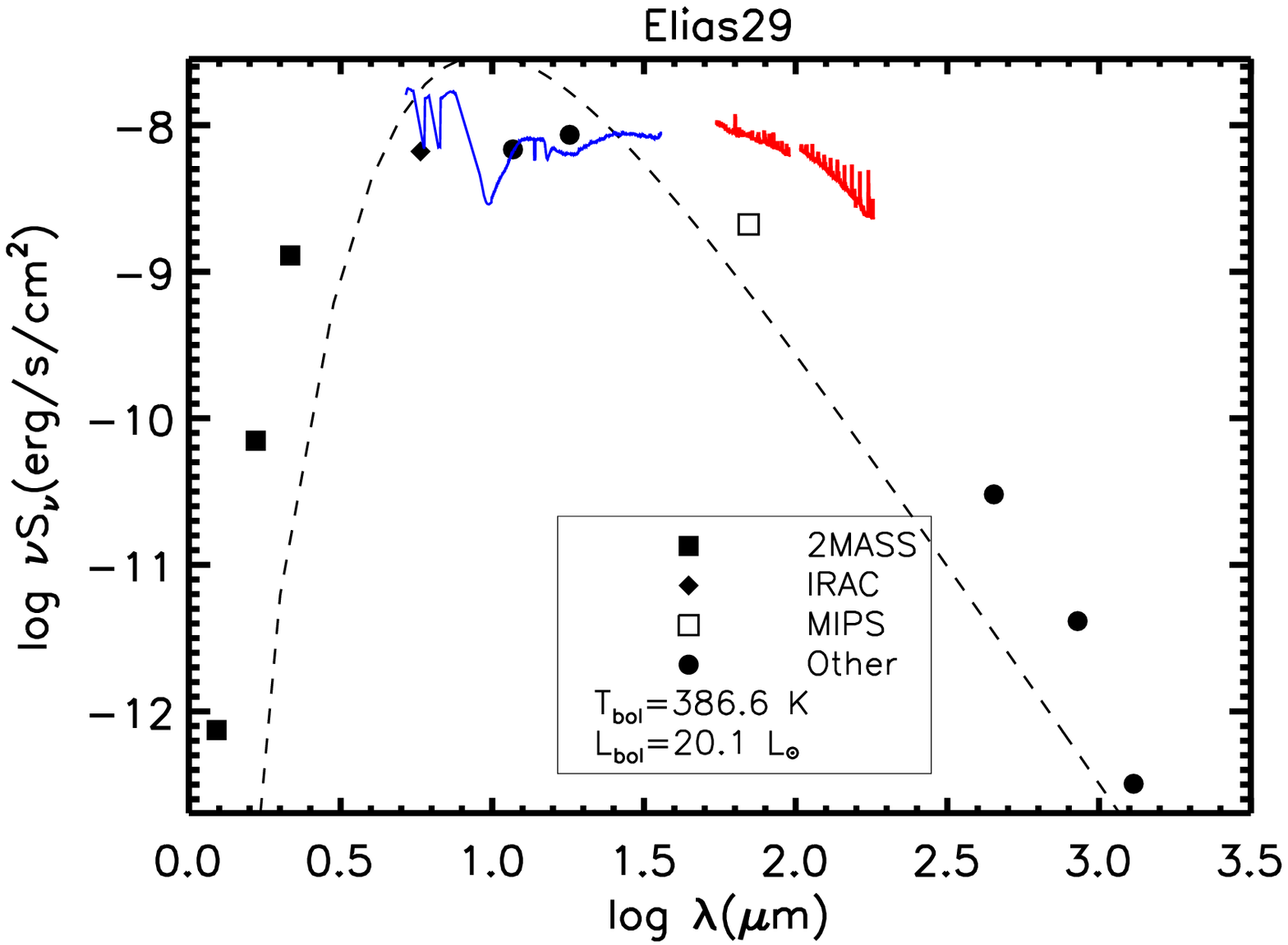}
\includegraphics[scale=0.45]{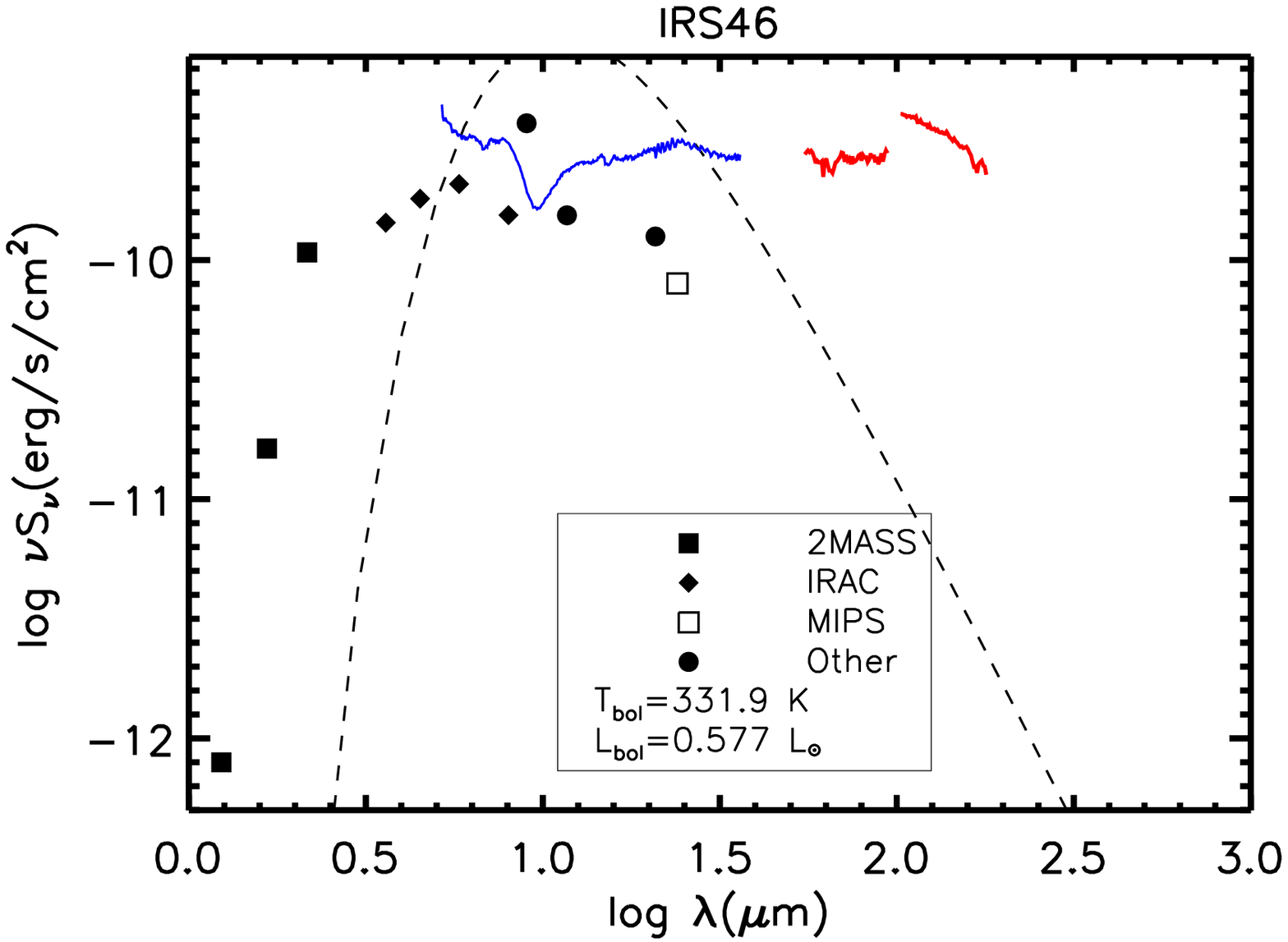}
\includegraphics[scale=0.45]{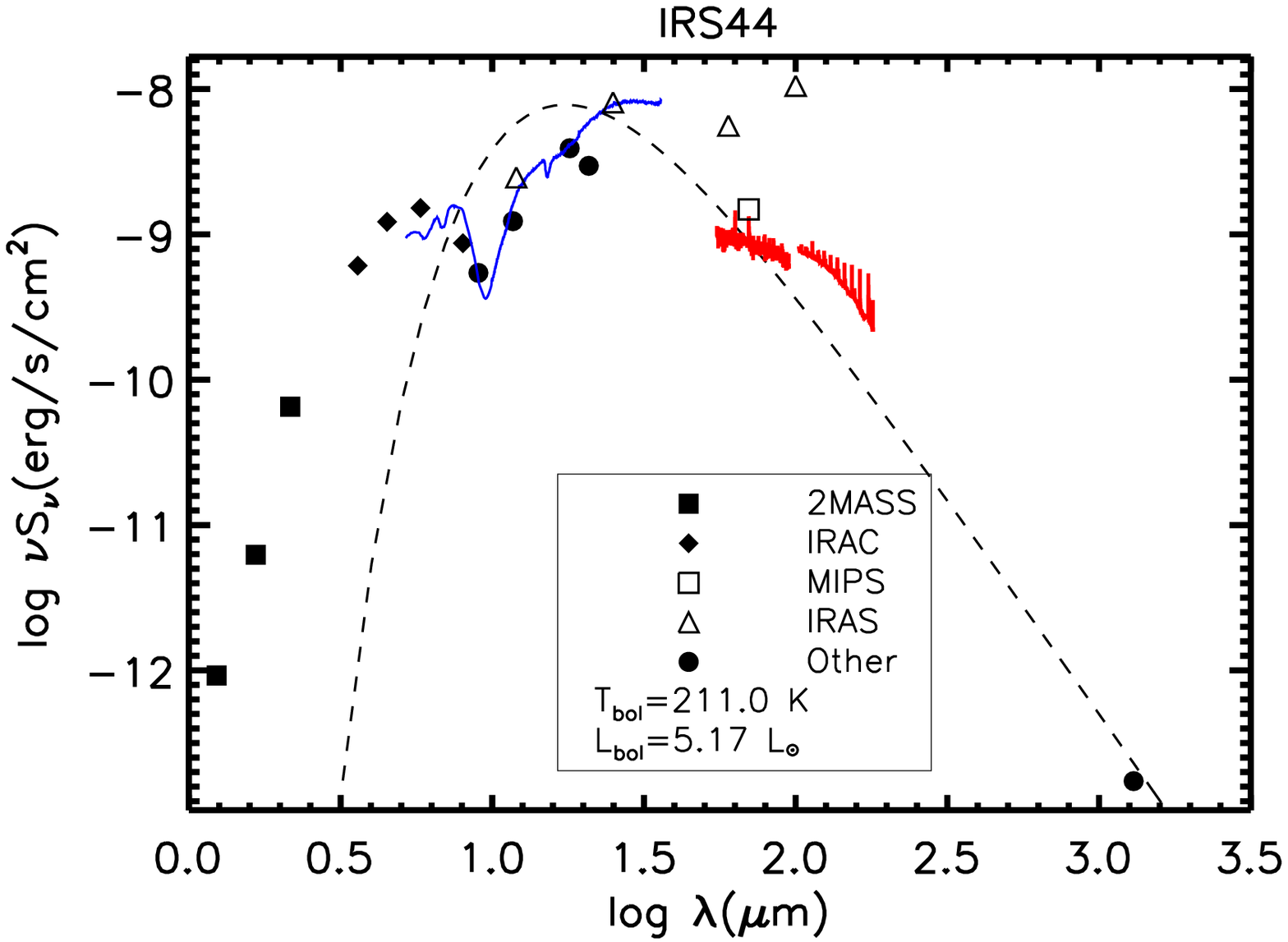}
\includegraphics[scale=0.45]{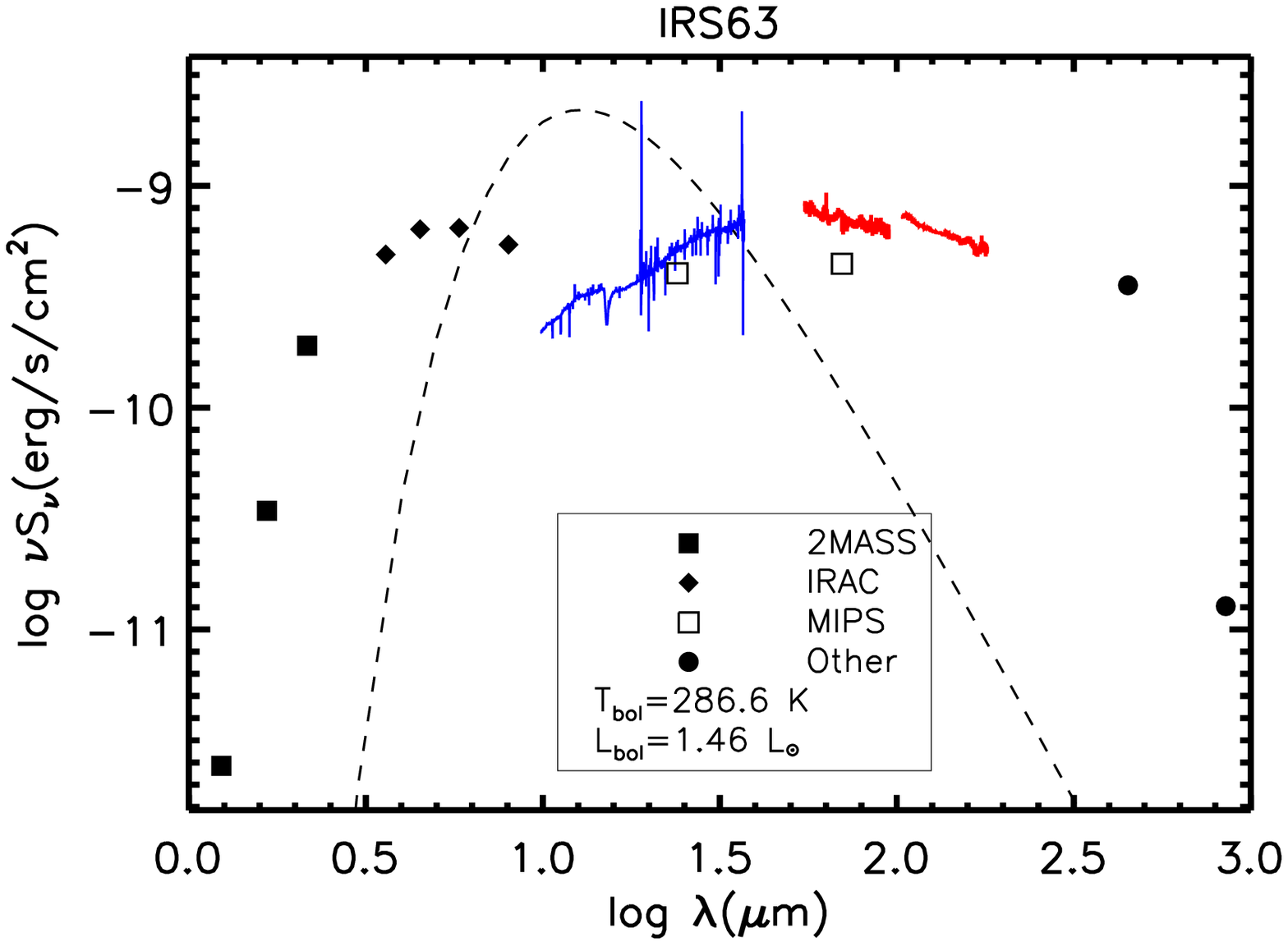}
\includegraphics[scale=0.45]{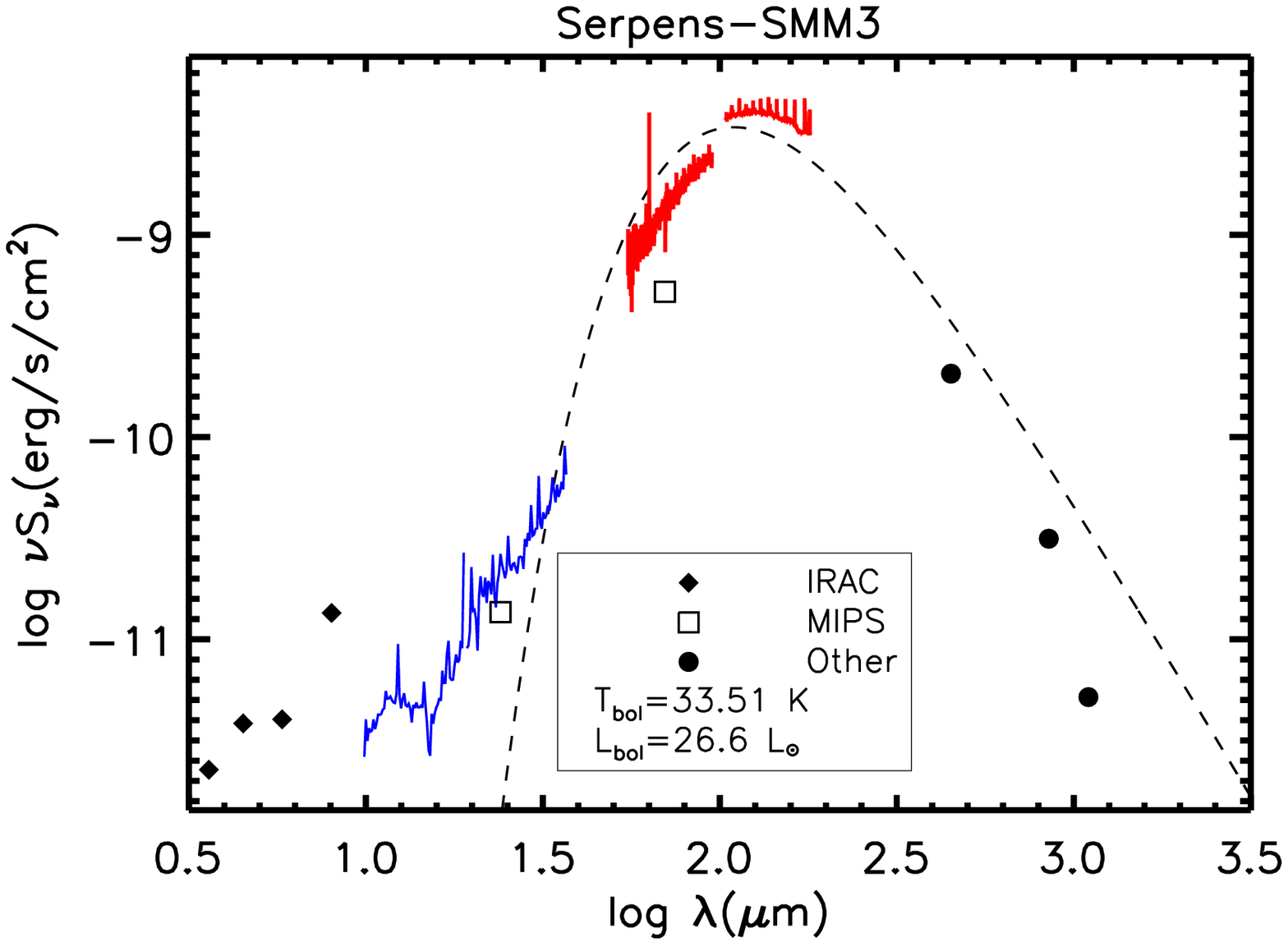}
\includegraphics[scale=0.45]{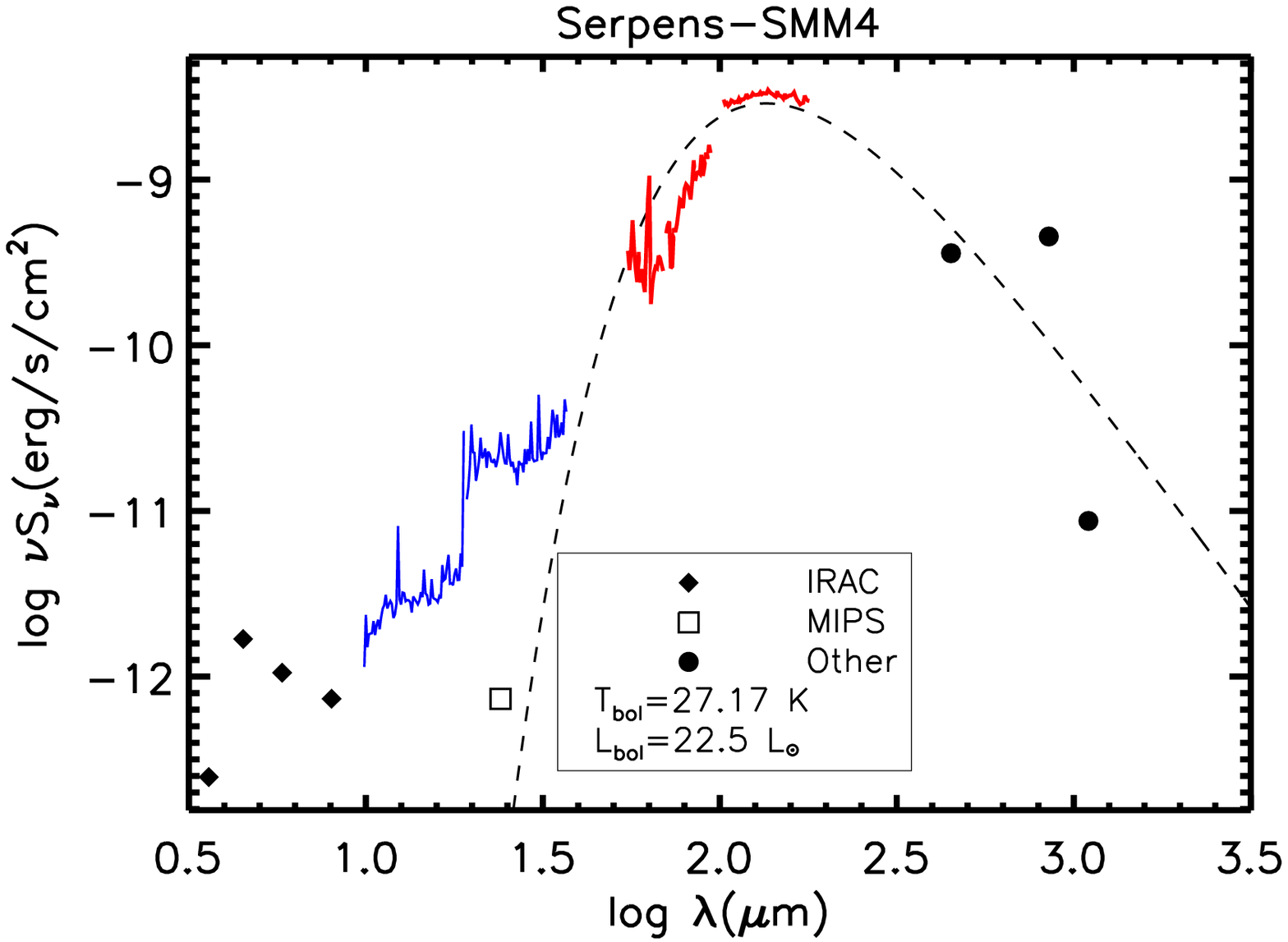}
\caption{0.5-1000 $\mu$m SEDs, as in Figure \ref{seds1}.   IRS44 is not corrected for PSF 
or its location on the edge of the PACS array.}
\label{seds4}
\end{center}
\end{figure}

\clearpage

\begin{figure}
\begin{center}
\includegraphics[scale=0.45]{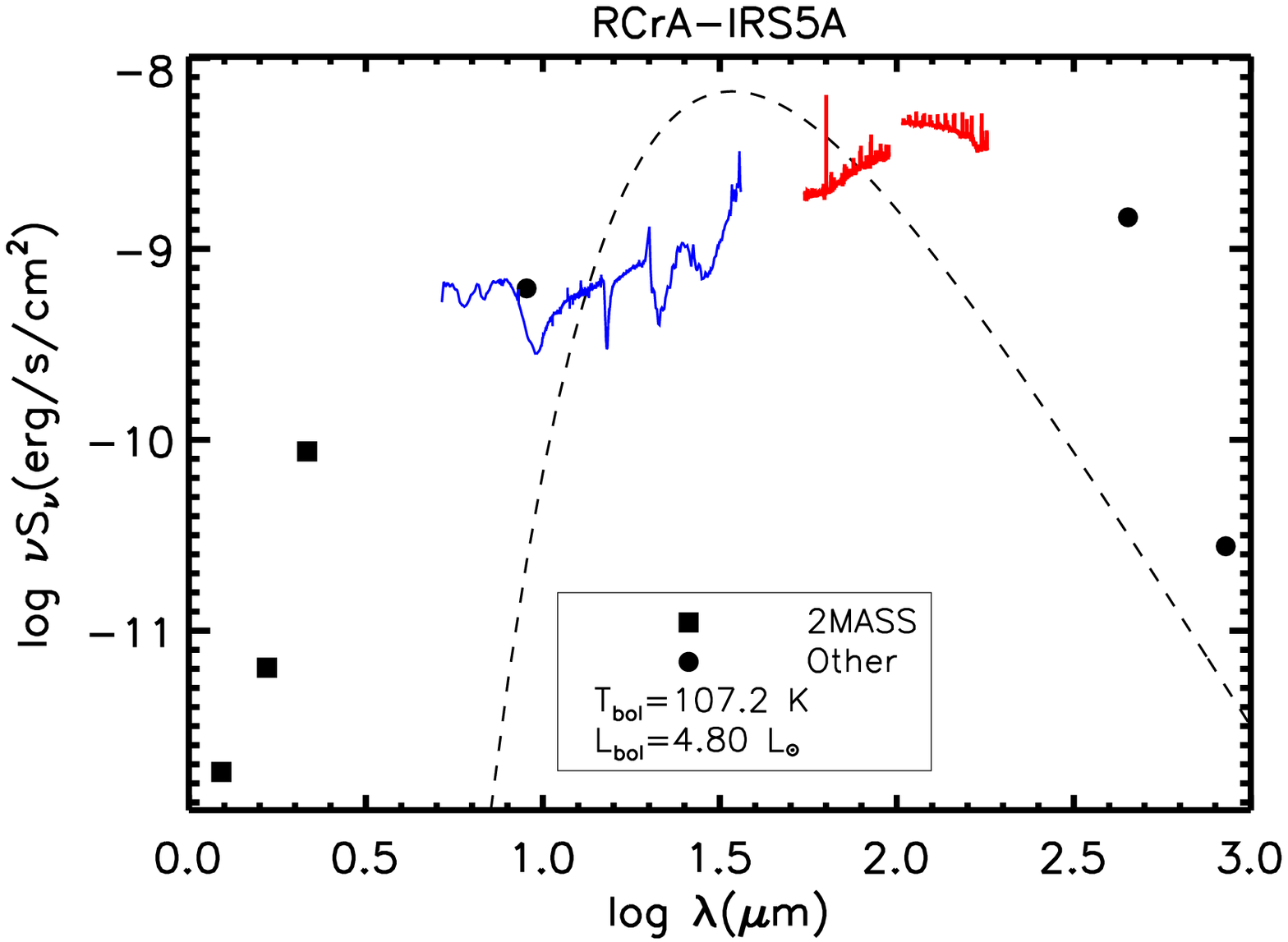}
\includegraphics[scale=0.45]{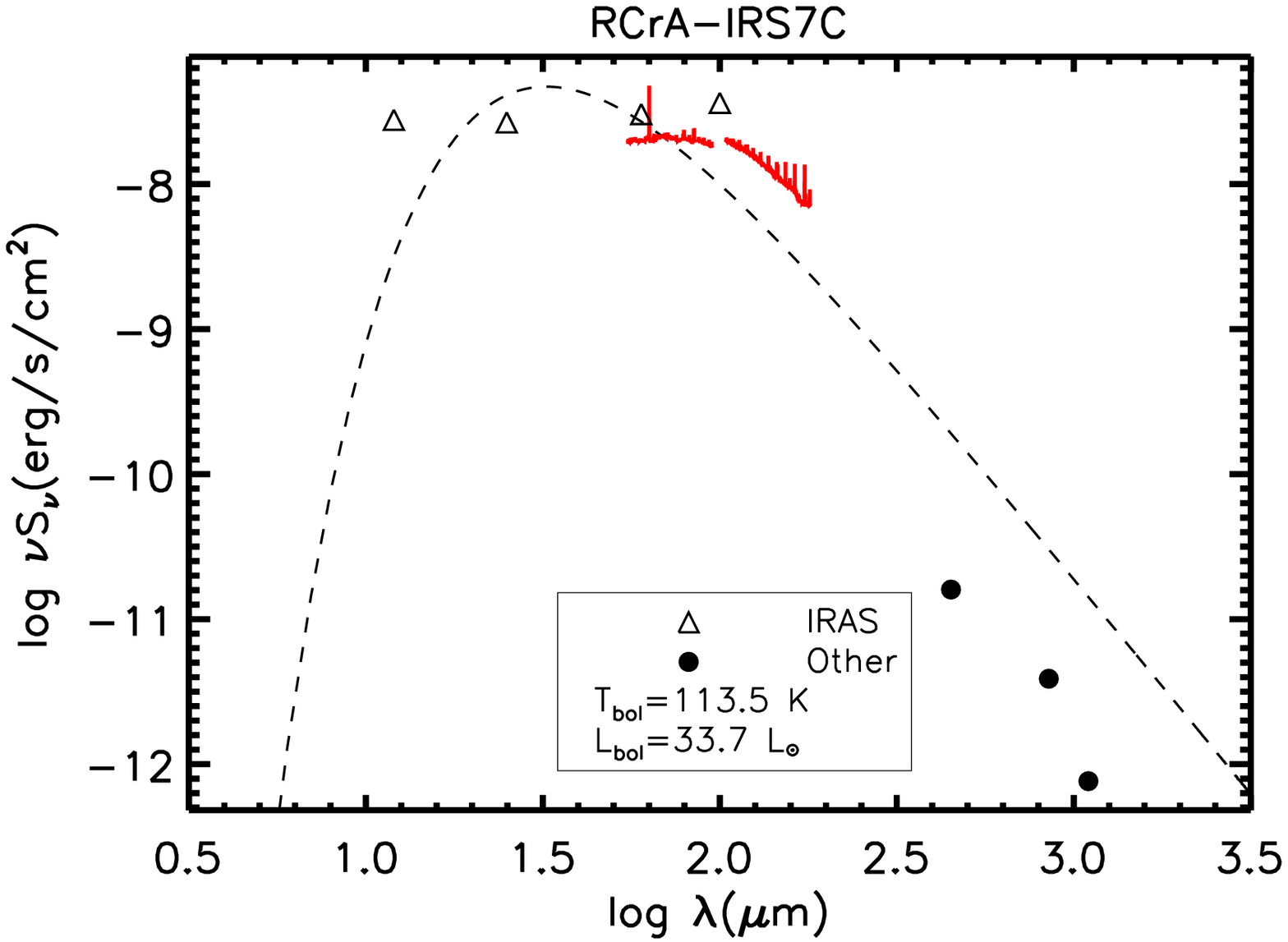}
\includegraphics[scale=0.45]{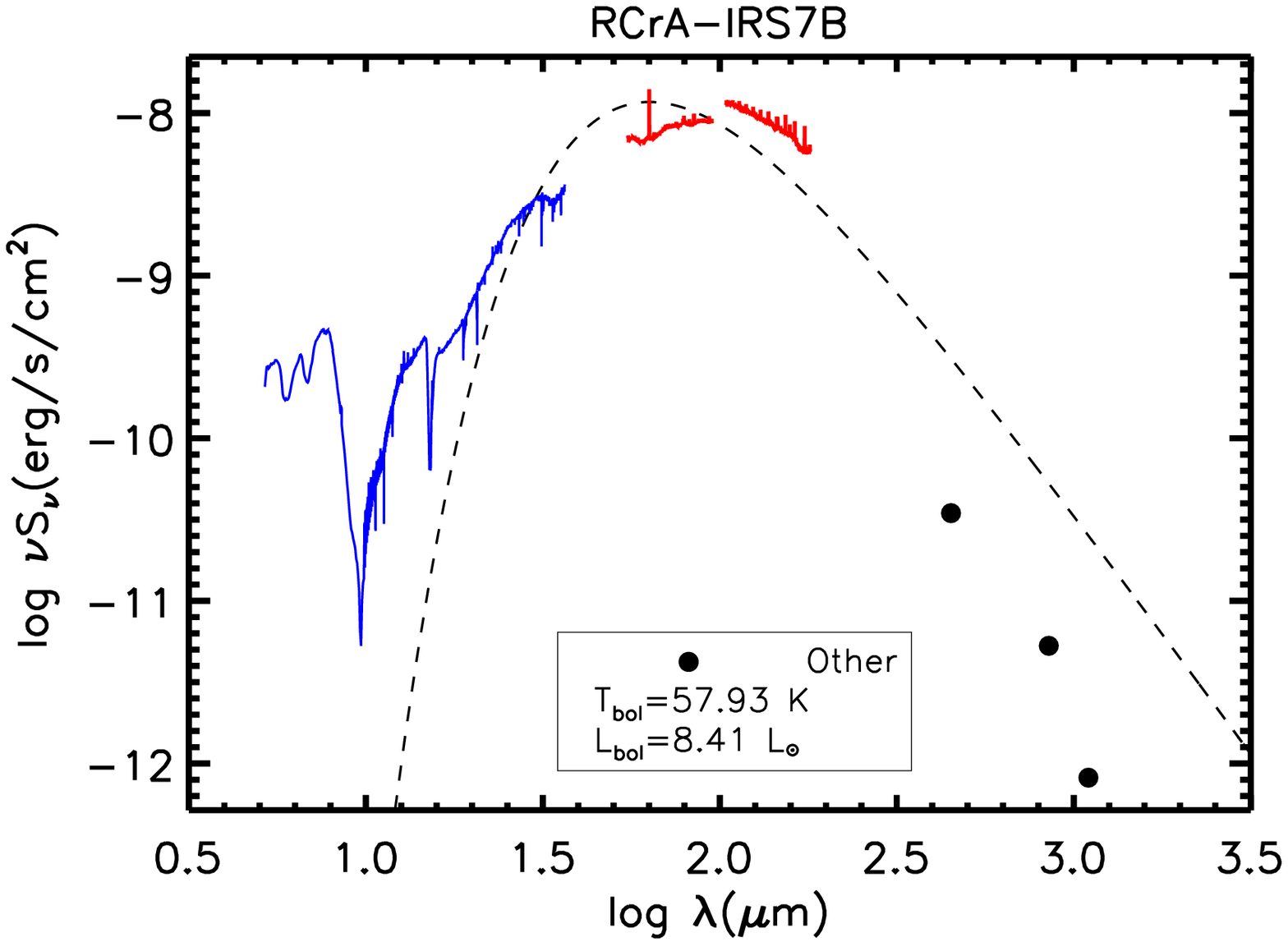}
\includegraphics[scale=0.45]{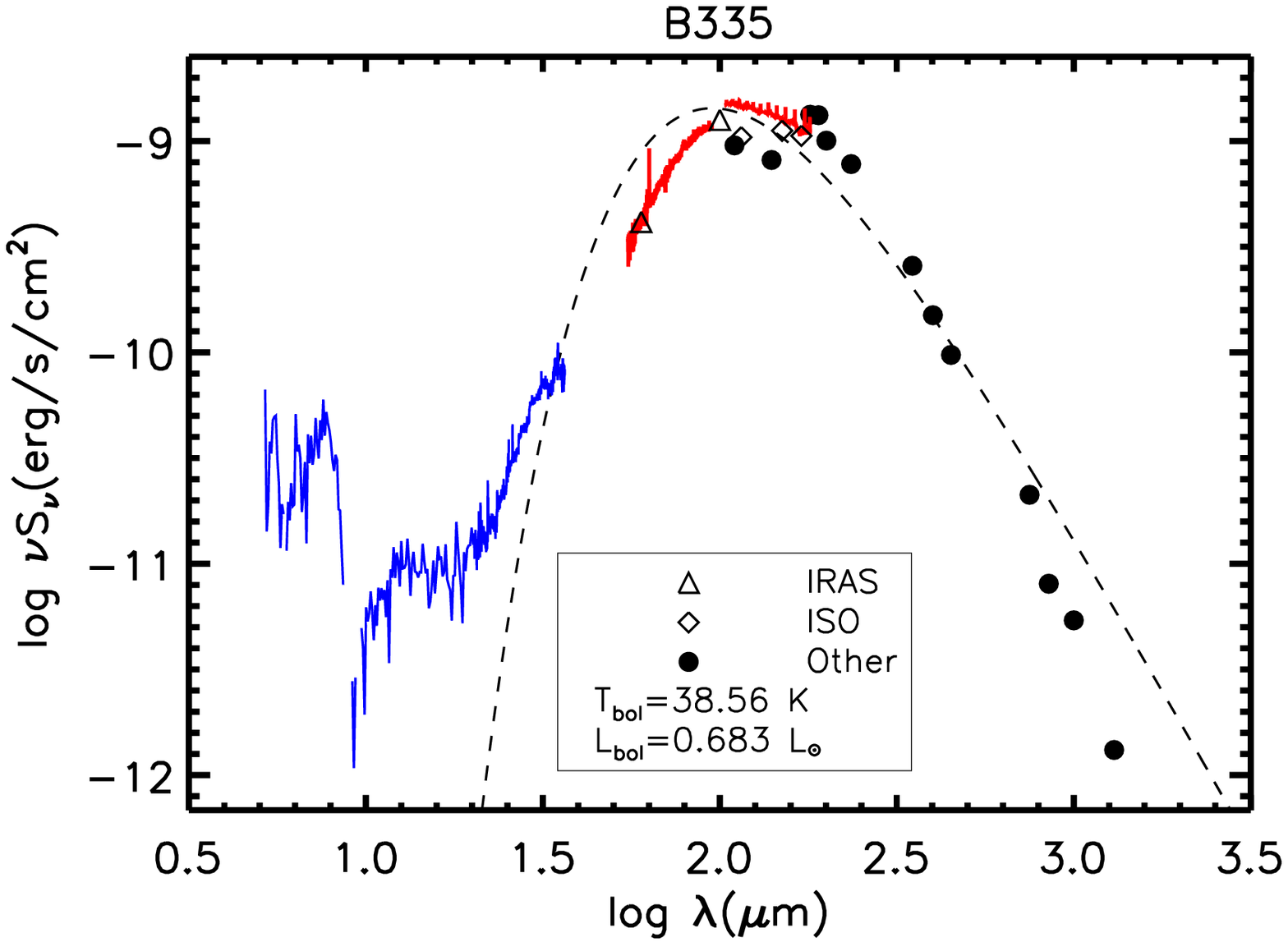}
\includegraphics[scale=0.45]{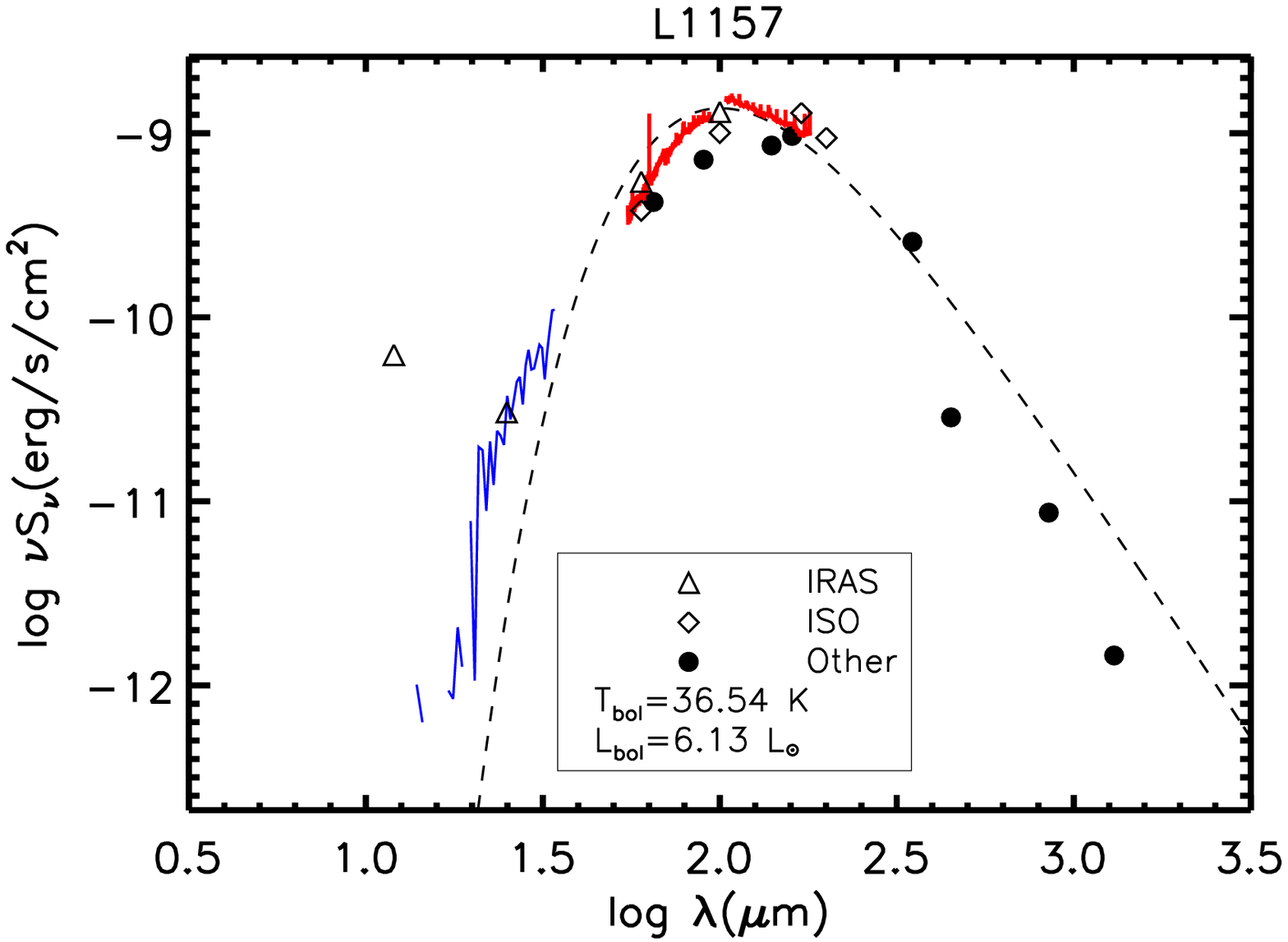}
\includegraphics[scale=0.45]{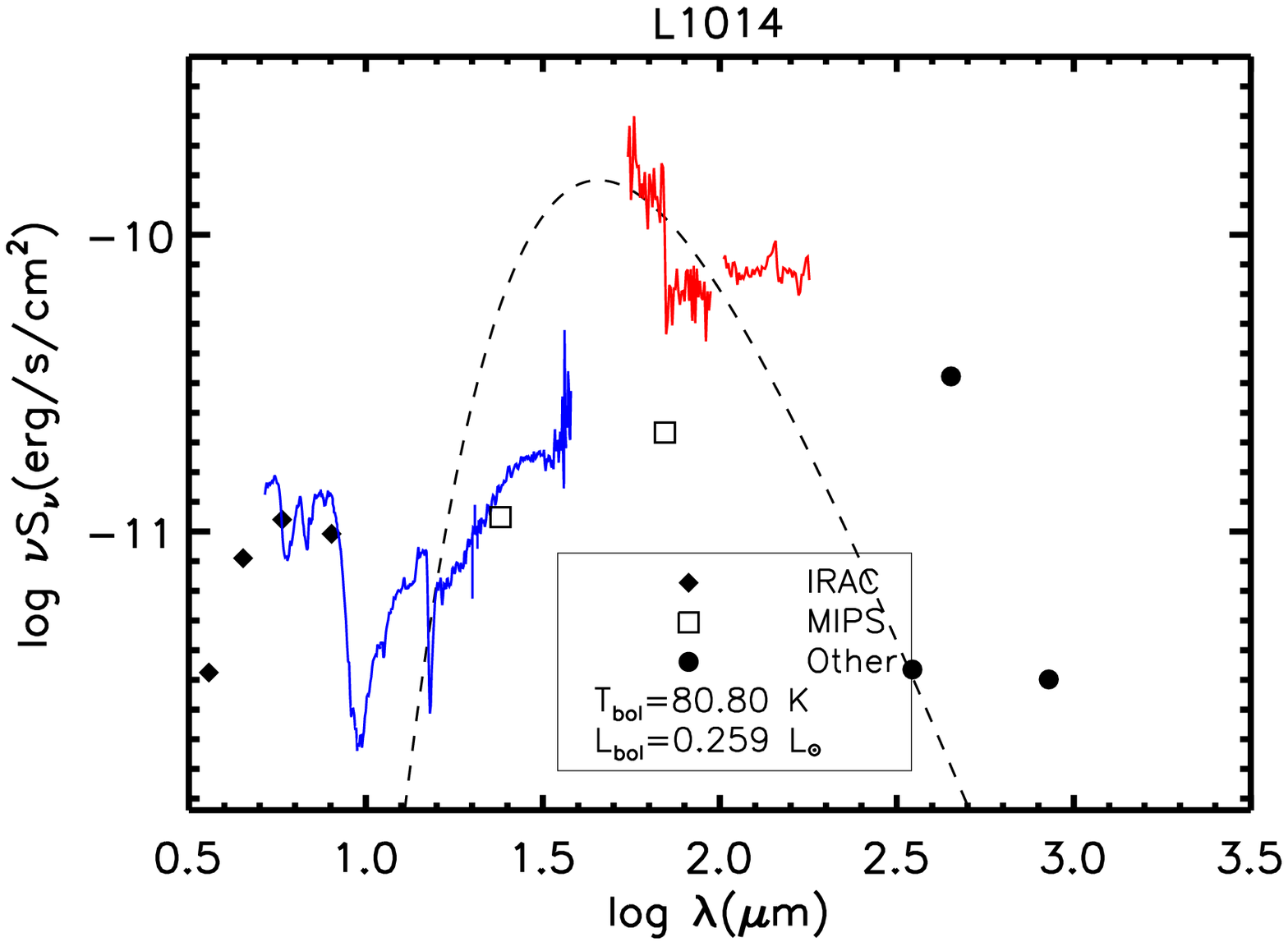}
\caption{0.5-1000 $\mu$m SEDs, as in Figure \ref{seds1}.  L1014 is not detected in continuum emission at $\lambda$ $<$ 75 $\mu$m.}
\label{seds5}
\end{center}
\end{figure}

\begin{figure}
\includegraphics[scale=0.45]{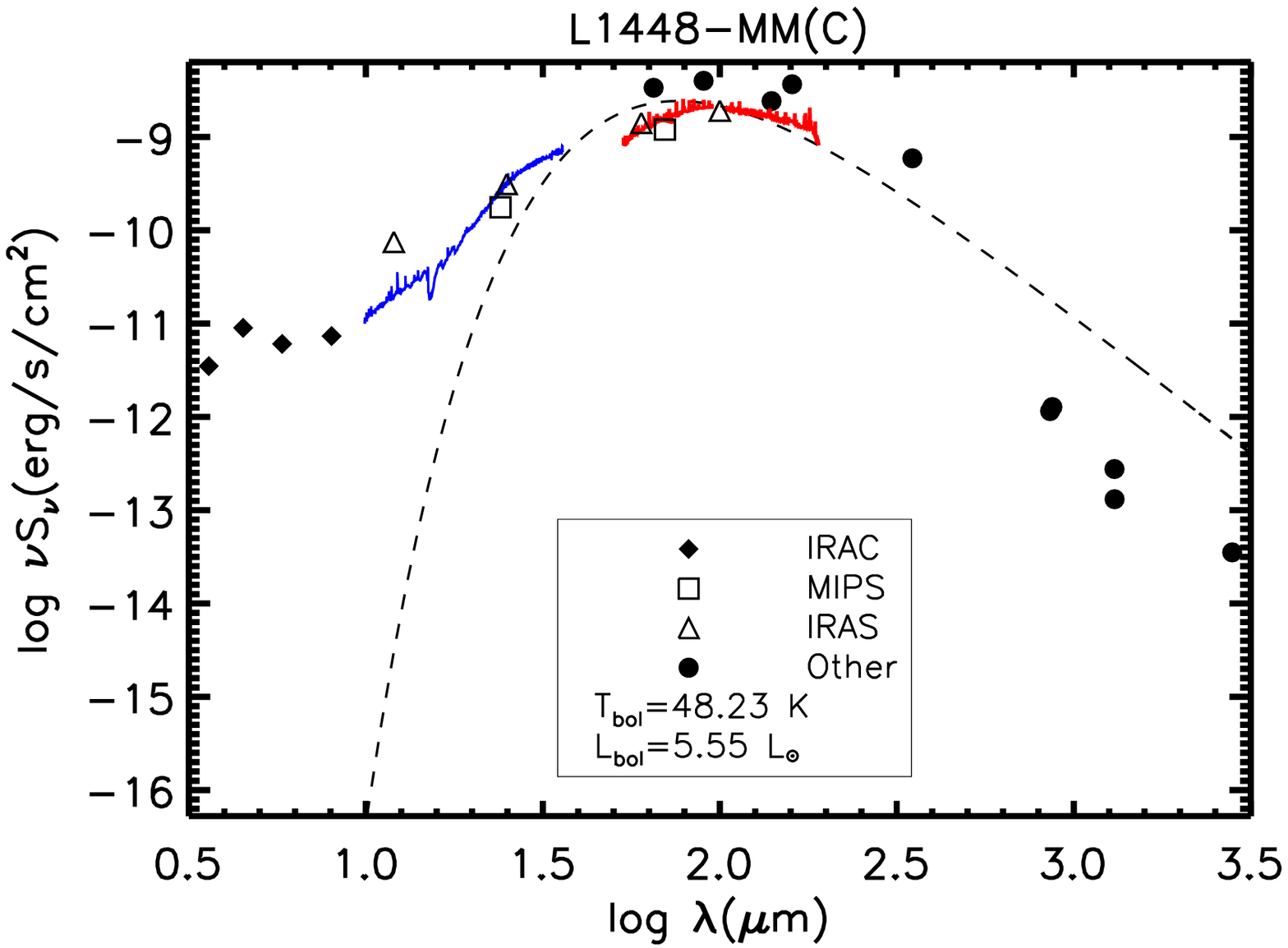}
\includegraphics[scale=0.45]{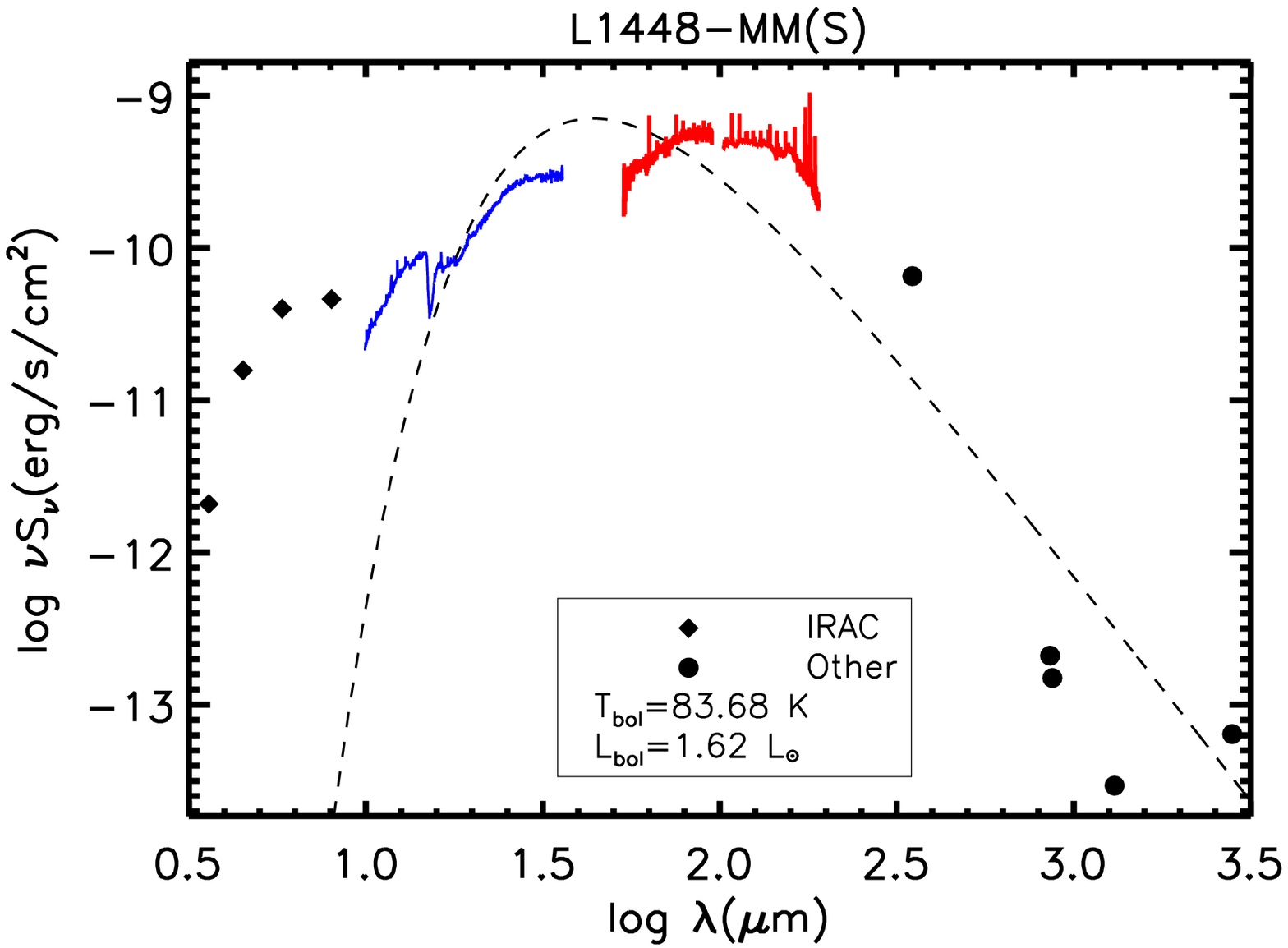}
\caption{0.5-3000 $\mu$m SED of L1448-MM, decomposed into sources C and S, 
with unresolved photometry ($\lambda$ $>$ 8 $\mu$m) weighted by the ratio of their 
IRAC 8 $\mu$m flux densities.  {\it Spitzer}-IRS spectra are extracted from separate pointings, 
and the SH (10-20 $\mu$m) spectrum is scaled up to match the LH (20-36 $\mu$m) spectrum.}
\label{l1448ab}
\end{figure}

\begin{figure}
\includegraphics[scale=0.8]{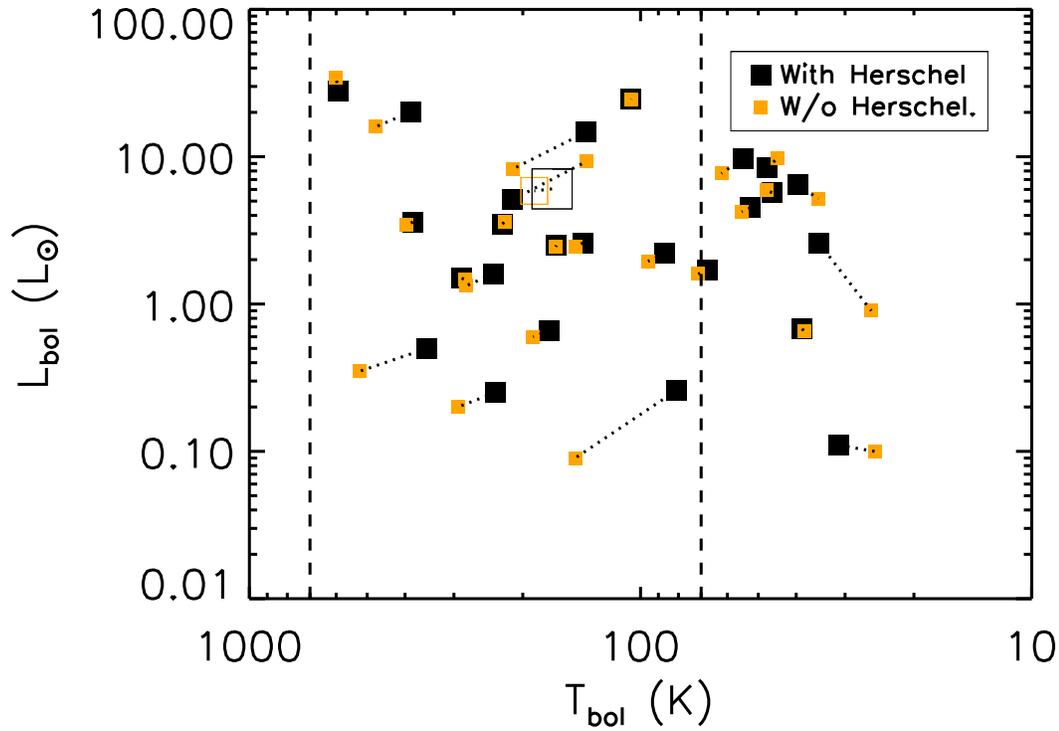}
\caption{The distribution in $L_{\rm bol}$ and $T_{\rm bol}$ before (orange small
squares) and after (black large squares) supplementing the SEDs with the DIGIT
PACS data.  The vertical dashed lines represent the Class 0/I and I/II boundaries.
The hollow squares are the sample means before (small orange) and after (large black).
Dotted lines connect each source in the distribution.
}
\label{lbolnew}
\end{figure}

\begin{figure}
\begin{center}
\includegraphics[scale=0.85]{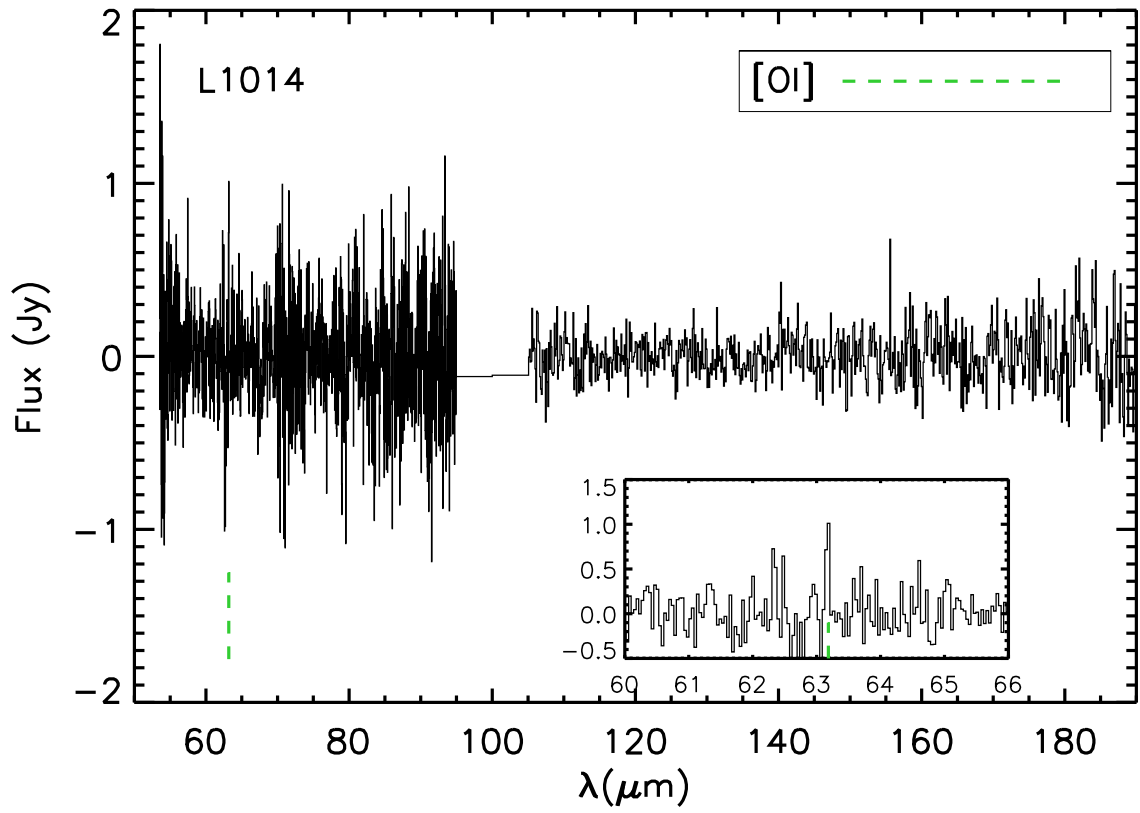}
\caption{Continuum-subtracted spectrum of L1014, representative of the ``cool'' subset.  The only detected line is \OI\ 63.18 \micron.
  {\bf Inset:} zoom-in on the detection of \OI.}
\label{quiescent}
\end{center}
\end{figure}

\begin{figure}
\begin{center}
\includegraphics[scale=0.85]{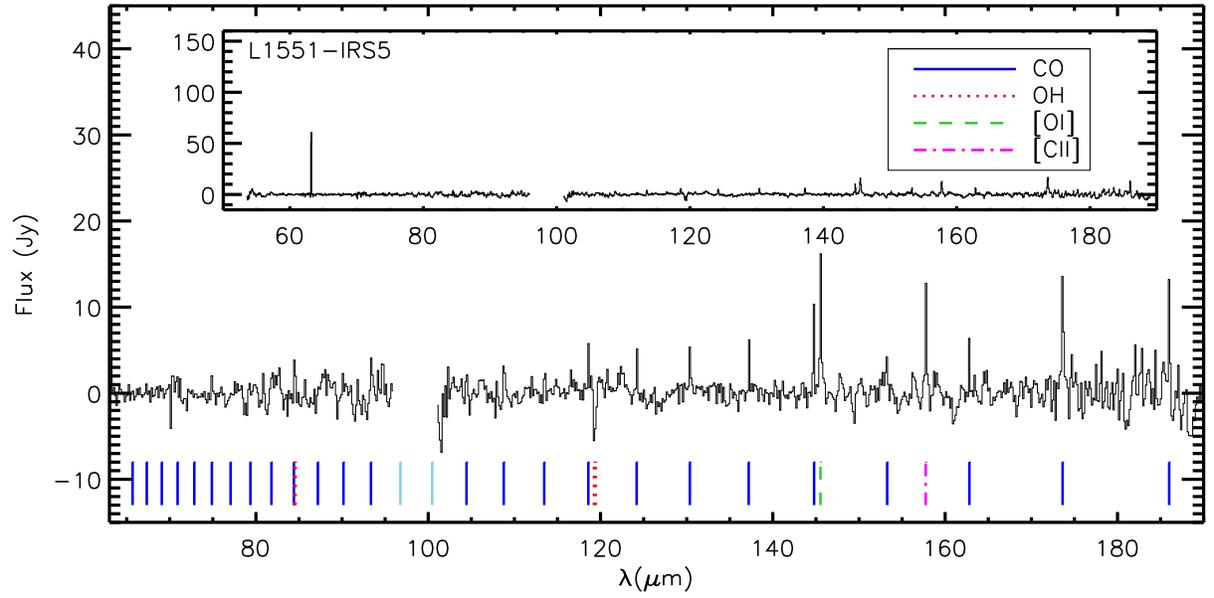}
\caption{Continuum-subtracted spectrum of L1551-IRS5, a representative of the
``warm''
subset of the DIGIT embedded sources sample.  The spectrum is rebinned to lower resolution at longer wavelengths to clarify line peaks.  {\bf Inset:} a zoomed-out image revealing the contrast
between the
strength of the \OI\ 63.2 $\mu$m line and the remaining spectral lines.}
\label{warm}
\end{center}
\end{figure}

\begin{figure*}
\includegraphics[scale=0.85]{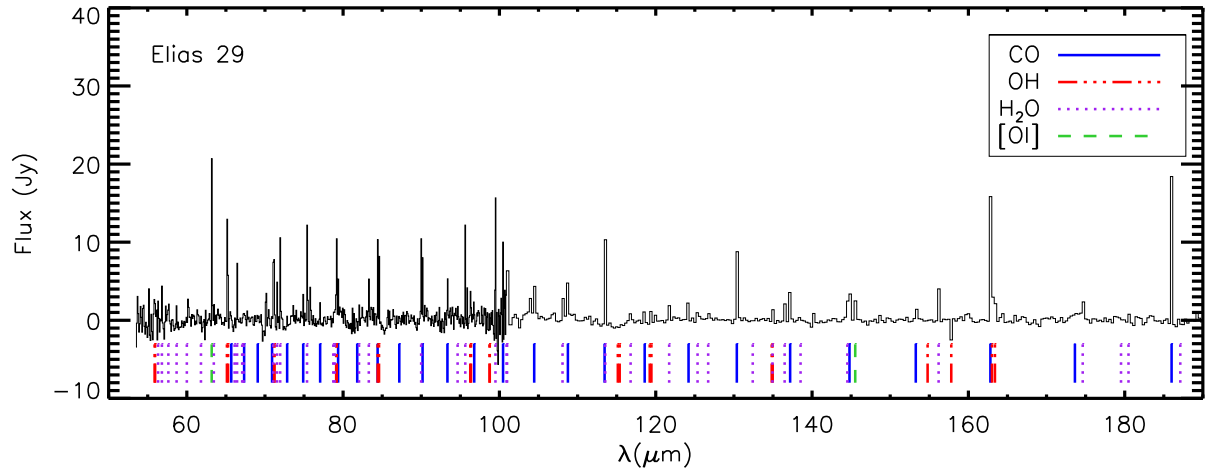}
\caption{Continuum-subtracted spectrum of Elias 29, a representative of the
``hot'' subset
and the most line-rich spectrum in the DIGIT sample.  The spectrum is rebinned to lower resolution at longer wavelengths to clarify line peaks.}
\label{hot}
\end{figure*}

\begin{figure}
\includegraphics[scale=0.6]{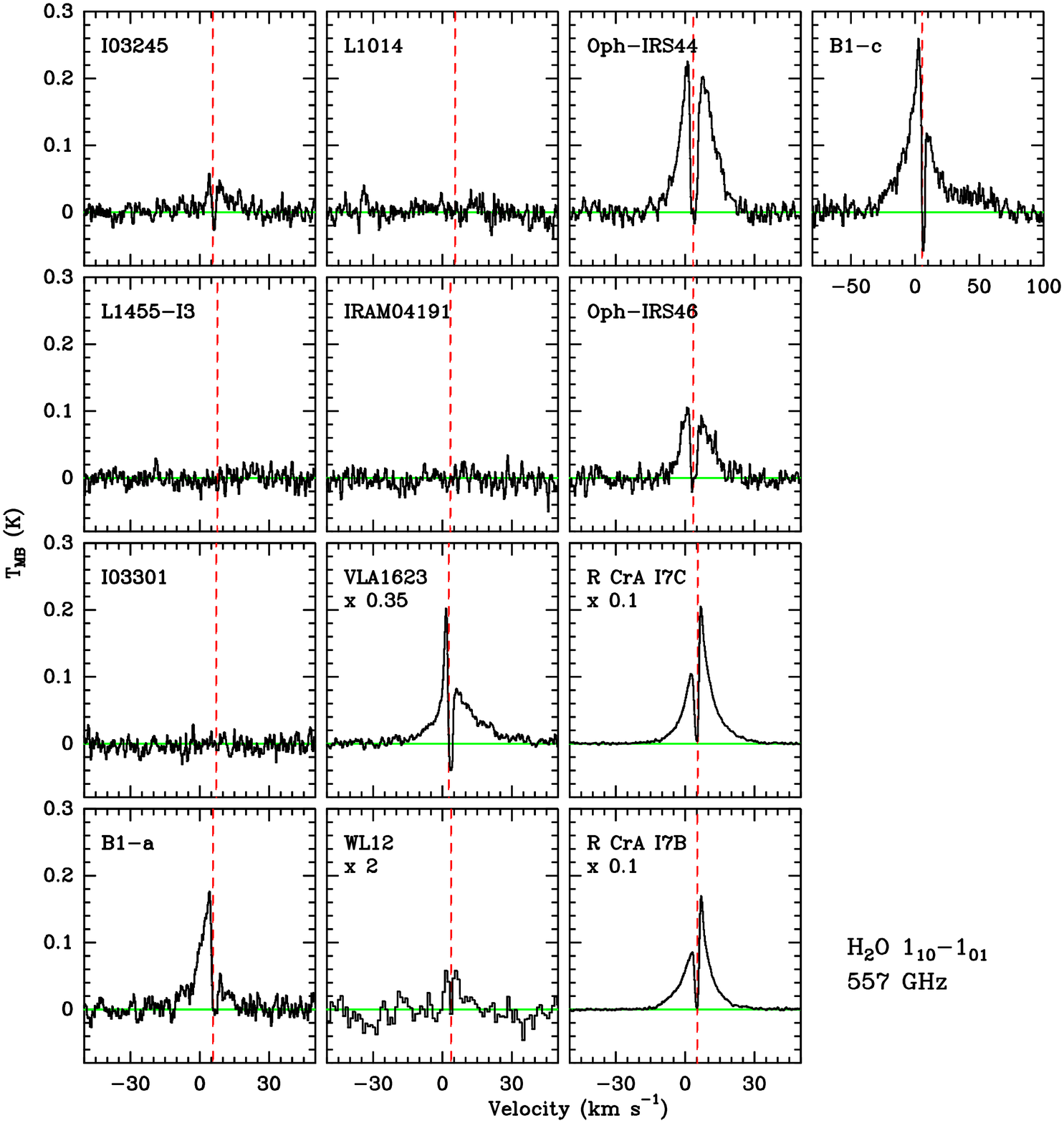}
\caption{HIFI observations of the 557 GHz H$_2$O line for the DIGIT sample.
The source velocity is marked with a dashed red line, and the zero-point with a
solid green line.}
\label{hifi}
\end{figure}

\begin{figure}
\begin{center}
\includegraphics[scale=0.4]{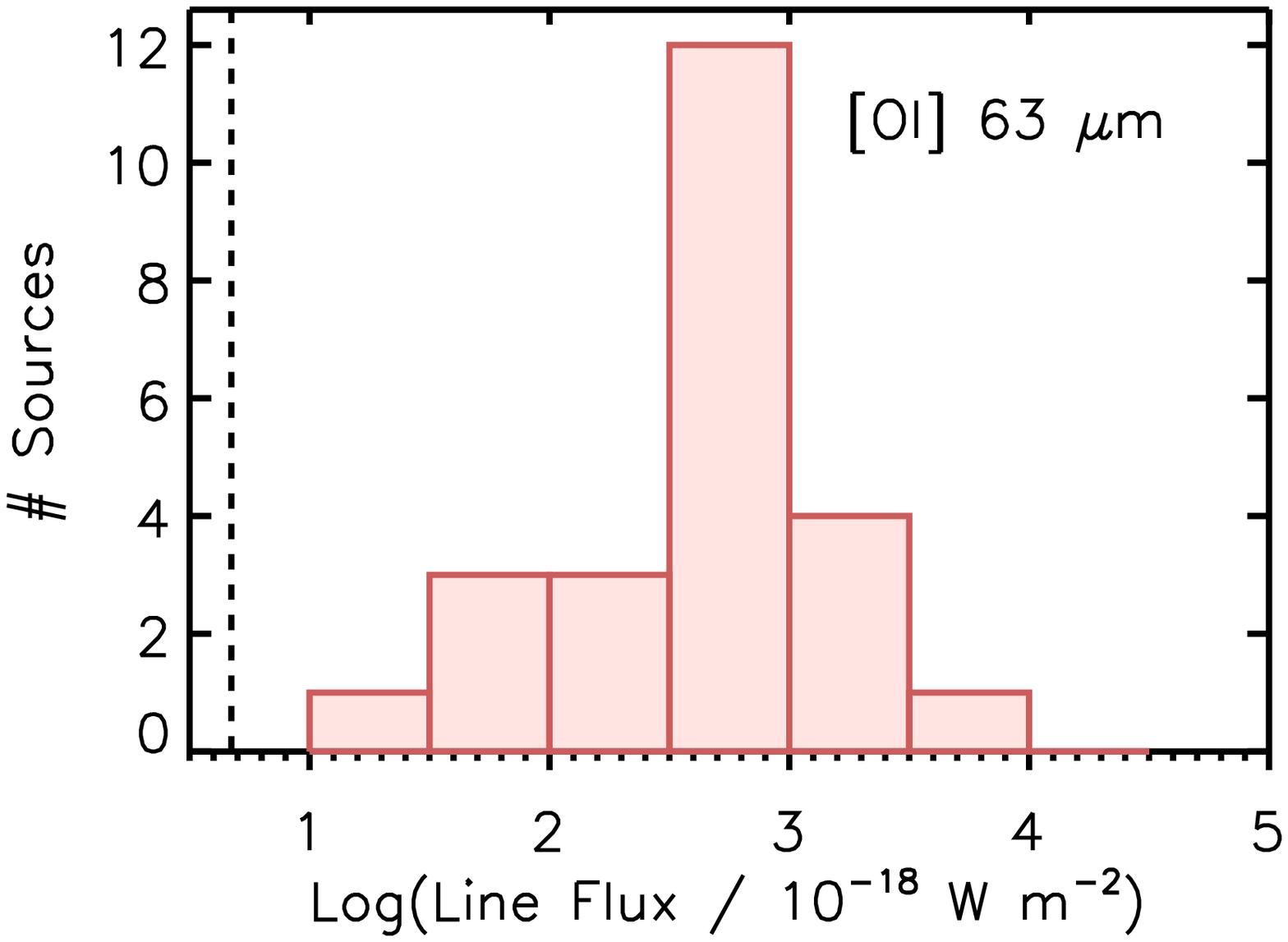}
\includegraphics[scale=0.4]{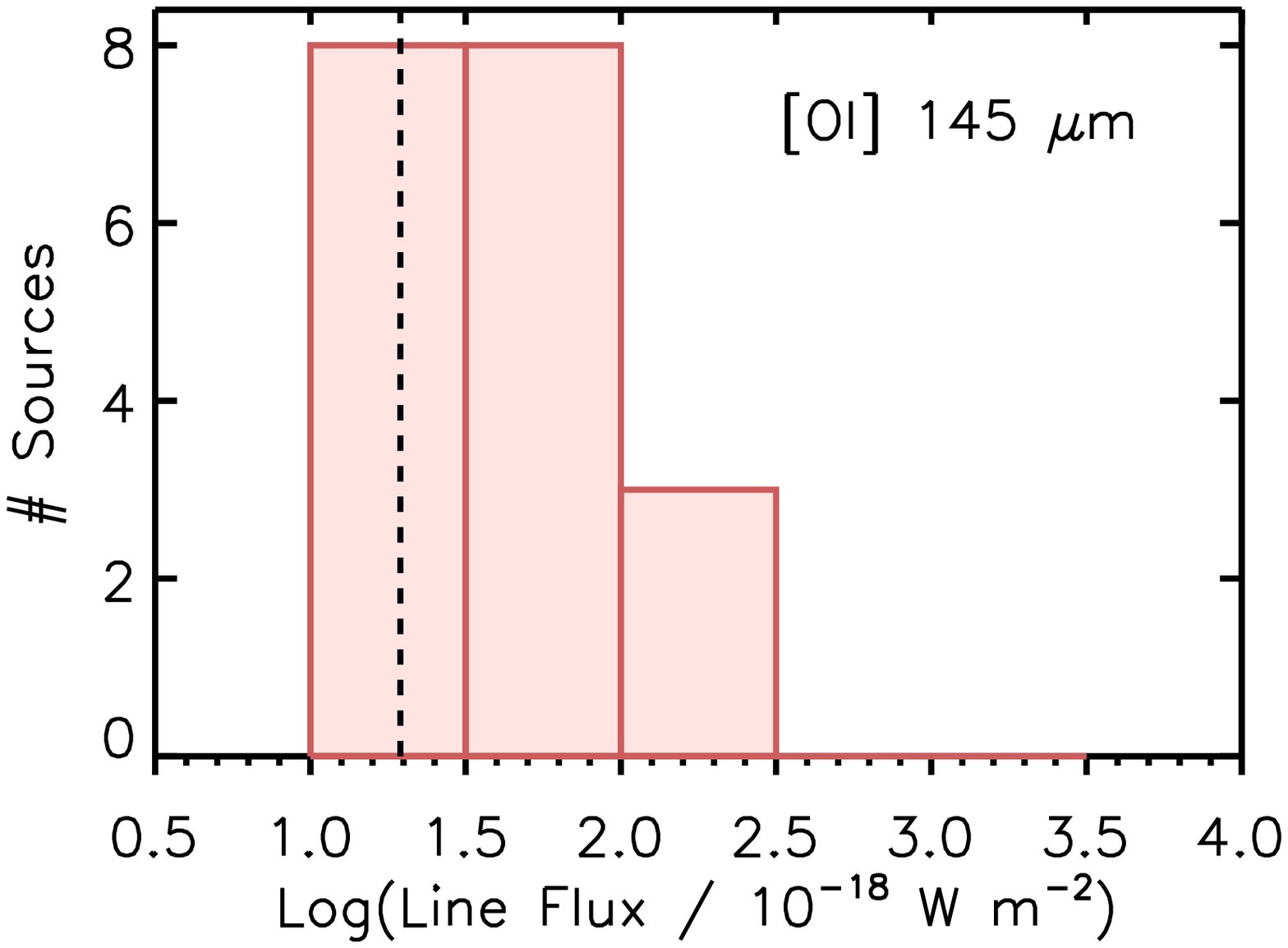}
\includegraphics[scale=0.4]{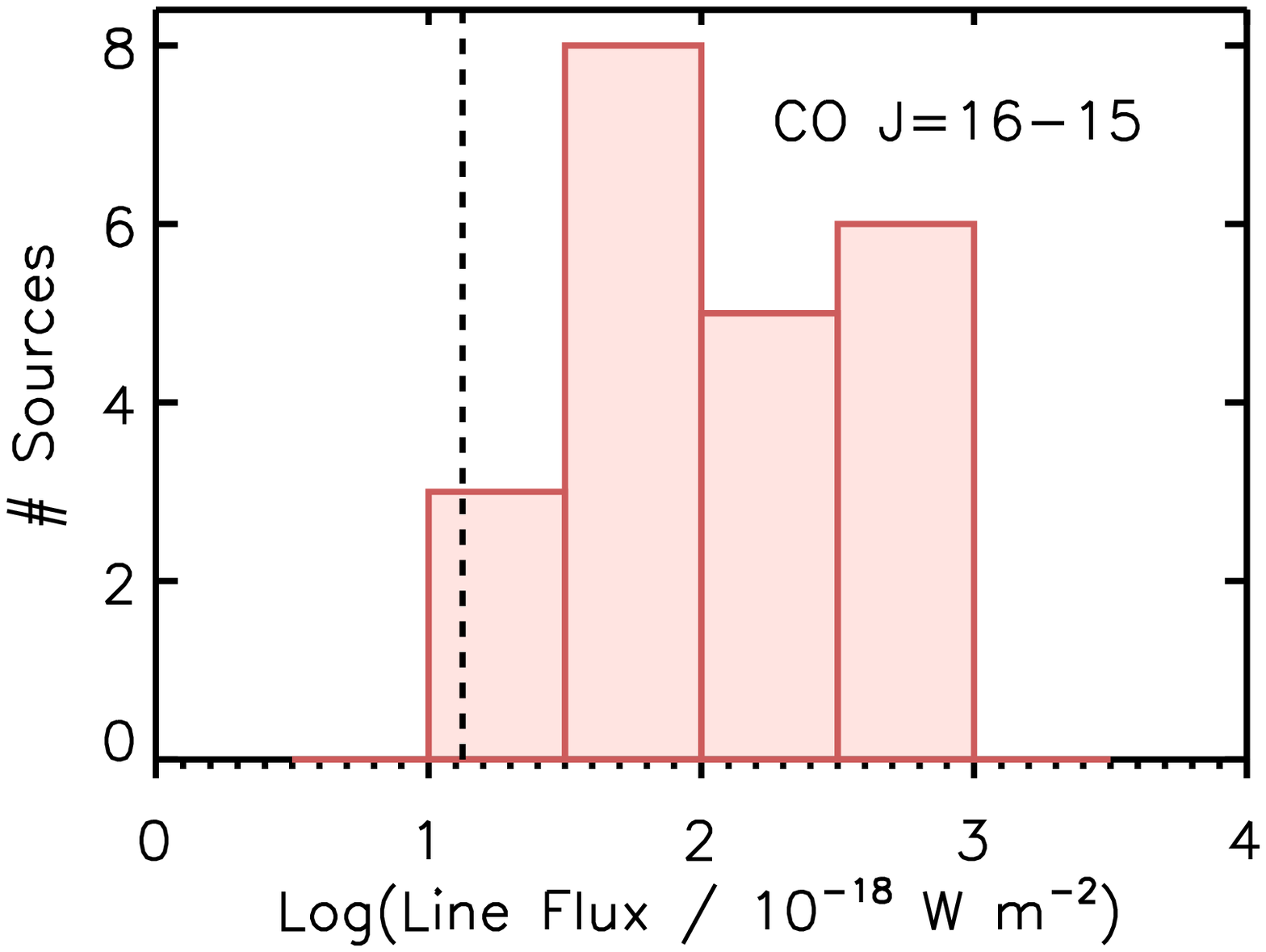}
\includegraphics[scale=0.4]{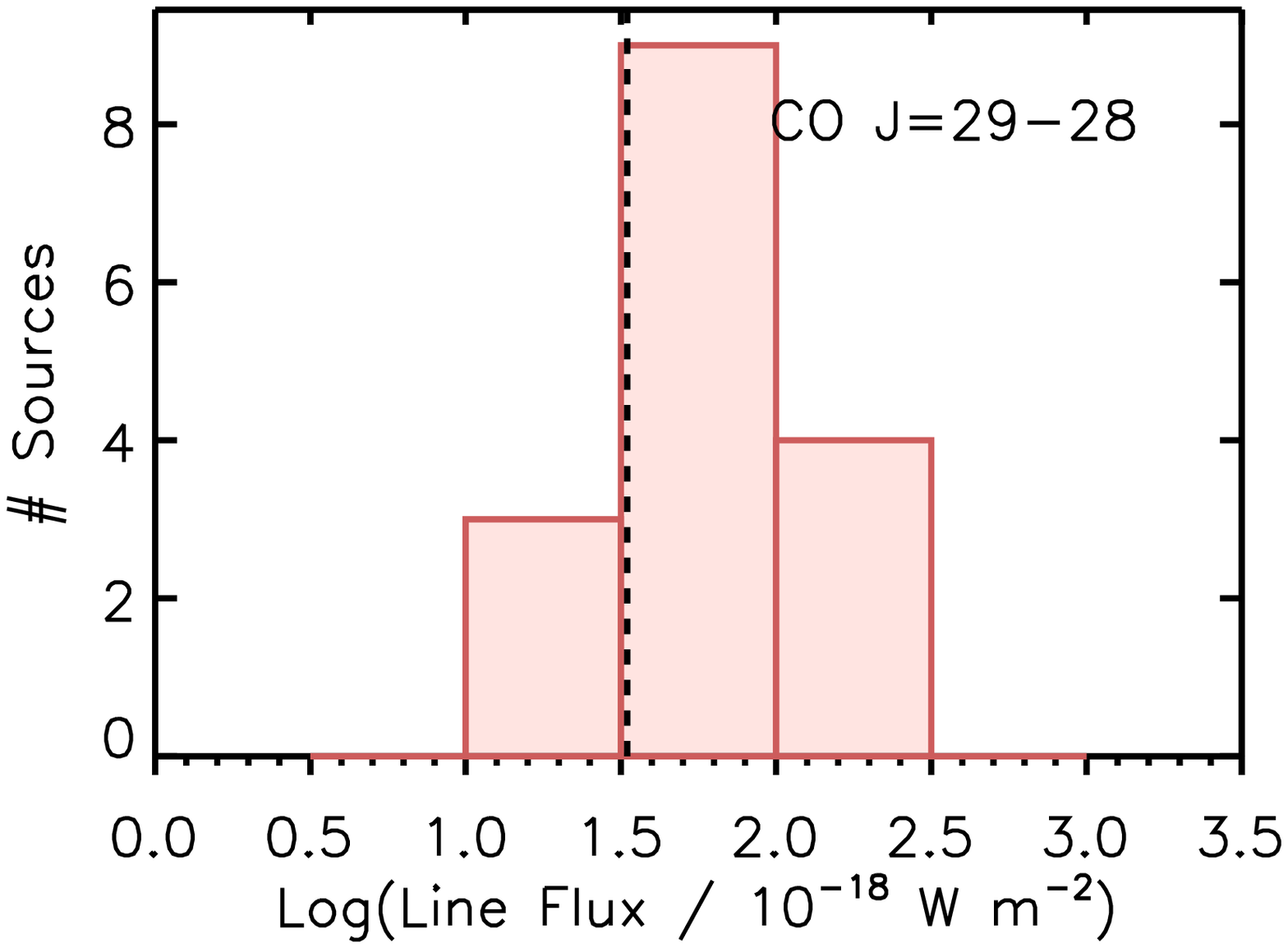}
\includegraphics[scale=0.4]{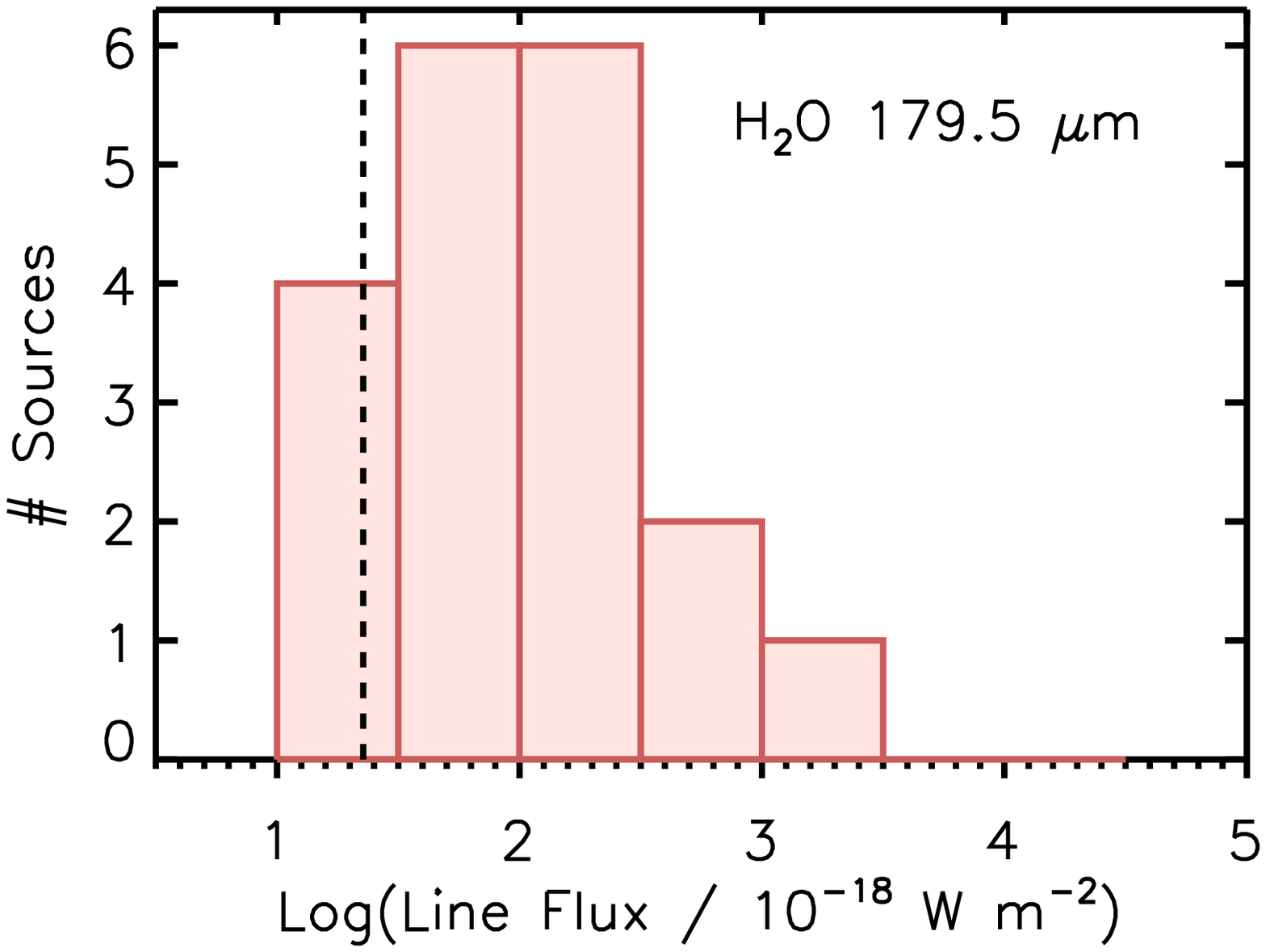}
\includegraphics[scale=0.4]{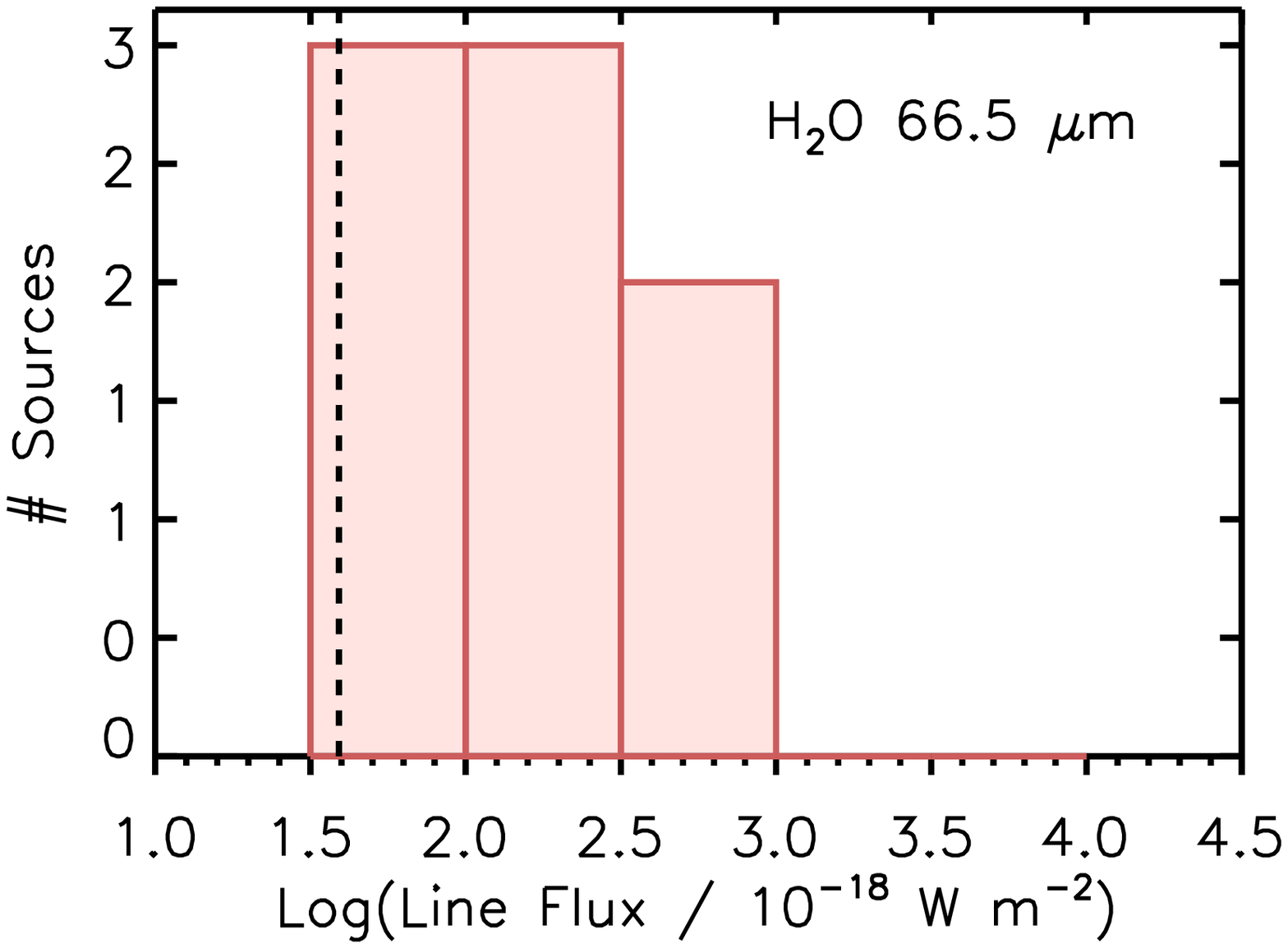}
\caption{Histograms of linefluxes for characteristic lines.   The dashed 
line represents the
1$\sigma$ uncertainty in lineflux for BHR71, one of the brighter sources in
the sample, and
can be taken as a conservative $1 \sigma$ noise.  The number of sources with
upper limits
is (25 - the total number plotted), for each histogram (ie. plots do not include the 5 confused 
sources in our sample).}
\label{hist1}
\end{center}
\end{figure}

\clearpage

\begin{figure}
\begin{center}
\includegraphics[scale=0.4]{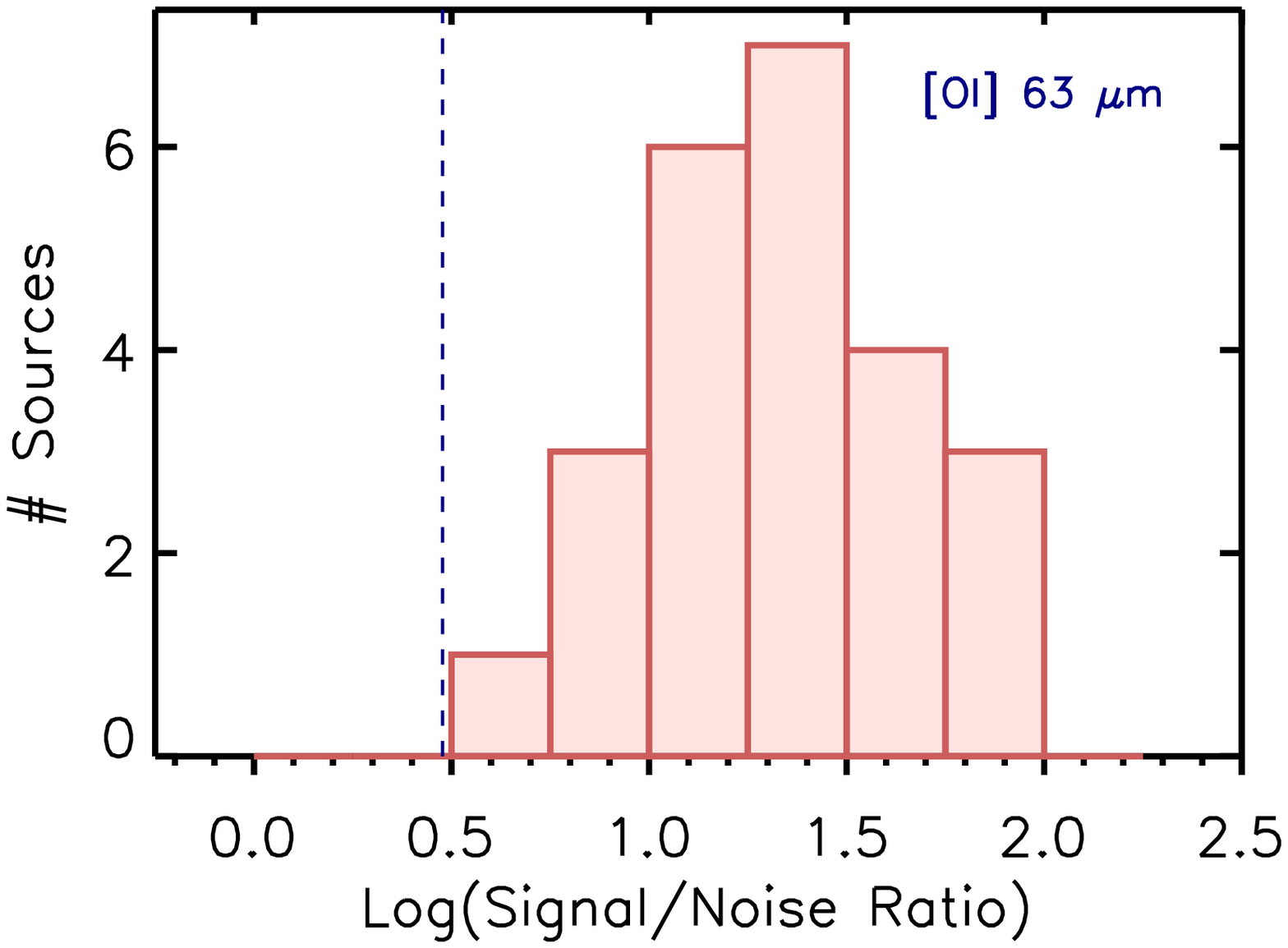}
\includegraphics[scale=0.4]{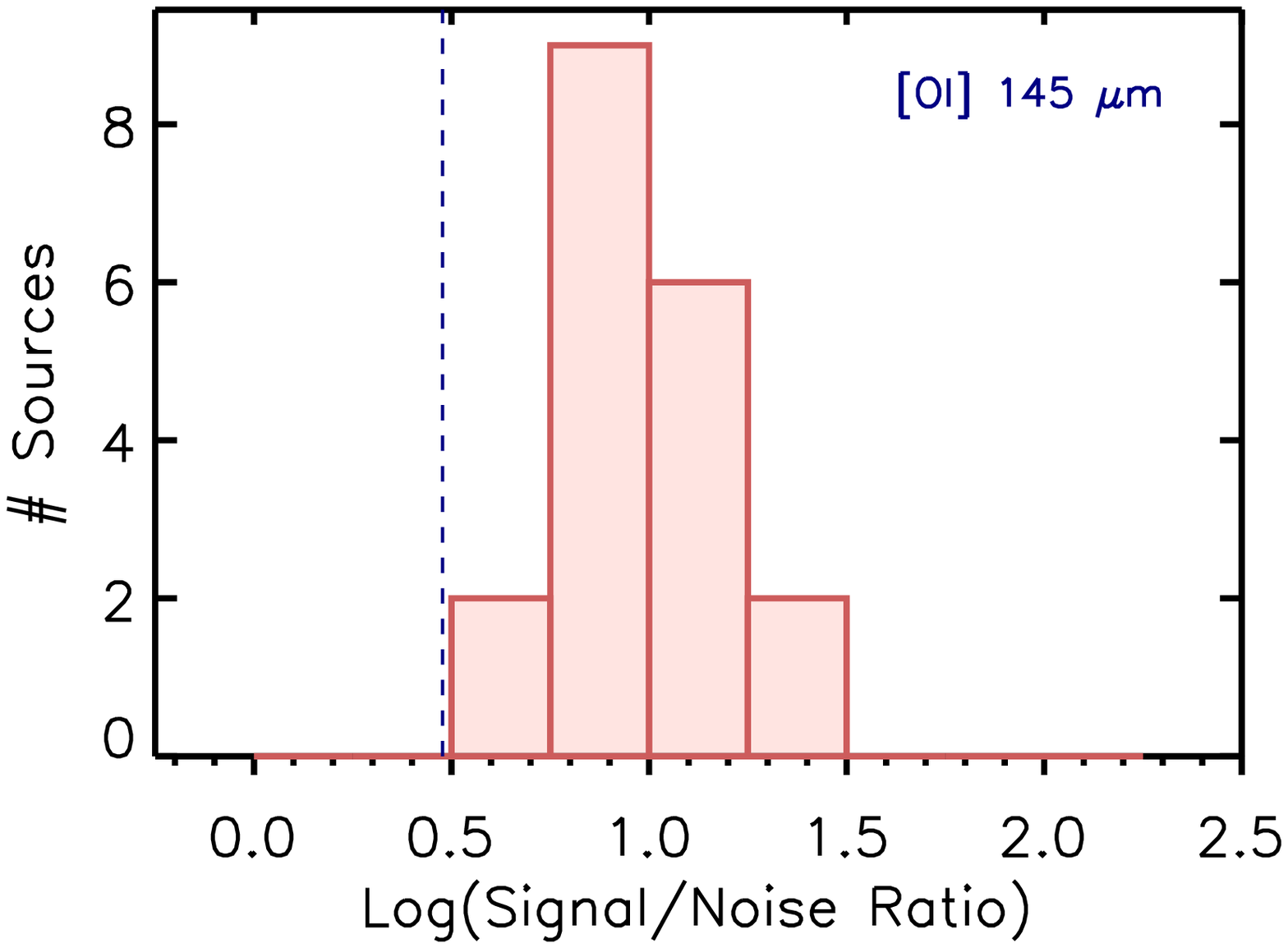}
\includegraphics[scale=0.4]{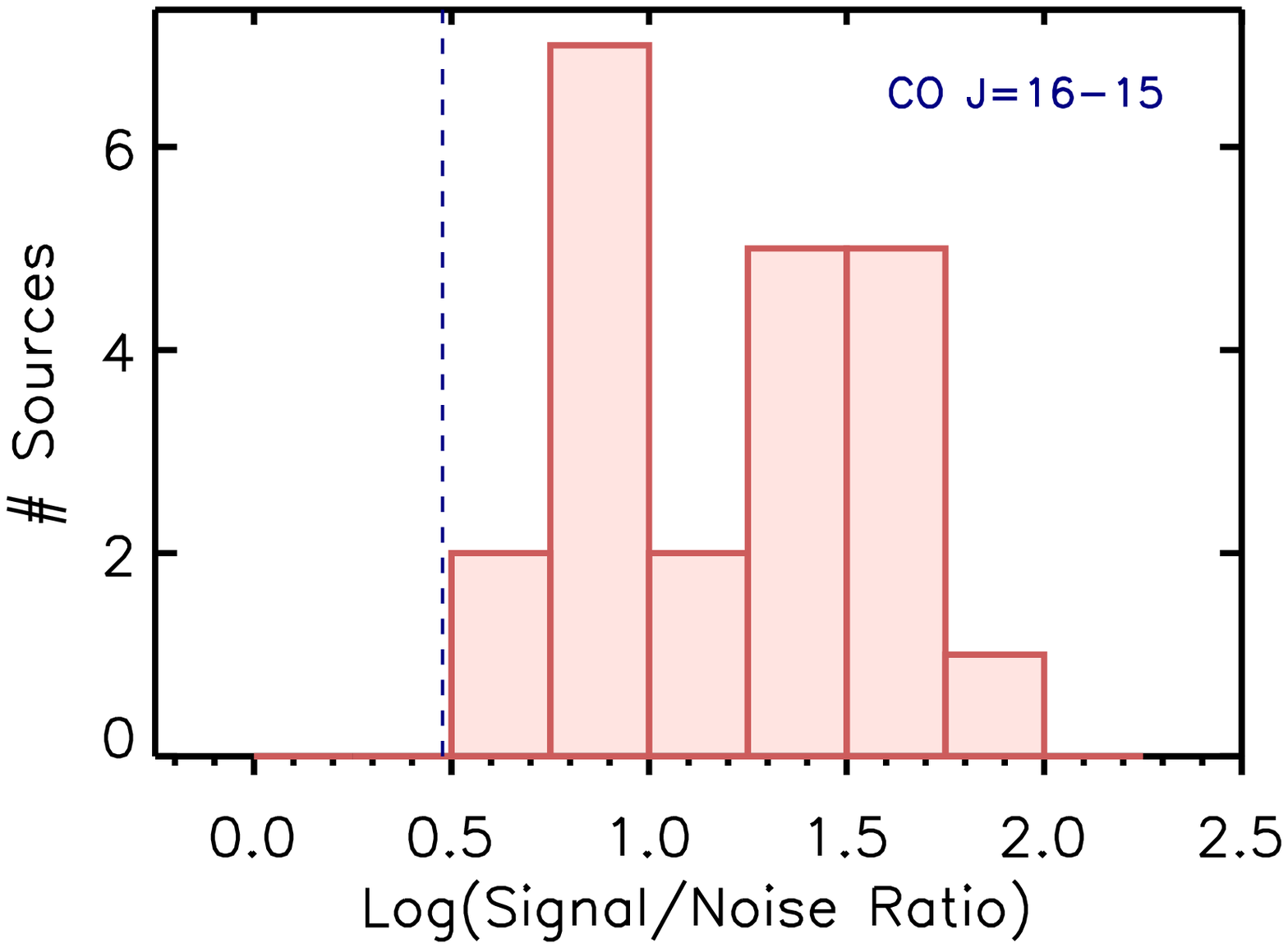}
\includegraphics[scale=0.4]{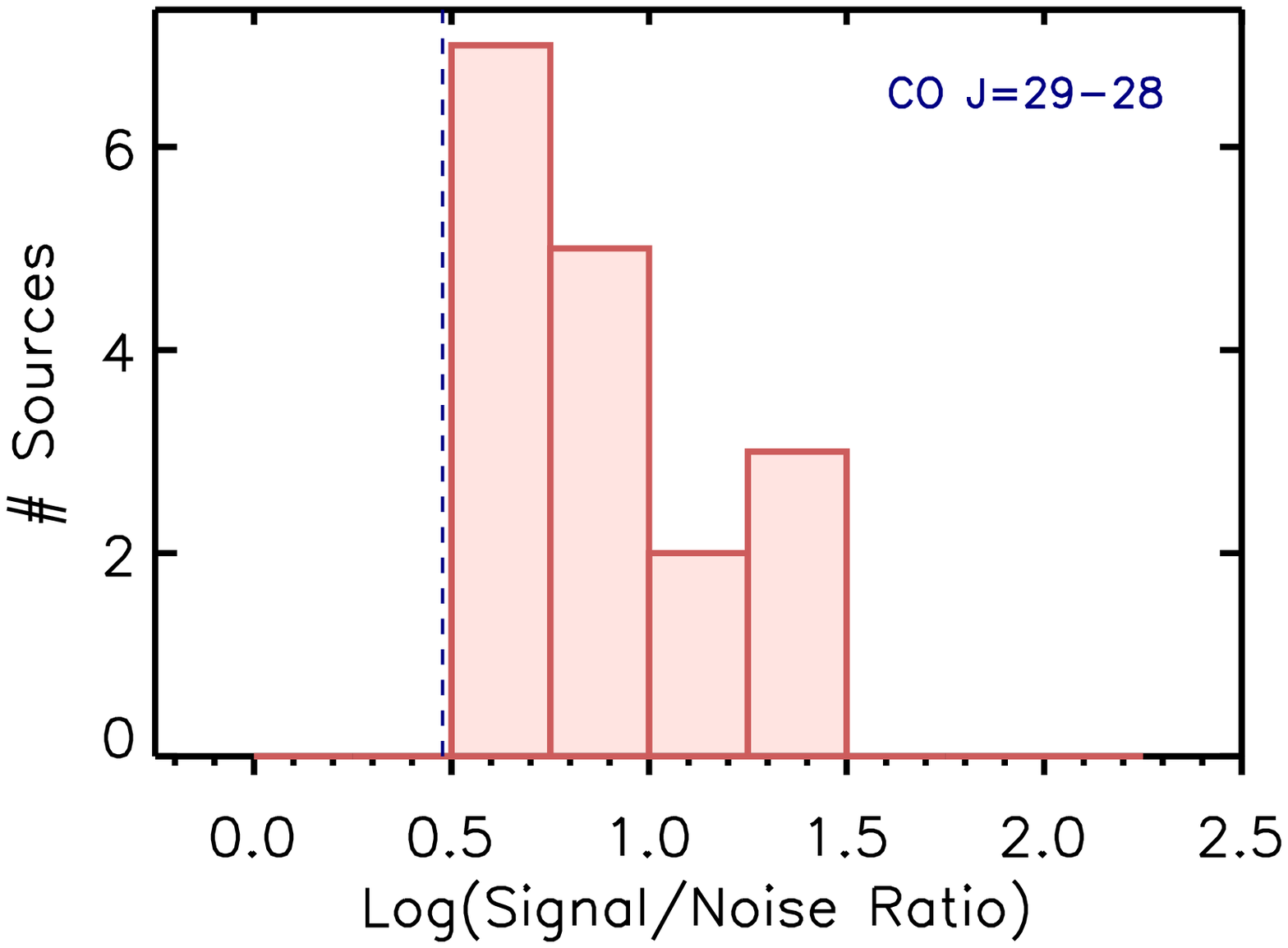}
\caption{Histograms of the S/N ratio for characteristic lines.  The blue vertical dashed line indicates a S/N of 3.  
We are complete in \OI\ 63 $\mu$m and CO \jj{16}{15}, but less so in \OI\ 145 $\mu$m and CO \jj{29}{28}.}
\label{histsn}
\end{center}
\end{figure}

\begin{figure}
\includegraphics[scale=0.43]{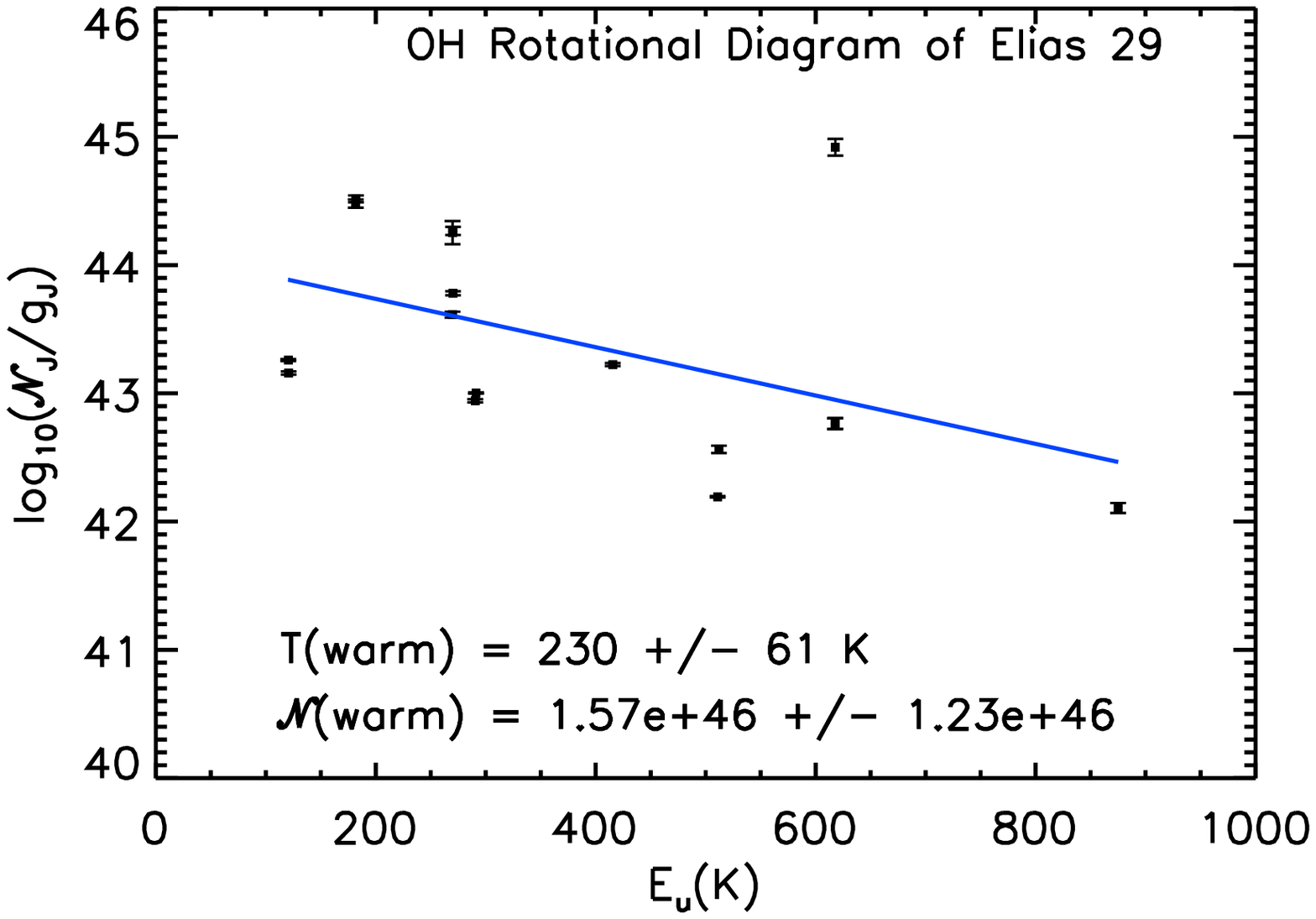}
\includegraphics[scale=0.43]{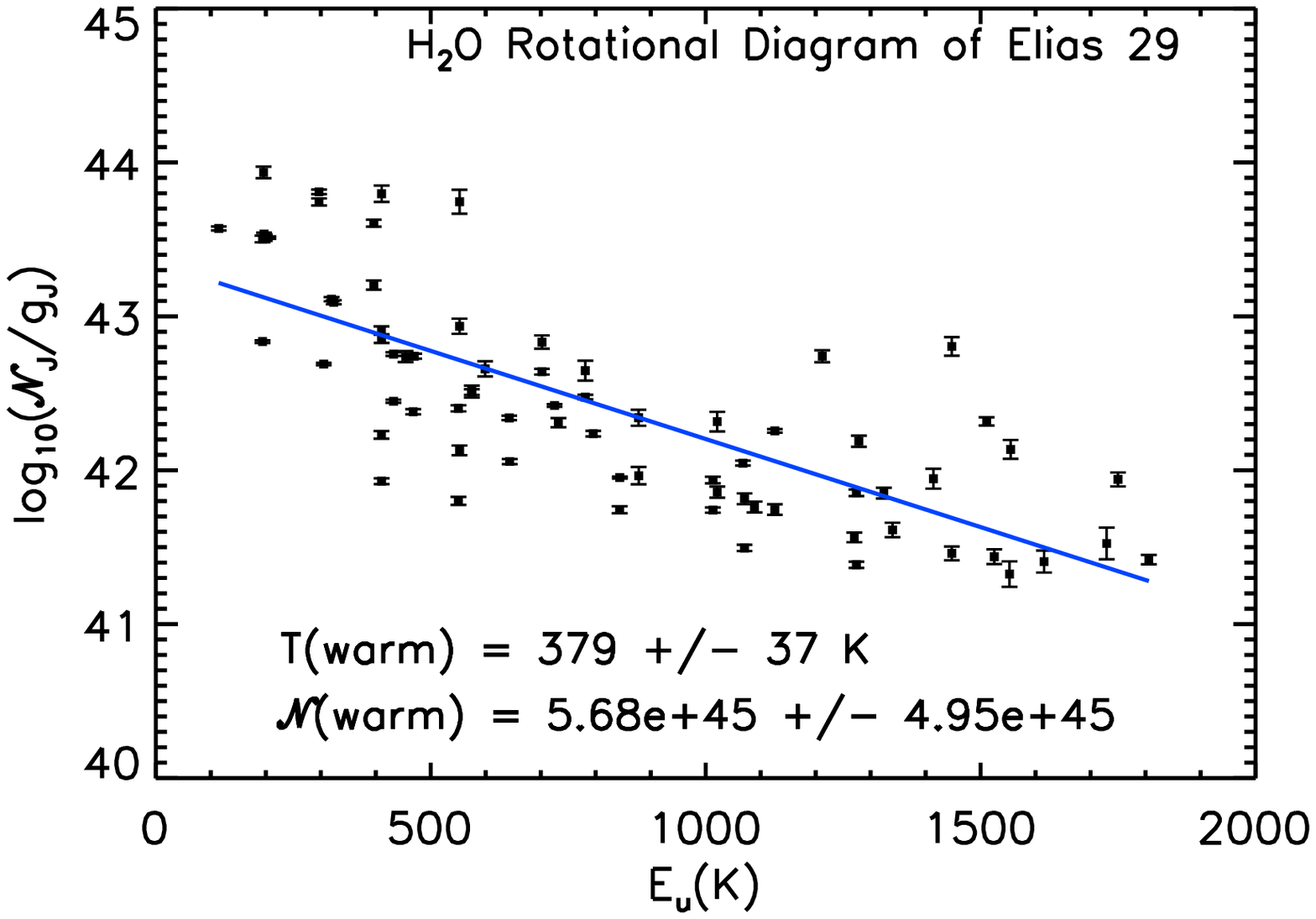}
\includegraphics[scale=0.43]{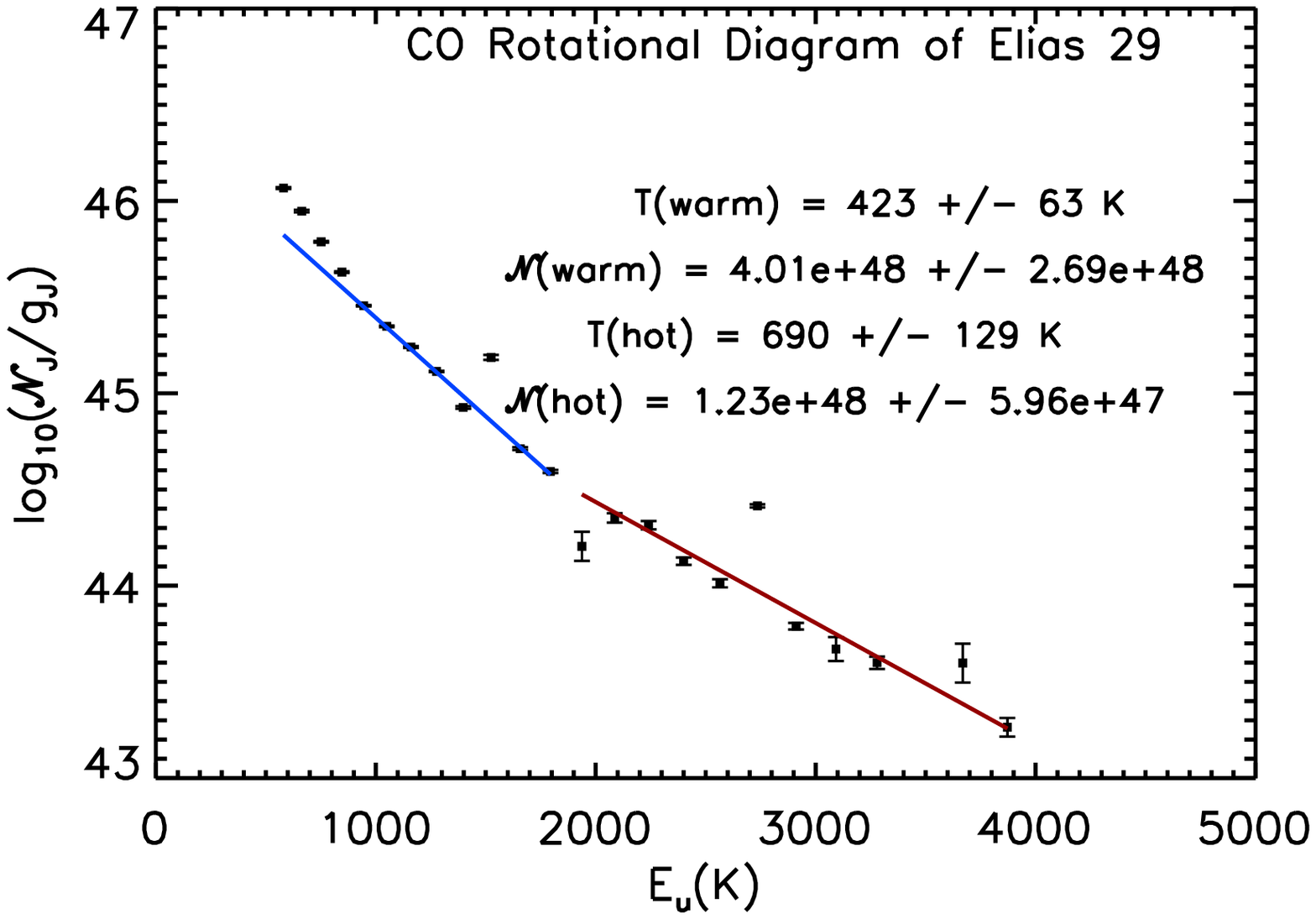}
\includegraphics[scale=0.43]{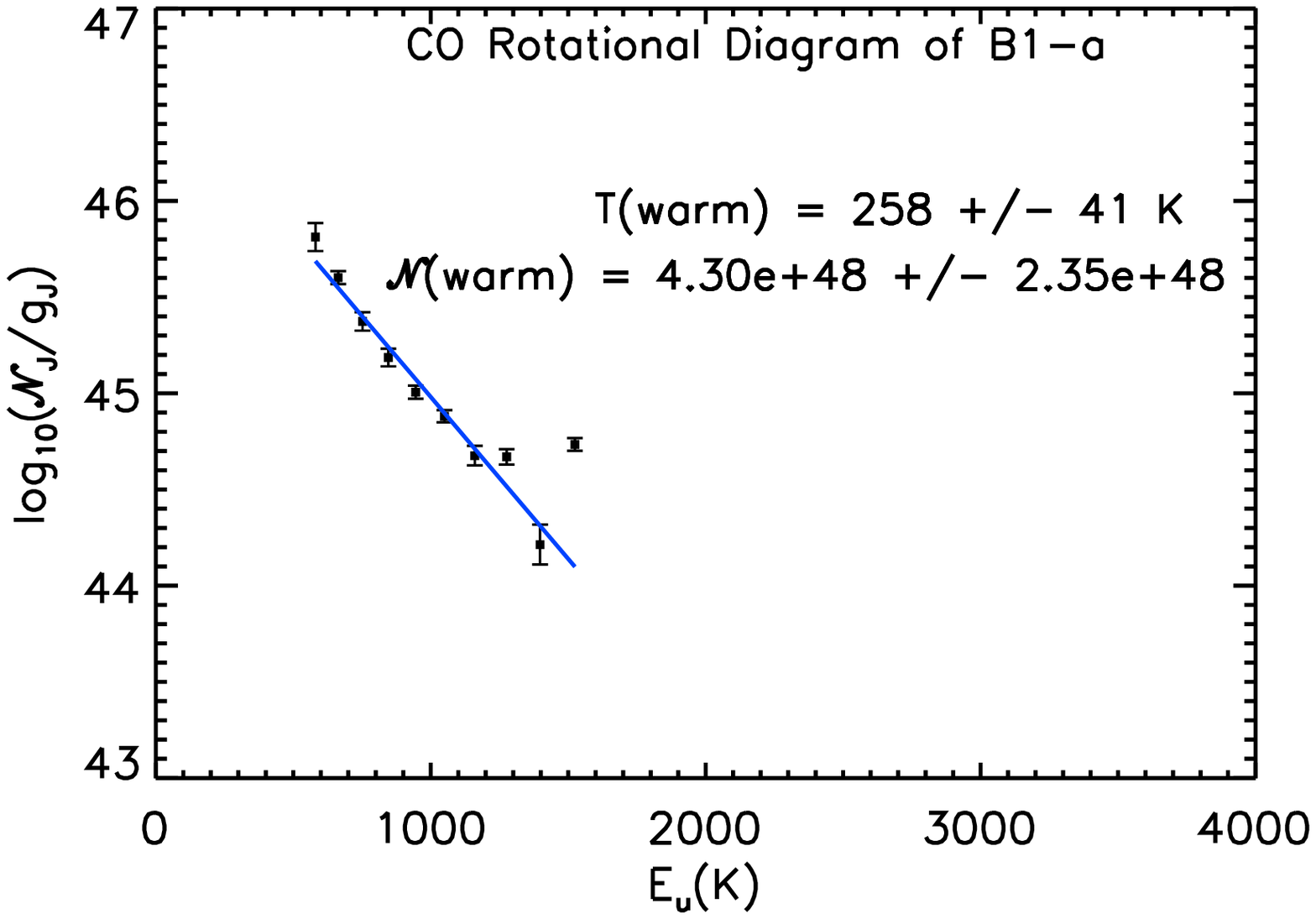}
\caption{Rotational diagrams for OH (upper left), H$_2$O (upper right), 
and CO (lower left) for
Elias 29, showing the results of typical fits.  We also show the rotational 
diagram for a source with only a detected warm component, B1-a (lower right).
In all CO rotational diagram fits, we 
ignore the CO \jj{23}{22} ($\eup\ = 1524$ K) and CO \jj{31}{30} ($\eup\ = 2735$ K) 
datapoints, which are blended with 
H$_2$O and OH respectively. 
}
\label{rotdiag}
\end{figure}
\begin{figure}
\includegraphics[scale=1.0,angle=0]{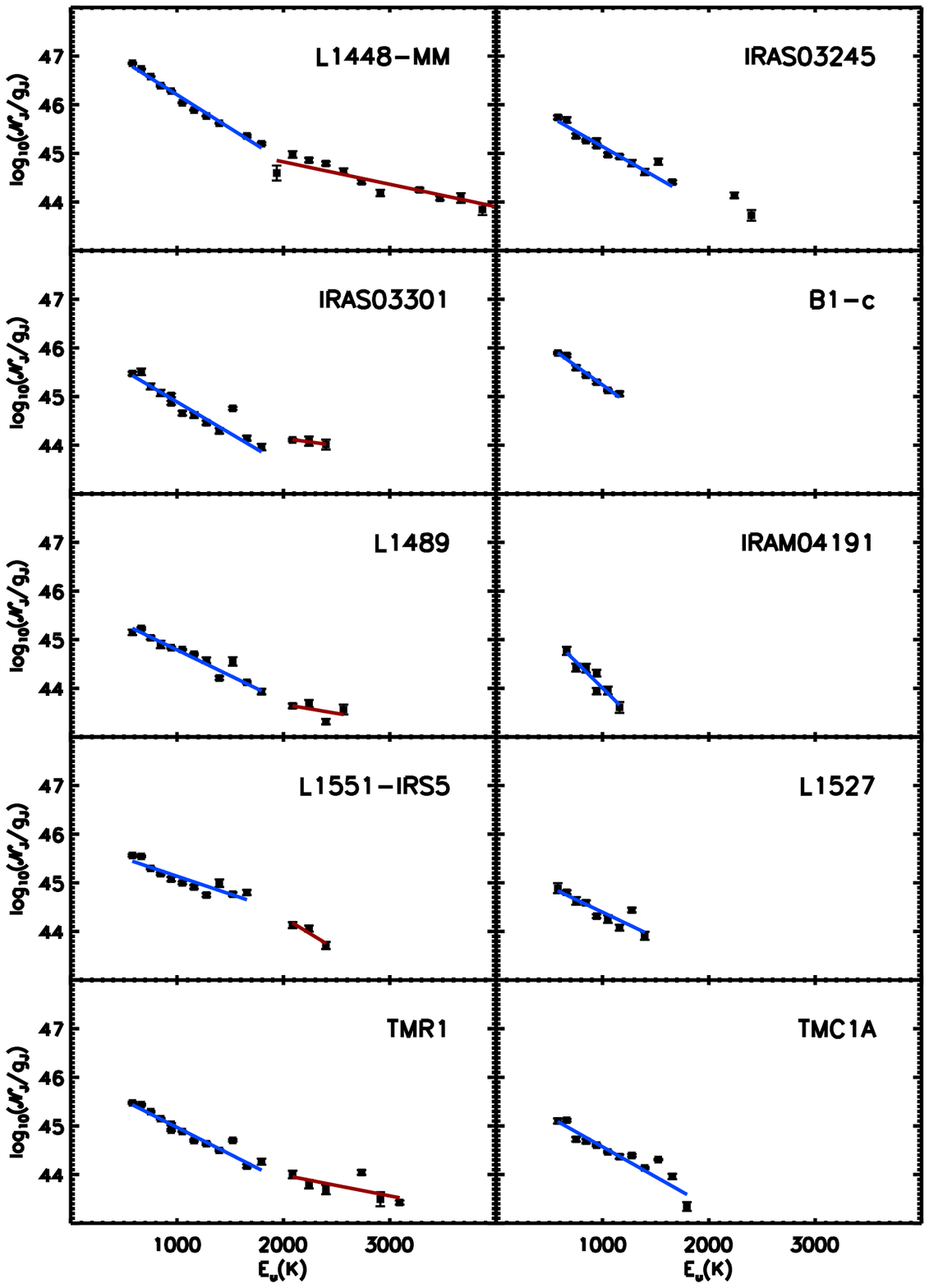}
\caption{CO rotational diagrams for the full sample.}
\label{rotdiag1}
\end{figure}

\begin{figure}
\includegraphics[scale=1.0,angle=0]{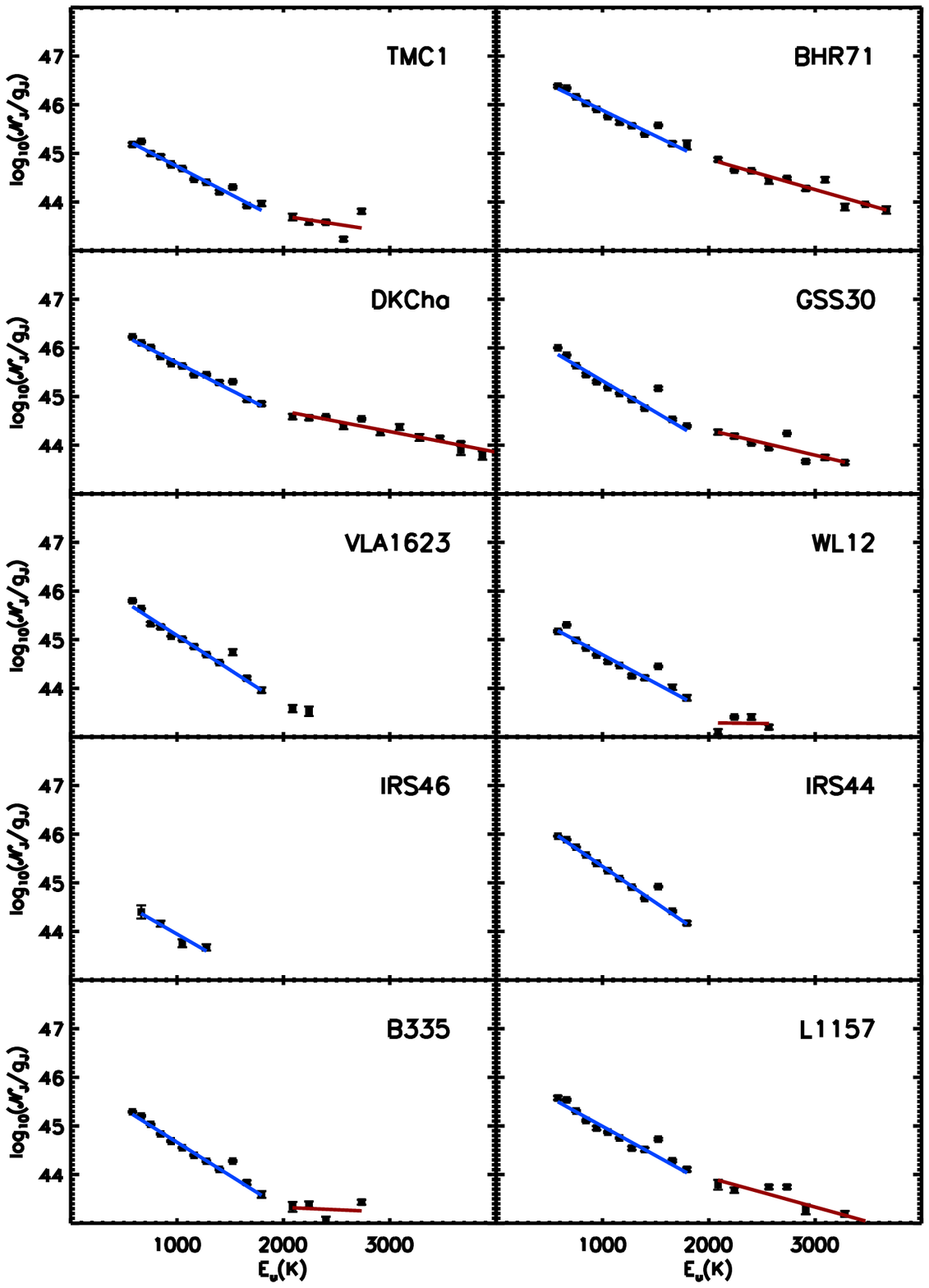}
\caption{CO rotational diagrams for the full sample (cont.).}
\label{rotdiag2}
\end{figure}

\begin{figure}
\begin{center}
\includegraphics[scale=0.4]{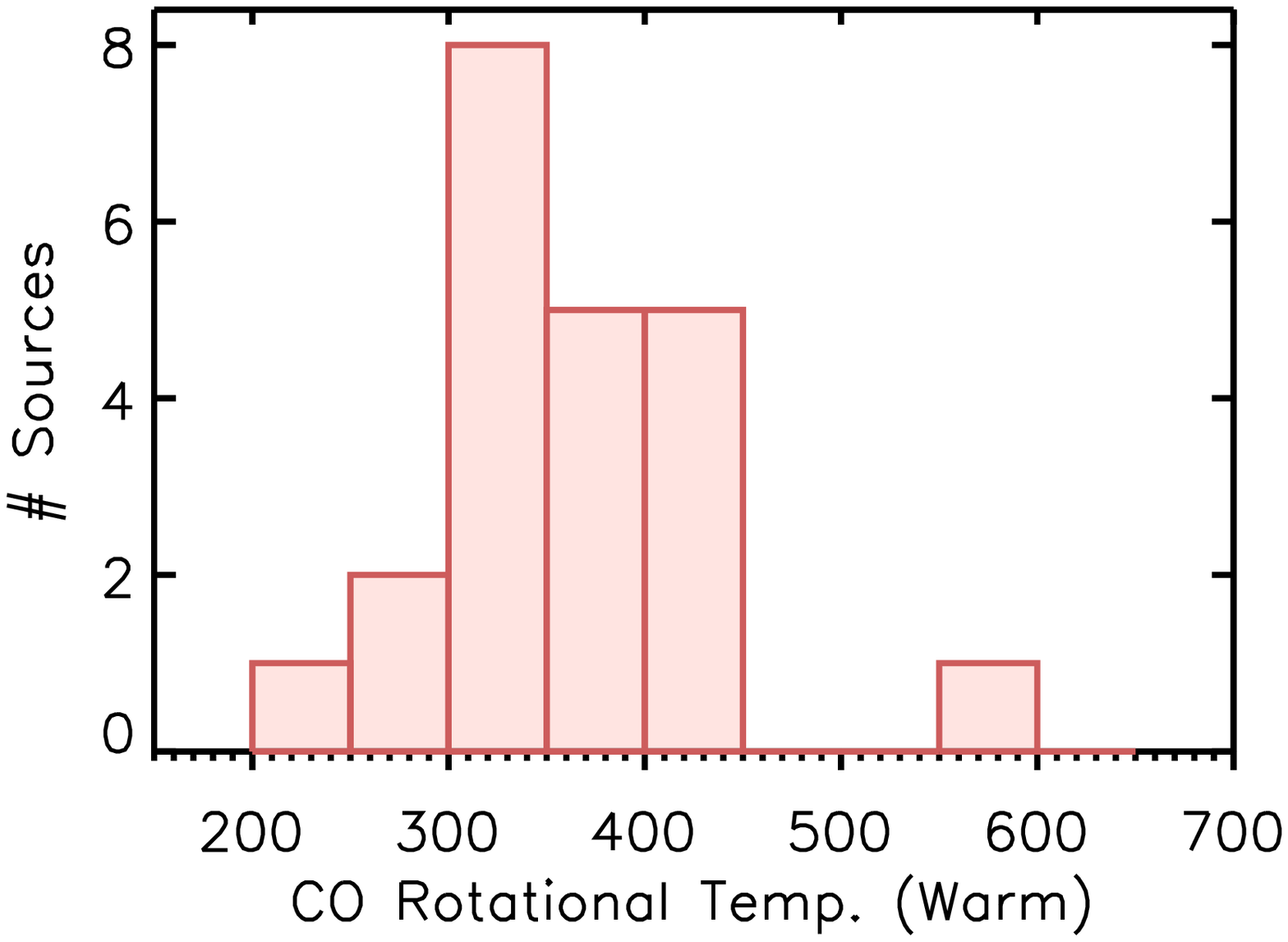}
\includegraphics[scale=0.4]{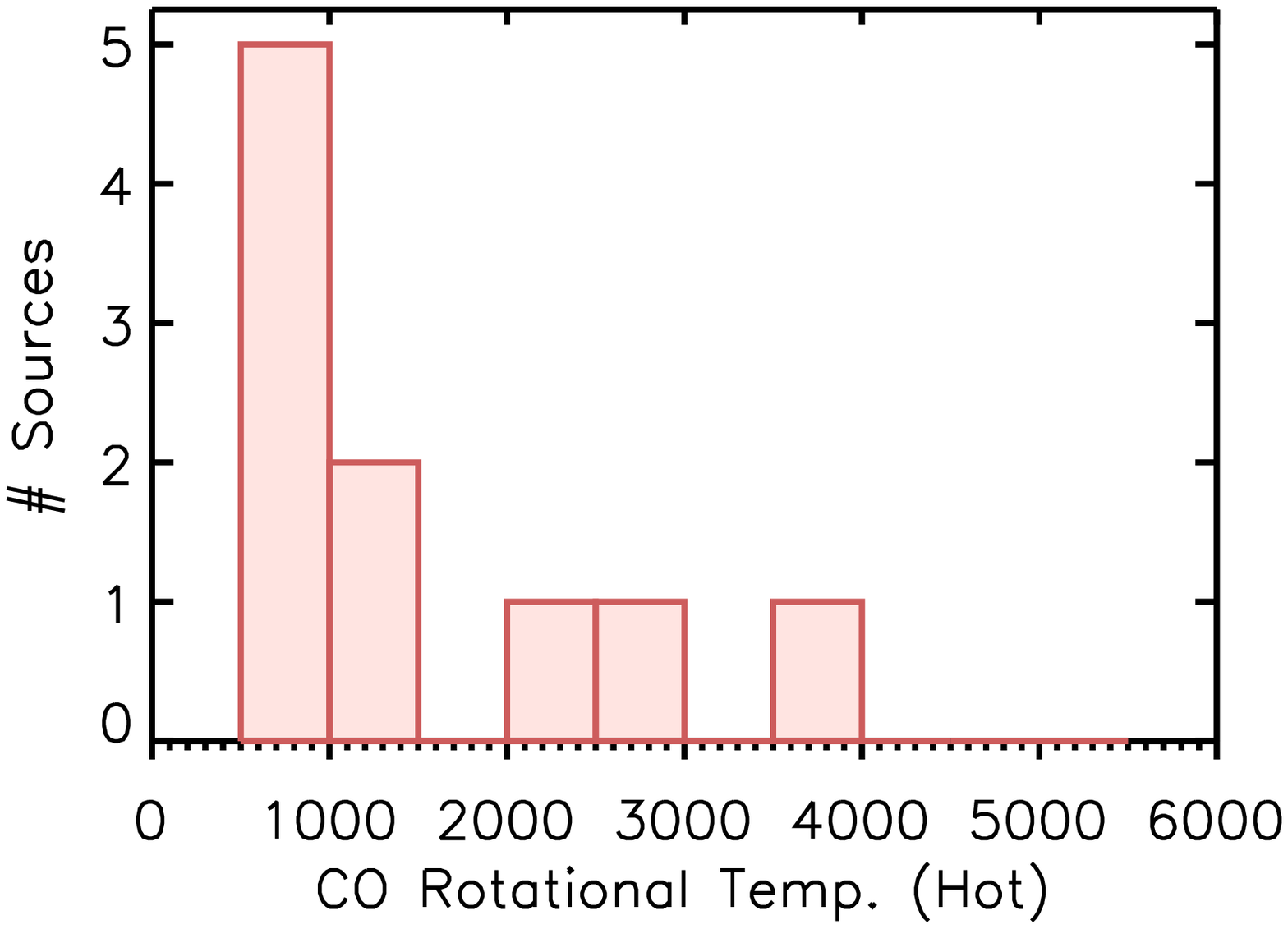}
\includegraphics[scale=0.4]{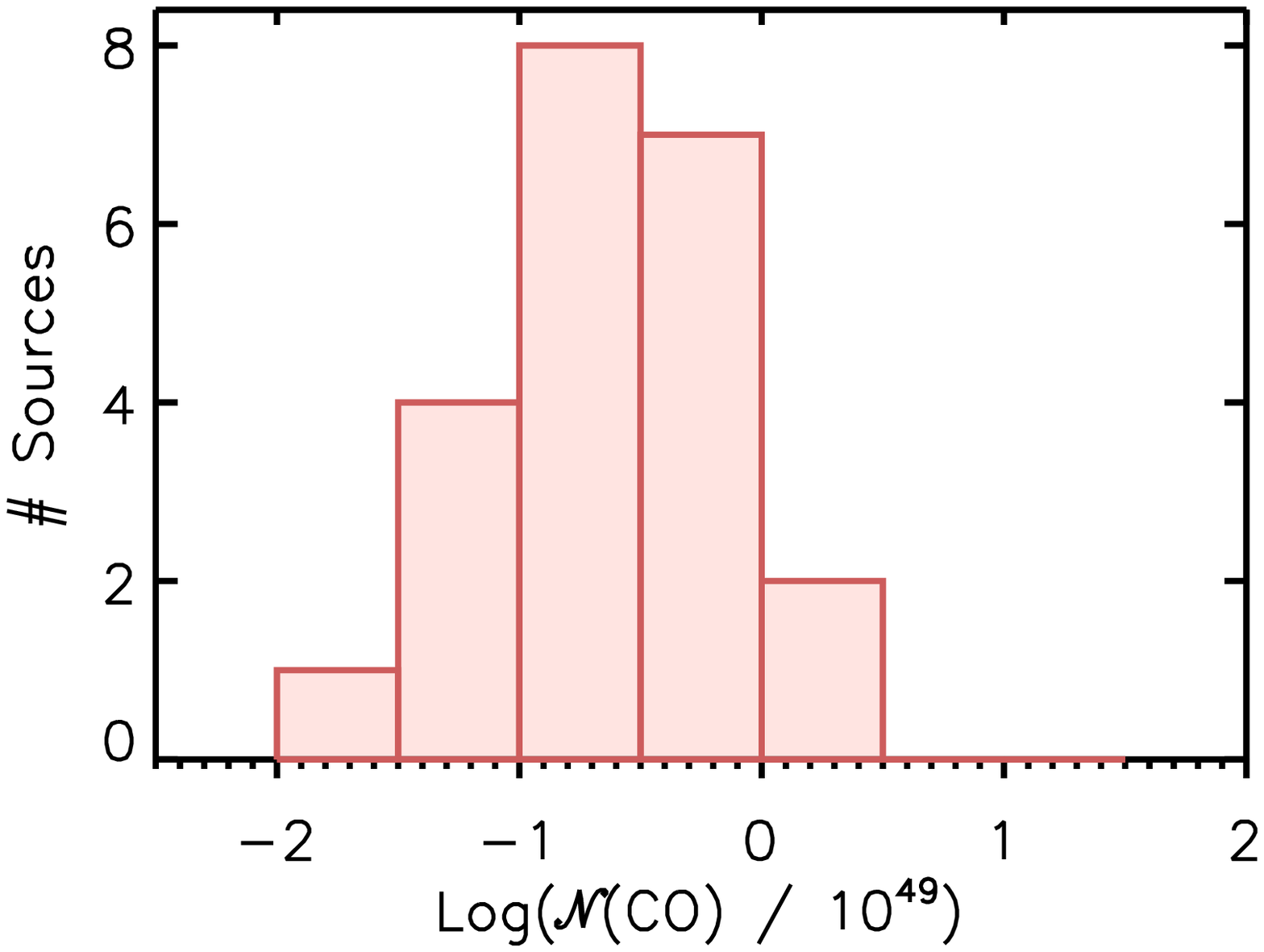}
\includegraphics[scale=0.4]{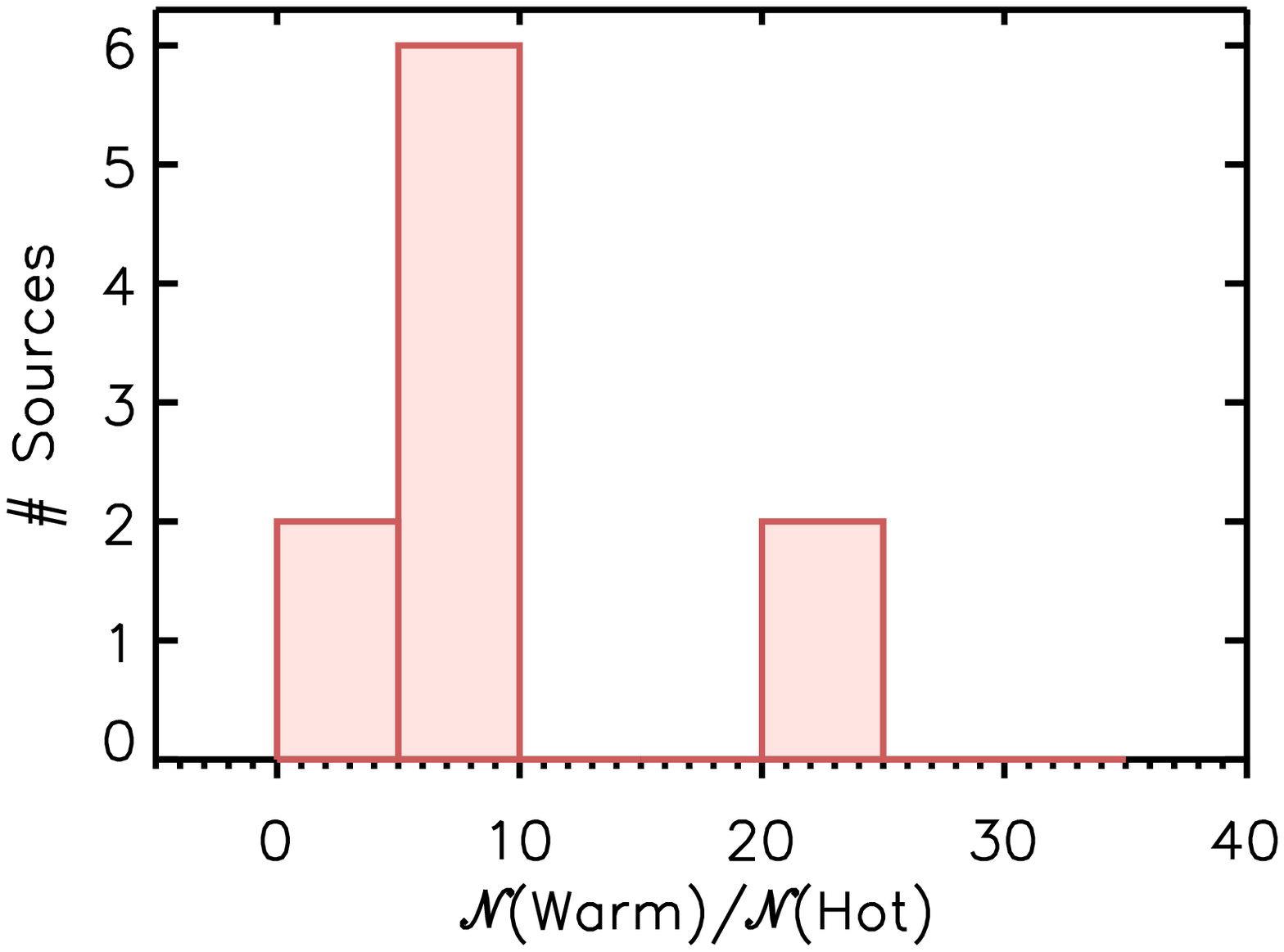}
\includegraphics[scale=0.4]{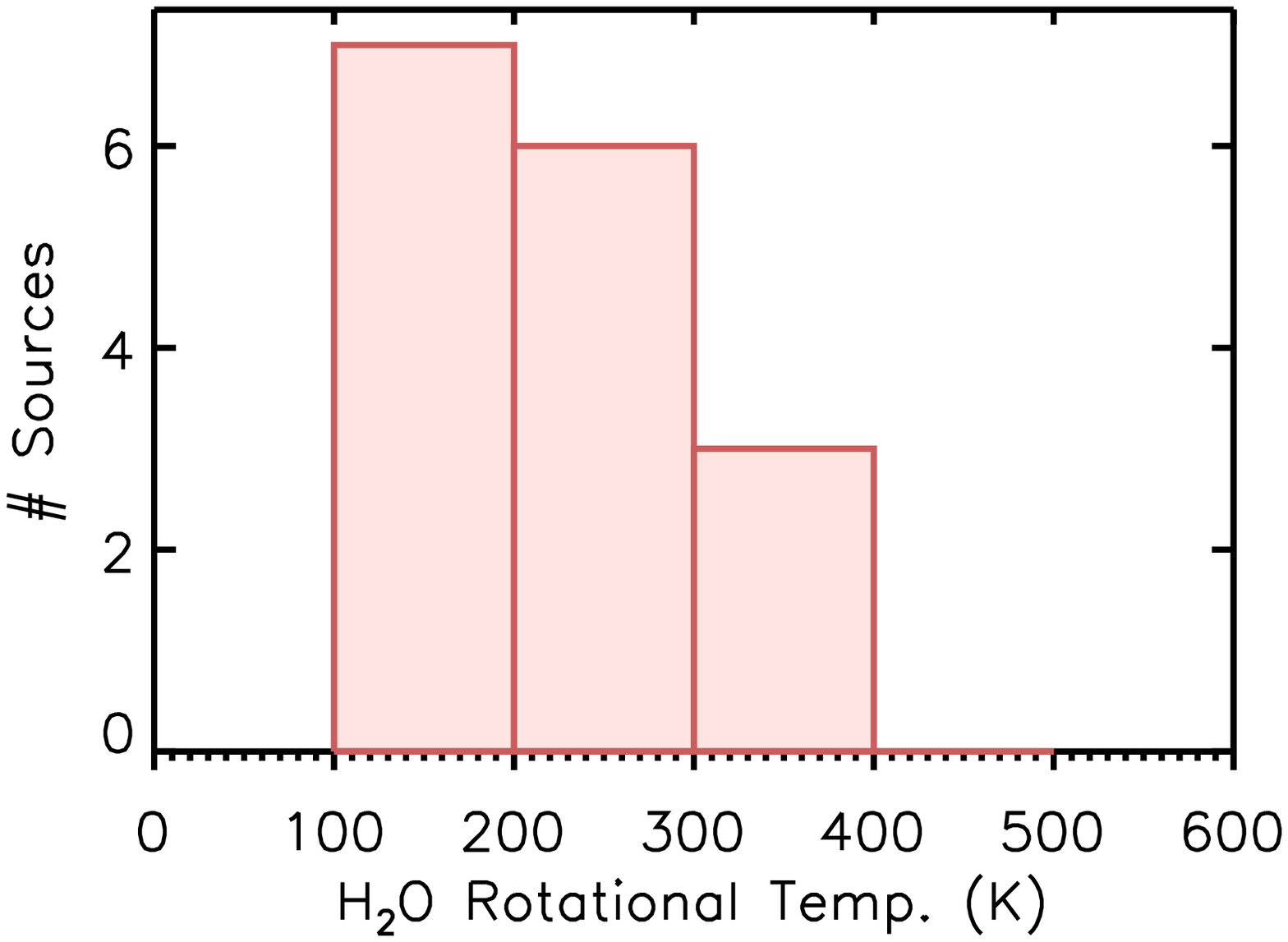}
\includegraphics[scale=0.4]{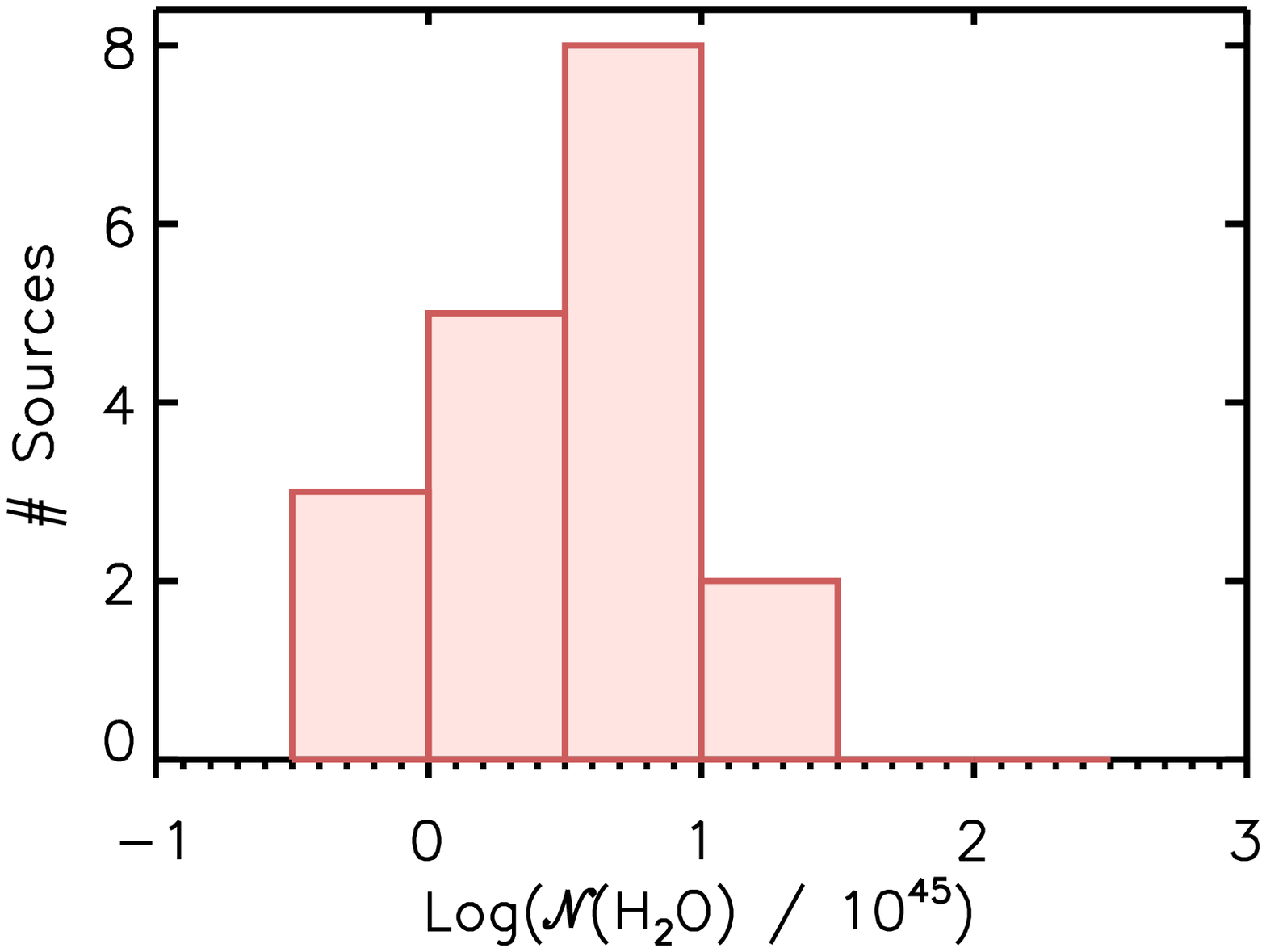}
\caption{Histograms of \funnyN\ and \trot\ for CO and H$_2$O.  {\bf Top:} 
CO rotational temperatures for the warm and hot components.  {\bf Middle:} 
Total number of CO molecules for the warm component, and ratio of molecules 
in the warm/hot populations.  {\bf Bottom:} Temperature and \funnyN\ for H$_2$O.}
\label{hist2}
\end{center}
\end{figure}

\begin{figure}
\begin{center}
\includegraphics[scale=0.5]{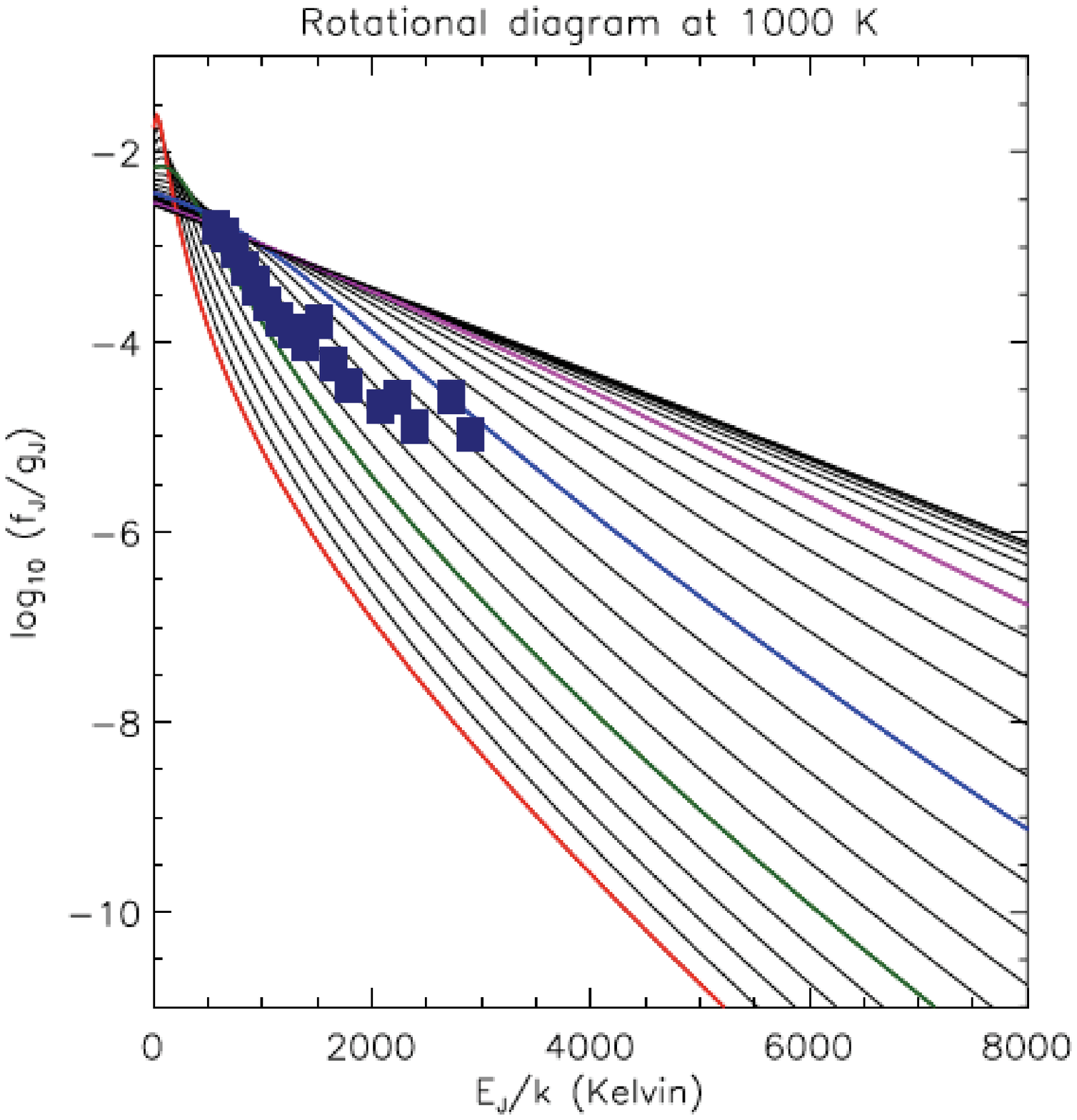}
\includegraphics[scale=0.55]{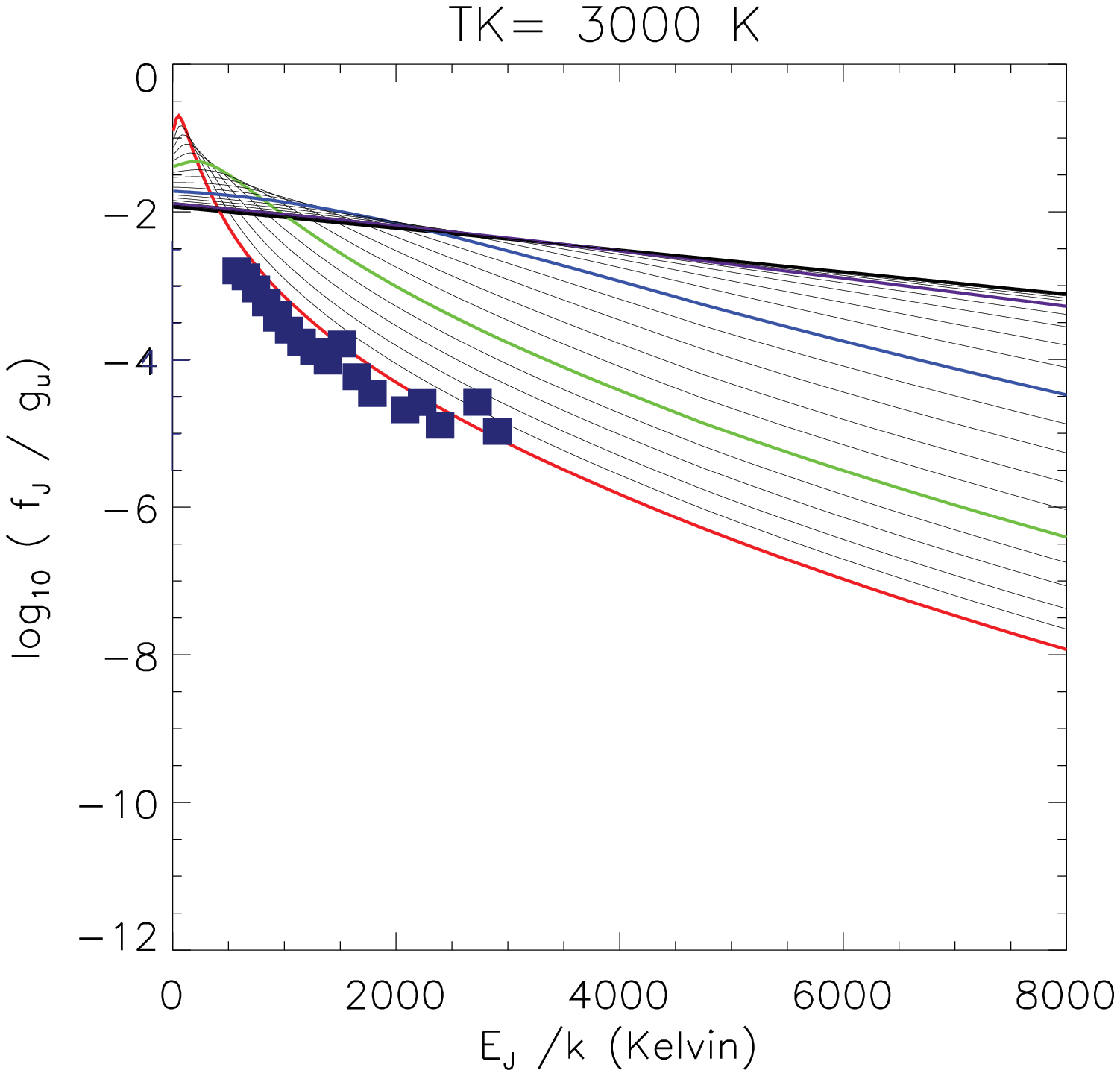}
\caption{{\bf Top:} Comparison of the CO state populations (blue squares) in
B335 overplotted on  models for a
1000 K isothermal, optically thin emitting region, from Fig. 5 in
\citet{neufeld12}.  We use the same axis labels for clarity; $E_{\rm J}$ is the energy of the 
upper state J.  The solid lines are constant density; the purple line is the solution for n$=$ 10$^7$ 
cm$^{-3}$,  the blue line is for 10$^6$, the green line is for 10$^5$, and the red (lowest) line is for 10$^4$. 
 {\bf Bottom:} The same data compared to  a 3000 K model.}
\label{neufeld}
\end{center}
\end{figure}

\begin{figure}
\includegraphics[scale=0.43]{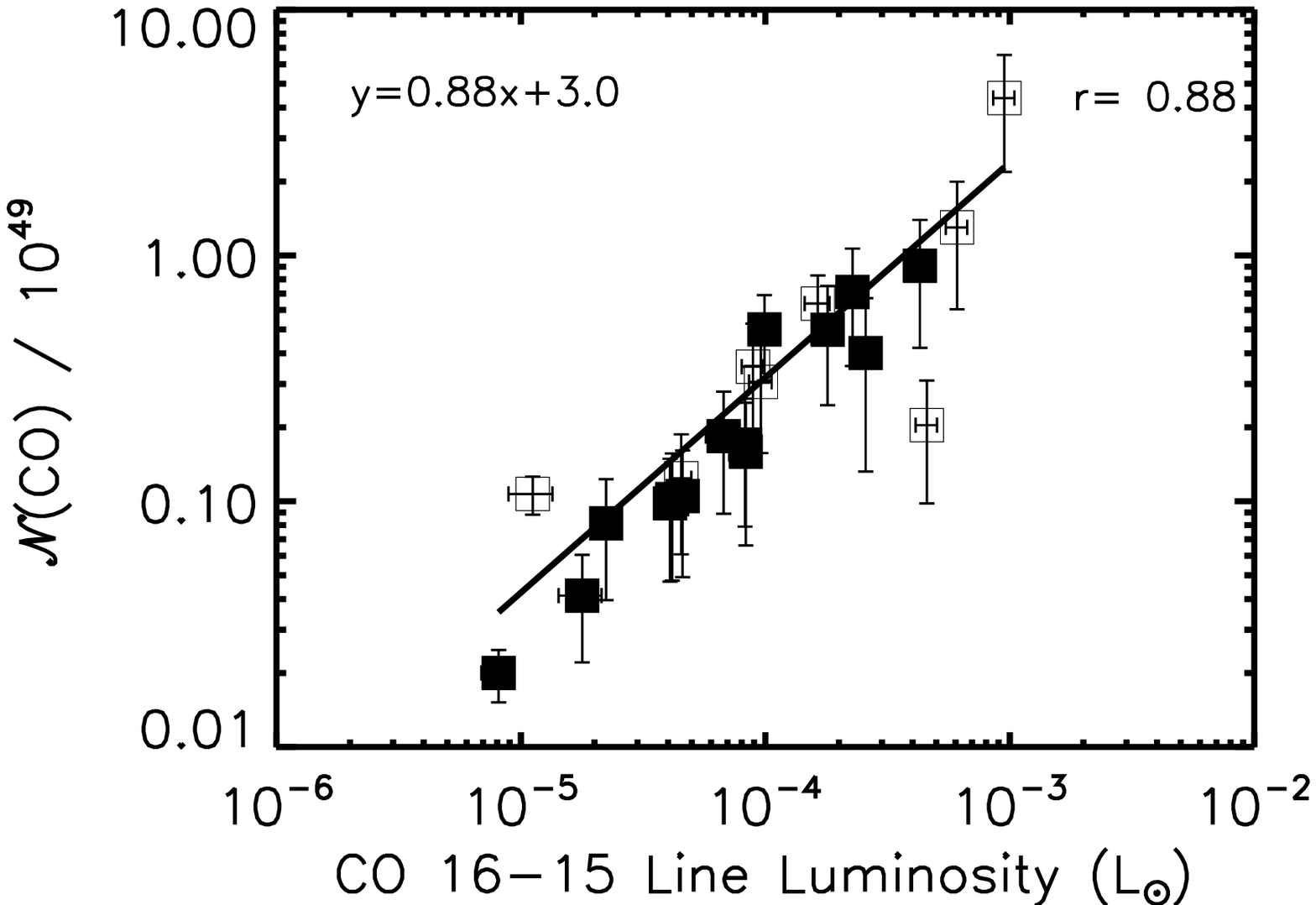}
\includegraphics[scale=0.43]{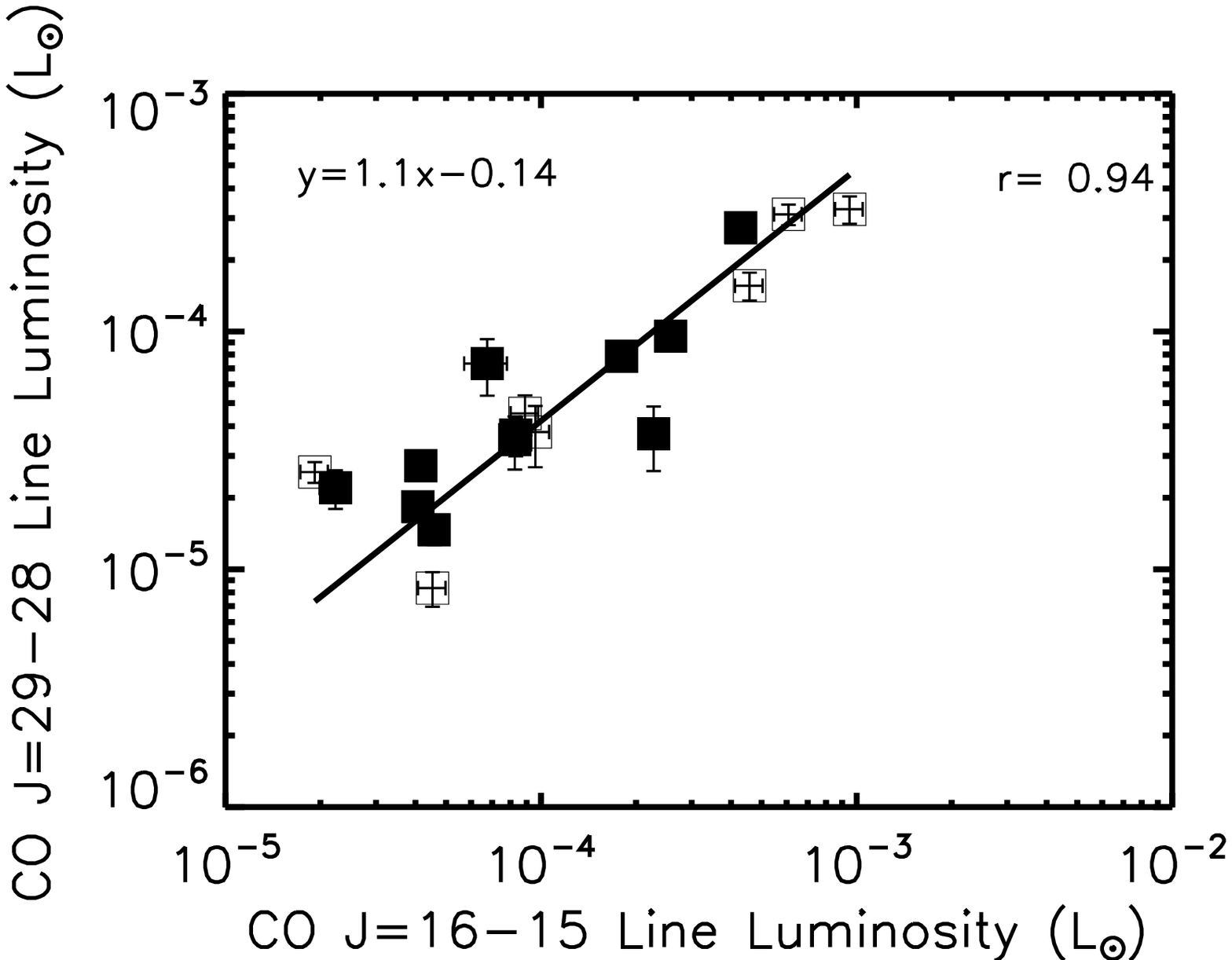}
\includegraphics[scale=0.43]{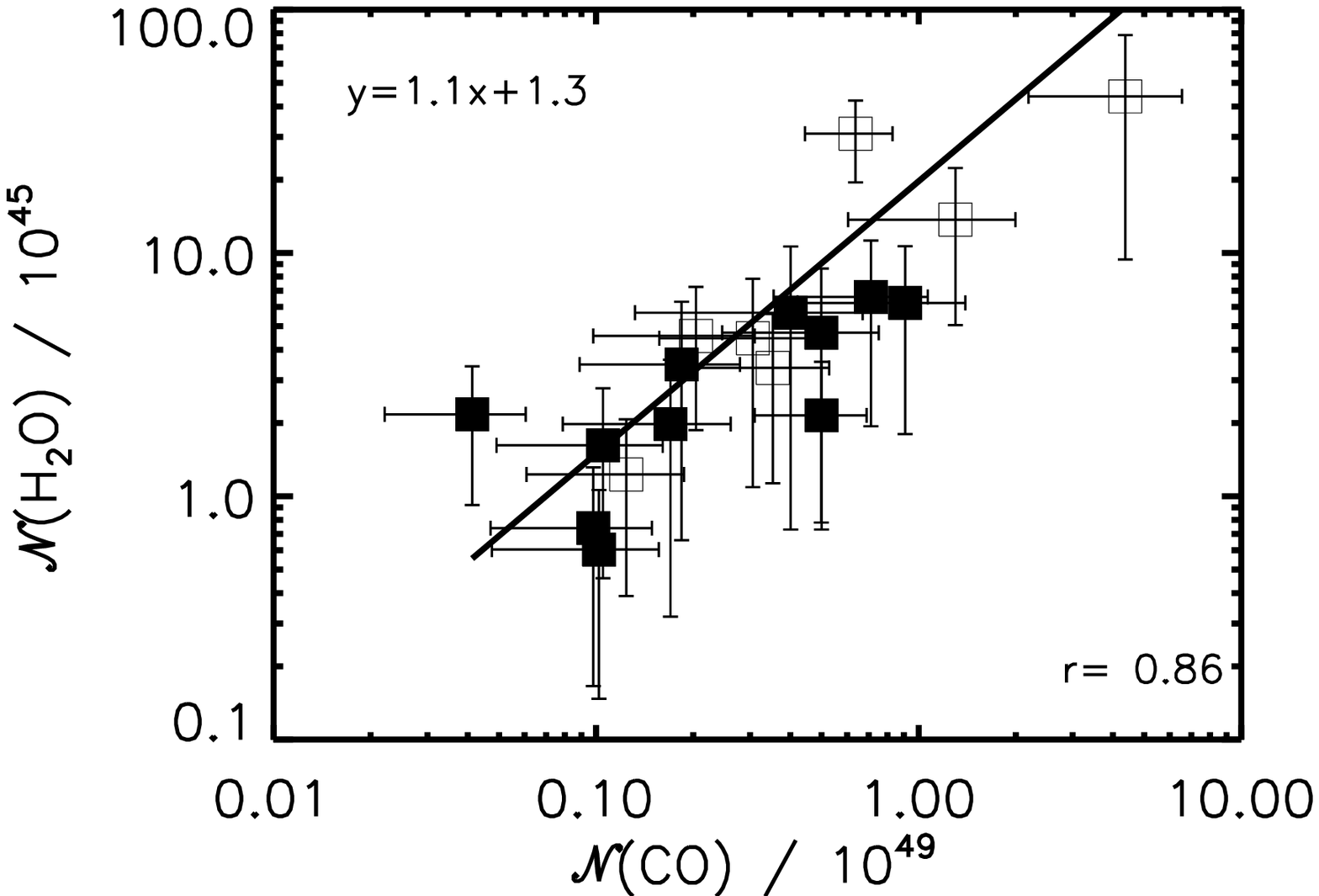}
\includegraphics[scale=0.43]{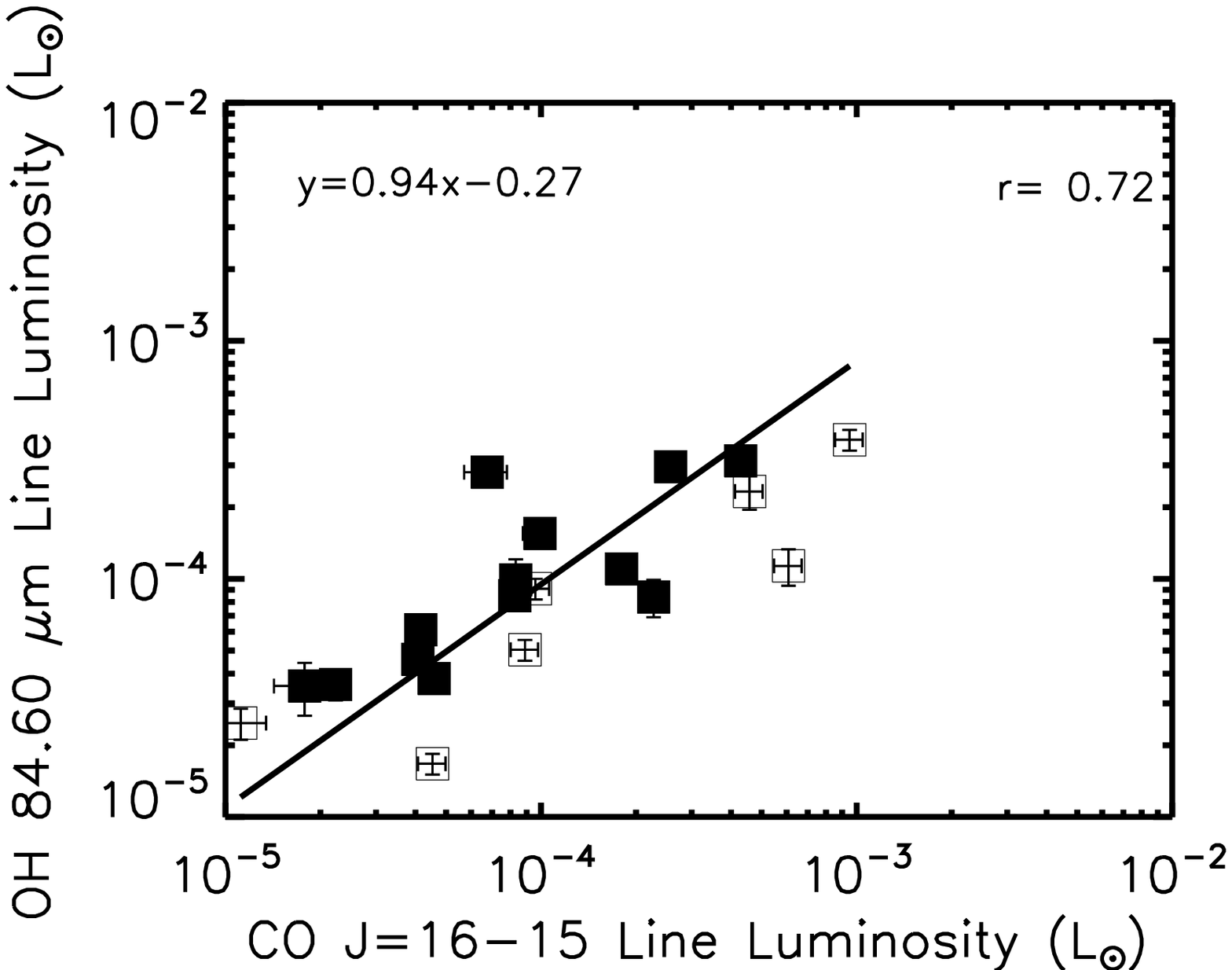}
\includegraphics[scale=0.43]{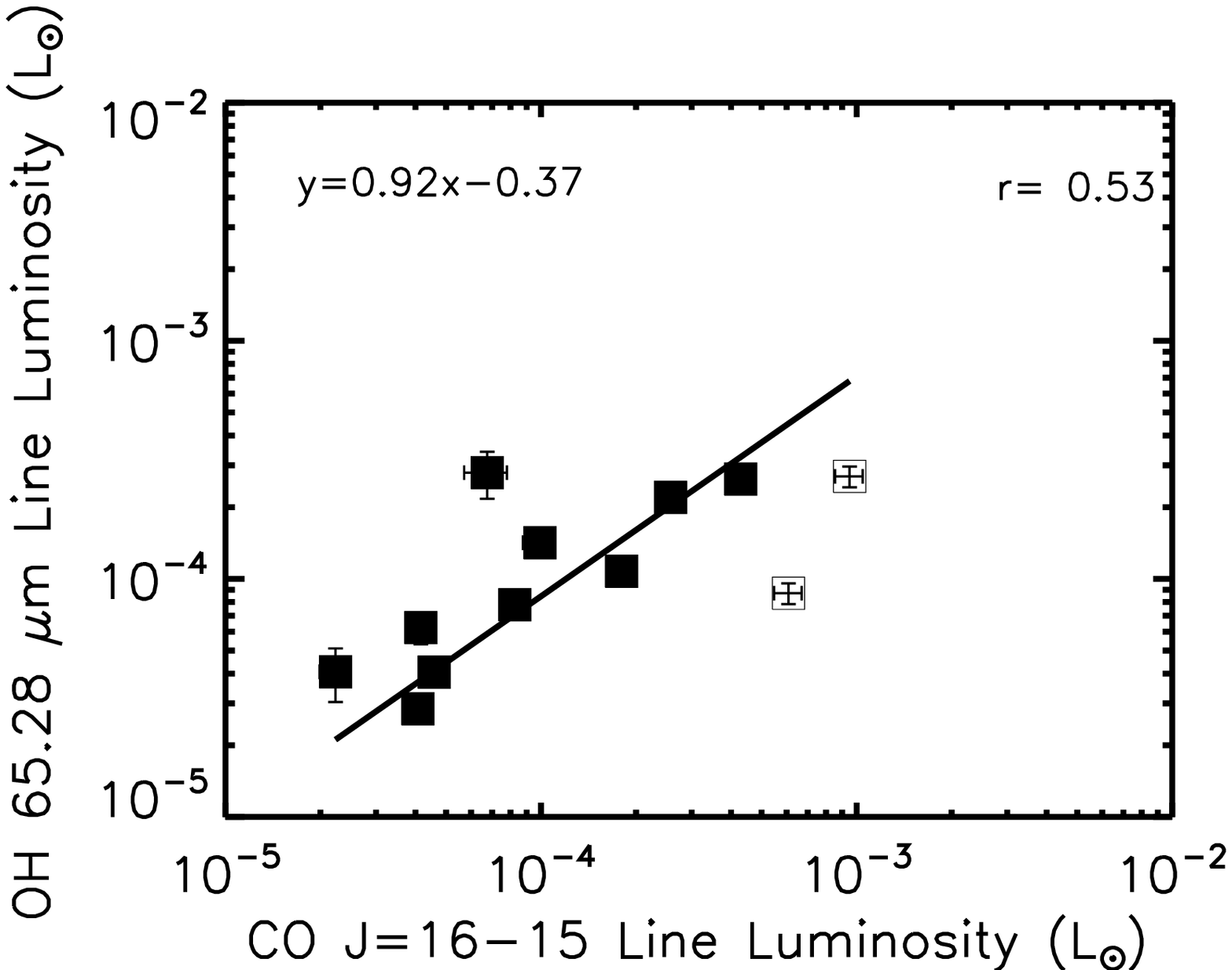}
\caption{{\bf Top Left:} 
$\mathcal{N}_{\rm CO}$ plotted versus a single line of CO (the \jj{16}{15}).
This line (and other nearby transitions) closely predict $\mathcal{N}_{\rm CO}$.  
{\bf Top Right:} CO \jj{29}{28} vs. CO \jj{16}{15} line luminosity.
{\bf Middle Left:} 
 $\mathcal{N}_{\rm CO}$ vs.  $\mathcal{N}_{\rm H_2O}$, showing a close correlation.
 {\bf Middle Right:} OH 84.60 $\mu$m vs. CO \jj{16}{15} line luminosity.
 {\bf Bottom:} OH 65.28 $\mu$m vs. CO \jj{16}{15} line luminosity.
 In all panels, Class 0 sources (according to \tbol) are plotted as open
symbols and Class I sources are plotted as filled symbols.  Plotted values without 
error bars have errors less than the symbol size.
}
\label{co1615}
\end{figure}

\begin{figure}
\includegraphics[scale=0.43]{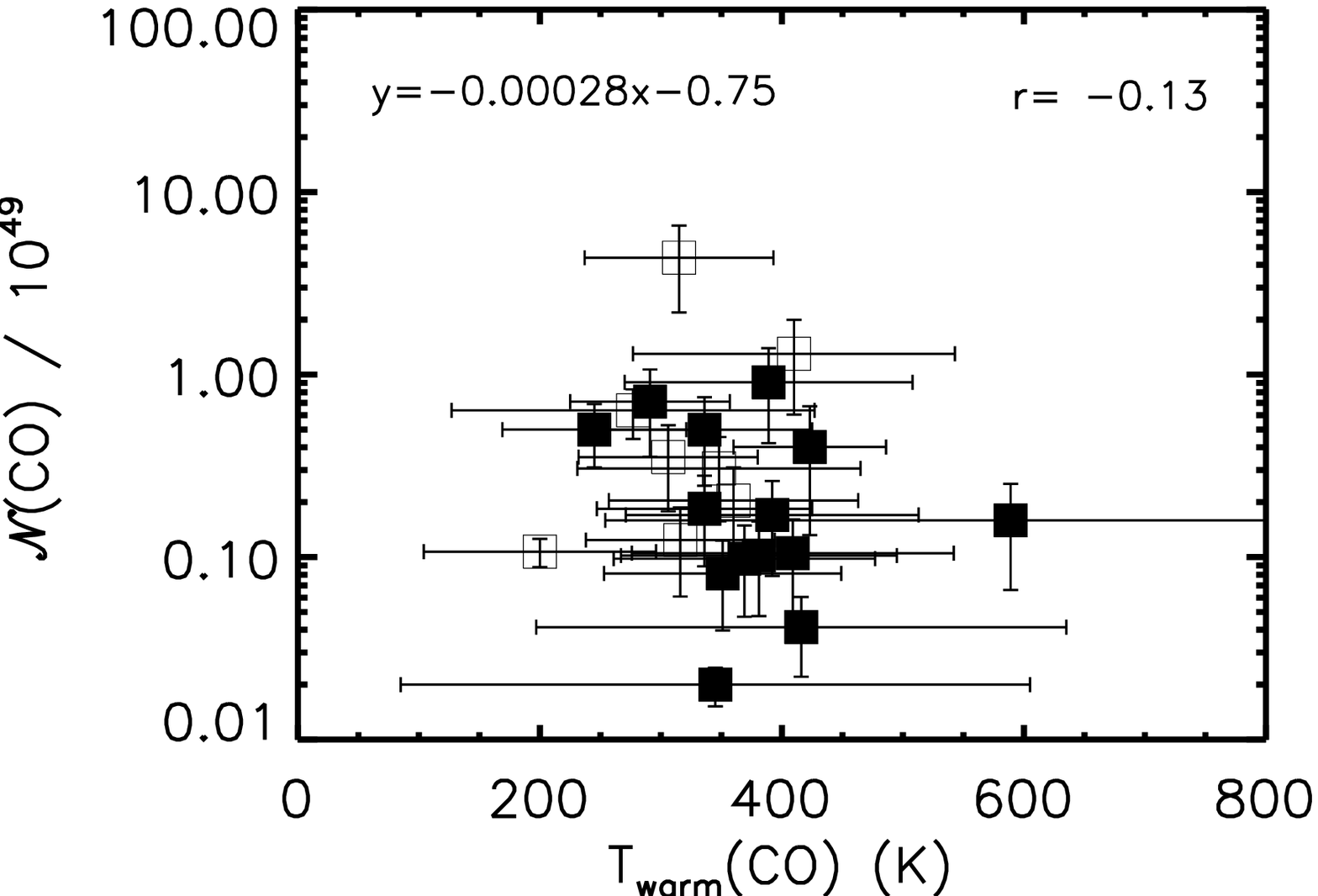}
\includegraphics[scale=0.43]{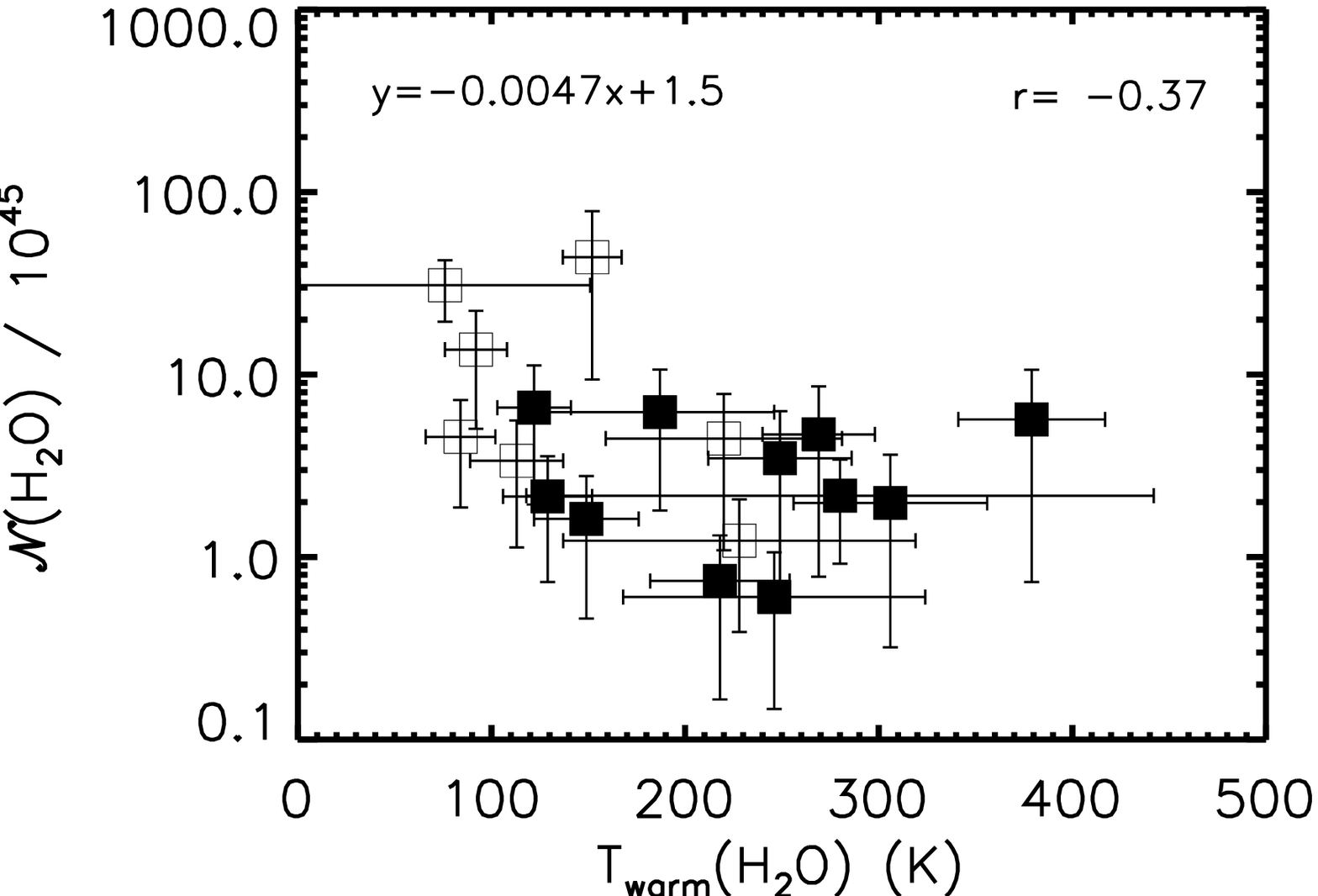}
\caption{{\bf Left:} 
$\mathcal{N}_{\rm CO}$ vs. $T_{\rm warm}$(CO), 
illustrating a lack of
correlation between other properties and rotational temperature.  {\bf Right:} 
Same comparison for H$_2$O.
In all panels, Class 0 sources (according to \tbol) are plotted as open
symbols and Class I sources are plotted as filled symbols.
}
\label{trotcorr}
\end{figure}

\begin{figure}
\includegraphics[scale=0.43]{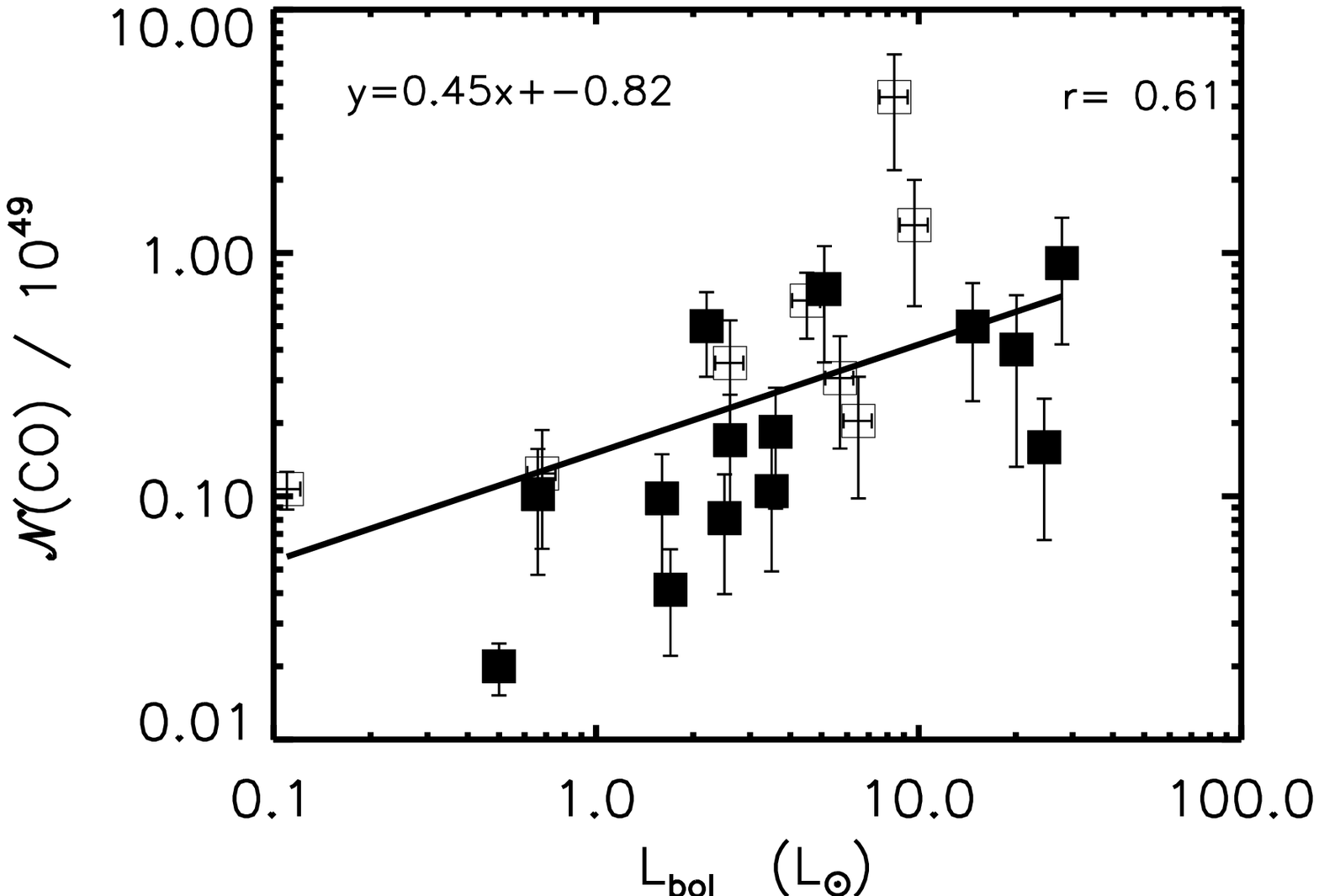}
\includegraphics[scale=0.43]{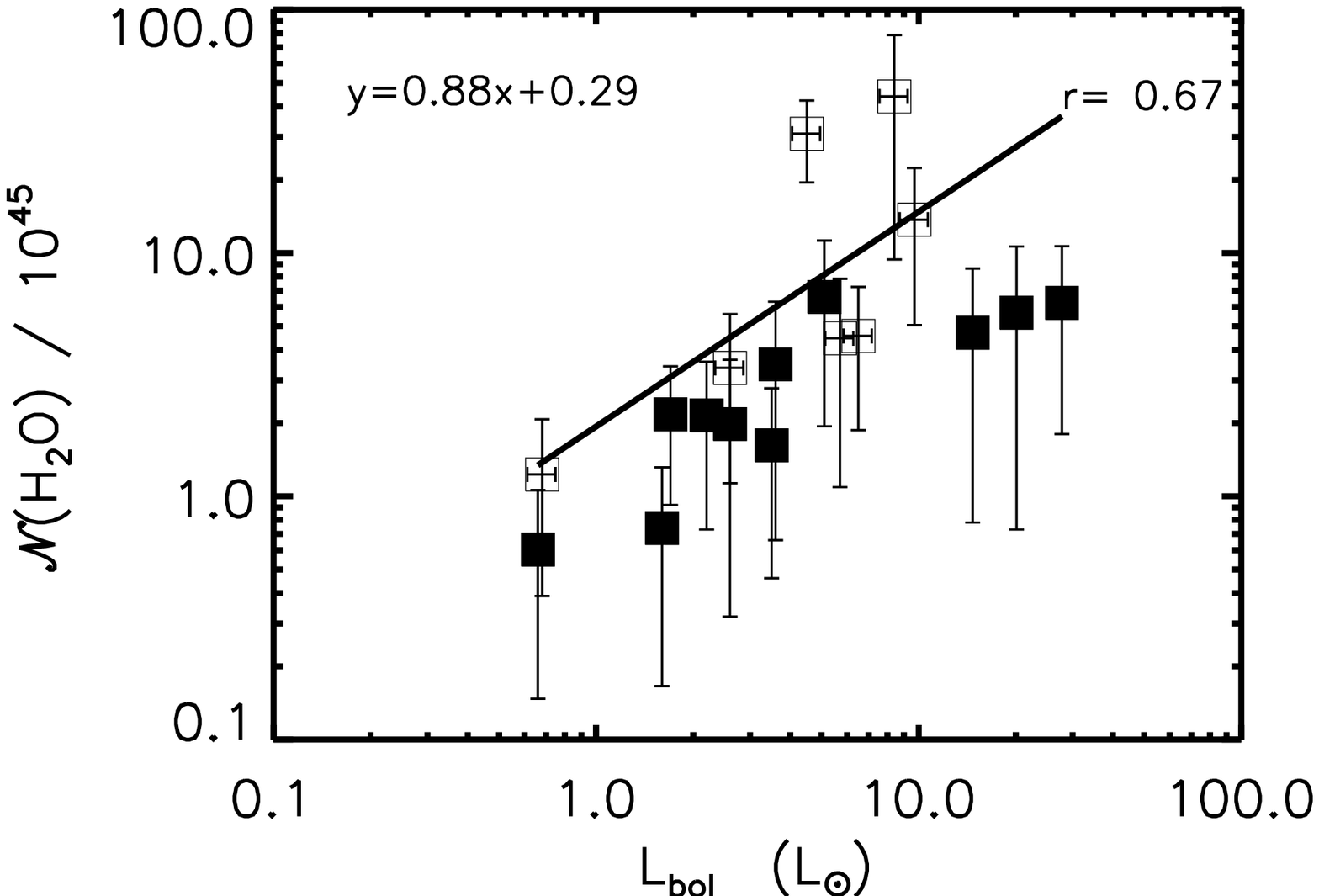}
\includegraphics[scale=0.43]{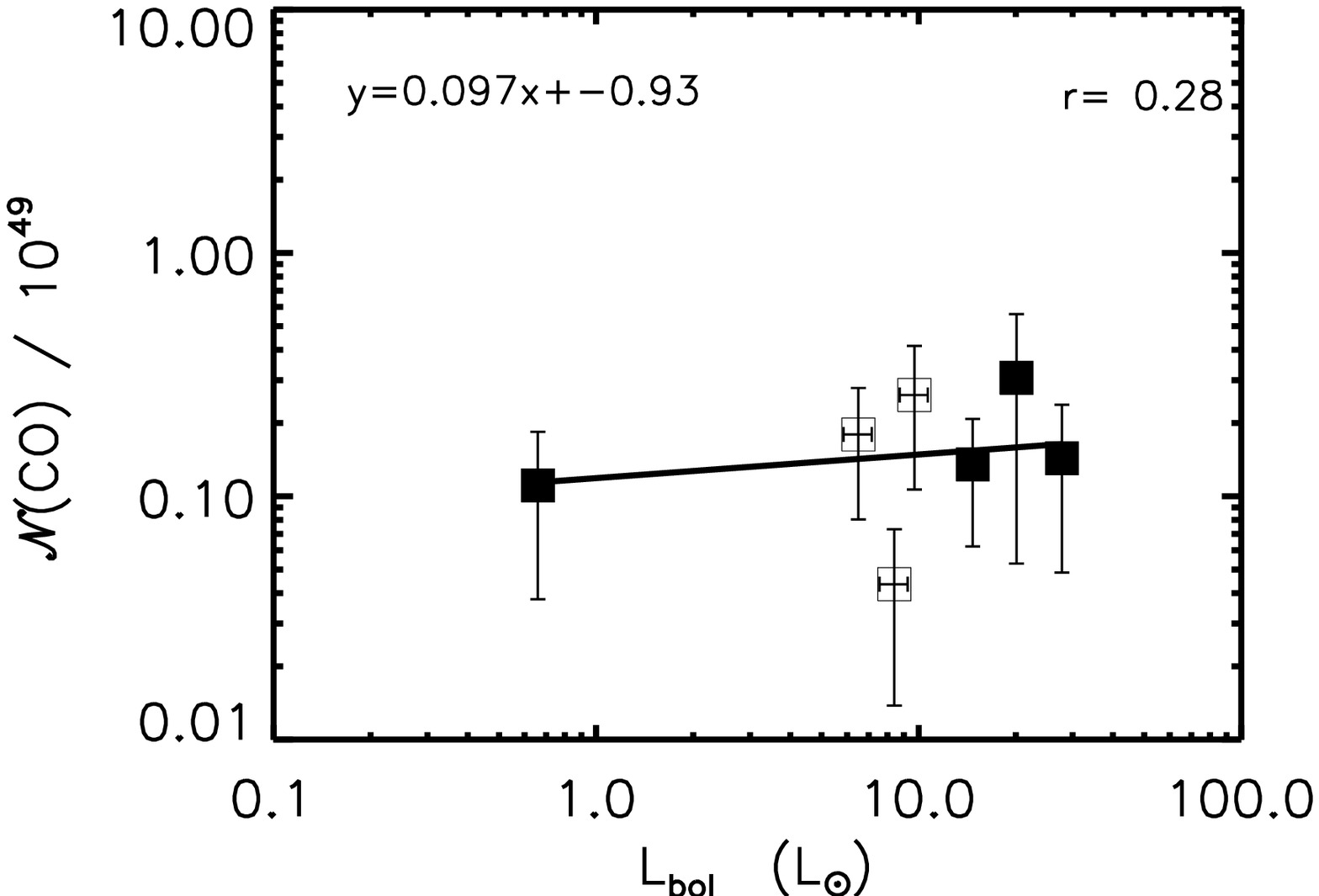}
\caption{{\bf Top Left:} $\funnyN_{\rm CO}$(warm) vs. \lbol.  
{\bf Top Right:} $\funnyN_{\rm H_2O}$ vs. \lbol.  The values are 
correlated over two orders of magnitude, for both CO
and H$_2$O.  B1-c exhibits an order of magnitude higher $\funnyN_{\rm H_2O}$ than the 
sample correlation would suggest, but this is likely 
due to large uncertainties in the fit, taken from only 4 lines.  The greatest outlier is 
L1551-IRS5, which exhibits from low luminosity in
CO, and no detected H$_2$O.  {\bf Bottom left:} 
$\mathcal{N}_{\rm CO}$ (hot)/(cold) vs.  $L_{\rm bol}$. 
Only the seven sources with both components detected 
(ie. at least five CO lines above \jj{24}{23} in excitation energy) are shown.
In all panels, Class 0 sources (according to \tbol) are plotted as open
symbols and Class I sources are plotted as filled symbols.
In each case, we take the uncertainty in \lbol\ to be dominated by the overall 
calibration uncertainty of 10\%.
}
\label{lbol_ntot}
\end{figure}

\begin{figure}
\includegraphics[scale=0.43]{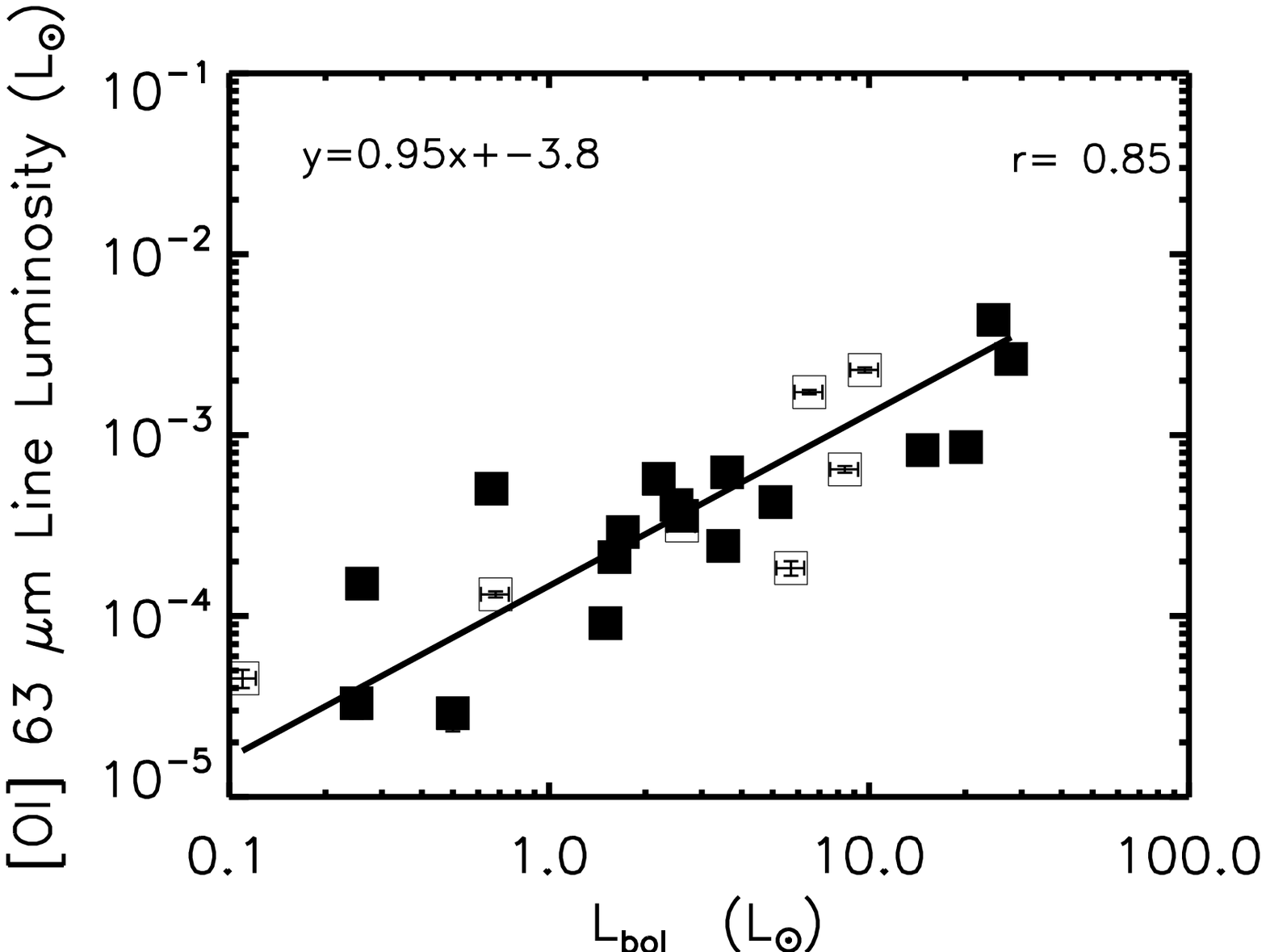}
\includegraphics[scale=0.43]{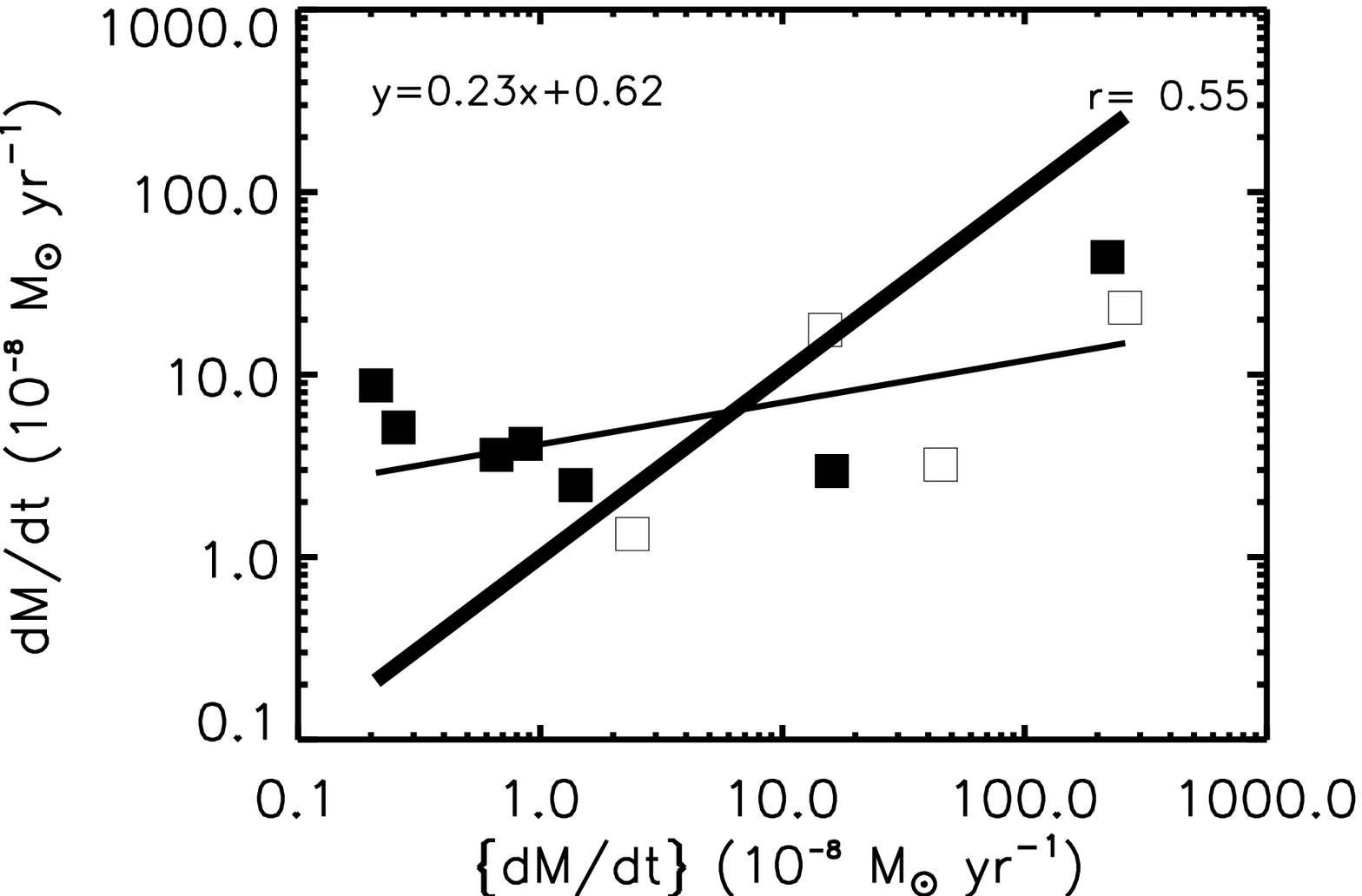}
\caption{{\bf Left:} 
\OI\ 63 $\mu$m luminosity (\lsun) vs. $L_{\rm bol}$ ($\lsun$) for the
DIGIT embedded sources sample.  {\bf Right:} Outflow rate derived from \OI\ 63 $\mu$m 
luminosity (10$^{-8}$ $\msunyr$) vs. 
outflow force derived from CO \jj{3}{2} (10$^{-5}$ $\msunyr$ km s$^{-1}$) 
divided 
 by a velocity of 1000 km s$^{-1}$, 
for sources in the DIGIT embedded sources sample.  The outflow rate is derived from 
ground-based CO \jj{3}{2} maps.  The thin solid line is a fit to the data, while the 
thick line assumes a match between the current and time-averaged values, at a 
velocity of 1000 km s$^{-1}$.
In all panels, Class 0 sources (according to \tbol) are plotted as open
symbols and Class I sources are plotted as filled symbols.
}
\label{lbol_o1}
\end{figure}

\clearpage

\appendix

\section{Data Reduction Recipe}

First, we describe the initial features of reduction that are common to
both methods.
Each segment is first reduced using a modified pipeline based on the standard
spectral reduction script with the relevant version of HIPE; these scripts 
are described in detail in separate subsections for HIPE 6.0 and 8.0.
The `simplecube' pipeline product is a collection of all of the 16
wavelength subarrays with 25 spaxels for each wavelength (16 $\times$ 5
$\times$ 5 cubic array).
The simplecube is non-uniform in wavelength, and next the
pipeline rebins the cube to a uniform wavelength coverage, user-selected.
The parameter to determine the uniform wavelength coverage is referred to as
``oversampling'' -- the pixels in the reduced data covering each resolution
element.
We used an oversampling rate of 2, closely matching the sampling rate of the
raw
PACS data, with an exception in the case of our science demonstration
phase data for DK Cha.  This object was observed using a slightly
different setup not sufficiently sampled in
wavelength, and we reduced our oversample to 1.  This process yields
two spectral cubes
(`rebinned cubes') consisting of 5 $\times$ 5 spaxels for each wavelength, over
a uniform set of wavelengths covering the entire module, for each nod
position.  The nod averaged product (`finalcube') is the average of the
two nods and is the final standard data product from the pipeline.  The
relative
spectral response function
was determined before launch but modified using calibration data taken in
the early phases of the mission, from observations of Neptune, to yield a flux 
density in Jansky (Jy).

After this point, we used two different procedures, optimized to produce
the best results for continuum and lines, respectively.  This dual procedure 
was necessary to produce reliable absolute flux calibration, while also 
obtaining the best S/N on continuum and lines.
For convenience, we refer to these two reductions as the ``bgcal'' (Background
Calibrated; best for absolute flux calibration) and ``calblock'' (Calibration Block; best 
for S/N) spectra.

\subsection{Continuum}

First, we determined the centroid position
of the continuum (using the HIPE 6.0 reduction) for two wavelength regions
in each segment, selected to avoid detectable line emission.
Using maps of these regions, we were able to interpolate a peak
position to within 0.2 spaxel widths. 
We find by this method that most of the sources have continuum centroids 
that fall between 0\farcs2 and 3.0\arcsec\ of the center coordinates, with the majority 
landing within 2\arcsec\ (0.2 spaxel widths) of the center.  For those sources, 
the pointing is consistent between the ``blue'' and ``red'' observations, independent 
observations executed at different times.

Second, in order to produce a spectrum with the best S/N and an accurate absolute
flux density, we chose two apertures for each spectrum: a smaller aperture ($S_{\rm sm}$) 
chosen for the best S/N, 
and a larger aperture ($S_{\rm lg}$) that included all of the source flux for calibration.  To do this, we
determined by inspection which spaxels contained high-S/N spectra of the source 
continuum and 
summed the flux in those spaxels, producing a single spectrum:

 \begin{equation}
 S_{\rm sm} = \Sigma_i S_{\rm i}(high~S/N)
 \end{equation}
 
In most cases, we 
simply use the central spaxel for this purpose.  For some sources we measured a
centroid offset from the center of greater
than 0.2 spaxels; in these cases we include additional spaxels with significant continuum.

Third, we determined the largest set of spaxels containing the source without
contamination from other sources in the field-of-view; in most cases this was
the
full 5$\times$5 array, but in some cases we selected a smaller set of spaxels (see 
section on extended emission, below).  We then computed the flux density summed 
over these spaxels (the total flux), $S_{\rm lg}$:

 \begin{equation}
 S_{\rm lg} = \Sigma_{\rm i} S_{\rm i}
 \end{equation}
 
Fourth, we fitted a 2nd (or smaller) order polynomial scaling factor in wavelength, 
determined from the ratio of the flux density in the larger aperture (S$_{\rm lg}$)
to that in the smaller aperture (S$_{\rm sm}$):

\begin{equation}
 S_{\rm lg}/S_{\rm sm}  \approx a\lambda^2+b\lambda+c
 \end{equation}
 
 where $\lambda$ is wavelength, and a, b, and c are fitted constants.

Fifth, we scale $S_{\rm sm}$ to $S_{\rm lg}$ using this correction 
factor:

\begin{equation}
S_{\rm final} = S_{\rm sm} \times (a\lambda^2 +b\lambda + c)
\end{equation}

This correction is applied  
module-by-module (e.g. a separate polynomial for each range of 50-75, 70-95, 
100-145, and 145-190 $\mu$m).
After scaling $S_{\rm sm}$ to $S_{\rm lg}$, the two ``blue'' (50-75 and 70-95 $\mu$m) 
and the two ``red'' (100-145 and 145-190 $\mu$m) 
modules were generally well-aligned with each other 
(less than 5\% difference on average), but retained the S/N of $S_{\rm sm}$.  
Unfortunately, 
even after applying this entire procedure,  $S_{\rm lg}$(red) $>$ $S_{\rm lg}$(blue) in 
all cases, a systematic discontinuity at 100 $\mu$m for every source.  

\subsection{Lines}

Our reduction technique for lines utilizes HIPE 6.0 {\it and} 
HIPE v8.0.2489 with the
``calibration block'', referred to as the ``calblock'' reduction,
which multiplies the chopped spectrum by a
similar spectrum of an on-board calibration source.  It works in a 
similar fashion to the continuum procedure, but with several key differences.

Our objective here is to measure line equivalent widths (\ewidth)
for gas phase lines,
which are spectrally unresolved, or in a few cases, perhaps slightly
resolved.
This reduction produced the lowest local (point-to-point) noise but larger
shifts between modules and an overall calibration less consistent
with photometry. Tests indicate that the equivalent width of lines
is independent of reduction. Consequently, we use the calblock reduction to
define the
equivalent width of lines and then use the bgcal reduction of the same
spectral region to convert to linefluxes. Mathematically, the process
is as follows:

\begin{equation}
\ewidth = \lineflux{\rm (calblock)}/ \contflux{\rm (calblock)},
\label{eweqn}
\end{equation}
where \contflux\ is the adjacent continuum flux density, with
units of \wmmicron, and \lineflux\ is the lineflux, with units
of \wmeter,  followed by

\begin{equation}
\lineflux{\rm (Final)} = \ewidth  \times  \contflux{\rm (bgcal)}
\label{linefluxeq}
\end{equation}

The first step is to extract linefluxes from the HIPE 8 calblock reduction  using
a narrow region to define the continuum (\contflux) and calculate the
\ewidth. We
utilized a modified version of the SMART reduction package \citep{higdon04}
to fit Gaussians and first or second-order baselines to the spectra
to remove local continuum features.
The LAMDA database of lines \citep{schoier05}, supplemented by HITRAN 
\citep{rothman05}, was
used to identify the features.
The fit for the line center varies only by 0.01 $\mu$m or less for
our different extraction methods (10-20\% of a resolution element);
therefore the fit uncertainty in velocity space is $\sim$ 30-50 km s$^{-1}$.
Line centers usually lie within 50 km s$^{-1}$
of the theoretical line center from laboratory measurements.  At our level 
of precision, we do not observe believable shifts in our data; nonetheless, 
this is still consistent with shifts observed in other datasets \citep[e.g.][]{karska13}.
However, the location of the
source within each slit also affects the velocity.
The uncertainty in \ewidth\ caused by the fit is quite small (S/N ratios as
high as 300 were obtained on bright
lines, particularly \OI), but for faint lines, the
uncertainty rises to 20\%.
In this treatment we do not include lines within the masked wavelength regions 
($<$ 55 $\mu$m, 95--102.5 $\mu$m, $>$ 190 $\mu$m) as they 
exhibit large calibration uncertainties.  The one exception is the 
CO  \jj{27}{26} line at 96.77 $\mu$m; the local RMS uncertainty is somewhat 
larger than, but still comparable to, the local continuum surrounding the CO  \jj{28}{27} 
line at 93.3 $\mu$m, 
and we only include this line in sources with confirmed detections of CO 
lines at shorter wavelengths.  The line does not significantly affect the derived 
source properties in the following sections.

\subsubsection{Corrections for spatial extent, and comparison to the empirical PSF}

Second, the linefluxes were corrected for PSF and extended emission.
As with the continuum, simply adding all the spaxels caused a serious loss  
of signal-to-noise and
a diminished rate of line detection, and the PSF correction
provided in the standard pipeline does not
account for extended emission.  Thus 
we do {\it not} use the PSF correction function, but instead compare the flux
over different sized apertures to determine an empirical correction for each
source.  In order to do this, we measured the lineflux, \lineflux(sm), in
the central spaxel (or a sum over a few spaxels for poorly centered sources)
that gave the best signal-to-noise on relatively strong lines.  Mathematically, 
the process is identical to the continuum, except for lineflux rather than flux 
density:

\begin{equation}
\lineflux(sm) = \Sigma_i  ~\lineflux(i)
\end{equation}

where i is the set of spaxels with good signal-to-noise on all lines.

Additionally, 
we correct the fluxes by the ratio between the bgcal and calblock local 
continuum for each individual line.
Because the spatial distribution of line emission could in principle
differ from that of the continuum, we did not simply scale the value
in the central spaxel by the continuum ratios.
To determine the total linefluxes, \lineflux(lg), we applied a polynomial 
correction, developed by comparing the spectral lineflux
detected in the central spaxel compared to that in the surrounding spaxels
for strong lines -- mostly CO lines, but also including strong H$_2$O lines.  
We did not include the 63 and 145 $\mu$m \OI\ lines in this correction, because they  
were frequently extended compared to these species and to the PSF (see Section \ref{reduccont}).
Weaker lines of all species were then scaled with the same factors.

Mathematically, the procedure is similar to the correction to the continuum.
First, the linefluxes, measured originally 
from the calblock reduction, are scaled to the bgcal level.   Then, as for 
continuum we derive the small aperture lineflux ($\lineflux(sm)$, typically the 
central spaxel) and the large aperture ($\lineflux(lg)$).  Then we correct 
the final linefluxes ($\lineflux(final)$) by a first-order polynomial fitted to the 
ratio of the fluxes in the apertures.  We only summed over the central 
3$\times$3 spaxels to calculate \lineflux(lg), 
rather than the full (5$\times$5) array.  Mathematically:

\begin{equation}
\lineflux(lg) = \Sigma_{\rm i} ~ \lineflux_{\rm i}
\end{equation}

where i is the index of the spaxels falling within the large aperture.

\begin{equation}
\lineflux(lg)/\lineflux(sm)  \approx a\lambda+b
\end{equation}

\begin{equation}
\lineflux(final) = \lineflux(sm) \times (a\lambda+b)
\end{equation}

This
method assumes that weak and strong lines have similar spatial distributions.
In Figure \ref{extendedlines}, we plot linefluxes of all
species measured from 3$\times$3 spaxel regions vs. linefluxes contained within
just the central spaxel, for four 
well-pointed sources.  For each source we fit a constant ratio from
50-100 $\mu$m, and a first order polynomial from 100-200 $\mu$m, a simplified
fitted version of the PSF correction, to the bright CO and H$_2$O lines.  For this 
exercise we set a minimum correction factor of 1.4 to the 50-100 $\mu$m fit, 
approximately the PSF correction.  Although 
additional lines are plotted, they are not considered in the fit. In cases for which 
we lack sufficient strong lines to determine the correction, we 
use a linear approximation to the PSF correction. 

\subsection{Detailed Steps in the Pipeline procedure for HIPE6 (Continuum)}

The HIPE pipeline is a Python script that executes Java modules line-by-line
in the interactive environment.  The ``flux calibrated'' version of our script
uses
calibration sets that were part of HIPE 6.0.  The steps executed in the pipeline
are
as follows, through Level 0 (raw data), Level 0.5 (sliced data frames), Level 1 
(the datacubes at the raw, non-uniform wavelength sampling) and Level 2 (the 
final datacubes after wavelength calibration, rebinning, nod averaging).

First, we step the data from Level 0 to 0.5.  Saturated frames are flagged and
removed from the ramps.  The data is converted from digital flux units to
volts.
The timestamp of the observation is reconstructed from the header, and pointing
information is added.  Next the details of the chopper throw are reconstructed
(6$\arcmin$ in the case of all DIGIT observations in this paper, or  the
``large chopper
throw'' setting).  Then a correction is applied for {\it Herschel}'s
velocity based on the pointing, UTC time, and location of  {\it Herschel}.  Next,
noisy/bad
pixels are flagged.  Finally the data is split into manageable ``slices'' to
conserve
RAM, and the Level 0.5 products are produced.

Next the data is reduced to Level 1.  First the masks are removed to apply the
glitch-flagging routines, producing a glitch mask.  The signal level is
converted
to the reference integration capacitance using a calibration file.  A
differential signal
is computed for each on/off-position pair of datapoints, for each chopper
cycle.
Then the spectrum is divided by the relative spectral response function (RSRF), and
the
nominal response function, in effect a flatfielding correction utilizing
calibration
observations of Neptune and Vesta.  This is then recorded as a ``PacsCube''
dataproduct, which contains the spectrum {\it prior} to wavelength calibration,
rebinning and nod averaging; this is the Level 1 product.

The last set of pipeline subroutines takes the Level 1 product to Level 2.  A user-selected
``oversampling'' parameter is used to rebin to a uniform spectral resolution.
 An oversample of 2 indicates that the data
are binned so that each resolution element is covered by 2 pixels in the
reduced data. 
Outliers are flagged with sigma-clipping in the wavelength domain.  The
``RebinnedCube'' product is the output of this routine, a spectrum for each
nod-difference spectrum.  In the standard chop-nod setting, this produces
two spectra.  The final step is to take the difference of the two nods by
combining
the two RebinnedCube products into a single ``FinalCube'' product, for each
of the 25 spaxels.  A post-HIPE routine is used to convert this to a datacube
for
final analysis.

\subsection{Detailed Steps in the Pipeline procedure for HIPE8 (Lines)}

The best signal-to-noise is in most cases generated using the current pipeline
calibration set from HIPE 8.0.2489. The steps executed in this version of
the
pipeline are similar to the HIPE 6.0 procedure, but with significant differences.

To step from Level 0 to 0.5, the pipeline begins by flagging and removing
saturated frames from the ramps.  Then the signal is converted from digital
flux
units to volts.  Next the timestamp and spatial coordinates of each pixel are
reconstructed from the header.  The wavelength calibration is determined.  A
correction is applied for  {\it Herschel}'s velocity based on the pointing, the location of
 {\it Herschel}, and UTC time.  Noisy/bad pixels are flagged, as are data affected by
chopper and grating movement.

Next the data are processed to Level 1.  The masks are removed to apply
the glitch-flagging routine, to make a glitch mask, and then all masks are
activated.  The signal is converted to reference integration capacitance.  {\it
In the 
case of the DIGIT pipeline, we derive the detector response from the standard
calibration block to improve signal-to-noise, rather than the standard pipeline
procedure.}  Then the on-off position difference signal is computed for each
pair of datapoints per chopper cycle.  The signal is divided by the relative
spectral
response function; an independent function is applied for each module.  The
flatfield is refined using a sub-procedure.  Finally the frames are converted
to a
set of PacsCubes, the level 1 dataproduct.

The last subsection processes the data to the Level 2 stage, where they are
ready for
analysis.  First the good (unmasked) data is located, and an upsample of 1
and oversampling parameter of 2 are used, in all cases except for the Science
Demonstration Phase observations of DK Cha (which was executed in a slightly
different engineering mode that lacked sufficient spectral sampling for an oversample of
2;
in this case we select oversample$=$1).  A clarification on terminology: in the context 
of PACS, upsampling refers to the addition of artificial bins between datapoints; 
an upsample 
of 1 does not use this interpolation.  Oversampling indicates a rebinning to a user-defined 
wavelength scale: the oversampling parameter indicates the number of pixels covering 
each element in the reduced data.  The wavelength grid is set by these
parameters to rebin the data.  A 5-sigma clipped mean flags and removes any
remaining outliers.  The data is rebinned at the selected sampling to create a
pair
of RebinnedCube products (one each for the source difference with Nod A, and
the source difference with Nod B).  Finally the RebinnedCubes are averaged to
form
a single ``FinalCube'' product.

As an additional post-processing step, we apply a PSF correction (derived from
the PACS
manual) to the well-pointed (centroid within $\sim$ 2$\arcsec$) of the center
coordinates
of the central spaxel), pointlike (non-extended)
sources in our sample.  This does not apply to any sources in the DIGIT
embedded
sample, which show some indication of extended or diffuse emission.

\section{Comparison of Spatial Extent for Lines and Continuum Across the DIGIT 
Sample}

In this section, we present the spatial distribution of all detected 
spectral lines in the DIGIT embedded sample (black), plotted against their local continuum (red) 
emission, drawn from linefree portions of spectra within 1 $\mu$m of the target line.  We plot 10\% contours, cutting off at twice the RMS noise of the outer ring of spaxels, except where noted below.

\begin{figure}
\includegraphics[scale=1.05]{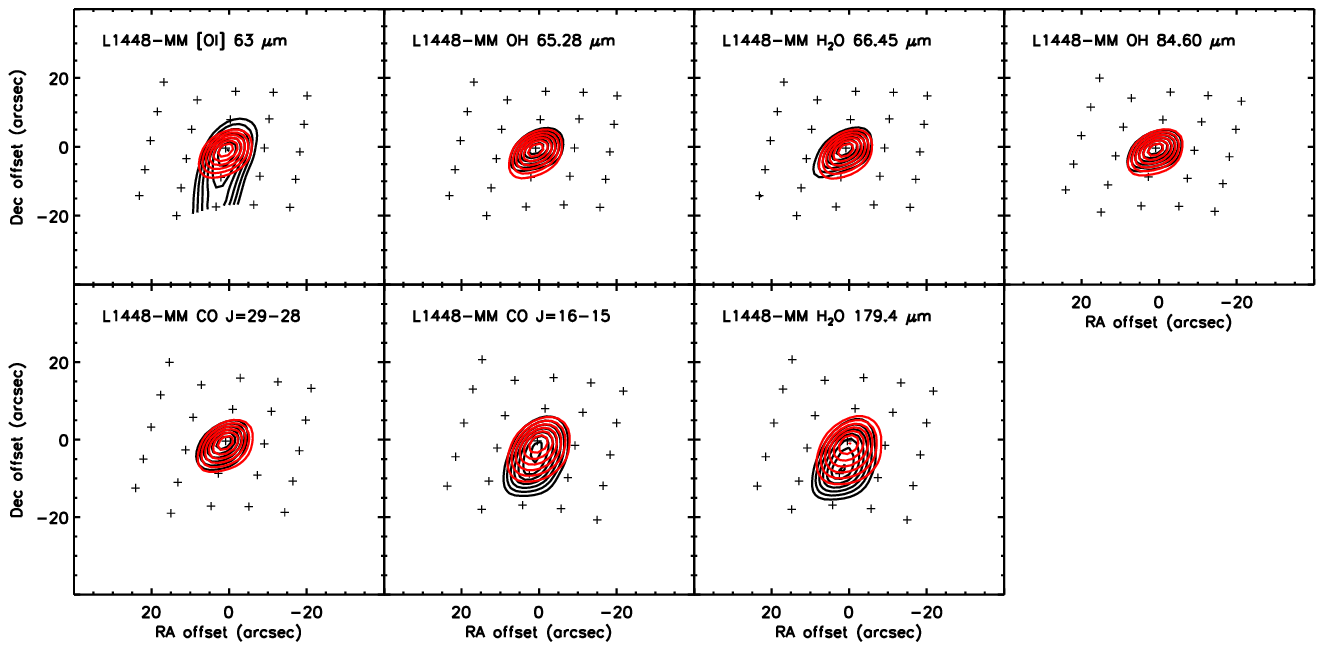}
\includegraphics[scale=1.05]{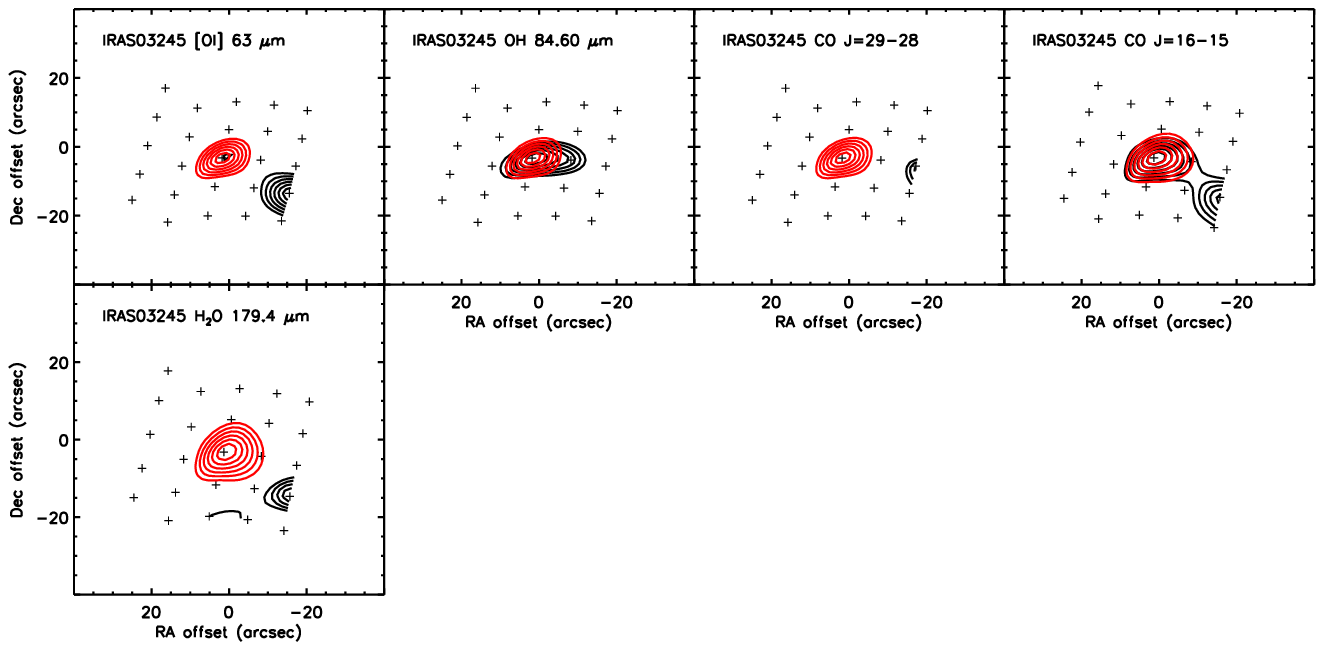}
\includegraphics[scale=1.05]{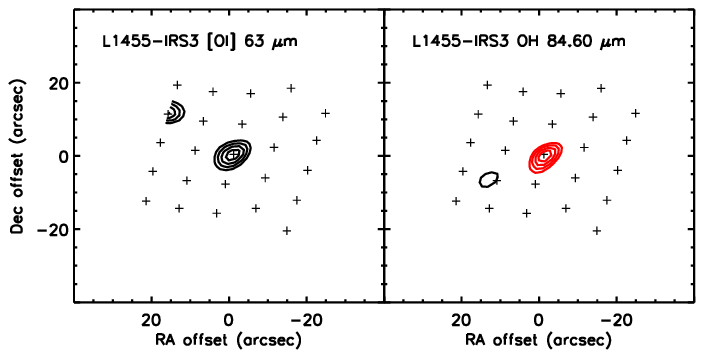}
\caption{Spatial distribution of lines (black) vs. local continuum (red).  The 
contours are in increments of 10\% from the peak flux (usually in the central 
spaxel), plotted down to the noise limit.  The noise limit is computed as 2$\times$ 
the average flux of the outer ring of 16 spaxels, with the exception of the IRS 46/44 
map, for which we avoid the edge spaxels containing IRS 44.}
\label{spatial1}
\end{figure}

\begin{figure}

\includegraphics[scale=1.05]{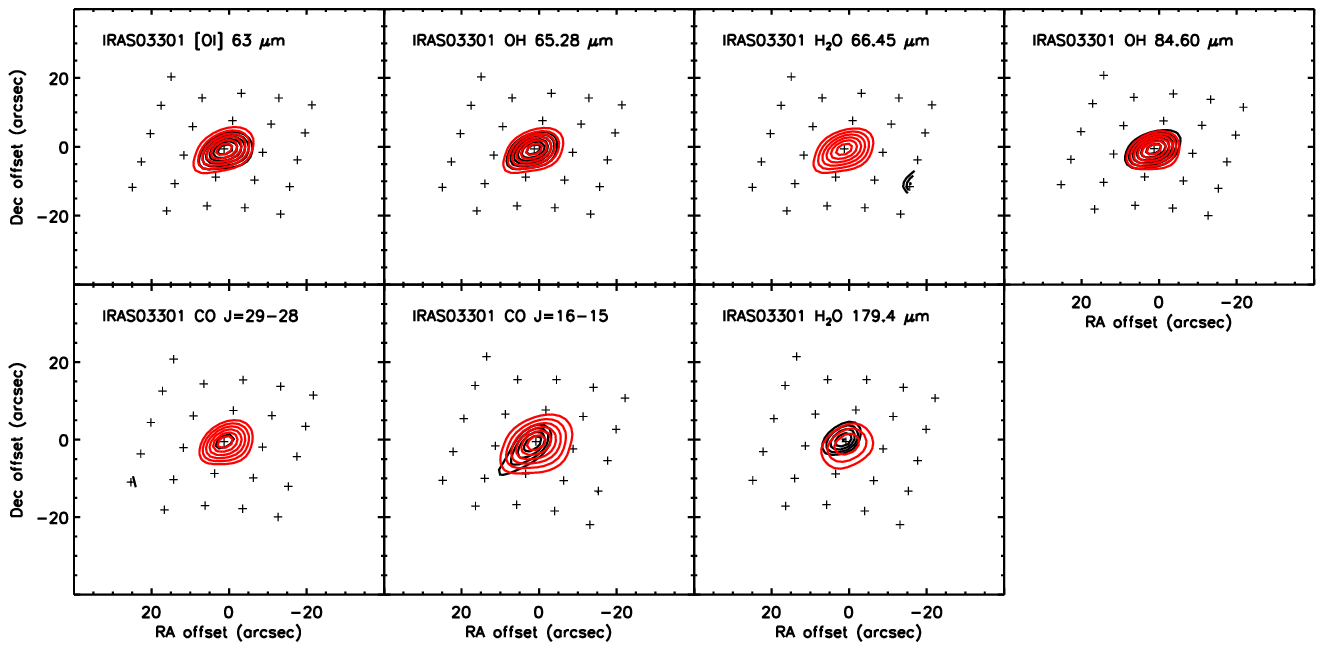}
\includegraphics[scale=1.05]{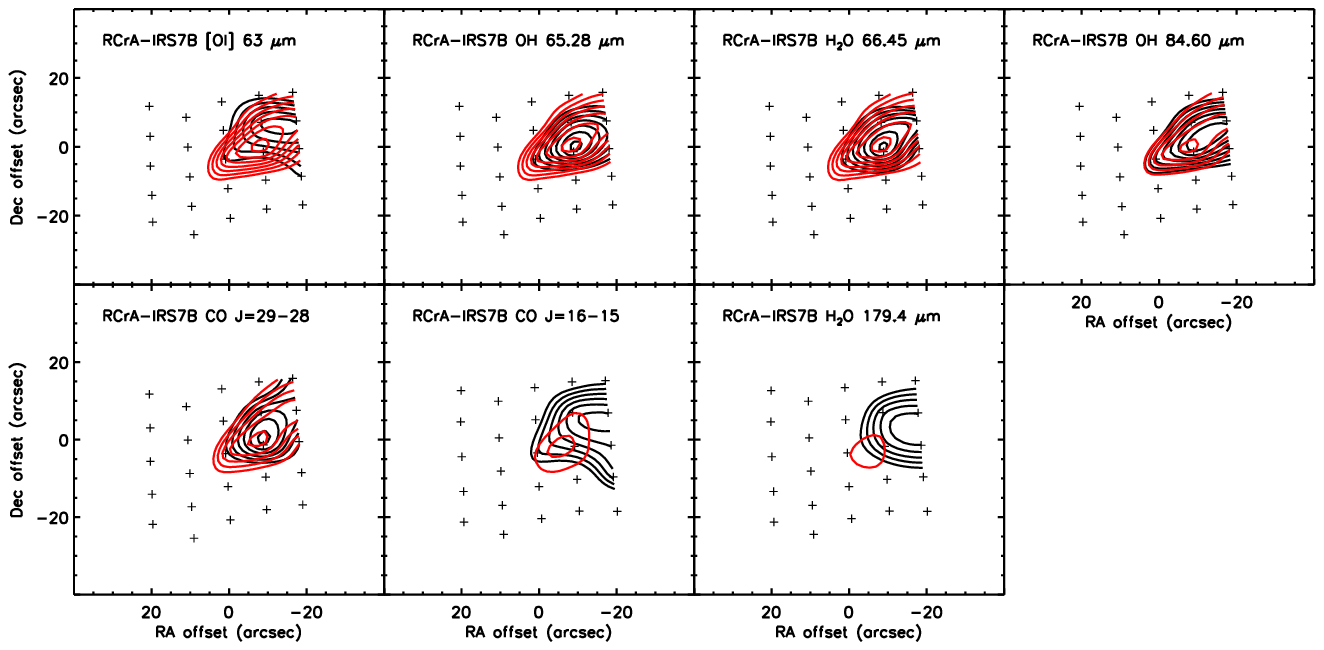}
\includegraphics[scale=1.05]{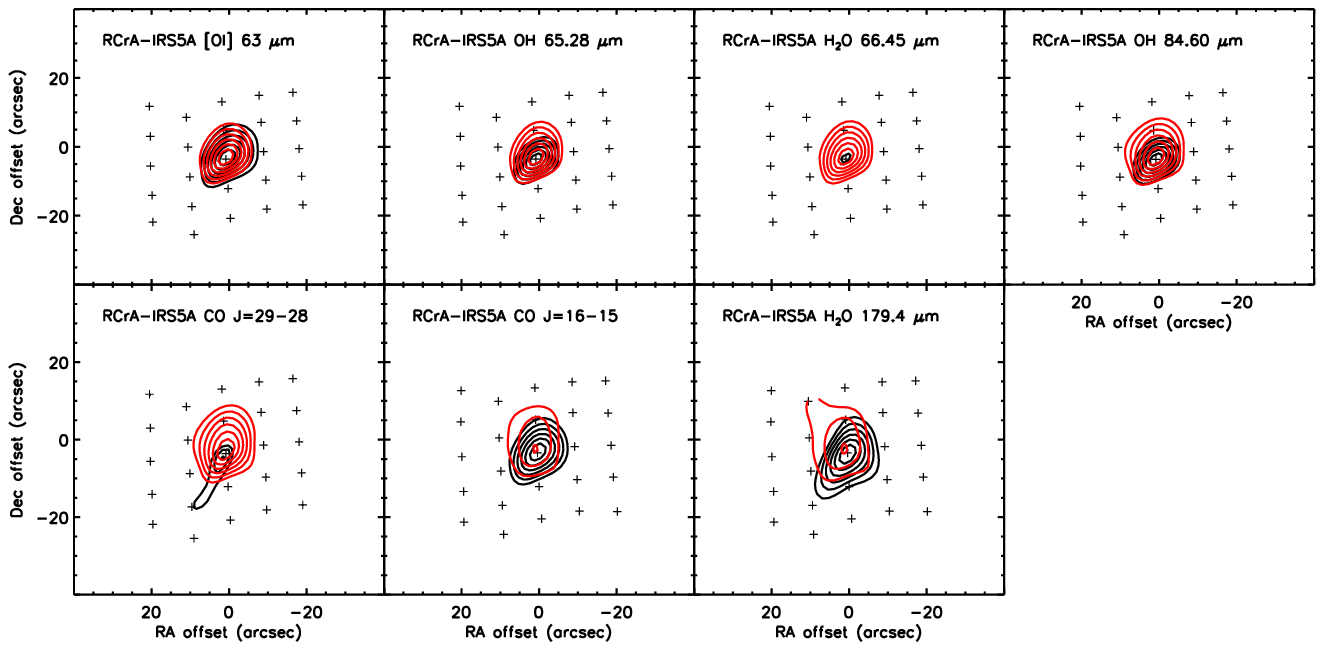}
\caption{See Figure \ref{spatial1} for a description of this figure.}
\label{spatial2}
\end{figure}

\begin{figure}
\includegraphics[scale=1.05]{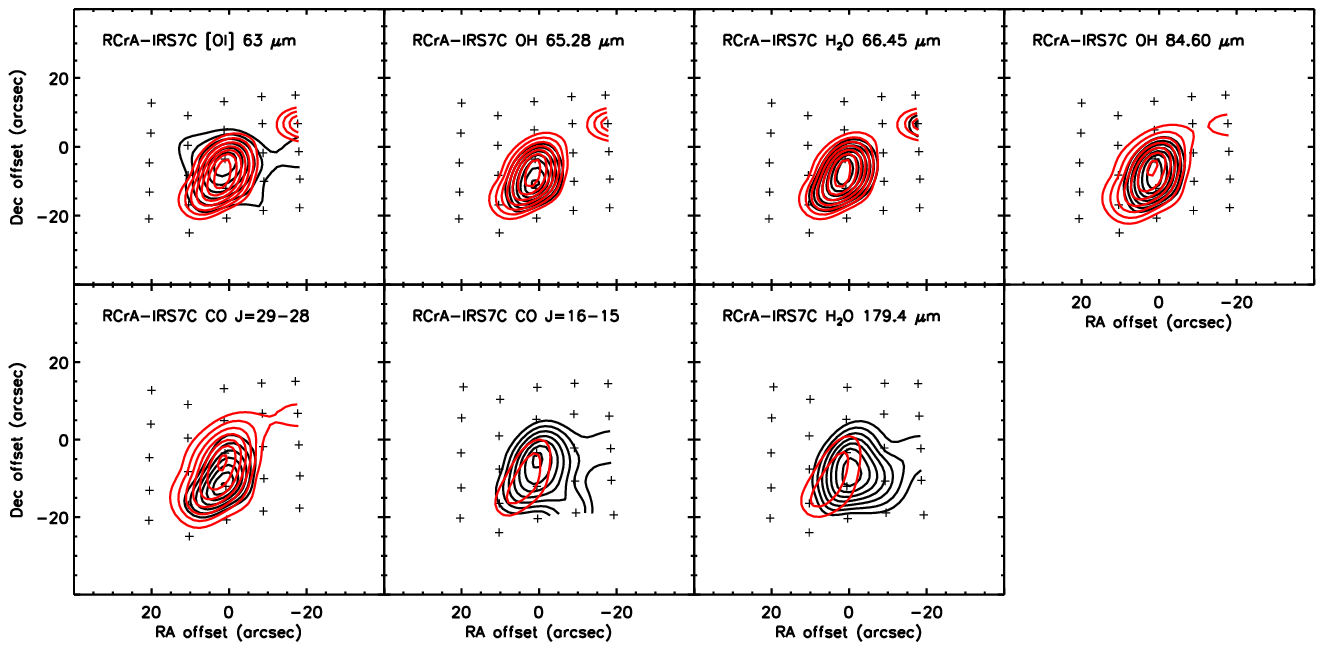}
\includegraphics[scale=1.05]{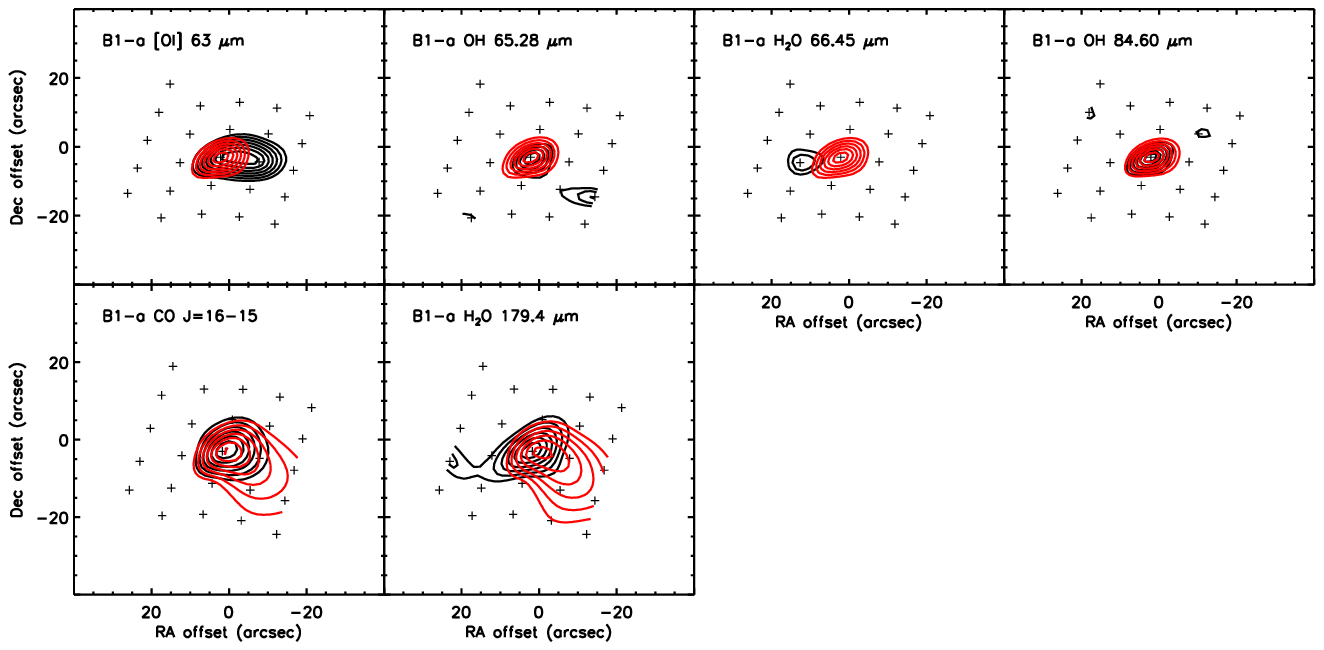}
\includegraphics[scale=1.05]{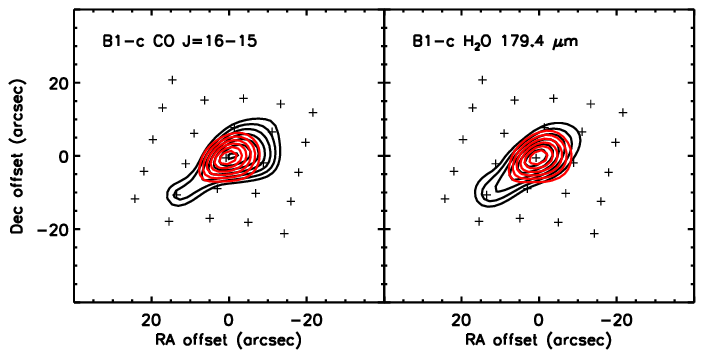}
\caption{See Figure \ref{spatial1} for a description of this figure.}
\label{spatial3}
\end{figure}

\begin{figure}
\includegraphics[scale=1.05]{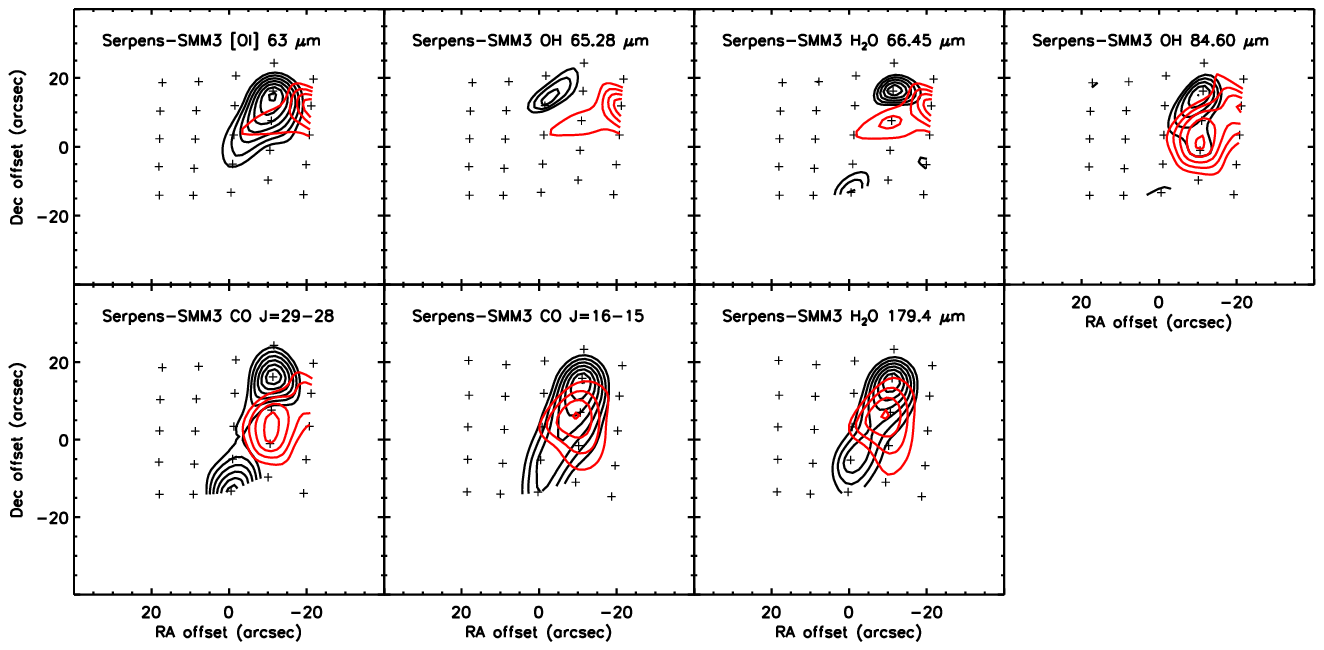}
\includegraphics[scale=1.05]{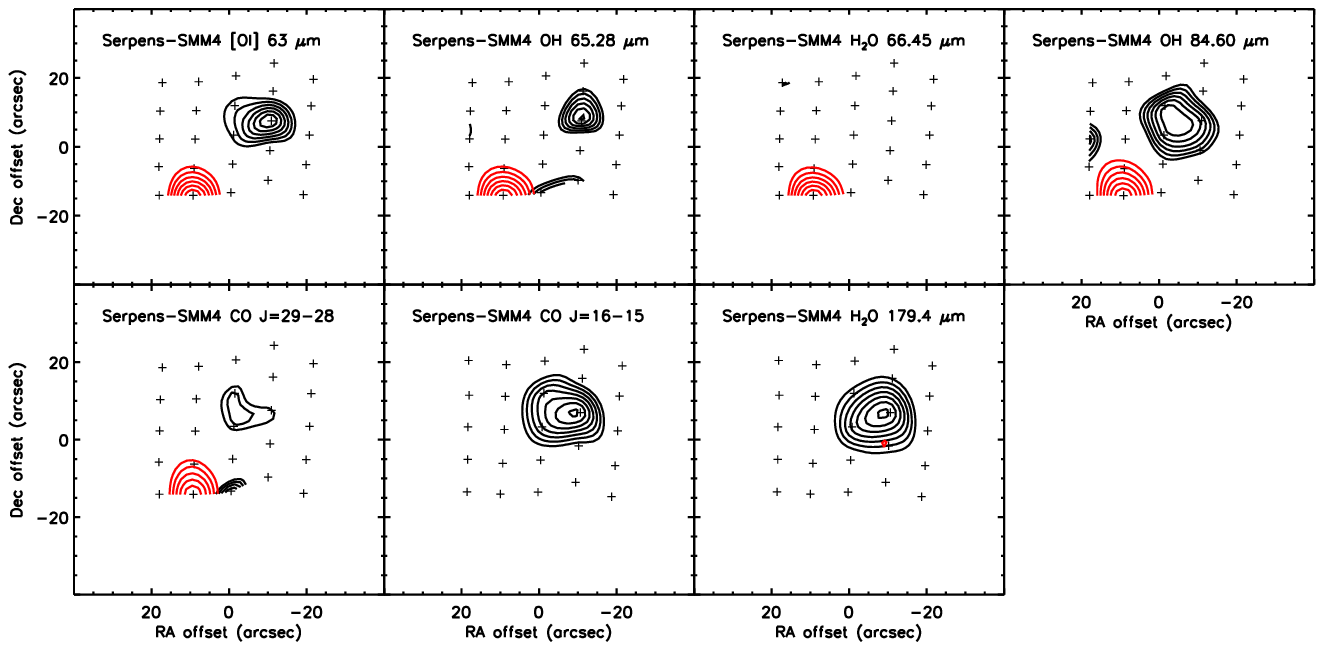}
\caption{See Figure \ref{spatial1} for a description of this figure.}
\label{spatial4}
\end{figure}

\begin{figure}
\includegraphics[scale=1.05]{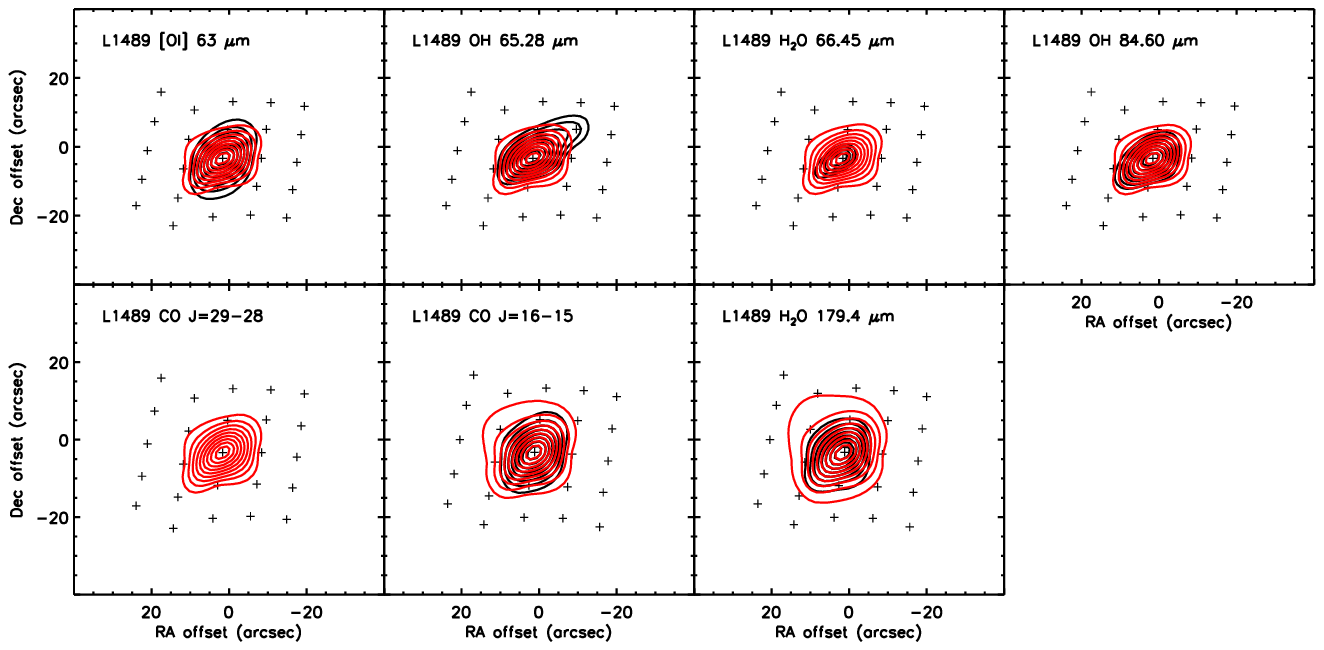}  
\includegraphics[scale=1.05]{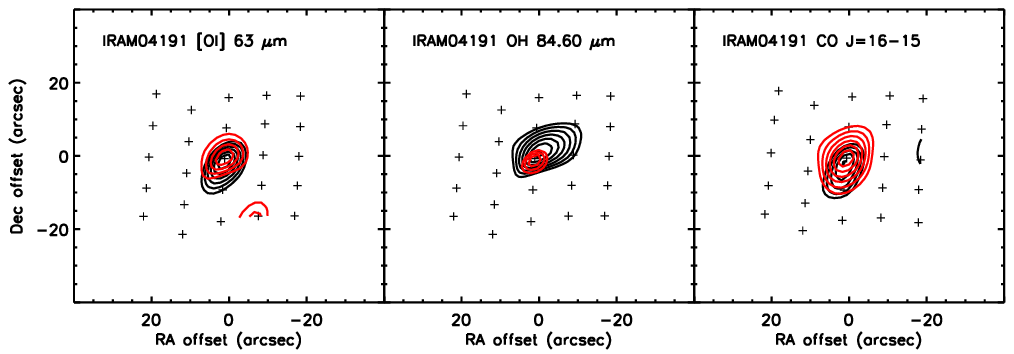} 
\includegraphics[scale=1.05]{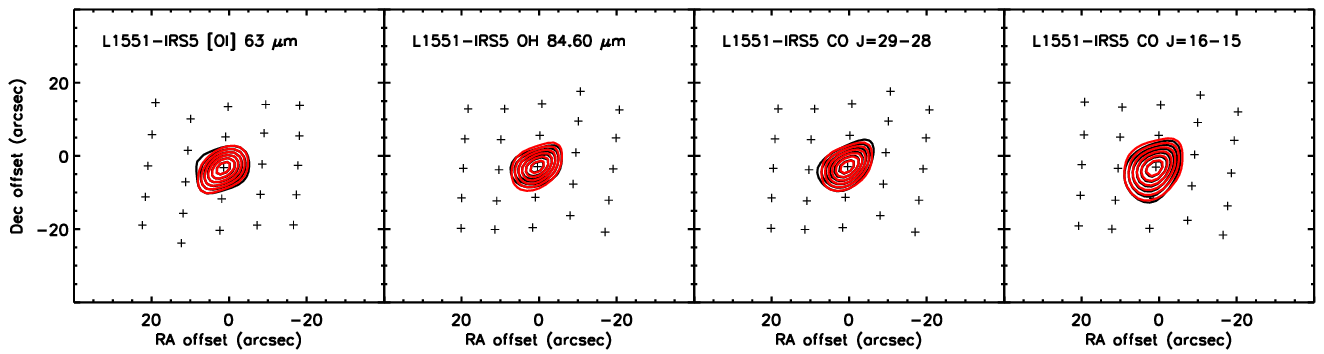} 
\caption{See Figure \ref{spatial1} for a description of this figure.}
\label{spatial5}
\end{figure}

\begin{figure}
\includegraphics[scale=1.05]{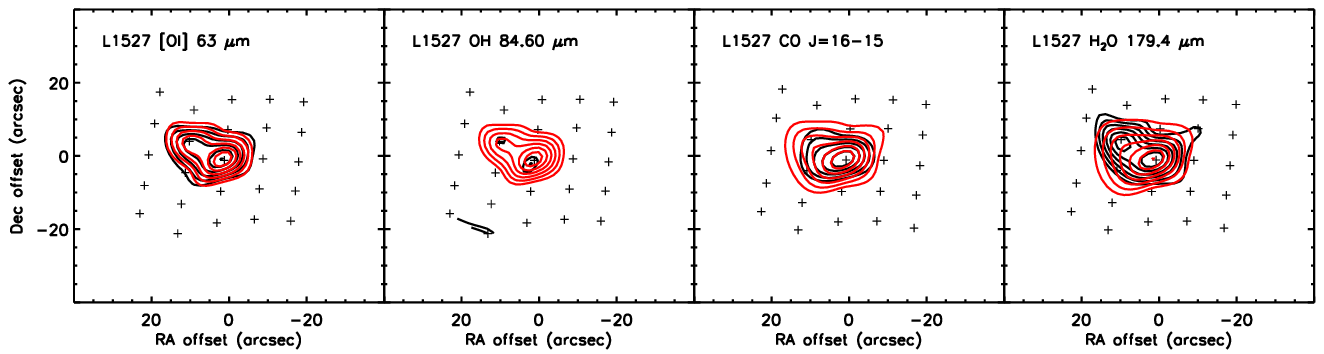}
\includegraphics[scale=1.05]{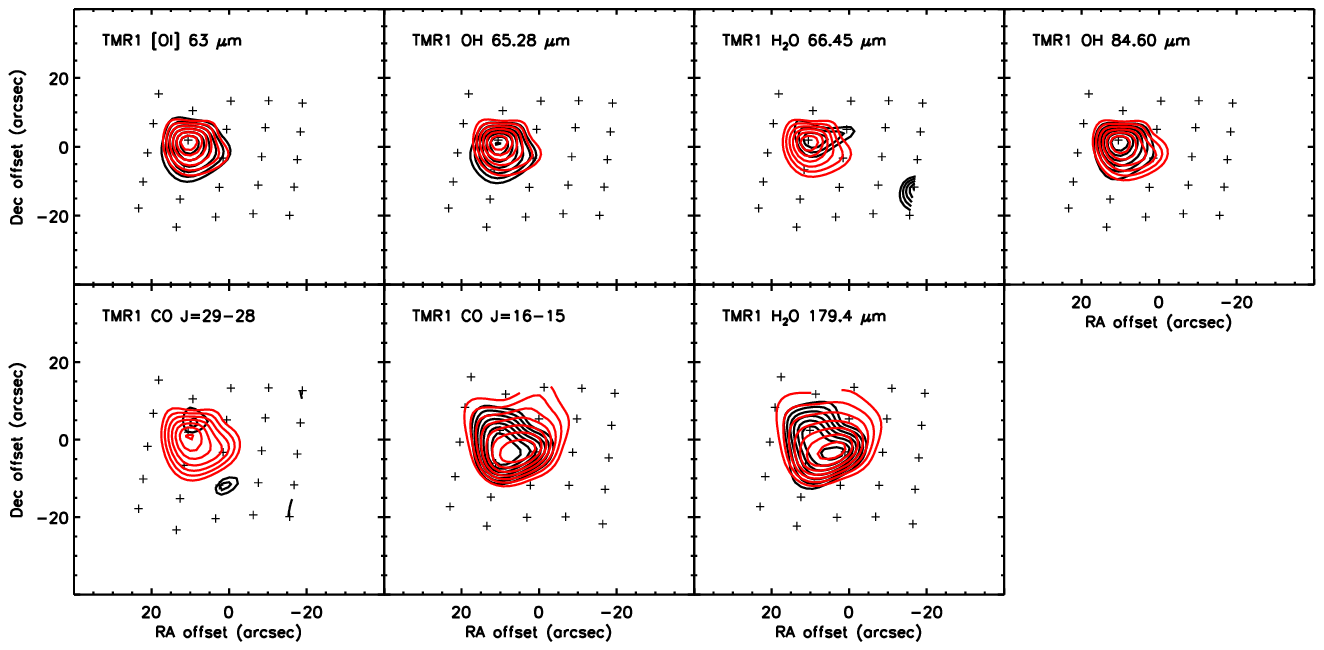}
\includegraphics[scale=1.05]{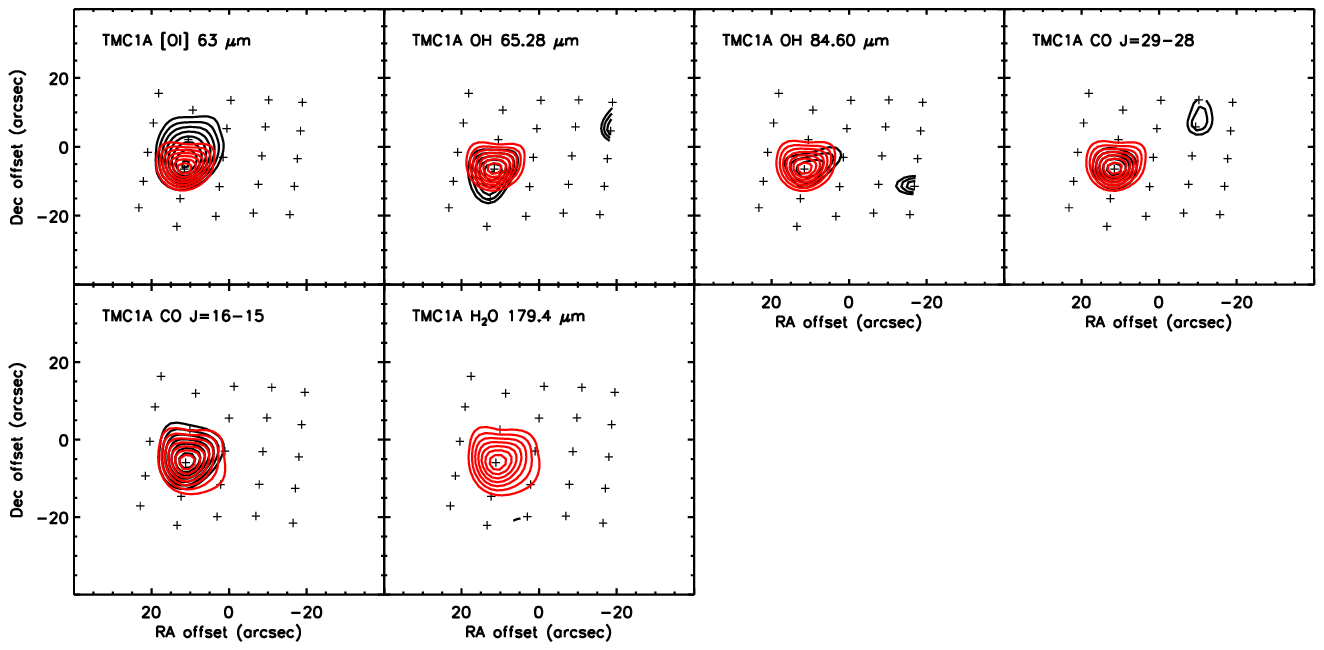}
\caption{See Figure \ref{spatial1} for a description of this figure.}
\label{spatial6}
\end{figure}

\begin{figure}
\includegraphics[scale=1.05]{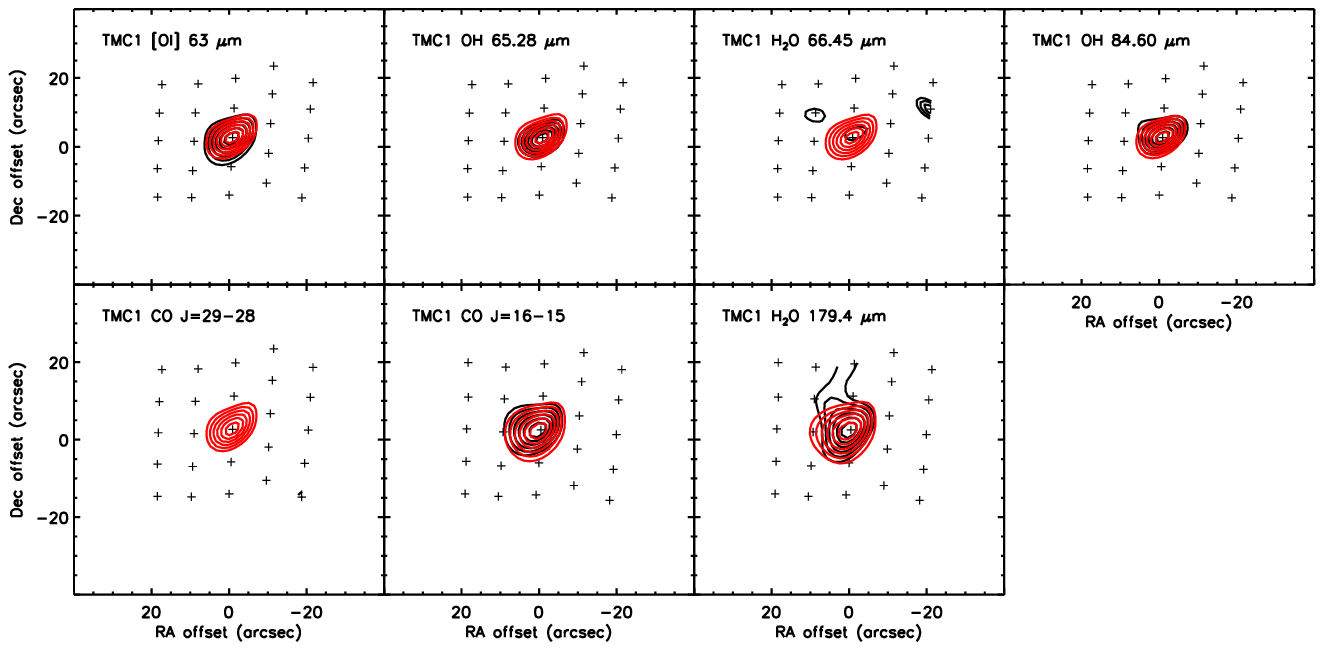}
\includegraphics[scale=1.05]{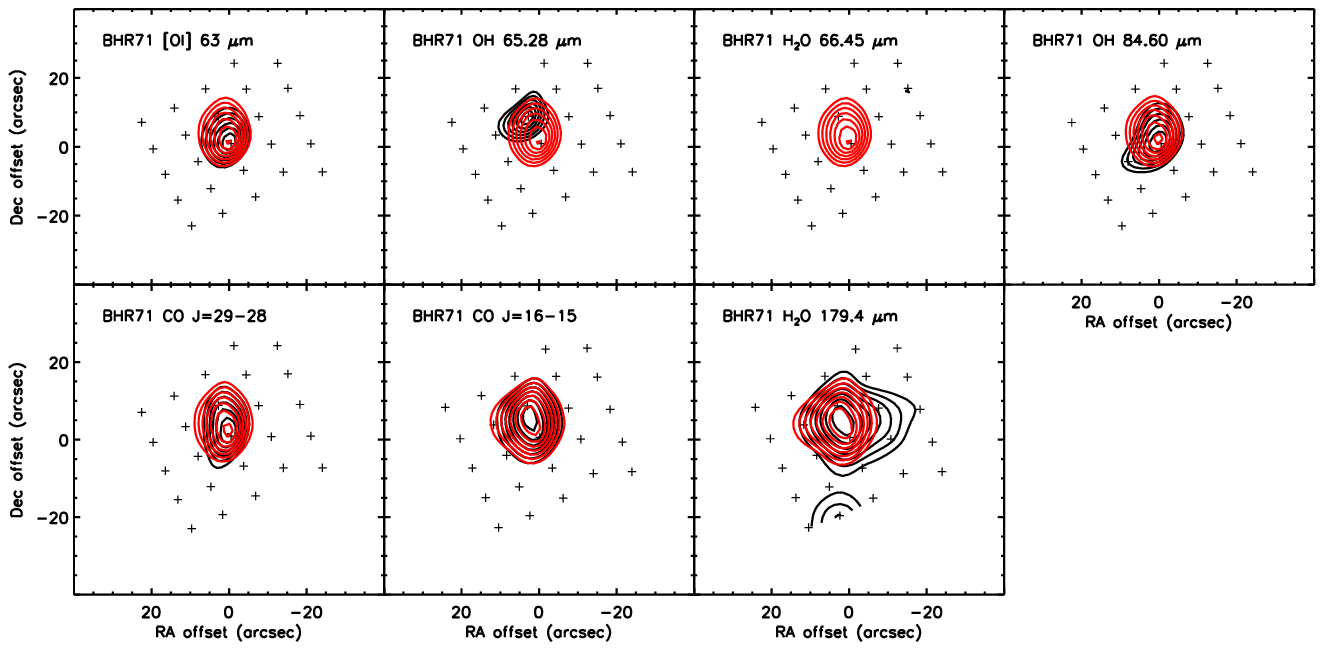}
\includegraphics[scale=1.05]{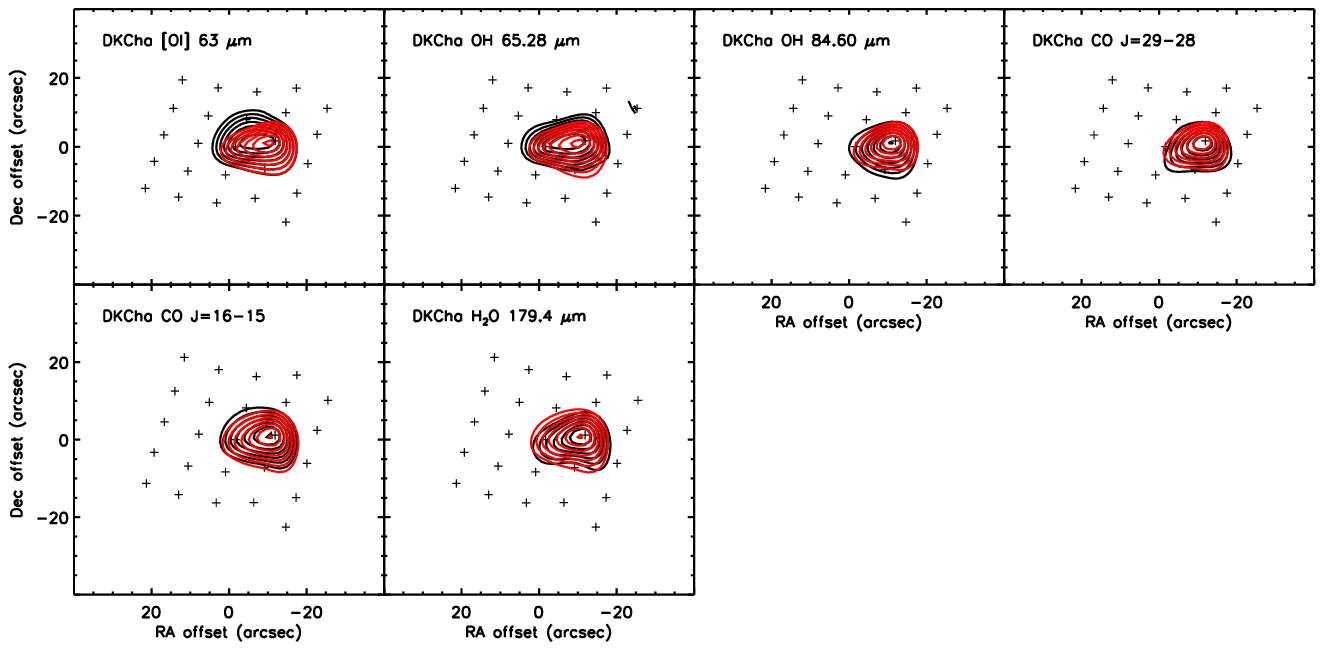}
\caption{See Figure \ref{spatial1} for a description of this figure.}
\label{spatial7}
\end{figure}

\begin{figure}
\includegraphics[scale=1.05]{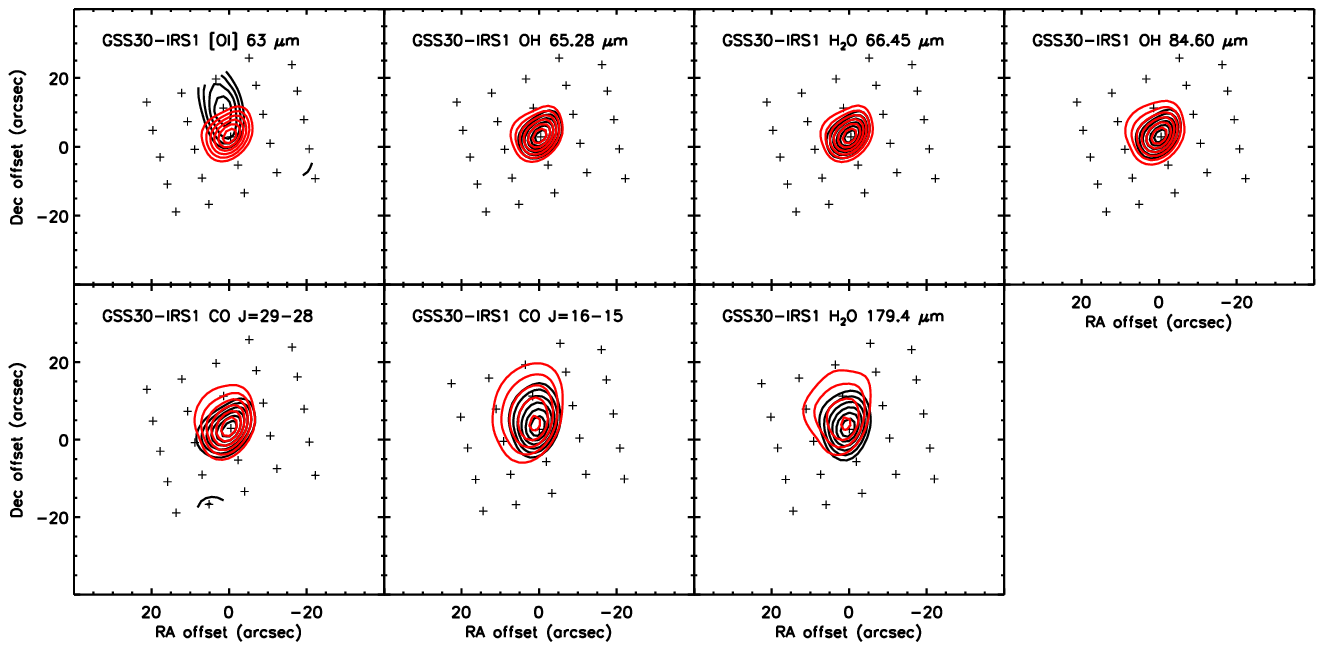}
\includegraphics[scale=1.05]{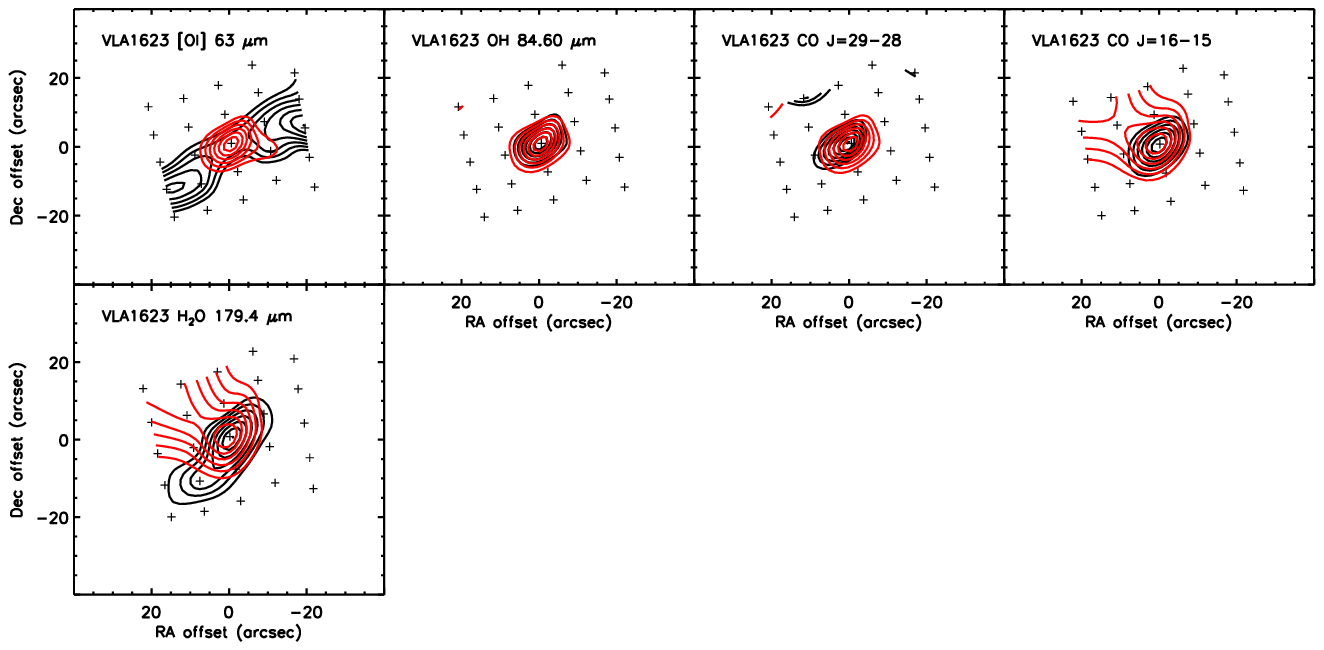}
\includegraphics[scale=1.05]{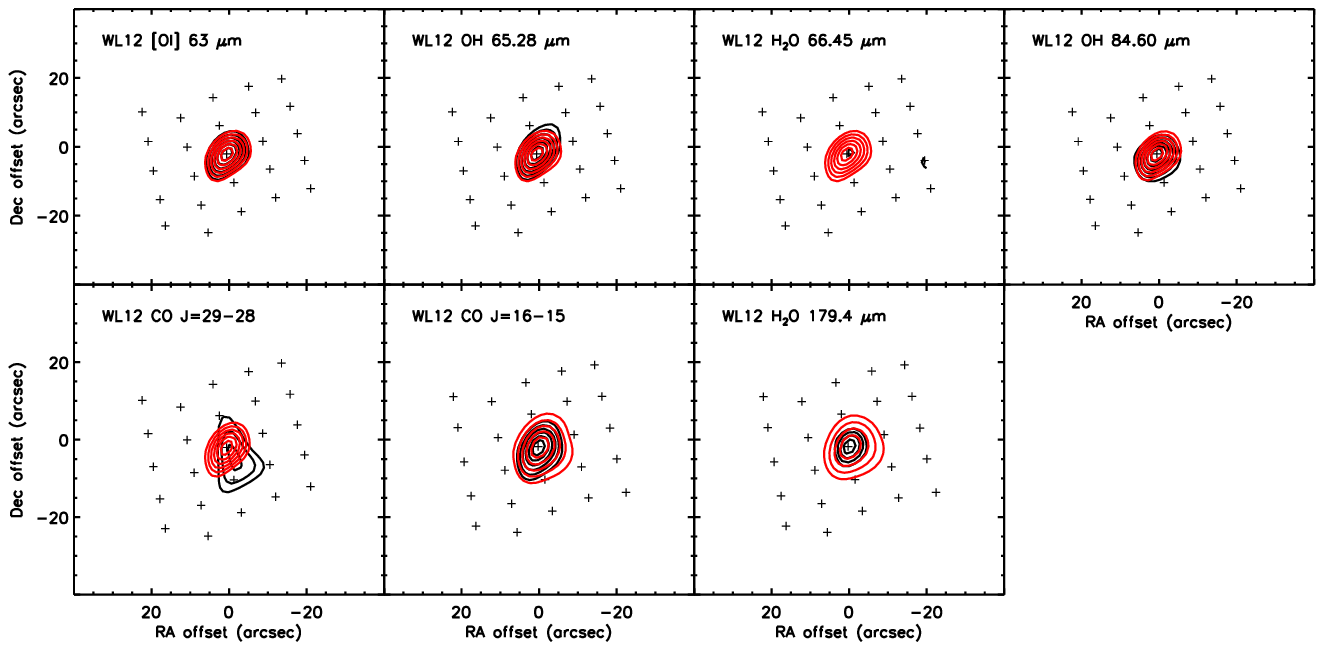}
\caption{See Figure \ref{spatial1} for a description of this figure.}
\label{spatial8}
\end{figure}

\begin{figure}
\includegraphics[scale=1.05]{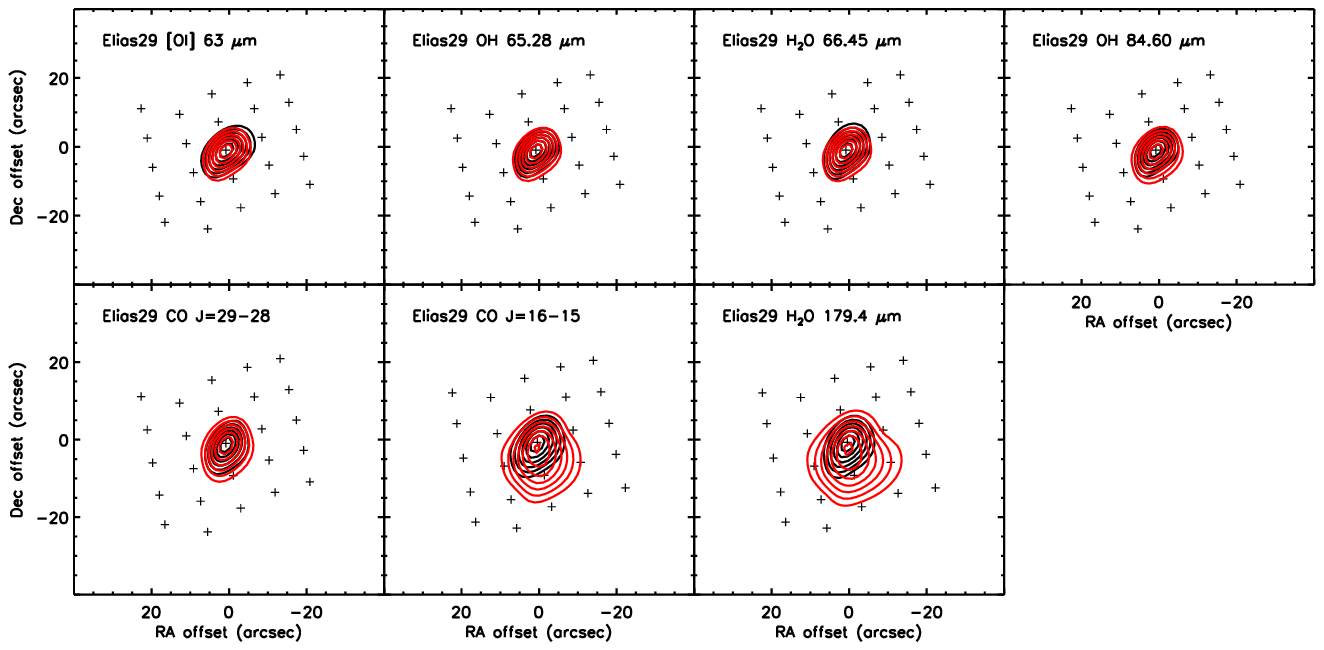}
\includegraphics[scale=1.05]{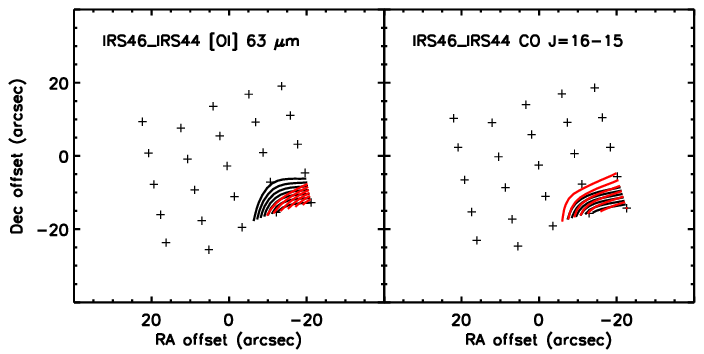}
\includegraphics[scale=1.05]{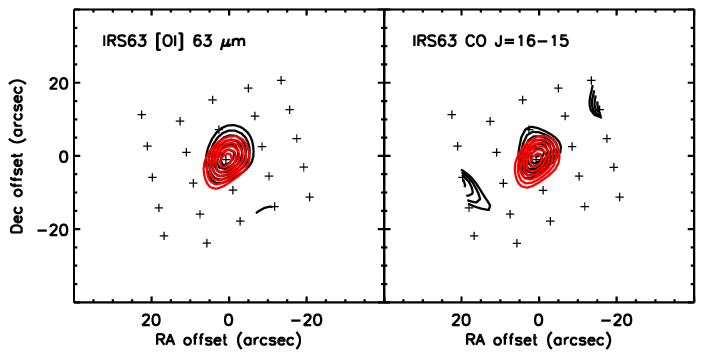}
\caption{See Figure \ref{spatial1} for a description of this figure.  Note that in the case of 
IRS 46, the \OI\ and CO lines are detected in the central spaxel, but at less than 10\% the peak value from IRS 44 (see Table \ref{fluxtable}); thus it does not appear at the 10\% significance level compared to contours derived from IRS 44.}
\label{spatial9}
\end{figure}

\begin{figure}
\includegraphics[scale=1.05]{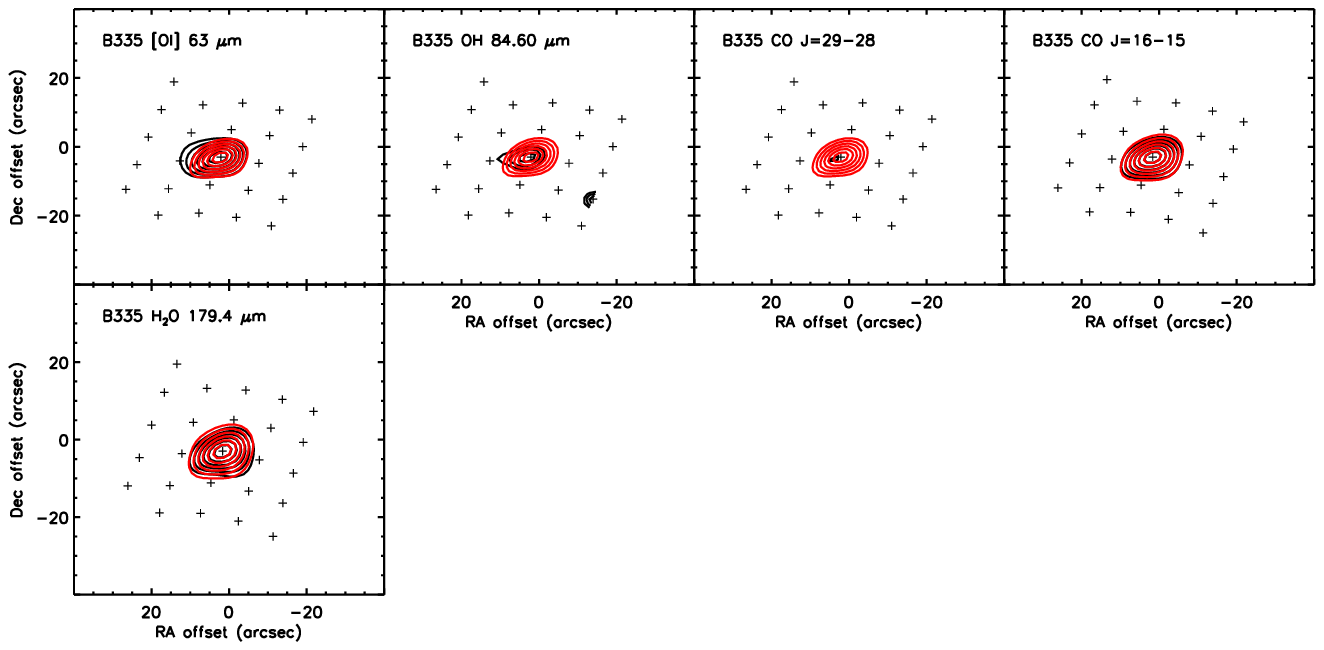}
\includegraphics[scale=1.05]{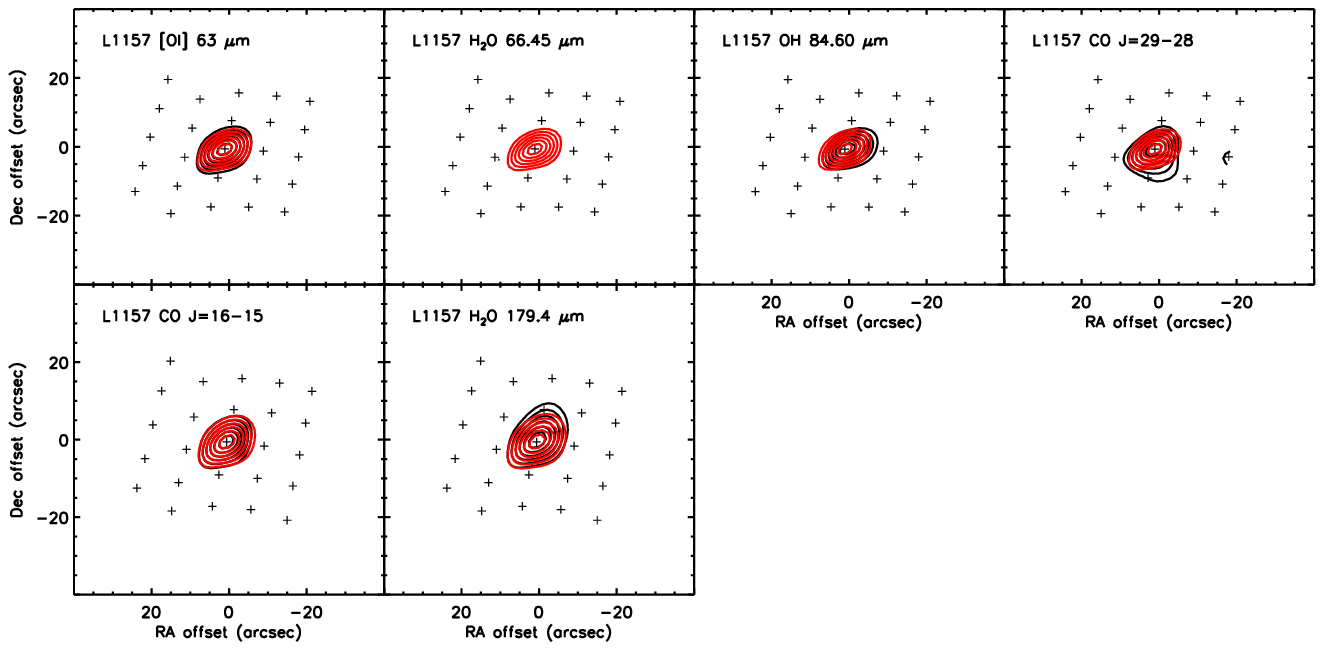}
\includegraphics[scale=1.05]{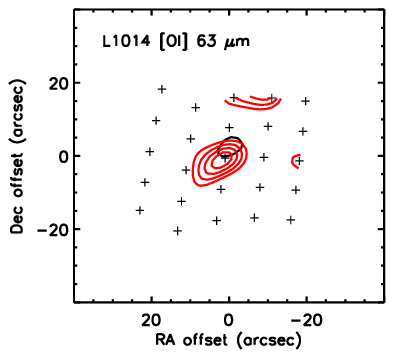}
\caption{See Figure \ref{spatial1} for a description of this figure.}
\label{spatial10}
\end{figure}

\section{Procedure to create a rotational diagram}

The line luminosities are simply computed from the linefluxes:
\begin{equation}
L_{\rm J} = \lineflux \times 4 \pi D^2,
\end{equation}
where $D$ is the distance to the source.
We use the subscript $J$ as shorthand for the full state descriptor
for the more complex species characterized by multiple quantum numbers.

If the emission lines are optically thin, 
the total numbers of molecules in the upper state of the transition, \funnyNJ,
are related to the line luminosities by
\begin{equation}
\mathcal{N}_{\rm J} = L_{\rm J}  /  (A(J) h \nu(J)),
\end{equation}
where $A(J)$ and $\nu(J)$ are the Einstein coefficient and frequency 
of the transition.

These populations can also be characterized by a Boltzmann distribution:
\begin{equation}
\frac{\mathcal{N}_{\rm J}}{g_{\rm J}} = \frac{\mathcal{N}}{Q} exp(\frac{-\eup}{T_{rot}}),
\end{equation}
where $Q$ is the partition function and \funnyN, $g_J$, and \eup\ are the
total number of molecules, degeneracy of the upper level,
and excitation energy (in K), respectively, of upper state $J$.  

When the logarithm of $\funnyNJ/g_J$ is plotted versus \eup, this
rotation diagram can provide a first, rough diagnosis of the physical
conditions in the emitting gas.
If the lines are optically thin and the energy level populations are characterized 
by a single effective temperature (\trot, which may or may not
correspond to the kinetic temperature \tkin)
the points can be fitted by a straight line, with slope of $-\trot^{-1}$.
If the rotational temperature is in equilibrium with the kinetic
(translational) temperature ($\trot = \tk$), the temperature of the emitting
gas is determined. The intercept of the fit, together with an evaluation
of $Q$, provides a measure of the total number of molecules in all levels
(\funnyN).  For simple, linear molecules \citep{kassel36},

\begin{equation}
Q = kT/hcB, hcB << kT
\end{equation}
where $h$ is Planck's constant, $c$ is the speed of light, $k$ is Boltzmann's
constant, and $B$ is the rotational constant for the species in question.
$B=192.25$ m$^{-1}$ for CO.
For H$_2$O we used  a polynomial function in
terms of \trot\ to find $Q$ \citep{fischer02}:

\begin{eqnarray}
Q=-4.9589+0.28147 \times T_{\rm rot}+0.0012848 \times T_{\rm rot}^2 & & \nonumber\\
-5.8343 \times 10^{-7} \times T_{\rm rot}^3 & &
\end{eqnarray}

and for OH:

\begin{eqnarray}
Q=7.7363+0.17152 \times T_{\rm rot}+0.00034886 \times T_{\rm rot}^2 & & \nonumber\\
-3.3504 \times 10^{-7} \times T_{\rm rot}^3 & &
\end{eqnarray}
Note that the equations for H$_2$O and OH apply if a single \trot\ between 
100 and 450 K describes the populations.

If the lower excitation lines become optically thick,
the state populations derived from the optically thin approximation
would exhibit curvature and diverge from the fitted line.
If the higher excitation lines are sub-thermally populated, the derived
state populations may fall below the fitted line.
Neither of these effects is seen in CO rotation diagrams for this sample,
for which the direct correspondence between \eup\ and the wavelength 
of the transition simplifies the analysis, but they are likely to cause
some of the larger scatter in the diagrams for OH and \water.

\clearpage

\end{document}